Katholieke Universiteit Leuven
Faculteit Wetenschappen
Instituut voor Theoretische Fysica

# An Algorithmic Approach to Operator Product Expansions, $\mathcal{W}$-Algebras and $\mathcal{W}$-Strings

Promotor: **Dr. W. Troost**

Proefschrift ingediend tot
het behalen van de graad van
Doctor in de Wetenschappen
door **Kris Thielemans**

Leuven, juni 1994.

# Abstract

String theory is currently the most promising theory to explain the spectrum of the elementary particles and their interactions. One of its most important features is its large symmetry group, which contains the conformal transformations in two dimensions as a subgroup. At quantum level, the symmetry group of a theory gives rise to differential equations between correlation functions of observables. We show that these Ward-identities are equivalent to Operator Product Expansions (OPEs), which encode the short-distance singularities of correlation functions with symmetry generators. The OPEs allow us to determine algebraically many properties of the theory under study. We analyse the calculational rules for OPEs, give an algorithm to compute OPEs, and discuss an implementation in *Mathematica*.
There exist different string theories, based on extensions of the conformal algebra to so-called $W$-algebras. These algebras are generically nonlinear. We study their OPEs, with as main results an efficient algorithm to compute the $\beta$-coefficients in the OPEs, the first explicit construction of the $WB_2$-algebra, and criteria for the factorisation of free fields in a $W$-algebra.
An important technique to construct realisations of $W$-algebras is Drinfel'd-Sokolov reduction. The method consists of imposing certain constraints on the elements of an affine Lie algebra. We quantise this reduction via gauged WZNW-models. This enables us in a theory with a gauged $W$-symmetry, to compute exactly the correlation functions of the effective theory.
Finally, we investigate the (critical) $W$-string theories based on an extension of the conformal algebra with one extra symmetry generator of dimension $N$. We clarify how the spectrum of this theory forms a minimal model of the $W_N$-algebra.

# Notations

$\partial,\ \bar\partial$ : $\frac{d}{dz}, \frac{d}{d\bar z}$.

$\delta_\varepsilon$ : infinitesimal transformation with $\varepsilon(x)$ an infinitesimal parameter.

$:T_1T_2:$ : normal ordening.

**OPA** : Operator Product Algebra, subsection 2.3.3.

**OPE** : Operator Product Expansions, section 2.3.

$[T_1(z)T_2(z_0)]$ : the complete OPE.

$\underbracket{T_1(z)T_2}(z_0)$ : singular part of an OPE.

$[T_1T_2]_n$ : coefficient of $(z-z_0)^{-n}$ in the OPE $[T_1(z)T_2(z_0)]$, eq. (2.3.3).

$\ll \frac{c}{2} \mid 0 \mid 2T \mid \partial T \gg$ : list of the operators at the singular poles in an OPE. Highest order pole is given first. First order pole occurs last in the list.

$\{T_1T_2\}_n(x_0)$ : coefficient of $\frac{(-1)^{n-1}}{(n-1)!}\,\partial^{n-1}\delta^{(2)}(x-x_0)$ in the Poisson bracket $\{T_1(x)T_2(x_0)\}$ subsection 2.3.5.

$\widehat{A}_m$ : mode of the operator $A$, section 2.4.

$L_{\{n\}}$ : is $L_{n_p}\cdots L_{n_1}$, $L_{\{-n\}}$ reversed order, chapter 4.

$\bar g$ : Lie algebra, see appendix B for additional conventions.

$\hat g$ : Kač-Moody algebra.

$e_0, e_+, e_-$ : generators of $sl(2)$.

$\mathcal{K}_\pm$ : kernel of the adjoint of $e_\pm$.

$\Pi$ : projection operator in $\bar g$, appendix B.

$\Pi_{hw}$ : projection on $\mathcal{K}_+$.

$\underline{c},\ \overline{c}$ : index limitation to generators of strictly negative, resp. positive grading, section 6.4.

$(a)_n$ : Pochhammer symbol, $(a)_n = \frac{\Gamma(a+n)}{\Gamma(a)} = \prod_{j=0}^{n-1}(a+j)$ ; $a \in \mathbb{R}$ and $n \in \mathbb{N}$, appendix 2.A.

# Contents

# Chapter 1

# Introduction and outline

The first chapter of a recent PhD. thesis in contemporary high-energy physics necessarily stresses the importance of symmetry [42, 51, 45]. The reason for this is that symmetry is the most powerful organising principle available, and a theoretical physicist wants to assume as little as possible. This has the peculiar consequence that he or she ends up making the far-reaching assumption that "nature" has the largest symmetry we are able to find.

A striking example is provided by string theory. The universe seems to contain a large number of "elementary" particles. It is an appealing idea to think of these particles as different states of one single object. This would enable us to treat them in a symmetric way. The simplest objects in every-day experience which have such different eigenstates are (violin) strings. One then has to find which action governs a string-object moving through space-time. The simplest (in a certain sense) action, was found by Polyakov [158]. It is a generalisation of the action for a free relativistic particle. For a bosonic string in $D$ dimensions the action is given by:

$$S[X^\mu, g^{ij}] = -\frac{1}{4\pi T}\int dx^2\, \sqrt{g(x)}g^{ij}(x)\partial_i X^\mu(x)\partial_j X_\mu(x)\,, \qquad (1.1)$$

where the fields $X^\mu$ describe the position of the string, and $x$ are coordinates which parametrise the two-dimensional surface ("world-sheet") which is swept out by the string as it moves in $D$-dimensional (flat) space-time. $g^{ij}$ is the metric on the world-sheet, with inverse determinant $g$. $T$ is a parameter that is related to the string tension.

The action (1.1) has (classically) a very large symmetry group, corresponding to reparametrisations of the world-sheet, and rescalings of the metric $g^{ij}$. These invariances are quite natural from the point of view of string theory. When viewing the theory defined by eq. (1.1) as a field theory in two dimensions, a first surprise awaits us. The field theory has an infinite dimensional symmetry group, which was quite uncommon in those days. A second surprise arises when we quantise the bosonic string theory. Requiring that the symmetry survives quantisation fixes the number of space-time dimensions to 26. Somehow, this makes one hope that a more realistic string theory would "explain" why we are living in a four-dimensional world. The third surprise is that, while we started with a free theory, interactions seem also to be fixed by the action (1.1), with only one parameter $T$. This is in contrast with the grand-unified theories where interactions have to be put in by hand, requiring the introduction of a number of parameters that have to be fixed by comparing with experiments. A last surprise which we wish to mention, is that the spectrum of the physical states contains a particle with the correct properties for a graviton. String theory thus seems to incorporate quantum gravity. This is particularly fortunate because no other theory has been found yet which provides a consistent quantisation of gravity (in four dimensions).

These four features – symmetry, fixing the number of dimensions, "automatic" interactions and quantum gravity – were so attractive that many physicists decided to put the book of Popper [166] back on the shelf for a while. Indeed, although string theory certainly looks like a "good" theory, it still does not produce any results which are falsifiable, *i.e.* which can be contradicted by an experiment.

To make any chance of being a realistic theory, a number of flaws of the original bosonic string theory (like apparently giving the wrong space-time dimension) had to be resolved. Several roads can be followed, but we will concentrate here on the one which is most related to this thesis: enlarging the symmetry of the theory. In fact, the Polyakov action has many more symmetries than we alluded to. However, they are only global symmetries, *i.e.* generated by transformations with constant parameters. To make some of these symmetries local, one has to introduce extra gauge fields, which can be viewed as generalisations of the metric $g^{ij}$ in eq. (1.1). In some cases, extra fields comparable to $X^\mu(x)$ are added to the theory. For instance, in superstrings one uses fermionic fields describing coordinates in a Grassmann manifold. The resulting string theories are called "$\mathcal{W}$–strings", and the (infinite dimensional) algebra formed by the infinitesimal transformations of the enlarged group is called an extended conformal algebra or, loosely speaking, a "$\mathcal{W}$–algebra".

Unfortunately, during the process of enhancing the original bosonic string, one of its attractive features has been lost, namely its uniqueness. This is due to a number of reasons, but we will only mention two. Because an infinite number of $\mathcal{W}$–algebras exist, an infinite number of $\mathcal{W}$–string theories can be found (although certainly not all of them are candidates for a realistic theory). A second reason is that by adding an extra action for the metric to eq. (1.1), one can make a consistent quantum theory for other dimensions of space-time than 26, the "noncritical" strings. As it is well-known since Einstein that the metric is related to gravity, the study of consistent quantum actions for the metric provides a quantisation of gravity in two dimensions. Two-dimensional $\mathcal{W}$–gravity is interesting in its own respect because one hopes to gain some insight in how to construct consistent quantum gravity in four dimensions.

Although uniqueness has been lost, the other attractive features of string theory still survive. In particular, the symmetry group of the Polyakov action has even been enlarged. The study of this new kind of symmetry has influenced, and has been influenced by, many other branches of physics and mathematics. This has happened quite often in the history of string theory, and is sometimes regarded as an important motivation for studying strings. To be able to discuss the relation of $\mathcal{W}$–algebras to other fields in physics, we have to be somewhat more precise.

The Polyakov action (1.1) is invariant under general coordinate transformations $x^i \to f^i(x)$ and local Weyl rescalings of the metric $g_{ij}(x) \to \Lambda(x)g_{ij}(x)$. A generalisation of the latter is to allow other fields $\Phi(x)$ to rescale as $\Phi(x) \to \Lambda(x)^h\Phi(x)$. $h$ is called the scaling dimension of the field $\Phi(x)$. The combination of the general coordinate and Weyl transformations can be used to gauge away components of the metric. In two dimensions one has exactly enough parameters to put the metric equal, at least locally, to the flat metric $\eta_{ij}$ (the conformal gauge). However, this does not yet completely fix the gauge. Obviously, conformal transformations (coordinate transformations which scale the metric) combined with the appropriate Weyl rescaling form a residual symmetry group. Therefore, field theories which have general coordinate and Weyl invariance are called conformal field theories.

The situation in two dimensions is rather special. In light-cone coordinates, $x_\pm = x_0 \pm x_1$, every transformation $x_\pm \to f^\pm(x_\pm)$ is conformal. We see that the group formed by the conformal transformations is infinite dimensional in two dimensions. This makes clear why the symmetry group of string theory is so exceptionally large. The conformal transformations are generated by the energy–momentum tensor $T^{ij}$ of the theory, which has scaling dimension $h = 2$. In fact, the algebra splits in two copies of the Virasoro algebra, related to the $x_+$ and $x_-$ transformations, and generated by $T^{++}$ and $T^{--}$. Similarly, an extended conformal algebra is formed by two copies of what is called a $\mathcal{W}$–algebra.

The symmetries of a theory have direct consequences for its correlation functions. The Ward identities are relations between $N$-point correlation functions where one of the fields is a symmetry generator, and $(N-1)$-point functions. Usually, this does not yet fix the $N$-point function, and correlation functions have to be calculated tediously. In two-dimensional conformal field theory, the consistency conditions imposed by the symmetries are so strong that the Ward identities determine all correlation functions with a symmetry generator in terms of those without any symmetry generators. This means that once the Ward identities have been found (which requires some regularisation procedure), all correlation functions with symmetry generators can be recursively computed.

In renormalisable field theories, "Operator Product Expansions" (OPEs) are introduced to calculate the short-distance behaviour of correlation functions [208]. This formalism has been extended by Belavin, Polyakov and Zamolodchikov [13] to two-dimensional conformal field theory. The Ward identities fix OPEs. Moreover, they impose a set of consistency conditions on the OPEs such that computing with OPEs amounts to applying a set of algebraic rules. They even almost determine the form of the OPEs. As an example, we give the OPE of one of the components of the energy–momentum tensor, which for any conformal field theory is:

$$T^{++}(x)T^{++}(y) = \frac{c/2}{(x^- - y^-)^4} + \frac{2T^{++}(y)}{(x^- - y^-)^2} + \frac{\partial_- T^{++}(y)}{(x^- - y^-)} + O(x^- - y^-)^0\,. \tag{1.2}$$

Here, the "central charge" $c$ is a number which can be determined by computing the two-point function $< T^{++}(x)T^{++}(y) >$, and depends on the theory we are considering. The important point here is that once the central charge is known, all correlation functions of $T^{++}$ can be algebraically computed. From the OPE eq. (1.2), the Virasoro algebra can be derived and vice versa. Similarly, if the symmetry algebra of the theory forms an extended conformal algebra, the generators form an Operator Product Algebra. This contains exactly the same information as the $\mathcal{W}$–algebra, and indeed is often called a $\mathcal{W}$–algebra.

There exists by now a wealth of examples of $\mathcal{W}$–algebras. Among the best known are the affine Lie algebras and the linear superconformal algebras. When a $\mathcal{W}$–algebra contains a generator with scaling dimension larger than two, the $\mathcal{W}$–algebra is (in most cases) nonlinear. Some examples of such $\mathcal{W}$–algebras with only one extra generator are $\mathcal{W}_3$ [211], the spin 4 algebra [29, 108] and the spin 6 algebra [74]. The Bershadsky-Knizhnik algebras [20, 133] have $N$ supersymmetry generators and an affine $so(N)$ subalgebra. Many other examples exist and no classification of $\mathcal{W}$–algebras seems as yet within reach.

The fields of a conformal field theory form a representation of its symmetry algebra. Belavin, Polyakov and Zamolodchikov [13] showed that under certain assumptions all fields of the theory are descendants of a set of primary fields. For a certain subclass of conformal field theories, the "minimal" models, the Ward identities fix all correlation functions. They also showed that the simplest minimal model corresponds to the Ising model at criticality. This connection with statistical mechanics of two-dimensional systems is due to the fact that a system becomes invariant under scaling transformations at the critical point of a phase transition. By hypothesising local conformal invariance, various authors (see [4, 118]) found the critical exponents of many two-dimensional models. Some examples of statistical models are the Ising ($m = 3$), tricritical Ising ($m = 4$), 3-state Potts ($m = 5$), tricritical 3-state Potts ($m = 6$), and the Restricted Solid-on-Solid (any $m$) models, where we denoted the number of the corresponding unitary Virasoro minimal model in brackets.

Another important connection was found, not in statistical mechanics, but in the study of integrable models in mathematics and physics. These models have two different Hamiltonian structures, whose Poisson brackets form examples of classical $\mathcal{W}$–algebras. For example, the Korteweg-de Vries (KdV) equation gives rise to a Virasoro Poisson bracket, while the Boussinesq equation has a Hamiltonian structure which corresponds to the classical $\mathcal{W}_3$ algebra. Moreover, the relation between the two Hamiltonian structures gives rise to a powerful method of constructing classical $\mathcal{W}$–algebras. Drinfeld and Sokolov [60] found a hierarchy of equations of the KdV type based on the Lie algebras $sl(N)$. The Hamiltonian structures of these equations provide explicit realisations of the classical $\mathcal{W}_N$ algebras, whose generators have scaling dimensions $2, 3, \ldots N$. An extension of this method is still the most powerful way at our disposal to find (realisations of) $\mathcal{W}$–algebras.

This thesis is organised as follows. Chapter 2 gives an introduction to conformal field theory. We discuss how the Ward identities are derived. For this purpose, the Operator Product Expansion formalism is introduced. We determine the complete set of consistency conditions on the OPEs. We then show how an infinite dimensional Lie algebra, corresponding to the symmetry algebra of the conformal theory, can be found using OPEs. We define the generating functional of the correlation functions of symmetry generators. The Ward identities can be used to find functional equations for the generating functional (or induced action). We conclude the chapter with some important examples of conformal field theories: free-field theories and WZNW–models.

In chapter 3, the consistency conditions on OPEs are converted to a set of algorithms to compute with OPEs, suitable for implementation in a symbolic manipulation program. We then describe the *Mathematica* package *OPEdefs* we developed. This package completely automates the computation of OPEs (and thus of correlation functions), given the set of OPEs of the generators of the $\mathcal{W}$–algebra.

The next chapter discusses $\mathcal{W}$–algebras using the Operator Product Expansion formalism. We first give some basic notions on highest weight representations of $\mathcal{W}$–algebras, of which minimal models are particular examples. The fields of a conformal field theory assemble themselves in highest weight representations. We then analyse the structure of $\mathcal{W}$–algebras using the consistency conditions found in chapter 2. The global conformal transformations fix the form of OPEs of quasiprimary fields, which are special examples of highest weight fields with respect to the global conformal algebra. A similar analysis is done for the local conformal transformations and primary fields. We then discuss the different methods which are used to construct $\mathcal{W}$–algebras and comment on the classification of the $\mathcal{W}$–algebras. Finally, as an example of the ideas in this chapter, the $\mathcal{W}_c B_2$ algebra is studied in detail. The complexity of the calculations shows the usefulness of *OPEdefs*.

Goddard and Schwimmer [103] proved that free fermions can always be factored out of a $\mathcal{W}$–algebra. Chapter 5 extends this result to arbitrary free fields. This is an important result as it shows that a classification of $\mathcal{W}$–algebras need not be concerned with free fields. We provide explicit algorithms to perform this factorisation. We then show how the generating functionals of the $\mathcal{W}$–algebra obtained via factorisation are related to the generating functional of the original $\mathcal{W}$–algebra. The $N = 3$ and $N = 4$ linear superconformal algebras are treated as examples.

The Drinfeld-Sokolov method constructs a realisation for a classical $\mathcal{W}$–algebra via imposing constraints on the currents of a Kač–Moody algebra. In particular, for any embedding of $sl(2)$ in a semi-simple (super)Lie algebra a different $\mathcal{W}$–algebra results. In chapter 6, Drinfeld-Sokolov reduction is extended to the quantum case. The reduction is implemented in an action formalism using a gauged WZNW–model, which enables us to find a path integral formulation for the induced action of the $\mathcal{W}$–algebra. The gauge fixing is performed using the Batalin-Vilkovisky [11] method. In a special gauge, the BV procedure reduces to a BRST approach [138]. Operator Product Expansions are used to perform a BRST quantisation. The cohomology of the BRST operator is determined in both the classical and quantum case. The results are then used to show that we indeed constructed a realisation of a quantum $\mathcal{W}$–algebra. The $N$-extended $so(N)$ superconformal algebras [20, 133] are used as an illustration of the general method.

The results of chapter 6 are then used in chapter 7 to study $\mathcal{W}$–gravity theories in the light-cone gauge. The gauged WZNW–model is used as a particular matter sector for the coupling to $\mathcal{W}$–gravity. Using the path integral formulation of the previous chapter, the effective action can be computed. It is shown that the effective action can be obtained from its classical limit by simply inserting some renormalisation factors. Explicit expressions for these renormalisation factors are given. They contain the central charge of the $\mathcal{W}$–algebra and some parameters related to the (super)Lie algebra and the particular $sl(2)$–embedding for which a realisation of the $\mathcal{W}$–algebra can be found. The example of the previous chapter is used to construct the effective action of $so(N)$ supergravity. We then check the results using the correspondence between the linear superconformal algebras and the $so(N)$ $\mathcal{W}$–algebras for $N \leq 4$, established in chapter 5. This is done using a semiclassical evaluation of the effective action for the linear superconformal algebras.

The last chapter contains a discussion of critical $\mathcal{W}$–strings. After a short review of $\mathcal{W}$–string theory, we concentrate to the case where the classical $\mathcal{W}$–algebra is formed by the energy–momentum tensor and a dimension $s$ generator. These $\mathcal{W}_{2,s}$ strings provide examples which can be analysed in considerably more detail than string theories based on more complicated $\mathcal{W}$–algebras. In particular, operator product expansions (and *OPEdefs*) are used to provide some insight in the appearance of $\mathcal{W}$–minimal models in the spectrum of $\mathcal{W}$–strings. Finally, some comments are made on the recent developments initiated by Berkovits and Vafa [14]. They showed how the bosonic string can be viewed as an $N = 1$ superstring with a particular choice of vacuum, and a similar embedding of $N = 1$ into $N = 2$ superstrings. The hope arises that a hierarchy of string embeddings exists, restoring the uniqueness of string theory in some sense.

# Chapter 2

# Conformal Field Theory and Operator Product Expansions

This chapter gives an introduction to conformal field theory with special attention to Operator Product Expansions (OPEs). It is of course not complete as conformal field theory is a very wide subject, and many excellent reviews exist, *e.g.* [97, 122, 190]. Because it forms an introduction to the subject, some points are probably trivial for someone who feels at home in conformal field theory. However, some topics are presented from a new standpoint, a few new results (on the associativity of OPEs) are given, and notations are fixed for the rest of the work.

We start by introducing the conformal transformations. Then, we study the consequences of a symmetry of a conformal field theory on its correlation functions. In the quantum case, this information is contained in the Ward identities. These identities are then used in the third section to develop the OPE formalism. We study the consistency conditions for OPEs in detail and introduce the notion of an Operator Product Algebra (OPA). We then draw attention to the close analogy between OPEs and Poisson brackets. Section 2.4 defines the mode algebra of the symmetry generators. In the next section, we define the generating functionals of the theory and show how they are determined by the Ward identities. We conclude with some important examples of conformal field theories, free field theories and WZNW–models.

## 2.1 Conformal transformations

A conformal field theory is a field theory which is invariant under general coordinate transformations $x^i \to f^i(x)$ and under the additional symmetry of Weyl invariance. The latter transformations correspond to local scale transformations of the metric $g_{ij}(x) \to \Lambda(x) g_{ij}(x)$ and fields $\Phi(x) \to \Lambda(x)^h \Phi(x)$, where $h$ is the scaling dimension of the field $\Phi$. The combination of these symmetries can be used to gauge away components of the metric. In two dimensions one has exactly enough parameters to put the metric equal, at least locally, to the Minkowski metric $\eta_{ij}$ (the conformal gauge). Obviously, conformal transformations (coordinate transformations which scale the metric) combined with the appropriate Weyl rescaling form a residual symmetry group. Therefore, we will study this conformal group first.

Conformal transformations are coordinate transformations which change the metric with a local scale factor. In a space-time of signature $(p, q)$ they form a group isomorphic with $SO(p+1, q+1)$. However, in the complex plane it is well-known that all (anti-)analytic transformations are conformal. This extends to the Minkowski plane where in light-cone coordinates, the conformal transformations are given by:

$$x^{\pm} \longrightarrow x'^{\pm}(x^{\pm})\,. \tag{2.1.1}$$

In a space-time of signature $(-1, 1)$, it is customary to perform a Wick rotation. We will always assume this has been done, and treat only the Euclidean case.

For a space of Euclidean signature, it is advantageous to use a complex basis $(\tau + i\sigma, \tau - i\sigma)$. In string theory, the space in which these coordinates live is a cylinder, as $\sigma$ is used as a periodic coordinate. This also applies to two-dimensional statistical systems with periodic boundary conditions in one dimension. One then maps this cylinder to the full complex plane with coordinates

$$(z, \bar{z}) = (\exp(\tau + i\sigma), \exp(\tau - i\sigma)\,, \tag{2.1.2}$$

where we will take a flat metric proportional to $\delta_{ij}$ in the real coordinates, or in the complex coordinates:

$$ds^2 = 2\sqrt{g}dzd\bar{z}\,. \tag{2.1.3}$$

We will use the notation $x$ for a coordinate of a point in the complex plane. Note that the points with fixed time $\tau$ lie on a circle in the $(z, \bar{z})$ plane.

It is often convenient to consider $z$ and $\bar{z}$ as independent coordinates (*i.e.* not necessarily complex conjugate). We can then restrict to the Euclidean plane by imposing $\bar{z} = z^*$. The plane with a Minkowski metric corresponds to $z, \bar{z} \in \mathbb{R}$.

As we are working in Euclidean space, the complex plane can be compactified to the Riemann sphere. Conformal field theory can also be defined on arbitrary Riemann surfaces, but we will restrict ourselves in this work to the sphere.

In the complex coordinates, a conformal transformation is given by:

$$z \to z' = f(z), \qquad \bar{z} \to \bar{z}' = \bar{f}(\bar{z}), \tag{2.1.4}$$

where $f$ is an analytic function, and $\bar{f}$ is antianalytic.

**Definition 2.1.1** *A primary field transforms under the conformal transformation (2.1.4) as:*

$$\Phi(z,\bar{z}) \quad \to \quad \Phi'(f(z),\bar{f}(\bar{z})) = (\partial f(z))^{-h}(\bar{\partial}\bar{f}(\bar{z}))^{-\bar{h}}\Phi(z,\bar{z}), \tag{2.1.5}$$

*where $\partial \equiv \frac{d}{dz}$ and $\bar{\partial} \equiv \frac{d}{d\bar{z}}$. The numbers $h$ and $\bar{h}$ are called the conformal dimensions of the field $\Phi$.*

For infinitesimal transformations of the coordinates $f(z) = z - \varepsilon(z)$, we see that the primary fields transform as:

$$\delta_\varepsilon \Phi(z,\bar{z}) = \varepsilon(z)\partial\Phi(z,\bar{z}) + h\partial\varepsilon(z)\Phi(z,\bar{z}). \tag{2.1.6}$$

By choosing for $\varepsilon(z)$ any power of $z$ we see that the conformal transformations form an infinite algebra generated by:

$$l_m = -z^{m+1}\partial \text{ and } \bar{l}_m = -\bar{z}^{m+1}\bar{\partial}, \qquad m \in \mathbb{Z}, \tag{2.1.7}$$

which consists of two commuting copies of the Virasoro algebra, but without central extension (see further):

$$[l_m, l_n] = (m-n)l_{m+n}, \tag{2.1.8}$$

and analogous commutators for the $\bar{l}_m$.

Clearly, $l_0$ corresponds to scaling transformations in $z$. The combination $l_0 + \bar{l}_0$ generates scaling transformations in the complex plane $x \to \lambda x$, while $i(l_0 - \bar{l}_0)$ generates rotations. This means that a field $\Phi(x)$ with conformal dimensions $h$ and $\bar{h}$ has scaling dimension $h + \bar{h}$ and spin $|h - \bar{h}|$. $l_{-1}$ generates translations and $l_1$ "special" conformal transformations. The subalgebra formed by $\{l_{-1}, l_0, l_1\}$ corresponds to the globally defined and invertible conformal transformations:

$$z \to \frac{az+b}{cz+d}, \tag{2.1.9}$$

where $a, b, c, d \in \mathbb{C}$ and $ac - bd = 1$. These transformations form a group isomorphic to $SL(2,\mathbb{C}) \approx SO(3,1)$.[1]

**Definition 2.1.2** *A quasiprimary field transform as eqs. (2.1.5,2.1.6) for the global transformations. A scaling field transforms this way for translations and scaling transformations.*

[1] Although there are two commuting copies of this algebra, we should take the appropriate real section when we restrict to the real surface.

## 2.2 Correlation functions and symmetry

The physics of a quantum field theory is contained in the correlation functions. In a path integral formalism, a correlation function of fields $\Phi_i$ corresponding to observables can be symbolically written as:

$$< \mathcal{R}\{\Phi_1(x_1)\Phi_2(x_2)\cdots\} > = \frac{1}{\mathcal{N}}\int [d\varphi_k] \, \exp(-S[\varphi_j]) \, \Phi_1(x_1)\Phi_2(x_2)\cdots, \tag{2.2.1}$$

where $S$ is a suitable action, a functional of the fields $\varphi_j$ in the theory, and $[d\varphi_k]$ denotes an appropriate measure. $\mathcal{R}$ denotes time–ordering in a radial quantisation scheme [92], *i.e.* $|x_i| > |x_{i+1}|$. We will drop this symbol for ease of notation. In fact, we will assume that for a correlation function $< \mathcal{R}\{\Phi_1(x_1)\Phi_1(x_2)\ldots\} >$, the expression for $|x_1| < |x_2|$ can be obtained by analytic continuation from the expression for $|x_2| < |x_1|$. This property is commonly called "crossing symmetry". It can be checked for free fields, and we restrict ourselves to theories where it is true.

The normalisation constant in eq. (2.2.1) is given by:

$$\mathcal{N} = \int [d\varphi_k] \, \exp(-S[\varphi_j]). \tag{2.2.2}$$

We suppose that $\mathcal{N}$ is different from zero (vacuum normalised to one). In general, the pathintegral (2.2.1) has to be computed perturbatively.

Symmetries of the theory put certain restrictions on the form of the correlation functions. We will investigate this in the next subsections.

### 2.2.1 Global conformal invariance

In this subsection, we consider the consequences of the invariance of the correlation functions of quasiprimary fields under the global conformal transformations (2.1.9). The results can be extended to conformal field theories in an arbitrary number of dimensions, but we treat only the two-dimensional case.

In conformal field theory one generally requires translation, rotation and scaling invariance of the correlation functions. This restricts all one-point functions of fields with zero conformal dimensions ($h = \bar{h} = 0$) to be constant, and all others to be zero. Two-point functions have the following form:

$$< \Phi_1(z_1,\bar{z}_1)\,\Phi_2(z_2,\bar{z}_2) > = \frac{\mathrm{cst}_{12}}{(z_1-z_2)^{h_1+h_2}(\bar{z}_1-\bar{z}_2)^{\bar{h}_1+\bar{h}_2}}, \tag{2.2.3}$$

with $\mathrm{cst}_{12}$ a constant. If one requires global conformal invariance (*i.e.* invariance under the special transformations as well), only fields with the same conformal dimensions can have non-zero two-point functions, see also eq. (2.2.9). Under this assumption, three-point functions are restricted to:

$$< \Phi_1(z_1,\bar z_1)\,\Phi_2(z_2,\bar z_2)\,\Phi_3(z_3,\bar z_3) >=$$

$$\mathrm{cst}_{123} \prod_{\substack{\{ijk\}\\ i<j}} \frac{1}{(z_i-z_j)^{h_{ijk}}(\bar z_i-\bar z_j)^{\bar h_{ijk}}}\,, \tag{2.2.4}$$

where $h_{ijk} = h_i + h_j - h_k$ and the product goes over all permutations of $\{1, 2, 3\}$. It is also possible to find the restrictions from global conformal invariance on four (or more)-point functions. It turns out that the correlation function is a function of the harmonic ratios of the coordinates involved, apart from $(z_i - z_j)$ factors like in (2.2.4) to have the correct scaling law. Crossing symmetry has to be checked for a four-point function, while it is automatic for two- and three-point functions.

Note that imposing invariance under the local conformal transformations would make all correlation functions zero.

### 2.2.2 Ward identities

In this subsection, we will discuss the consequences of a global symmetry of the action using Ward identities.

We start by considering an infinitesimal transformation $\delta_\varepsilon$ of the fields, where $\varepsilon(x)$ is an infinitesimal parameter. Throughout this work, we will only consider local transformations, *i.e.* $\delta_\varepsilon\Phi(x)$ depends on a finite number of fields and their derivatives. If the transformation leaves the measure of the pathintegral (2.2.1) invariant, we find the following identity:

$$\int [d\varphi_k]\, \exp(-S[\varphi_j])\ \delta_\varepsilon\left(\Phi_1(x_1)\Phi_2(x_2)\cdots\right) \ =$$

$$\int [d\varphi_k]\, \exp(-S[\varphi_j])\ \left(\delta_\varepsilon S[\varphi]\right)\Phi_1(x_1)\Phi_2(x_2)\cdots\,. \tag{2.2.5}$$

This follows by considering a change of variables in the pathintegral. Using eq. (2.2.1), eq. (2.2.5) implies an identity between correlation functions. This Ward identity is useful in many cases, but at this moment we are interested in transformations $\delta_\varepsilon$ which leave the action invariant if $\delta_\varepsilon$ is constant, *i.e.* global symmetries.

As an example, we will derive the Ward identity for the energy–momentum tensor $T^{ij}$. Consider an infinitesimal transformation of the fields of the form:

$$\delta_\varepsilon\Phi(x) = \varepsilon^i(x)\partial_i\Phi(x) + \cdots\,, \tag{2.2.6}$$

where the dots denote corrections according to the tensorial nature of the field $\Phi$, see (2.1.6). The action is supposed to be invariant under these transformations if the $\varepsilon^i(x)$ are constants. Using Noether's law, one has a classically conserved current associated to this symmetry of the action $S$:

$$\delta_\varepsilon S = -\frac{1}{\pi}\int d^2x\,\sqrt{g}T^i{}_j(x)\partial_i\varepsilon^j(x)\,, \tag{2.2.7}$$

where $g$ is the absolute value of the determinant of $g_{ij}$.[2] This equation defines the energy–momentum tensor $T^{ij}$ up to functions whose divergence vanishes. One can show that the alternative definition

$$T_{ij} = \frac{-2\pi}{\sqrt{g}}\frac{\delta S}{\delta g^{ij}} \tag{2.2.8}$$

satisfies (2.2.7). It is obvious that the tensor (2.2.8) is symmetric. If the action is invariant under Weyl scaling, we immediately see that it is traceless. Hence, in a conformal field theory in two dimensions, the energy–momentum tensor has only two independent components. In the complex basis, the components that remain are $T^{zz}$ and $T^{\bar z\bar z}$. Using the metric eq. (2.1.3), we see from eq. (2.2.7) that Weyl invariance implies that the action is not only invariant under a global transformation, but also under a conformal transformation, $\bar\partial\varepsilon^z = 0, \varepsilon^{\bar z} = 0$. This transformation is sometimes called semi-local.

Let us now consider the expectation value of some fields $\Phi_k$. Combining eqs. (2.2.5) and (2.2.7), we find:

$$< (\delta_\varepsilon\Phi_1(x_1))\,\Phi_2(x_2)\cdots > \,+\, < \Phi_1(x_1)\,(\delta_\varepsilon\Phi_2(x_2))\cdots > + \cdots$$

$$= \ -\frac{\sqrt{g}}{\pi}\int d^2x\,\left(\partial_i\varepsilon^j(x)\right) < T^i{}_j(x)\Phi_1(x_1)\Phi_2(x_2)\cdots > \,. \tag{2.2.9}$$

The equation (2.2.9) is the Ward identity for the energy–momentum tensor. It shows that $T^{ij}$ is the generator of general coordinate transformations. Note that $T^{ij}$ has to be symmetric and traceless for the correlation functions to be rotation and scaling invariant.

We derived the Ward identity (2.2.9) in a formal way. In a given theory, it has to be checked using a regularised calculation. We will not do this in this work, and assume that eq. (2.2.9) holds, possibly with some quantum corrections. This enables us to make general statements for every theory where eq. (2.2.9) is valid. Similar Ward identities can be derived for any global symmetry of the action which leaves the measure invariant.

An important corrollary of the Ward identity for a symmetry generator, is that it is conserved "inside" correlation functions. We will again treat $T^{ij}$ as an example. We take $\varepsilon^i(x)$ functions which go to zero when $x$ goes to infinity, and which have no singularities. In this case, we can use partial integration. Now, the *lhs* of (2.2.9) depends only on the value of $\varepsilon$ (and its derivatives) in the $x_i$. This means that, after a partial integration, the coefficient of $\varepsilon$ in the integrandum in the *rhs* of (2.2.9) has to be zero except at these points. We find:

$$\frac{\partial}{\partial x^i} < T^{ij}(x)\Phi_1(x_1)\Phi_2(x_2)\cdots >= 0 \tag{2.2.10}$$

where $x \neq x_k$.

[2] We restrict ourselves to the case of a constant metric.

From now on, we consider the case of a conformal field theory in two dimensions. We already showed that $T^{ij}$ is symmetric and traceless, in the sense that all correlation functions vanish if one of the fields is the antisymmetric part of $T^{ij}$ or its trace. In the complex basis the two remaining components are $T^{\bar{z}\bar{z}}$ and $T^{zz}$. The conservation of $T^{ij}$ eq. (2.2.10) gives:

$$\frac{d}{d\bar{z}} < T^{\bar{z}\bar{z}}(z,\bar{z})\Phi_1(z_1,\bar{z}_1)\Phi_2(z_2,\bar{z}_2)\cdots >= 0\,, \qquad z \neq z_i\,, \tag{2.2.11}$$

and an analogous equation for $T^{zz}$. This shows that $< T^{\bar{z}\bar{z}}(z,\bar{z})\Phi_1\Phi_2\cdots >$ is a holomorphic function of $z$ with possible singularities in $z_i$. Using the assumption that the transformation of the fields $\delta_\varepsilon\Phi_i(x_i)$ is local, we see that the correlation function is a meromorphic function, with possible poles in $z = z_i$. We define:

$$T \equiv gT^{\bar{z}\bar{z}} = T_{zz} \quad \text{and} \quad \bar{T} \equiv gT^{zz} = T_{\bar{z}\bar{z}} \tag{2.2.12}$$

and we will write $T$ as a function of $z$ only.

For $\varepsilon^i(x)$ as specified above eq. (2.2.10), only the singularities $z = z_i$ contribute to the Ward identity (2.2.9). We now take $\varepsilon^z = \varepsilon(z)$, $\varepsilon^{\bar{z}} = 0$ in a neighbourhood of these points. Using eq. (A.5), the Ward identity (2.2.9) becomes for a conformal transformation:

$$\begin{aligned}&\sum_{j=1}^{N} < \Phi_1(z_1,\bar{z}_1)\ \cdots\ (\delta_\varepsilon\Phi_j(z_j,\bar{z}_j))\ \cdots > \\ &\quad = \oint_C \frac{dz}{2\pi i}\varepsilon(z) < T(z)\Phi_1(z_1,\bar{z}_1)\Phi_2(z_2,\bar{z}_2)\cdots >\,,\end{aligned} \tag{2.2.13}$$

where the contour $C$ encircles all $z_j$ exactly once anticlockwise (and an analogous formula for antianalytic variations).

If all $\Phi_j$ are primary fields, we can use (2.1.6) to compute the *lhs* of the Ward identity (2.2.13):

$$\begin{aligned}&\oint_C \frac{dz}{2\pi i}\varepsilon(z) < T(z)\Phi_1(z_1,\bar{z}_1)\Phi_2(z_2,\bar{z}_2)\cdots > \\ &\quad = \sum_{j=1}^{N}(\varepsilon(z_j)\partial_j + h_j\,\partial_j\varepsilon(z_j)) < \Phi_1(z_1,\bar{z}_1)\Phi_2(z_2,\bar{z}_2)\cdots >\,.\end{aligned} \tag{2.2.14}$$

This fixes the singular part in $(z = z_i)$ of the meromorphic $(N+1)$-point function $< T\Phi_1\cdots >$. We will assume that all correlation functions have no singularity at infinity. Hence, the regular terms in $(z - z_i)$ vanish, except for a possible constant.

To know the transformation law of $T(z)$ itself, we observe that since it is classically a rank two conformal tensor, its dimensions are $h = 2$, $\bar{h} = 0$. However, due to quantum anomalies, a Schwinger term can arise in its transformation law:

$$\delta_\varepsilon T(z) = \varepsilon(z)\partial T(z) + 2\partial\varepsilon(z)T(z) + \frac{c}{12}\partial^3\varepsilon(z)\,. \tag{2.2.15}$$

In the path integral formalism, the Schwinger term is non-zero if the measure is not invariant under the symmetry transformation generated by $T$, see eq. (2.2.9). From dimensional arguments, it follows that $c$ should have dimension zero. In fact, $c$ is in general a complex number. When $c$ is different from zero, $T$ is not a primary field, but only quasiprimary, see definition 2.1.2.

The transformation law (2.2.15) involves only $T$ and no other components of the energy–momentum tensor. In the classical case, this is a direct consequence of the conformal symmetry of the theory. We will assume that the same property holds in the quantum case.

Recall that eq. (2.2.14) determines the correlation function $< T\Phi_1\cdots >$ up to a constant. When we require invariance of the correlation function under scaling transformations ($\varepsilon(z) \sim z$), this constant has to be zero. Under these assumptions, the Ward identity completely fixes all correlation functions with one of the fields equal to $T$ and all other fields being primary:

$$\begin{aligned}&< T(z)\Phi_1(z_1,\bar{z}_1)\Phi_2(z_2,\bar{z}_2)\cdots >= \\ &\quad \sum_{j=1}^{N}\left(\frac{h_j}{(z-z_j)^2} + \frac{1}{(z-z_j)}\partial_j\right) < \Phi_1(z_1,\bar{z}_1)\Phi_2(z_2,\bar{z}_2)\cdots >\,.\end{aligned} \tag{2.2.16}$$

We can now look at correlation functions of $T$ only. Using eq. (2.2.15), we find in a completely analogous way to eq. (2.2.16):

$$\begin{aligned}&< T(z)T(z_1)T(z_2)\cdots >= \\ &\quad \sum_{j=1}^{N} < T(z_1)\cdots\left(\frac{c/2}{(z-z_j)^4} + \frac{2T(z_j)}{(z-z_j)^2} + \frac{\partial_j T(z_j)}{(z-z_j)}\right)\cdots T(z_N) >\,.\end{aligned} \tag{2.2.17}$$

Let us consider a few examples. Because of scaling invariance, one-point functions of fields with non-zero dimension vanish. Together with the fact that $< \mathbb{1} >= 1$, we see that eq. (2.2.15) implies:

$$< \delta_\varepsilon T >= \frac{c}{12}\partial^3\varepsilon(z)\,. \tag{2.2.18}$$

The Ward identity (2.2.13) fixes then the two-point function to:

$$< T(z)T(w) > = \frac{c/2}{(z-w)^4}\,, \tag{2.2.19}$$

in accordance with eq. (2.2.3). Similarly, the constant in the three-point function (2.2.4) of three $T$'s is fixed to $c$.

The analysis of the Ward identity for the energy–momentum tensor can be repeated for every generator of a global symmetry of the theory. In two dimensions, all tensors can be decomposed in symmetric tensors. If they are in addition traceless, only two components remain $W^{zzz\cdots}$ and $W^{\bar{z}\bar{z}\bar{z}\cdots}$. We again find that $\bar{\partial}W^{\bar{z}\bar{z}\bar{z}\cdots} = 0$, and the correlation functions of $W^{\bar{z}\bar{z}\bar{z}\cdots}$ are meromorphic functions. For generators with a spinorial character, or generators with non-integer conformal dimensions, we can only conclude that the correlation functions are holomorphic, *i.e.* fractional powers of $z - z_i$ could occur. We restrict ourselves in this work to the meromorphic sector of the conformal field theory. The symmetry algebra of the theory is a direct product[3] $\mathcal{A} \otimes \bar{\mathcal{A}}$ where $\mathcal{A}$ is the algebra formed by the "chiral" generators (with holomorphic correlation functions), and $\bar{\mathcal{A}}$ is its antichiral counterpart. From now on, we will concentrate on the *chiral* generators.

To conclude this section, we wish to stress that the Ward identities are regularisation dependent. In this work, we will not address the question of deriving the Ward identities for a given theory. However, once they have been determined, the Ward identities enable us to compute the singular part of any correlation function containing a symmetry generator in terms of (differential polynomials of) correlation functions of the fields of the theory. This determines those correlation functions up to a constant, because we required that a correlation function has no singularity at infinity. Furthermore, if the correlation functions are invariant under scale transformations, this extra constant can only appear when the sum of the conformal dimensions of all fields in the correlator is zero.

## 2.3 Operator Product Expansions

In this section, we discuss Operator Product Expansions (OPEs). In a first step, we show how they encode the information contained in the Ward identities. As such, OPEs provide an algebraic way of computing correlation functions. In the next subsection, we derive the consistency conditions on the OPEs. Subsection 2.3.3 introduces the concept of an Operator Product Algebra (OPA).

[3]This is only true when $z$ and $\bar{z}$ are regarded is independent coordinates.

### 2.3.1 OPEs and Ward identities

To every field in a conformal field theory, we assign a unique element of a vectorspace $\mathcal{V}$ of "operators". For the moment, we leave the precise correspondence open, but our goal is to define a bilinear operation in the vectorspace which enables us to compute correlation functions. We simply write the same symbol for the field and the corresponding operator. Similarly, if a field has conformal dimension $h$, we say that the corresponding operator has conformal dimension $h$.

Let us look at an example to clarify what we have in mind. Consider the Ward identities of the chiral component of the energy–momentum tensor $T(z)$. Eq. (2.2.16) suggests that we assign the OPE:

$$T(z)\Phi(z_0,\bar{z}_0) = \frac{h\Phi(z_0,\bar{z}_0)}{(z-z_0)^2} + \frac{\partial\Phi(z_0,\bar{z}_0)}{z-z_0} + O(z-z_0)^0 \tag{2.3.1}$$

for a primary field $\Phi$ with conformal dimension $h$. We do not specify the regular part yet. Similarly, eq. (2.2.17) imposes the following OPE for $T$ with itself:

$$T(z)T(z_0) = \frac{c/2}{(z-z_0)^4} + \frac{2T(z_0)}{(z-z_0)^2} + \frac{\partial T(z_0)}{z-z_0} + O(z-z_0)^0\,. \tag{2.3.2}$$

We see that an OPE is a bilinear map from $\mathcal{V}\otimes\mathcal{V}$ to the space of formal Laurent expansions in $\mathcal{V}$. We will use the following notation for OPEs:

$$T_1(z)T_2(z_0) = \sum_{n<=h(T_1,T_2)} \frac{[T_1T_2]_n(z_0)}{(z-z_0)^n}\,, \tag{2.3.3}$$

where $h(T_1,T_2)$ is some finite number. Because of the correspondence between OPEs and Ward identities, we see that all terms in the sum (2.3.3) have the same conformal dimension: $\dim([T_1T_2]_n) = \dim(T_1) + \dim(T_2) - n$. If no negative dimension fields are present in the field theory, we immediately infer that $h(T_1,T_2)$ is less than or equal to the sum of the conformal dimensions of $T_1$ and $T_2$. As noted before, we restrict ourselves in this work to the case where the correlation functions of the chiral symmetry generators $T_i(z_i)$ are meromorphic functions in $z_i$. In other words, the sum in eq. (2.3.3) runs over the integer numbers. We denote the singular part of an OPE by $T_1\underbrace{(z)T_2}(z_0)$. It is determined via:

$$\oint_{C_{z_0}} \frac{dz}{2\pi i}\varepsilon(z)T_1\underbrace{(z)T_2}(z_0) = \delta_\varepsilon^{T_1}T_2(z_0)\,, \tag{2.3.4}$$

where we denoted the transformation generated by $T_1$ with parameter $\varepsilon$ as $\delta_\varepsilon^{T_1}$. Note that we can assign an OPE only when $T_1$ is a symmetry generator.

The prescription we use for the moment to calculate correlation functions with OPEs, is simply an application of conformal Ward identities like eq. (2.2.13). As an example, to compute a correlation function with a symmetry generator $T_1$, we substitute the contractions of $T_1$ with the other fields in the correlator[4]:

$$
\begin{aligned}
&< T_1(z)\Phi_1(z_1,\bar{z}_1)\Phi_2(z_2,\bar{z}_2)\cdots > = \\
&\qquad < \underbrace{T_1(z)\Phi_1}(z_1,\bar{z}_1)\Phi_2(z_2,\bar{z}_2)\cdots > + \\
&\qquad < \underbrace{T_1(z)\Phi_1(z_1,\bar{z}_1)\Phi_2}(z_2,\bar{z}_2)\cdots > + \cdots
\end{aligned}
\tag{2.3.5}
$$

In this way, we can always reduce any correlation function, where symmetry generators are present to a differential polynomial of correlation functions of fields only. Of course, if only generators were present from the start, we will end up with a function of all arguments $z, w, \ldots$ containing one-point functions and $< \mathbb{1} >= 1$.

We now consider the map from the space of fields to the vectorspace $\mathcal{V}$ in more detail. The normal ordered product of two fields $T_1$ and $T_2$ is defined by considering a correlation function $< T_1(z)T_2(w)\,X >$, where $X$ denotes an arbitrary sequence of operators, and taking the limit of $z$ going to $w$ after substracting all singular terms:

$$
\begin{aligned}
&<: T_1T_2 : (w)\,X >= \\
&\qquad \lim_{z\to w} < \left( T_1(z)T_2(w) - \sum_{n>0} \frac{[T_1T_2]_n(w)}{(z-w)^n} \right) X > .
\end{aligned}
\tag{2.3.6}
$$

This definition corresponds to a point-splitting regularisation prescription. In the case of non-meromorphic correlation functions, extra fractional powers of $(z-w)$ are included in this definition. We will not treat this case here. In the notation of eq. (2.3.3), we see that the map from the space of the fields to $\mathcal{V}$ is in this case given by:

$$
: T_1T_2 : \to [T_1T_2]_0 \,. \tag{2.3.7}
$$

In fact, we can use the same procedure to define all operators in the regular part of the OPE[5], *i.e.*

$$
\begin{aligned}
&< \sum_{n\leq 0} \frac{[T_1T_2]_n(w)}{(z-w)^n}\, X > = \\
&\qquad < \left( T_1(z)T_2(w) - \sum_{n>0} \frac{[T_1T_2]_n(w)}{(z-w)^n} \right) X > .
\end{aligned}
\tag{2.3.8}
$$

[4]In fact, this is only valid when the correlation function is zero at infinity, *i.e.* when no extra constant appears, see the discussion at the end of the previous section.

[5]It is sufficient to define an operator $A$ by specifying all correlators $< A(z)\,X >$. Crossing symmetry of the correlators gives then $< X\,A(z)\,Y >$, where $X$ and $Y$ stand for arbitrary sequences of operators.

The definition (2.3.8) implies that correlation functions can be computed by substituting for two operators their *complete* OPE. We write:

$$
< T_1(z_1)T_2(z_2)\ X >= < [T_1(z_1)T_2(z_2)]\ \ X > . \tag{2.3.9}
$$

In this way, an OPE is now an identity between two bilocal operators.

Because one-point functions are constants and we assumed that correlators have no singularity at infinity, the definition (2.3.8) implies:

$$
< [T_1(z)T_2(w)]_n > =\ 0\,, \qquad \forall n < 0\,. \tag{2.3.10}
$$

If all fields in the correlation function eq. (2.3.9) are symmetry generators, we end up with one-point functions $< T_i(z) >$. These can of course not be computed in the OPE formalism. As infinite sums are involved, one should pay attention to convergence, which will in general only be assured when certain inequalities are obeyed involving the $z_i$. However, the result should be a meromorphic function with poles in $(z_i - z_j)$. So analytic continuation can be used.

The definition (2.3.8) gives the same results as the straightforward application of the Ward identities eq. (2.3.5). This will be proved in the next subsection after determining the consistency requirements on the OPEs.

### 2.3.2 Consistency conditions for OPEs

In this subsection, we determine the consistency requirements on the OPEs by considering (contour integrals of) correlators. We first list the properties of correlation functions used in subsection 2.2.2:

**Assumption 2.3.1** *The correlators have the following properties:*

- *translation and scaling invariance*
- *no singularity at infinity*
- *the correlation functions involving a chiral symmetry generator $W(z)$ are meromorphic functions in $z$*
- *crossing symmetry*

We now suppose that there is a map from the space of fields of the conformal field theory to a vectorspace $\mathcal{V}$. A $\mathbb{Z}_2$ grading in $\mathcal{V}$ should exist, corresponding to bosonic and fermionic operators. We denote it with $|A|$. There is an even linear map $\partial$ from $\mathcal{V}$ to $\mathcal{V}$, and a sequence of bilinear operations for every $n \in \mathbb{Z}$ which we denote by $[AB]_n$, see eq. (2.3.3). For $n > 0$ the bilinear operations are defined in eq. (2.3.4). We require that correlation functions can be computed by substituting the complete OPE as done for eq. (2.3.9). From this definition, we determine the properties of these maps. In the remainder of this subsection $X$ and $Y$ denote arbitrary sequences of operators.

We first determine $[\partial A\, B]_n$ by requiring that $\partial$ corresponds to the derivative on the fields. Because we have:

$$< X\ \partial A(z)\ Y >= \frac{d}{dz} < X\ A(z) Y > , \tag{2.3.11}$$

we find that the OPE $\partial A(z)B(w)$ is given by taking the derivative of the OPE $A(z)B(w)$ with respect to $z$:

$$\partial A(z)B(w) = \sum_{n<=h(A,B)} -n\frac{[AB]_n(w)}{(z-w)^{n+1}} , \tag{2.3.12}$$

and hence:

$$[\partial A\, B]_{n+1} = -n[AB]_n . \tag{2.3.13}$$

This equation enables us to write all the terms in the regular part in terms of normal ordered operators:

$$[AB]_{-n} = \frac{1}{n!} : (\partial^n A)\ B : , \quad n \in \mathbb{N} . \tag{2.3.14}$$

By applying derivatives with respect to $w$ we find:

$$[A\, \partial B]_{n+1} = n[AB]_n + \partial[AB]_{n+1} . \tag{2.3.15}$$

We now investigate the consequences of the crossing symmetry of the correlators.

First, consider a correlator $< A(z)B(w)\, X >$. As $X$ is completely arbitrary, we see that the OPE $A(z)B(w)$ must be equal to the OPE $(-1)^{|A|\,|B|}B(w)A(z)$. Looking at eq. (2.3.3) for both OPEs, one sees that a Taylor expansion of $[AB]_n(z)$ around $w$ has to be performed to compare the OPEs. The result is:

$$[BA]_q = (-1)^{|A|\,|B|} \sum_{l\geq q} \frac{(-1)^l}{(l-q)!} \partial^{(l-q)} [AB]_l \tag{2.3.16}$$

for arbitrary $q$.

**Intermezzo 2.3.1**

Eq. (2.3.16) leads to a consistency equation when $A = B$:

$$[AA]_q = -\sum_{l>0} \frac{(-1)^l}{2l!} \partial^l [AA]_{q+l} \quad \text{if} |A| + q \text{ odd} , \tag{2.3.17}$$

where we shifted the summation index $l$ over $q$ with respect to eq. (2.3.16). If $A$ is bosonic (fermionic), this relation determines an odd (even) pole of the OPE $A(z)A(w)$ in terms of derivatives of the higher poles, and thus in terms of the higher even (odd) poles. We can write:

$$[AA]_q = \sum_{k\geq 0} a_k \partial^{2k+1} [AA]_{q+2k+1} ,$$

where the constants $a_k$ satisfy the following recursion relation:

$$a_k = \frac{1}{2}\left( \frac{1}{(2k+1)!} - \sum_{0\leq l<k} \frac{a_l}{(2k-2l)!} \right) ,$$

which can be solved by using the generating function:

$$f(x) = \sum_{k\geq 0} a_k\, x^{2k} = \frac{\tanh(x/2)}{x} ,$$

for which the Taylor expansion is well known. Hence:

$$a_k = 2B_{2k+2} \frac{2^{2k+2}-1}{(2k+2)!} ,$$

where $B_n$ are the Bernoulli numbers. The first values in this series are:

$$a_0 = \frac{1}{2} ,\ a_1 = -\frac{1}{24} ,\ a_2 = \frac{1}{240} ,\ a_3 = -\frac{17}{40320} .$$

Applying the rule (2.3.17) for $q = 0$ and $A$ fermionic, expresses the normal ordered product of a fermionic field with itself in terms of the operators occuring in the singular part of its OPE with itself. An example is found in the $N = 1$ superconformal algebra. This algebra contains a supersymmetry generator $G$ which has the following OPE:

$$G(z)G(w) = \frac{2c/3}{(z-w)^3} + \frac{2T(w)}{(z-w)} + O(z-w)^0 .$$

Applying the above formulas, we find:

$$[GG]_0 = \partial T .$$

---

It is clear that eq. (2.3.16) shows that (dropping sign factors):

$$\begin{aligned} &< [[A(z)B(w)]\, C(u)]\ X >_{\rm OPE} \\ &= \ < [[B(w)A(z)]\, C(u)]\ X >_{\rm OPE} \\ &= \ < [C(u)\, [A(z)B(w)]]\ X >_{\rm OPE} . \end{aligned} \tag{2.3.18}$$

This does not yet prove that the correlator is crossing symmetric. Indeed, for this we also need (dropping arguments as well):

$$< [[AB]C]\ X >_{\rm OPE} = < [[CA]B]\ X >_{\rm OPE} . \tag{2.3.19}$$

We see that crossing ssymmetry implies "associativity" of the OPEs: the order in which the OPEs are substituted should be irrelevant. This puts very stringent conditions on the OPEs. These conditions are most easily derived by using contour integrals of correlators. Indeed, we can isolate the contribution of a certain part of the OPE by taking appropriate contour integrals:

$$< [A[BC]_p]_q(u)\, X >= \oint_{C_u} \frac{dz}{2\pi i}\,(z-u)^{q-1}$$

$$\oint_{C_u} \frac{dw}{2\pi i}\,(w-u)^{p-1}\;< A(z)B(w)C(u)\, X >\,, \tag{2.3.20}$$

where $C_u$ denotes a contour which encircles $u$ once anti-clockwise, not including any other points involved in the correlator. We can now use a contour deformation argument relating the contour integral in eq. (2.3.20) to a contour integral where the integration over $w$ is performed last, see fig. 2.1. This integral has two terms: one where the $z$ contour is around $u$ (corresponding to the correlator $(-1)^{|A|\,|B|} <$ $[B\,[AC]]\; X >$, and one where it is around $w$ ($< [[AB]\,C]\; X >$). Using the definition

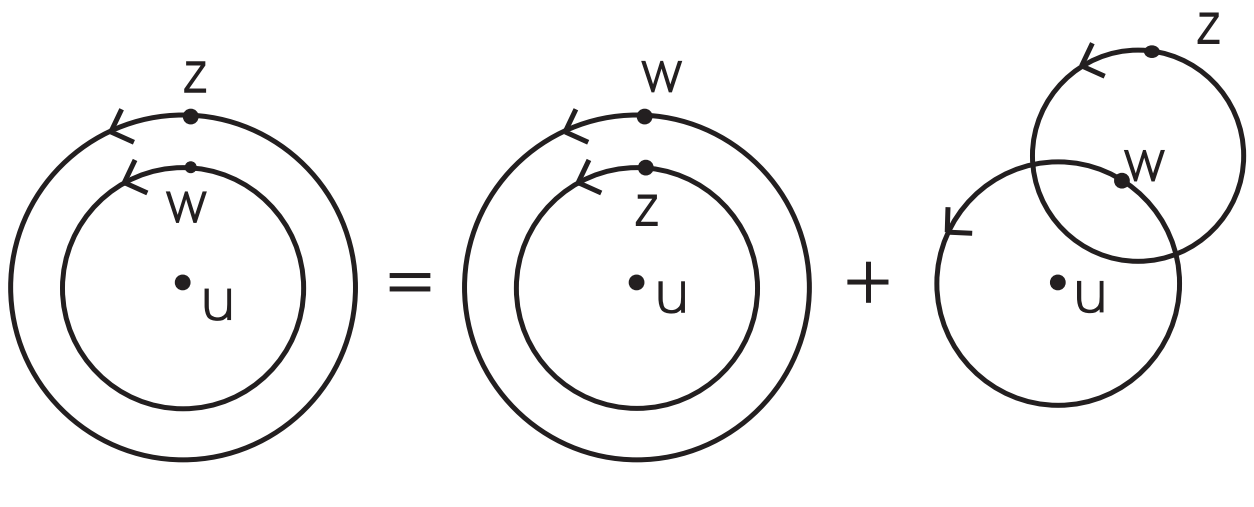

Figure 2.1: Contour deformations

(2.3.3) for the OPEs, and Cauchy's residue formula for contour integrals, we arrive at:

$$[A[BC]_p]_q = (-1)^{|A|\,|B|}[B[AC]_q]_p + \sum_{l>0} \binom{q-1}{l-1} [[AB]_l C]_{p+q-l}\,. \tag{2.3.21}$$

A second equation follows in the same manner by interchanging the role of $z$ and $w$ in fig. 2.1 and bringing the second contour integral of the *rhs* of eq. (2.3.20) to the left:

$$\begin{aligned}&[A[BC]_p]_q\\ =\;& (-1)^{|A|\,|B|}\left([B[AC]_q]_p - \sum_{l>0}\binom{p-1}{l-1}[[BA]_l C]_{p+q-l}\right)\,.\end{aligned} \tag{2.3.22}$$

Both eqs. (2.3.21),(2.3.22) have to be satisfied for any $p, q \in \mathbb{Z}$.[6]

Similarly, by starting with:

$$< [[AB]_p\; C]_q(u)\, X >= \oint_{C_u} \frac{dw}{2\pi i}\,(w-u)^{q-1}$$

$$\oint_{C_w} \frac{dz}{2\pi i}\,(z-w)^{p-1}\;< A(z)B(w)C(u)\, X >\,, \tag{2.3.23}$$

and using the same contour deformation of fig. 2.1, but now with the first term of the *rhs* brought to the left, we get:

$$\begin{aligned}[[AB]_p\; C]_q \;=\;& \sum_{l\geq q}(-1)^{q-l}\binom{p-1}{l-q}[A[BC]_l]_{p+q-l} - \\ & (-1)^{|A|\,|B|}\sum_{l>0}(-1)^{p-l}\binom{p-1}{l-1}[B[AC]_l]_{p+q-l}\,.\end{aligned} \tag{2.3.24}$$

which has again to be satisfied for all $p, q \in \mathbb{Z}$.

Eq. (2.3.16) and eq. (2.3.21) for $q = 0$ first appeared in [5] where they were derived using the mode algebra associated to OPEs (see section 2.4). In [186], contour deformation arguments were used to find eq. (2.3.21) and an equation related to eq. (2.3.22) for restricted ranges of $p, q$. Reference [186] also contains eq. (2.3.24) for $q > 1$.

At this point, we found the conditions on the OPEs of the operators in $\mathcal{V}$ such that we find "correlation functions" which satisfy the assumptions 2.3.1. We still need to show that this procedure indeed gives the correlation functions of the conformal field theory we started with, *i.e.* that we find correlation functions which satisfy the Ward identities of the theory. Before proceeding, we prove the following lemma:

**Lemma 2.3.1** *The associativity condition eq. (2.3.21) for strictly positive $q$ can be rewritten as:*

$$\underbrace{A(z)[B(w)}C(u)] = (-1)^{|A|\,|B|}\left[B(w)\underbrace{A(z)C}(u)\right] + \left[\underbrace{A(z)B}(w)C(u)\right]$$

*for $|w-u| < |z-u|$.*

**Proof :**

The proof of this lemma is in some sense the reverse of the derivation of eq. (2.3.21). We multiply eq. (2.3.21) with $(z-u)^q(w-u)^p$ and sum for $p$ and $q$ over the appropriate ranges. After substracting the result from the proposition of this lemma, we need to

[6] See appendix 2.A for the extended definition of the binomial coefficients.

prove:

$$\left[\underbrace{A(z)B}(w)C(u)\right] =$$

$$\sum_{q>0}\sum_{p}(z-u)^{-q}(w-u)^{-p}\sum_{l>0}\binom{q-1}{l-1}[[AB]_lC]_{p+q-l}(u)\,.$$

Due to the binomial coefficient, and because $q$ is strictly positive, the sum over $l$ is from one to $q$. Call $r=p+q-l$ and $s=q-l$. We find:

$$\begin{aligned} rhs &= \sum_{r}\sum_{l>0}[[AB]_lC]_r(u) \\ &\qquad \left(\sum_{s\geq 0}\binom{l+s-1}{l-1}(z-u)^{-l-s}(w-u)^{-r+s}\right) \\ &= \sum_{r}\sum_{l>0}(z-u)^{-l}(w-u)^{-r}(1-\frac{w-u}{z-u})^{-l}[[AB]_lC]_r(u)\,, \end{aligned}$$

which is exactly the *lhs*. In the last step we used the Taylor expansion of $(1-x)^{-l}$ which only converges for $|x|<1$. ■

The Ward identities have the form (2.2.13):

$$\begin{aligned} &\oint_C \frac{dz}{2\pi i}\varepsilon(z) < [T_0(z)T_1(z_1)]\cdots T_n(z_n) >= \\ &\quad \oint_{C_1}\frac{dz}{2\pi i}\varepsilon(z) < \underbrace{T_0(z)T_1}(z_1)\cdots T_n(z_n) > +\cdots \\ &\quad + \oint_{C_n}\frac{dz}{2\pi i}\varepsilon(z) < \underbrace{T_0(z)T_1(z_1)\cdots T_n}(z_n) >\,, \end{aligned} \tag{2.3.25}$$

where the contractions correspond by definition to the transformation generated by $T_0$ of the field $T_i$, eq. (2.3.4). The contour $C$ surrounds $z_1\cdots z_n$ once in the anti-clockwise direction, while the contours $C_i$ encircle only $z_i$. $\varepsilon(z)$ is analytic in a region containing $C$.

**Theorem 2.3.1** *The Ward identities eq. (2.3.25) are satisfied for correlation functions which we compute by substituting two operators with their complete OPE.*

**Proof :**

The $(n+1)$-point function $< T_0(z)T_1(z_1)\cdots T_n(z_n) >$ has only poles in $z$ when $z=z_i$. We will use this to deform the contour $C$.

Suppose that the theorem holds for $n$-point functions. We rearrange the operators $T_i$ such that $|z_1-z_2|$ is smaller than all other $|z_i-z_j|$. We substitute $T_1(z_1)T_2(z_2)$ by their OPE. Then we have to compute a $n$-point function.

We now split the contour $C$ in a contour $C_2'$ and the contours $C_i$, $i>2$, used in eq. (2.3.25). $C_2'$ encircles $z_1$ and $z_2$, but no other $z_i$. Because of the rearrangement, we can take $C_2'$ such that for any point $z$ on $C_2'$, $|z-z_2|>|z_1-z_2|$. By applying the recursion assumption we find:

$$\begin{aligned} &\oint_C \frac{dz}{2\pi i}\varepsilon(z) < T_0(z)T_1(z_1)\cdots T_n(z_n) >= \\ &\quad \oint_{C_2'}\frac{dz}{2\pi i}\varepsilon(z) < \underbrace{T_0(z)[T_1(z_1)}T_2(z_2)]\cdots T_n(z_n) > \\ &\quad + \oint_{C_3}\frac{dz}{2\pi i}\varepsilon(z) < \underbrace{T_0(z)[T_1(z_1)T_2(z_2)]T_3}(z_3)\cdots T_n(z_n) > +\cdots \end{aligned}$$

We can now use lemma 2.3.1 for the first term of the *rhs*. We indeed find the *rhs* of eq. (2.3.25). ■

We comment on when a correlation function can be computed by taking all contractions, which was the prescription we used to define OPEs in the previous subsection, see eq. (2.3.5). Computing a correlation function in this way, means that we drop the integrals in the Ward identity eq. (2.3.25). This can only be done if all one-point functions $< [T_1T_2]_0 >$ vanish (except if $T_i = \mathbb{1}$). Because of scaling invariance, this is true when all operators (except $\mathbb{1}$) have strictly positive dimension. Bowcock argued in [34] that correlation functions can be computed by substituting the complete OPE, or by using contractions. His argument is based on the claim that eq. (2.3.25) is true, but no proof is given.

To conclude this subsection, we wish to mention that Wilson Operator Product Expansions were already used outside the scope of two-dimensional conformal field theory [208, 191]. However, because the consistency requirements on the OPEs are especially strong in two-dimensional conformal field theory (due to the fact that the conformal algebra has infinite dimension), it is there where the full power of the formalism comes to fruition.

### 2.3.3 Operator Product Algebras

We can assemble the consistency conditions on OPEs in a definition, see also [28, 96].

**Definition 2.3.1** *An Operator Product Algebra (OPA) is a $\mathbb{Z}_2$ graded vectorspace $\mathcal{V}$ with elements $\mathbb{1}, A, B, C\cdots$, an even linear map $\partial$, and a bilinear binary operation:*

$$[.,.]_l : \mathcal{V}\otimes\mathcal{V}\to\mathcal{V}, \qquad l\in\mathbb{Z}\,,$$

*which is zero for $l$ sufficiently large. The following properties hold:*

- **unity**:

$$[\mathbb{1}A]_l = \delta_l\ A$$

- **commutation** : (eq. (2.3.16))

$$[BA]_n = (-1)^{|A|\,|B|} \sum_{l\geq n} \frac{(-1)^l}{(l-n)!} \partial^{(l-n)} [AB]_l \qquad \forall n \in \mathbb{Z}$$

- **associativity** : (eq. (2.3.21))

$$\begin{aligned} [A[BC]_p]_q &= (-1)^{|A|\,|B|} [B[AC]_q]_p \\ &\quad + \sum_{l>0} \binom{q-1}{l-1} [[AB]_l C]_{p+q-l} \qquad \forall p, q \in \mathbb{Z} \end{aligned}$$

The properties in this definition are sufficient to recover all consistency conditions of the previous subsection. Using $\partial A = [A\mathbb{1}]_1$ and the associativity condition, the equations for OPEs of derivatives (2.3.13,2.3.15) follow. Similarly, the other associativity conditions (2.3.22) and (2.3.24) can be derived from the properties of an OPA.

An alternative definition would be to impose eqs. (2.3.13, 2.3.15), requiring the associativity condition eq. (2.3.21) only for $p, q \geq 0$. One then usually considers eq. (2.3.21) for $p$ or $q$ equal to zero as being the definition of how we can calculate OPEs of normal ordered products. The set of consistency conditions eq. (2.3.21) for $p, q > 0$ has then to be checked for all generators of the OPA.

In some cases, the definition 2.3.1 is too strong. OPEs are intended to compute correlation functions. It is possible that there are some fields in the theory which have vanishing correlators with all other fields. These null fields should be taken into account in the definition of an OPA. Indeed, they could occur in every consistency equation for OPEs without affecting the results for the correlation functions. We need an algebraic definition of a null field. From the way we compute correlators using OPEs, we see that if $N$ is a null field, $[NA]_n$ should be again a null field, for any $A \in \mathcal{V}$ and $n \in \mathbb{Z}$. If this would not be true, we could write down a nonvanishing correlator with $N$. We see that the null fields form an ideal in the OPA.

**Definition 2.3.2** *Consider an OPA as defined in def. 2.3.1. Suppose there is an ideal $\mathcal{N}$ in the OPA, whose elements we call null operators (or null fields). We extend the definition of an OPA to algebras where the defining properties are only satisfied up to elements of $\mathcal{N}$.*

This extension of the definition of an OPA was not considered in [28, 96].

It is in general difficult to check if we can consider an operator $N$ to be null. We do not want that $\mathbb{1} \in \mathcal{N}$, because then all operators are null. Hence, a necessary condition for $N$ to be null is that we can find no operator $A$ in $\mathcal{V}$ such that $[NA]_n \sim \mathbb{1}$ (for some $n$). Usually, this is regarded as a sufficient condition, because one generally works with fields of strictly positive conformal dimension. Due to the scaling invariance of the correlation functions, all one-point functions are then zero, except $< \mathbb{1} >$. So, in this case, any field which does not produce the identity operator in some OPE has zero correlation functions.

### 2.3.4 OPA–terminology

In this subsection, we introduce some terminology which is continually used in the rest of this work.

**Definition 2.3.3** *We call $A$ a composite operator if it is equal to $[BC]_0$, for some $B, C \neq \mathbb{1}$, except when $B$ is equal to $C$ and fermionic.*

The condition when $B = C$ in this definition comes from the considerations in intermezzo 2.3.1.

**Definition 2.3.4** *A set of operators is said to generate the OPA if all elements of $\mathcal{V}$ can be constructed by using addition, scalar multiplication, derivation and taking composites.*

**Definition 2.3.5** *A Virasoro operator $T$ has the following non-zero poles in its OPE (2.3.2):*

$$[TT]_4 = \frac{c}{2}\mathbb{1}, \qquad [TT]_2 = 2T, \qquad [TT]_1 = \partial T\,.$$

**Definition 2.3.6** *An operator $A$ is a scaling operator with respect to $T$, with (conformal) dimension $h$ if:*

$$[TA]_2 = hA, \qquad [TA]_1 = \partial A\,.$$

*A quasiprimary operator is a scaling operator with in addition $[TA]_3 = 0$.*
*A primary operator is a scaling operator with in addition $[TA]_n = 0$ for $n \geq 3$.*
*A conformal OPA is an OPA with a Virasoro $T$. All other operators of the OPA (except $\mathbb{1}$) are required to be scaling operators with respect to $T$.*

**Intermezzo 2.3.2**
As an application of the above definitions, we wish to show that when $A$ and $B$, with dimension $a$ and $b$, are scaling operators with respect to $T$ then $[AB]_n$ is a scaling operator with dimension $a+b-n$. Let us compute the first order pole of this operator with $T$ using eq. (2.3.21):

$$\begin{aligned}[T[AB]_n]_1 &= [A[TB]_1]_n + [[TA]_1 B]_n \\ &= [A\,\partial B]_n + [\partial A\ B]_n \\ &= \partial[AB]_n\,,\end{aligned}$$

where we used the sum of eq. (2.3.13) and eq. (2.3.15) in the last step. For the second order pole we have:

$$\begin{aligned}[T[AB]_n]_2 &= [A[TB]_2]_n + [[TA]_2 B]_n + [[TA]_1 B]_{n+1} \\ &= (a+b)[AB]_n + [\partial A\ B]_{n+1} \\ &= (a+b-n)[AB]_n.\end{aligned}$$

---

**Definition 2.3.7** *A map from the OPA to the half-integer numbers is called a dimension if it has the properties:*

$$\begin{aligned}\dim(\mathbb{1}) &= 0 \\ \dim(\partial A) &= \dim(A)+1 \\ \dim([AB]_l) &= \dim(A)+\dim(B)-l\,.\end{aligned}$$

*If such a map exists, we call the OPA graded.*

**Definition 2.3.8** *A $\mathcal{W}$–algebra is a conformal OPA where one can find a set of generators which are quasiprimary.*

Different definitions of a $\mathcal{W}$–algebra exist in the literature. Sometimes one requires that the generators are primary (except $T$ itself). In this work, we will mainly consider $\mathcal{W}$–algebras of this subclass, and for which the number of generators is finite. The importance of $\mathcal{W}$–algebras lies in the fact that the chiral symmetry generators of a conformal field theory form a $\mathcal{W}$–algebra[7]. We will treat $\mathcal{W}$–algebras in more detail in chapter 4.

Finally, we introduce a notation for OPEs which lists only the operators in the singular terms, starting with the highest order pole. As an example, we will write a Virasoro OPE (2.3.2) as:

$$T\times T \;= \ll \frac{c}{2}\mid 0\mid 2T\mid \partial T\gg\ . \tag{2.3.26}$$

[7] In fact, the chiral symmetry generators form only a conformal OPA, but we know of no example in the literature where no quasiprimary generators can be found.

### 2.3.5 Poisson brackets

To conclude this section, we want to show the similarity between Poisson bracket calculations and OPEs.

In a light-cone quantisation scheme (choosing $\bar z$ as "time") [209], the symmetry generators in classical conformal field theory obey Poisson brackets of the form:

$$\{A(z),B(z_0)\}_{\rm PB} = \sum_{n>0}\frac{(-1)^{n-1}}{(n-1)!}\{AB\}_n(z_0)\partial^{n-1}\delta(z-z_0)\,, \tag{2.3.27}$$

where $\{AB\}_n$ are also symmetry generators of the theory. The derivative is with respect to the $z$–coordinate. We choose the normalisation factors such that:

$$\{AB\}_n(z) = \int dz\ (z-w)^{n-1}\{A(z),B(w)\}_{\rm PB}\,. \tag{2.3.28}$$

For convenience, we drop the subscript PB in the rest of this subsection. The Poisson brackets satisfy:

$$\begin{aligned}\{A(z),B(z_0)\} &= (-1)^{|A|\,|B|}\{B(z_0),A(z)\} \\ \{\partial A(z),B(z_0)\} &= \frac{d}{dz}\{A(z),B(z_0)\} \\ \{\{A(z_1),B(z_2)\},C(z_3)\} &= \{A(z_1),\{B(z_2),C(z_3)\}\} - \\ &\quad (-1)^{|A|\,|B|}\{B(z_2),\{A(z_1),C(z_3)\}\}\end{aligned} \tag{2.3.29}$$

These relations imply identities for the $\{AB\}_n$. We do not list the consequences of the first two, as they are exactly the same as eqs. (2.3.16) and (3.3.1). The Jacobi identities give:

$$\begin{aligned}&\{A\ \{BC\}_p\}_q(z_3)\ -\ (-1)^{|A|\,|B|}\{B\ \{AC\}_q\}_p(z_3) \\ &= \int dz_2\ (z_2-z_3)^{p-1}\int dz_1\ \sum_{l=1}^{q}\binom{q-1}{l-1} \\ &\qquad (z_1-z_2)^{l-1}(z_2-z_3)^{q-l}\{\{A(z_1),B(z_2)\},C(z_3)\} \\ &= \sum_{l=1}^{q}\binom{q-1}{l-1}\{\{AB\}_l\ C\}_{p+q-l}(z_3)\}\,.\end{aligned} \tag{2.3.30}$$

These equations are of exactly the same form as the associativity conditions for OPEs (see (2.3.21) with $q,p>0$).

An important difference with OPEs is that no "regular" part is defined for Poisson brackets. In particular, normal ordering is not necessary. A Poisson bracket where a product of fields is involved, is simply:

$$\{A(z), B(w)C(w)\} = \{A(z), B(w)\}C(w) + (-1)^{|A|\,|B|}B(w)\{A(z), C(w)\}\,. \quad (2.3.31)$$

When using the notation:

$$\{AB\}_{-n}(z) = \frac{1}{n!}(\partial^n A(z))\, B(z)\,, \qquad n \geq 0\,, \quad (2.3.32)$$

we see that eq. (2.3.31) corresponds to eq. (2.3.21) for $p = 0$ and with the double contractions $l < q$ dropped. This is also true for a classical version of eq. (2.3.24). We can conclude that when using the correspondence:

$$\{AB\}_n \leftrightarrow [AB]_n\,, \quad (2.3.33)$$

computing with Poisson brackets follows almost the same rules as used for OPEs: one should drop double contractions and use a graded-commutative and associative normal ordering. In particular, as the Jacobi–identities are the same, any linear PB–algebra corresponds to an operator product algebra and vice-versa. For nonlinear algebras, this is no longer true because of normal ordering.

Due to this correspondence, we will often write "classical OPEs" for Poisson brackets.

## 2.4 Mode algebra

In this section, we show that there is an infinite dimensional algebra with a graded–symmetric bracket associated to every OPA. For every operator $A$ we define the $m$-th mode of $A$ by specifying how it acts on an operator:

$$\widehat{A}_m B \equiv [AB]_{m+a}\,, \quad (2.4.1)$$

where $a$ is the conformal dimension of $A$. The shift in the index is made such that $\widehat{A}_m B$ has dimension $b - m$, independent of $A$ [8]. Hence for operators with (half-)integer dimension, $m$ is (half-)integer. An immediate consequence of this definition follows by considering (2.3.13):

$$\left(\widehat{\partial A}\right)_m = -(m+a)\widehat{A}_m\,. \quad (2.4.2)$$

We can now compute the graded commutator of two modes:

$$[\widehat{A}_m, \widehat{B}_n]C = [A[BC]_{n+b}]_{a+m} - (-1)^{|A|\,|B|}[B[AC]_{m+a}]_{n+b}\,. \quad (2.4.3)$$

[8] This definition assumes that the OPA is graded, def. 2.3.7. Of course, the mode algebra can also be defined without this concept.

Using the associativity condition (2.3.21) we see that:

$$[\widehat{A}_m, \widehat{B}_n] = \sum_{l>0} \binom{m+a-1}{l-1} (\widehat{[AB]_l})_{m+n}\,. \quad (2.4.4)$$

Note that this commutator is determined by the singular part of the OPE.

**Theorem 2.4.1** *Eq. (2.4.4) defines a graded commutator.*

**Proof :**
We can use the relation between the OPEs $AB$ and $BA$ (2.3.16) and eq. (2.4.2) to show that:

$$\begin{aligned} &-(-1)^{|A|\,|B|}[\widehat{B}_n, \widehat{A}_m] \\ &= \sum_{l>0}\binom{n+b-1}{l-1}\sum_{p>l}\frac{(-1)^p}{(p-l)!}(\widehat{\partial^{p-l}[AB]_p})_{m+n} \\ &= \sum_{l>0}\sum_{p>l}\binom{n+b-1}{l-1}\binom{m+n+a+b-l-1}{p-l}(\widehat{[AB]_p})_{m+n}\,. \end{aligned}$$

Using eq. (2.A.4), we see that this is equal to (2.4.4). ∎

A first example of a mode algebra is given by the modes of a Virasoro operator $T$ with OPE (2.3.2). For historic reasons, we denote these modes with $L$. We find using eq. (2.4.4):

$$[\widehat{L}_m, \widehat{L}_n] = (m-n)\widehat{L}_{m+n} + \frac{c}{2}\binom{m+1}{3}\delta_{m+n}\,, \quad (2.4.5)$$

where we used that the modes of the unit operator (which is implicit in the fourth order pole of (2.3.2)) are given by:

$$\widehat{\mathbb{1}}_m = \delta_m\,. \quad (2.4.6)$$

The infinite dimensional Lie algebra with commutator (2.4.5) is called the Virasoro algebra. We see that this algebra is a central extension of the algebra of the classical generators of conformal transformations (2.1.8). The modes corresponding to the global conformal transformations $\widehat{L}_1, \widehat{L}_0, \widehat{L}_{-1}$ form a finite dimensional subalgebra where the central extension drops out.

Similarly, the OPE of $T$ with a primary field $\Phi$ with dimension $h$ (2.3.1) gives the following commutator:

$$[\widehat{L}_m, \widehat{\Phi}_n] = ((h-1)m - n)\widehat{\Phi}_{m+n}\,. \quad (2.4.7)$$

We have the following important theorem.

**Theorem 2.4.2** *For a given OPA, the commutator of the corresponding mode algebra satisfies graded Jacobi identities modulo modes of null fields:*

$$[\widehat{A}_k, [\widehat{B}_l, \widehat{C}_m]] = (-1)^{|A|\,|B|}[\widehat{B}_l, [\widehat{A}_k, \widehat{C}_m]] + [[\widehat{A}_k, \widehat{B}_l], \widehat{C}_m]\,. \qquad (2.4.8)$$

**Proof :**
The *lhs* of eq. (2.4.8) is by definition (2.4.4) equal to:

$$[\widehat{A}_k, [\widehat{B}_l, \widehat{C}_m]] = \sum_{q>0}\sum_{p>0} \binom{k+a-1}{q-1} \binom{l+b-1}{p-1} ([A[\widehat{BC}]_p]_q)_{k+l+m}\,. \qquad (2.4.9)$$

We can now use the associativity condition (2.3.21). Calling the summation index in (2.3.21) $r$ and renaming $q = s + r - p$, it follows that the Jacobi identity (2.4.8) will be satisfied if:

$$\sum_{p>0} \binom{k+a-1}{r+s-p-1} \binom{l+b-1}{p-1} \binom{r+s-p-1}{r-1} =$$
$$\binom{k+a-1}{r-1} \binom{k+l+a+b-r-1}{s-1}\,. \qquad (2.4.10)$$

After cancelling out factors, one sees that this equation follows from eq. (2.A.5). ■

Moreover, from the above proof it is clear that the reverse is also true:

**Theorem 2.4.3** *If the mode algebra satisfies the Jacobi-identities (2.4.8) up to null fields, the associativity conditions (2.3.21) are satisfied.*

Modes of normal ordered operators are given by eq. (2.3.24):

$$([\widehat{AB}]_0)_m = \sum_l : \widehat{A}_l \widehat{B}_{m-l} :, \qquad (2.4.11)$$

where

$$: \widehat{A}_l \widehat{B}_m : \;\equiv \begin{cases} \widehat{A}_l \widehat{B}_m & \text{if } \; l \le -a \\ (-1)^{|A|\,|B|} \widehat{B}_m \widehat{A}_l & \text{if } \; l > -a \end{cases} \qquad (2.4.12)$$

Consider an OPA where the generators have OPEs whose singular part contains composite operators. In this case, the mode algebra is only an infinite dimensional (super-)Lie algebra when those composites are viewed as new elements of the algebra, such that the commutators close linearly. Otherwise, the commutators close only in the enveloping algebra of the modes of the generators.

As the mode algebra contains the same information as the OPEs, one can always choose which one uses in a certain computation. For linear algebras (where the OPEs close on a finite number of noncomposite operators) modes are very convenient. However, for nonlinear algebras the infinite sums in the modes of a composite are more difficult to handle.

The definition eq. (2.4.1) provides a realisation of the mode algebra. In a canonical quantisation scheme, another representation is found in terms of the creation- and annihilation operators. For different periodicity conditions in the coordinate $\sigma$ (relating $z$ and $\exp(2i\pi)z$) of the symmetry generators, formally the same algebra arises. However, the range of the indices differs. As an example, the $N = 1$ superconformal algebra consist of a Virasoro operator $T$ and a fermionic dimension 3/2 primary operator $G$ with OPE:

$$G \times G \;= \ll \frac{2c}{3}\mathbb{1} \mid 0 \mid 2T \gg \;. \qquad (2.4.13)$$

This gives for the anticommutator of the modes:

$$[\widehat{G}_m, \widehat{G}_n] = \frac{2c}{3}\binom{m+1/2}{2}\delta_{m+n} + 2\widehat{L}_{m+n}\,. \qquad (2.4.14)$$

In the representation eq. (2.4.1) defined via the OPEs, $m$ and $n$ in this commutator are half-integer numbers. However, the algebra is also well-defined if $m$ and $n$ are integer. This corresponds to different boundary conditions on $G(z)$. The relation between the different modings of the linear superconformal algebras is studied in [178, 50]. We will use the notation $\widehat{A}_m$ in the representation eq. (2.4.1) of the mode algebra, and drop the hats otherwise.

## 2.5 Generating functionals

In this section, we will define the generating functionals for the correlation functions of a conformal field theory and show that the Ward identities give a set of functional equations for these functionals.

Consider a conformal field theory with fields $\phi_i$ and action $S[\phi_i]$. We denote the generators of the chiral symmetries of the theory with $T_k$. The partition function $Z$ is defined by:

$$Z[\mu] \;=\; \frac{1}{\mathcal{N}}\int [d\varphi_i]\, \exp\left(-S[\varphi_j] - \frac{1}{\pi}\int d^2x\, \mu^k(x) T_k(x)\right) \qquad (2.5.1)$$
$$\;=\; < \exp\left(-\frac{1}{\pi}\int d^2x\, \mu^k(x) T_k(x)\right) >_{\rm OPE}\;. \qquad (2.5.2)$$

Here the normalisation constant $\mathcal{N}$ was defined in eq. (2.2.2) and $\mu^k$ (the "sources") are non-fluctuating fields. $Z$ is the generating functional for the correlation functions of the generators $T_k$. Indeed, by functional derivation with respect to the sources we can determine every correlation function:

$$< T_1(x_1)T_2(x_2)T_3(x_3) > = (-\pi)^3 \frac{\delta}{\delta\mu^1(x_1)}\frac{\delta}{\delta\mu^2(x_2)}\frac{\delta}{\delta\mu^3(x_3)} Z[\mu]\Big|_{\mu=0}. \qquad (2.5.3)$$

If a generator $T_k$ is fermionic, $\mu^k$ is Grasmann-odd, *i.e.* anticommuting. We will always use left-functional derivatives in this work.

It is often useful to define the induced action $\Gamma$ as the generating functional of all "connected" diagrams:

$$Z[\mu] = \exp(-\Gamma[\mu])\,. \qquad (2.5.4)$$

We can view the sources $\mu^k$ as gauge fields. By assigning appropriate transformation rules for $\mu^k$, we can try to make the global symmetries local at the classical level. As an example, in a conformal field theory, the action is invariant under the conformal transformations generated by $T$, when the parameter $\varepsilon^z$ is analytic. If we want to make the theory invariant for any $\varepsilon^z$, we have to couple the generator $T$ to a gauge field. We find that:

$$\delta_\varepsilon(S + \frac{1}{\pi}\int \mu T) = \frac{1}{\pi}\int \left(-\bar\partial\varepsilon T + \delta\mu T + \mu\delta T\right)\,, \qquad (2.5.5)$$

where we used the definition of $T$ as a Noether current (eq. (2.2.7)). Applying the transformation rule for $T$ (eq. (2.2.15)) in the classical case ($c = 0$), we see that if the source $\mu$ transforms as:

$$\delta_\varepsilon\mu = \bar\partial\varepsilon + \varepsilon\partial\mu - \partial\varepsilon\mu\,, \qquad (2.5.6)$$

the action $S + \frac{1}{\pi}\int \mu T$ is classically invariant. When gauging not only the conformal symmetry, we would expect that higher order terms in the sources have to be added to eq. (2.5.1) to obtain invariance. However, Hull [115] proved that minimal coupling (addition of only linear terms) is sufficient to gauge chiral symmetries.

It is possible that the resulting local symmetry does not survive at the quantum level. For example, the Schwinger term in the transformation law of $T$ (eq. (2.2.15)) breaks gauge invariance. In general, central terms in the transformation laws of the symmetry generators give rise to "universal" anomalies, which cannot be canceled by changing the transformation law of the gauge fields[9]. In this case, the induced action $\Gamma$ is a (in general nonlocal) functional of the gauge fields $\mu^k$ and is used in $\mathcal{W}$–gravity theories (see chapter 7) and non-critical $\mathcal{W}$–strings.

The Ward identities can be used to derive functional equations for the generating functionals $Z$ and $\Gamma$. As an example, consider the induced action where only the energy–momentum tensor is coupled to a source (*i.e.* the generating functional for correlation functions with only $T$). Under the variation of the source (eq. (2.5.6)), the partition function transforms as:

$$\begin{aligned} \delta Z[\mu] &= < \exp\left(-\frac{1}{\pi}\int(\mu+\delta\mu)\,T\right) > - Z[\mu] \\ &= -\frac{1}{\pi}\int(\bar\partial\varepsilon + \varepsilon\partial\mu - (\partial\varepsilon)\mu) < T\,\exp\left(-\frac{1}{\pi}\int\mu T\right) > \end{aligned} \qquad (2.5.7)$$

To compute the first integral with $\bar\partial\varepsilon T$, we note that in the complex basis, when $\varepsilon^{\bar z} = 0$ and all $\Phi_i = T$, eq. (2.2.9) becomes:

$$\begin{aligned} &-\frac{1}{\pi}\int d^2x\,\bar\partial\varepsilon(x)\ < T(x)T(x_1)\cdots T(x_N) > \\ &= \sum_{j=1}^{N} < T(x_1)\cdots\delta_\varepsilon T(x_j)\cdots T(x_N) > \,. \end{aligned} \qquad (2.5.8)$$

We can now multiply this equation with $\mu(x_1)\cdots\mu(x_N)$ and integrate over all $x_j$. Using crossing symmetry in the *rhs* of eq. (2.5.8), we get (in an obvious notation):

$$\begin{aligned} &\int d^2x\,\bar\partial\varepsilon(x)\ < T(x)\left(-\frac{1}{\pi}\int\mu T\right)^N > \\ &= N\int d^2x\,\mu(x) < \delta_\varepsilon T(x)\left(-\frac{1}{\pi}\int\mu T\right)^{N-1} > \,. \end{aligned} \qquad (2.5.9)$$

This gives the following result:

$$\begin{aligned} &\int d^2x\,\bar\partial\varepsilon(x)\ < T(x)\,\exp\left(-\frac{1}{\pi}\int\mu T\right) > = \\ &\int d^2x\,\mu(x) < \delta_\varepsilon T(x)\exp\left(-\frac{1}{\pi}\int\mu T\right) > \,. \end{aligned} \qquad (2.5.10)$$

Using eq. (2.2.15), the variation of the partition function (2.5.7) becomes:

$$\delta Z[\mu] = \frac{c}{12\pi}\left(\int \varepsilon\partial^3\mu\right) Z[\mu]\,. \qquad (2.5.11)$$

On the other hand, we can rewrite the *rhs* of eq. (2.5.7) using:

$$< T(x)\,\exp\left(-\frac{1}{\pi}\int\mu T\right) > = -\pi\frac{\delta Z}{\delta\mu(x)}[\mu] = \pi Z[\mu]\frac{\delta\Gamma}{\delta\mu(x)}[\mu]\,. \qquad (2.5.12)$$

[9] The universal anomalies could be canceled by including a gauge field for the "generator" $\mathbb{1}$.

Combining (2.5.7) and (2.5.12) gives us a functional equation for $Z$, and thus for $\Gamma$:

$$\frac{c}{12\pi}\partial^3\mu = \bar\partial t - \mu\partial t - 2(\partial\mu)t \tag{2.5.13}$$

where:

$$t = \frac{\delta\Gamma}{\delta\mu}\,. \tag{2.5.14}$$

We see that the Ward identities fix the generating functionals. This is of course no surprise, as we already knew that the Ward identities determine the correlation functions. The underlying theory $S[\varphi_j]$ determines the central charge $c$.

By rescaling $t$ in eq. (2.5.13), we can make the functional equation $c$-independent. We see that $\Gamma = c\Gamma^{(0)}$, where $\Gamma^{(0)}$ is $c$-independent. In fact, the solution of the Ward identity (2.5.13) was given by Polyakov [159]:

$$\Gamma[\mu] = \frac{c}{24\pi}\int \mu\partial^2\left(1-\bar\partial^{-1}\mu\partial\right)^{-1}\frac{\partial}{\bar\partial}\mu\,, \tag{2.5.15}$$

where the inverse derivative is defined in eq. (A.7). For a general set of symmetry generators, it will not be possible to solve the functional equation in closed form.

We now show how OPEs can be used to derive the functional equation on the partition function $Z$ (eq. (2.5.1)). We start by computing:

$$-\pi\bar\partial\frac{\delta Z}{\delta\mu^i(z,\bar z)} = \bar\partial < T_i(z)\exp\left(-\frac{1}{\pi}\int d^2x\,\mu^k(x)T_k(x)\right) > \,. \tag{2.5.16}$$

Let us look at order $N$ in the sources:

$$\begin{aligned}
&\bar\partial < T_i(z)\left(-\int\mu^kT_k\right)^N > \\
&= -N\int d^2x_0\,\mu^j(x_0)\bar\partial < \sum_{n>0}\frac{[T_iT_j]_n(z_0)}{(z-z_0)^n}\left(-\int\mu^kT_k\right)^{N-1} > \\
&= \pi N\sum_{n>0}\frac{(-1)^n}{(n-1)!} \\
&\qquad \partial^{n-1}\left(\mu^j(z,\bar z) < [T_iT_j]_n(z)\left(-\int\mu^kT_k\right)^{N-1} >\right) \\
&= \pi N\sum_{n>0}\frac{\partial^{n-1}\mu^j(z,\bar z)}{(n-1)!} < [T_jT_i]_n(z)\left(-\int\mu^kT_k\right)^{N-1} >\,,
\end{aligned} \tag{2.5.17}$$

where we used eq. (A.3) in the second step, and the last step [42] relies on eq. (2.3.16). Hence,

$$-\pi\bar\partial\frac{\delta Z}{\delta\mu^i(z,\bar z)} = \sum_{n>0}\frac{\left(\partial^{n-1}\mu^j(z,\bar z)\right)}{(n-1)!} < [T_jT_i]_n(z)\exp\left(-\frac{1}{\pi}\int\mu^kT_k\right) >, \tag{2.5.18}$$

Note that only the singular parts of the OPEs contribute. One can easily check that eq. (2.5.18) reproduces the functional equation (2.5.13).

Consider now the case of a linear algebra, *i.e.* the singular parts of the OPEs contain only the generators $T_j$ or their derivatives. The result (2.5.18) can then be expressed in terms of functional derivatives with respect to $\mu^j$ of $Z$ as in eq. (2.5.12). Central extension terms in eq. (2.5.18) are simply proportional to $Z$. This means an overall factor of $Z$ can be divided out. If all central extension terms are proportional to $c$, we again infer that $\Gamma = c\Gamma^{(0)}$.

When the OPEs close nonlinearly, *i.e.* contain also normal ordered expressions of the generators, it is still possible to write down a functional equation for $Z$ or $\Gamma$ by using eq. (2.3.6). An explicit calculation of such a case is given in section 5.3. The resulting functional equations have not been solved up to now. We will treat this problem at several points in this work, but especially in chapter 7.

## 2.6 A few examples

### 2.6.1 The free massless scalar

The action for a massless scalar propagating in a space with metric $g_{ij}$ is given by:

$$S_s[X,g^{ij}] = -\frac{1}{4\pi\lambda}\int dx^2\,\sqrt{g}g^{ij}\partial_iX(x)\partial_jX(x)\,, \tag{2.6.1}$$

with $g$ the absolute value of the determinant of $g_{ij}$ and $\lambda$ a normalisation constant. For $D$ scalars, this gives an action for the bosonic string in $D$ dimensions with flat target space, see chapter 8.

Before discussing the Ward identity of the energy–momentum tensor for this action, we first observe that the action (2.6.1) has a symmetry:

$$\delta X(x) = \varepsilon(x)\,. \tag{2.6.2}$$

We will follow the reasoning of subsection 2.2.2 to find the Ward identity corresponding to this symmetry. The Noether current for this symmetry is $\frac{1}{2\lambda}\partial_iX(x)$ (see eq. (2.2.7)). It satisfies the Ward identity (see eq. (2.2.9)):

$$\begin{aligned}
&-\frac{1}{2\pi\lambda}\int d^2x\,\sqrt{g}g^{ij}\partial_i\varepsilon(x) < \partial_jX(x)X(x_1)X(x_2)\cdots >= \\
&\qquad < \varepsilon(x_1)X(x_2)\cdots > + < X(x_1)\varepsilon(x_2)\cdots > + \cdots
\end{aligned} \tag{2.6.3}$$

This symmetry has a chiral component $\partial X$ for which the equation corresponding to (2.2.16) is:

$$< \partial X(z,\bar{z}) X(z_1,\bar{z}_1) X(z_2,\bar{z}_2)\cdots >=$$
$$\frac{\lambda}{z-z_1} < X(z_2,\bar{z}_2)\cdots > + \frac{\lambda}{z-z_2} < X(z_1,\bar{z}_1)\cdots > + \cdots \tag{2.6.4}$$

corresponding to the OPE[10]:

$$\partial X(z) X(z_0,\bar{z}_0) = \frac{\lambda}{z-z_0} + O(z-z_0)^0 \,. \tag{2.6.5}$$

This agrees with the propagator for $X$ which is given by the Green's function for the Laplacian (see appendix A):

$$< X(z,\bar{z}) X(z_0,\bar{z}_0) >= \lambda \log(z-z_0)(\bar{z}-\bar{z}_0)\,. \tag{2.6.6}$$

Note that $X$ itself is a field whose two-point function is not a holomorphic function of $z$ or $\bar{z}$. This in fact makes it impossible to use the OPE formalism with the field $X$. This need not surprise us, as $X$ not a symmetry generator. We will only work with the chiral symmetry generator $\partial X$. For easy reference we give its OPE, which follows from eq. (2.6.5) by taking an additional derivative:

$$\partial X(z) \partial X(z_0) \;=\; \frac{\lambda}{(z-z_0)^2} + O(z-z_0)^0 \tag{2.6.7}$$

The action (2.6.1) is invariant under the transformation $\delta_\varepsilon X = \varepsilon^i \partial_i X$. Hence, we have that the energy–momentum tensor eq. (2.2.8) is classically conserved. It is given by:

$$T_{ij} = \frac{1}{2\lambda}\partial_i X(x)\partial_j X(x) \;-\frac{1}{4\lambda} g_{ij} g^{kl} \partial_k X(x) \partial_l X(x)\,. \tag{2.6.8}$$

In the complex basis, $T_{ij}$ has only two non-vanishing components:

$$T_{zz} = \frac{1}{2\lambda}\partial X \partial X\,, \qquad\qquad T_{\bar{z}\bar{z}} = \frac{1}{2\lambda}\bar{\partial} X \bar{\partial} X\,. \tag{2.6.9}$$

However, these expression are not well-defined in the quantum case, due to short-distance singularities. We use point-splitting regularisation to define the quantum operator. In the OPE formalism, this becomes $T = \frac{1}{2\lambda}[\partial X \partial X]_0$. We can now use the OPE (2.6.7), and the rules of subsection 2.3.2 to compute the OPE of $T$ with itself. One ends up with the correct Virasoro OPE (2.3.2), with a central charge $c = 1$. Furthermore, it is easy to check that $X$ is a primary field with respect to $T$ of dimension zero. Moreover, $\partial X$ is also primary, having dimension one. In the rest of this subsection we set $\lambda = 1$.

[10] As before, we use the misleading notation $\partial X(z)$.

Let us now check for the mode algebra eq. (2.4.4) corresponding to the OPE of $\partial X$ with itself eq. (2.6.7). The modes of $\partial X$ are traditionally denoted with $\alpha_m$. We find:

$$[\alpha_m, \alpha_n] \;=\; m\delta_{m+n}\,, \tag{2.6.10}$$

which is related to the standard harmonic oscillator commutation relations via a rescaling with $\sqrt{|m|}$.

In string theory, computing scattering amplitudes is done by inserting local operators of the correct momentum in the path integral and integrating over the coordinates [127, 105]. These local operators are (composites with) normal ordered exponentials of the scalar field. Because $X$ cannot be treated in the present OPE scheme, we should resort to different techniques to define these exponentials. This is most conveniently done using a mode expansion of $X$. One shows that one can define a chiral operator which we write symbolically as:

$$V_a(z) \;=\, :\exp(aX(z)):\,, \qquad\qquad a\in\mathbb{C}\,, \tag{2.6.11}$$

for which the following identities hold:

$$\partial V_a \;=\; a :\partial X\, V_a: \tag{2.6.12}$$
$$\partial X(z)\, V_a(w) \;=\; a\frac{V_a(w)}{(z-w)} + O(z-w)^0 \tag{2.6.13}$$

and

$$\begin{aligned} & V_a(z)\, V_b(w) \\ &\quad = \;(z-w)^{ab} : \exp(aX(z))\;\exp(bX(w)): \\ &\quad = \;(z-w)^{ab} : \Big(1 + (z-w) a\partial X(w) + . \\ &\qquad\qquad \frac{(z-w)^2}{2}(a\partial^2 X(w) + a^2 \partial X(w)\partial X(w))\ldots\Big) V_{a+b}(w):\,, \end{aligned} \tag{2.6.14}$$

where the last line is a well-defined expression where normal ordering, as defined in eq. (2.3.6), from right to left is understood. We will use these equations as the definition for the vertex operators.

Using the eqs. (2.6.12) and (2.6.13), it is easy to check that $V_a$ is primary with conformal dimension $a^2/2$ with respect to $T$ (2.6.9). This differs from the classical dimension which is zero as $X$ has dimension zero.

The OPE (2.6.14) is distinctly different from the ones considered before. Indeed, noninteger poles are possible. The rules constructed in section 2.3.2 are not valid for such a case. However, every OPE with vertex operators we will need, will follow immediately from the above definitions. Another peculiarity is that when $ab \in \mathbb{Z}$, the definition (2.6.14) fixes the regular part of the OPE of two vertex operators in terms of normal ordered products of $\partial X$ and a vertex operator. An interesting example of this is for $V_{\pm 1}$:

$$\begin{aligned} V_{\pm 1}(z)\, V_{\pm 1}(w) &= (z-w)V_2(w) + \ldots \\ V_{\pm 1}(z)\, V_{\mp 1}(w) &= \frac{1}{(z-w)} \pm \partial X(w) + \ldots \end{aligned} \tag{2.6.15}$$

We see that we have two operators:

$$\frac{V_1 + V_{-1}}{\sqrt{2}} = \sqrt{2}\cosh X \quad \text{and} \quad \frac{V_1 - V_{-1}}{\sqrt{2}i} = -i\sqrt{2}\sinh X \tag{2.6.16}$$

which satisfy the OPEs of two free fermions (see eq. (2.6.26) below). Vertex operators can also be used to construct realisations of affine Lie algebras, as we will see in section 4.6.

When considering realisations of $\mathcal{W}$–algebras using scalars, it is obviously a problem that the energy–momentum tensor of the free scalars is restricted to an integer central charge $n$ by considering $n$ free bosons. However, we can modify the eq. (2.6.9) to:

$$T = \frac{1}{2} : \partial X \partial X : - q\partial^2 X\,, \tag{2.6.17}$$

which has a central charge $1 - 12q^2$. This corresponds to adding a background charge to the free field action [73, 91, 59].

### 2.6.2 The free Majorana fermion

The free Majorana fermion in two dimensions has the following action in the conformal gauge:

$$S_f[\psi] = \frac{1}{2\pi\lambda} \int d^2x\, \psi(x)\bar{\partial}\psi(x)\,, \tag{2.6.18}$$

where $\lambda$ is a normalisation constant. The equation of motion for $\psi$ is:

$$\bar{\partial}\psi = 0\,. \tag{2.6.19}$$

Hence, $\psi$ is a chiral field and we will write $\psi(z)$. Note that under conformal transformations it has to transform as a primary field of dimension $h = 1/2$, $\bar{h} = 0$ to have a conformal invariant action. $-\frac{1}{\lambda}\psi(x)$ is the Noether current corresponding to the transformation:

$$\delta_\varepsilon \psi(x) = \varepsilon(x)\,, \tag{2.6.20}$$

where $\varepsilon$ is a Grasmann–odd field. Although we can proceed as before, we will derive the OPEs of $\psi$ using a different technique. Consider the generating functional (see eq. (2.5.1)):

$$Z[\mu] = \frac{1}{\mathcal{N}} \int [d\psi]\, \exp\left( -S_f[\psi] - \frac{1}{\pi} \int d^2x\, \mu(x)\psi(x) \right)\,, \tag{2.6.21}$$

where $\mu$ is Grasmann–odd. Such a pathintegral can be computed by converting it to a Gaussian path integral, as we will now show. We start by rewriting the exponent in the following way:

$$\begin{aligned} Z[\mu] &= \frac{1}{\mathcal{N}} \int [d\psi]\, \exp\left( -\frac{1}{2\pi\lambda} \int (\psi + \lambda(\bar{\partial}^{-1}\mu))\bar{\partial}(\psi + \lambda\bar{\partial}^{-1}\mu) \right) \\ &\quad \exp\left( \frac{\lambda}{2\pi} \int (\bar{\partial}^{-1}\mu)\mu \right)\,, \end{aligned} \tag{2.6.22}$$

where we used the definition of the inverse derivative of appendix A, supposing that $\mu$ decays sufficiently fast at infinity. One can now shift the integration variables to $\tilde{\psi} = \psi + \lambda\bar{\partial}^{-1}\mu$. This shift has a Jacobian 1. One arrives at:

$$Z[\mu] = \frac{1}{\mathcal{N}} \int [d\tilde{\psi}]\, \exp\left( \frac{-1}{2\lambda\pi} \int \tilde{\psi}\bar{\partial}\tilde{\psi} \right) \exp\left( \frac{\lambda}{2\pi} \int (\bar{\partial}^{-1}\mu)\mu \right)\,. \tag{2.6.23}$$

The remaining path integral cancels $\mathcal{N}$ (see eq. (2.2.2)) and we have:

$$Z[\mu] = \exp\left( -\frac{\lambda}{2\pi^2} \int dx^2 dx_0^2\, \mu(z,\bar{z}) \frac{1}{z - z_0} \mu(z_0, \bar{z}_0) \right)\,. \tag{2.6.24}$$

From this result we immediately see that the one-point function $< \psi(x) >$ is zero, while the two-point function is given by:

$$\begin{aligned} < \psi(z)\psi(z_0) > &= \pi^2 \frac{\delta}{\delta\mu(z,\bar{z})} \frac{\delta}{\delta\mu(z_0,\bar{z}_0)} Z[\mu]\Big|_{\mu = 0} \\ &= \frac{\lambda}{z - z_0}\,, \end{aligned} \tag{2.6.25}$$

where we used left-functional derivatives. The corresponding OPE is[11]:

$$\psi(z)\,\psi(w) = \frac{\lambda}{z - w} + O(z-w)^0\,. \tag{2.6.26}$$

[11] It is easy to prove that this OPE gives rise to a functional equation for $Z[\mu]$ as in subsection 2.5, which is satisfied by eq. (2.6.24).

To find the energy–momentum tensor for the action (2.6.18), we use the definition (2.2.7) with $\varepsilon^i$ decaying at infinity, together with the transformation law (2.1.6) for $\psi$. We find:

$$\begin{aligned}\delta S_f &= \frac{1}{\lambda\pi}\int d^2x\,\left(\varepsilon^z\partial\psi + 1/2\partial\varepsilon^z\psi + \varepsilon^{\bar z}\bar\partial\psi\right)\bar\partial\psi \\ &= \frac{1}{2\lambda\pi}\int d^2x\,\bar\partial\varepsilon^z\psi\partial\psi\,. \end{aligned} \tag{2.6.27}$$

Hence, we find only one component $T^{\bar z\bar z}$ non-zero, according to (2.2.12) we have:

$$T \;=\; \frac{1}{2\lambda} :\partial\psi\,\psi:, \tag{2.6.28}$$

which has central charge $c = 1/2$.

### 2.6.3 Other first order systems

We will need two other first order systems, the fermionic $b, c$ and the bosonic $\beta, \gamma$ system [91]. They will arise as the ghosts in the BRST-quantisation of systems with conformal invariance. We will combine both using a supermatrix notation (see appendix B). The action is:

$$S_{\rm bc}[b,c] = \frac{1}{\pi}\int d^2x\, b^i(x)\,{}_iA^j\,\bar\partial({}_jc)(x)\,, \tag{2.6.29}$$

where ${}_iA^j$ is a constant bosonic supermatrix which is invertible. We assume that it does not mix fermions and bosons. We also take $b$ and $c$ to be bosonic matrices[12]. The equations of motion again show that $b^i, c_i$ are chiral fields. Conformal invariance requires the $b^i$ to be primary fields with dimension $h^i$ and $c_i$ also to be primary such that $\dim({}_iA^j\;{}_jc) = 1 - h^i$ (and zero for $\bar h$). We proceed now as in the previous subsection:

$$\begin{aligned} Z[\mu,\nu] &= \frac{1}{\mathcal{N}}\int[db^i][dc_j]\,\exp-\left(S_{\rm bc}[b,c] + \frac{1}{\pi}\int \mu_k\,{}^kb + \nu^k\,{}_kc\right) \\ &= \frac{1}{\mathcal{N}}\int[db^i][dc_j]\,\exp\left(\frac{1}{\pi}\int(b+\bar\partial^{-1}\nu A^{-1})A\bar\partial(c-(A\bar\partial)^{-1}\mu)\right) \\ &\qquad \exp\left(\frac{1}{\pi}\int(\bar\partial^{-1}\nu)A^{-1}\mu\right) \\ &= \exp\left(-\frac{1}{\pi}\int d^2xd^2x_0\;\nu^i(z,\bar z)\frac{{}_i(A^{-1})^j}{z-z_0}\;{}_j\mu(z_0,\bar z_0)\right)\,. \end{aligned} \tag{2.6.30}$$

[12]One can of course take $b$ and $c$ to be fermionic matrices. The two-point function (2.6.31) does not depend on this convention.

Carefully keeping track of the signs, we get for the two-point functions:

$$\begin{aligned} <{}_ic(z)\;b^j(z_0)> &= \pi^2\frac{\delta}{\delta\nu^i(z,\bar z)}\frac{\delta}{\delta\mu_j(z_0,\bar z_0)}Z[\mu,\nu]\Big|_{\mu,\nu\,=\,0} \\ &= \frac{-{}_i(A^{-1})^j}{z-z_0}\,. \end{aligned} \tag{2.6.31}$$

This OPE does not depend on the dimensions of the fields, while the energy–momentum tensor of course does. To find the energy–momentum tensor, we proceed as in the previous subsection. From:

$$\delta_\varepsilon\left({}_i(Ac)\right) = \varepsilon^z\partial({}_i(Ac)) + (1-h^i)\partial\varepsilon^z\;{}_i(Ac)\,, \tag{2.6.32}$$

we get:

$$T_{\rm bc} =: bA\partial c - (1-h^i)\partial(b^i\;{}_iA^j\;{}_jc):, \tag{2.6.33}$$

with central charge:

$$c = \sum_i(-1)^i 2(6(h^i)^2 - 6h^i + 1))\,, \tag{2.6.34}$$

where the phase factor is $-1$ for $b^i$ fermionic. Note that the second term of eq. (2.6.33) is the derivative of the ghost current. It is not present in eq. (2.6.28) because $:\psi\psi := 0$.

### 2.6.4 Wess-Zumino-Novikov-Witten models

A final example of a conformal field theory is the WZNW–model [156, 209]. It is a nonlinear sigma model with as target space a group manifold. The (super) Lie group $\mathcal{G}$ is required to be semisimple (see [154, 81] for a weakening of this condition). We denote its (super) Lie algebra with $\bar g$, see appendix B for conventions.

The WZWN action $\kappa S^+[g]$ is a functional of a $\mathcal{G}$ -valued field $g$ and is given by:

$$\begin{aligned} \kappa S^+[g] &= \frac{\kappa}{4\pi x}\int_{\partial\Omega} d^2x\; str\left\{\partial g^{-1}\bar\partial g\right\} \\ &\quad + \frac{\kappa}{12\pi x}\int_\Omega d^3x\;\varepsilon^{\alpha\beta\gamma}\,str\left\{g_{,\alpha}g^{-1}g_{,\beta}g^{-1}g_{,\gamma}g^{-1}\right\}\,, \end{aligned} \tag{2.6.35}$$

where $\Omega$ is a three-manifold with boundary $\partial\Omega$. It satisfies the Polyakov-Wiegman identity [161]:

$$S^+[hg] = S^+[h] + S^+[g] - \frac{1}{2\pi x}\int str\left(h^{-1}\partial h\bar\partial g g^{-1}\right)\,, \tag{2.6.36}$$

which is obtained through direct computation. We also introduce a functional $S^-[g]$ which is defined by:

$$S^-[g] = S^+[g^{-1}]. \tag{2.6.37}$$

It is with this functional that we now continue. Using the Polyakov-Wiegman identity (2.6.36), we can show that the action $S^-[g]$ transforms under:

$$\delta_\eta g = \eta g \tag{2.6.38}$$

as

$$\delta_\eta S^-[g] = -\frac{\kappa}{2\pi x}\int str\left(\left(\partial g\, g^{-1}\right)\bar\partial\eta\right) . \tag{2.6.39}$$

We see that $S^-[g]$ is invariant when $\bar\partial\eta = 0$. The corresponding conserved current is:

$$J_z = \frac{\kappa}{2}\partial g\, g^{-1} , \tag{2.6.40}$$

which is chiral, $\bar\partial J_z = 0$. Similarly, for $\delta_{\bar\eta} g = g\bar\eta$ with $\partial\bar\eta = 0$, we find a conserved current:

$$J_{\bar z} = -\frac{\kappa}{2} g^{-1}\bar\partial g . \tag{2.6.41}$$

$J_z$ transforms under eq. (2.6.38) as:

$$\delta_\eta J_z = \frac{\kappa}{2}\partial\eta + [\eta, J_z] , \tag{2.6.42}$$

$$= -\int dy \Big\{ str\left(\eta(y) J_z(y)\right) , J_z(x) \Big\}_{\rm PB} \tag{2.6.43}$$

where the Poisson bracket is given by[13]:

$$J_z^a \times J_z^b = \ll -\frac{\kappa}{2} g^{ab} \mid {}^a g^d J_z^c{}_{cd} f^b \gg , \tag{2.6.44}$$

which defines a current algebra of level $\kappa$. It can be argued that the relation eq. (2.6.44) does not renormalise when going to the quantum theory, and we will take (2.6.44) as the definition of the OPE. However, the relation (2.6.40) can be renormalised to:

$$J_z = \frac{\alpha_\kappa}{2}\partial g\, g^{-1} . \tag{2.6.45}$$

Using OPE techniques, it is argued in [135] that for a current algebra of level $\kappa$, normalising the currents as in (2.6.44) gives:

$$\alpha_\kappa = \kappa + \tilde h , \tag{2.6.46}$$

where the dual Coxeter number $\tilde h$ is the eigenvalue of the quadratic Casimir in the adjoint representation (see also appendix B). This follows from consistency requirements in the OPA of the currents with $g$. We will need this renormalisation in chapter 7.

[13] We use the "classical OPE" notation by virtue of the correspondence discussed in subsection 2.3.5.

The OPA generated by the currents $J_z^a$ with the OPEs (2.6.44) is known as a Kač–Moody algebra or affine Lie algebra. Kač–Moody algebras were studied in the mathematical literature [126, 151] much earlier than WZNW–models. For a review on the algebraic aspects of Kač–Moody algebras, see [101]. We will denote the Kač–Moody algebra corresponding to $\bar g$ as $\hat g$.

One can check that the Sugawara tensor:

$$T_S = \frac{1}{x\left(\kappa + \tilde h\right)} str[J_z J_z]_0 , \tag{2.6.47}$$

satisfies the Virasoro algebra with the central extension given by:

$$c = \frac{k(d_B - d_F)}{k + \tilde h} . \tag{2.6.48}$$

The currents $J_z^a$ have conformal dimension 1 with respect to $T_S$.

We now briefly review some basic formulas for gauged WZNW–models [156, 209, 161, 3, 58] which we will need in chapter 6. We can gauge the symmetries generated by $J_z$ by introducing gauge fields $A_{\bar z}$. The action:

$$\kappa S^-[g] + \frac{1}{\pi x}\int d^2x\; str\Big(J_z(x) A_{\bar z}(x)\Big) \tag{2.6.49}$$

is (classically) invariant when the gauge fields transform as:

$$\delta_\eta A_{\bar z} = \bar\partial\eta + [\eta, A_{\bar z}] . \tag{2.6.50}$$

The induced action (2.5.4) $\Gamma[A_{\bar z}]$ for the gauge fields $A_{\bar z}$ is:

$$\exp\left(-\Gamma[A_{\bar z}]\right) = \langle \exp -\frac{1}{\pi x}\int d^2x\; str\Big(J_z(x) A_{\bar z}(x)\Big)\rangle . \tag{2.6.51}$$

By using the OPEs eq. (2.6.44), and the general formula for the functional equation for $\Gamma$ (2.5.18) we find the following Ward identity:

$$\bar\partial u_z - [A_{\bar z}, u_z] = \partial A_{\bar z} , \tag{2.6.52}$$

where:

$$u_z^a(x) = -\frac{2\pi}{\kappa} g^{ab}\frac{\delta\Gamma[A_{\bar z}]}{\delta\, {}^b A_{\bar z}(x)} . \tag{2.6.53}$$

The Ward identity is independent of $\kappa$, therefore:

$$\Gamma[A_{\bar z}] = \kappa\Gamma^{(0)}[A_{\bar z}], \tag{2.6.54}$$

where $\Gamma^{(0)}[A_{\bar z}]$ is independent of $\kappa$. In [161, 3], it was observed that eq. (2.6.52) states that the curvature for the Yang-Mills fields $\{A, u\}$ vanishes.

## 2.A Appendix : Combinatorics

This appendix combines some formulas for binomial coefficients and related definitions. We define first the Pochhammer symbol:

$$(a)_n = \frac{\Gamma(a+n)}{\Gamma(a)} = \prod_{j=0}^{n-1}(a+j)\,, \qquad a \in \mathbb{R},\, n \in \mathbb{N} \tag{2.A.1}$$

where the last definition is also valid if $a$ is a negative integer. We then define the binomial as:

$$\binom{a}{n} = \frac{(a)_n}{n!} \qquad a \in \mathbb{R},\; n \in \mathbb{N}\,. \tag{2.A.2}$$

In particular,

$$\binom{-n}{m} = (-1)^m \binom{m+n-1}{m}\,, \qquad m, n \in \mathbb{N}\,. \tag{2.A.3}$$

We now list some identities for sums of binomial coefficients which are used in section 2.4.

$$\sum_{l\in\mathbb{N}} (-1)^l \binom{r}{l} \binom{r+s-l}{p-l} = \binom{s}{p}\,, \quad r, s \in \mathbb{Z},\, p \in \mathbb{N} \tag{2.A.4}$$

which can be proven by multiplying with $x^p$ and summing over $p$.

$$\sum_{p\in\mathbb{N}} \binom{n}{p} \binom{m}{s-p} = \binom{m+n}{s}\,, \quad m, n \in \mathbb{Z},\, s \in \mathbb{N} \tag{2.A.5}$$

which is proven by multiplying with $x^s$ and summing over $s$.

# Chapter 3

# Operator Product Expansions in Mathematica

The previous chapter indicates that the method of Operator Product Expansions (OPEs) gives us a very useful algebraic tool to study a conformal field theory. However, the relevant formulas show that calculations can become quite tedious and error-prone, especially when nested normal ordered products are involved. This was the motivation to implement an algorithm to handle OPEs and normal ordered products. By now, the *OPEdefs* package is widely used, showing that the need for automated OPE calculations was (and is) present among conformal field theorists.

This chapter discusses the *OPEdefs* package. In the first section, the desired capabilities of the package are discussed and a brief history of the development of *OPEdefs* is given. Section 3.2 is devoted to design considerations. The implementation we used is explained in section 3.3. We first give the algorithm, and prove that it is finite. Then some crucial points in the actual code of *OPEdefs* are highlighted. Section 3.4 is a user's guide to *OPEdefs*, and the next section presents an explicit example of a calculation. Section 3.6 gives an outlook on possible extensions. We end this chapter with a review of other symbolic manipulation packages used in conformal field theory.

Some of the material found in this chapter was published in [192] and [193].

## 3.1 Intention and history

We want to construct a package which, given the OPEs of a set of generators, is able to compute any OPE in the Operator Product Algebra (OPA)[1]. Moreover, as the definition of normal ordering is noncommutative and nonassociative, the package should bring composite operators in a standard form. This can be done by introducing an order for the generators and reordering composites such that generators are ordered from left to right, using the associativity conditions of the previous chapter.

[1]When we say, "compute an OPE", we mean computing all operators of the singular part.

We will restrict ourselves to meromorphic conformal field theory. As we use only the concept of an OPA (see def. 2.3.2), we will be able to use operators with any (also negative) conformal dimension.

An important step in the calculations with OPEs is to check that the Jacobi identitities (2.3.21) discussed in the previous chapter are satisfied. The package should be able to verify this. This will also enable us to fix some free parameters in the OPEs of the generators by requiring associativity.

A first version of my package was used to perform the calculations on the Casimir algebra of $B_2$ [78] (see section 4.6), where OPEs of composites of up to four fields are involved. After some optimisations, *OPEdefs* 2.0 [192] was made available to the public. Several well-known results were checked with this version, *e.g.* the Sugawara construction for $\widehat{SU(n)}$, the free field realisations and homological constructions of $W_3$ based on a $\widehat{SU}(3)$-Kač–Moody algebra [56]. The power of having all calculations automated was already clear at this point, as we found a small misprint in [56][2], while this calculation was highly nontrivial by hand.

Important improvements (version 3.0) were published in [193]. It was necessary to introduce a different syntax, to be able to make extensive changes in the future. Since then, many other small changes have been made to the package. The latest version, which will be discussed in this chapter, is 3.1.

The package is freely available from the author[3].

## 3.2 Design considerations

In this section, we comment on how *OPEdefs* was designed. Several aspects are important.

[2]The second term in the expression for $J_3^+$ (eq. 2.15) of [56] should read $-\beta_1\gamma_1\gamma_3$.

[3]It should be in the *MathSource* archive soon. See `http://www.wri.com/mathsource/index.html`.

- How will the program present itself to the user ? (interface)
  Preferably, concepts and terminology that the user has to handle should be familiar to any conformal field theorist.

- How will the data be represented ?
  The representation has to be chosen such that storage space and computation time are minimised. Furthermore, it should be possible to incorporate any related problems (for instance Poisson brackets) without changing the data representation.
  Whatever representation is chosen, it is important to hide this choice for the user. Ideally, one should be able to change the representation (and thus the internal working of the program) without affecting the interface.

- Make sure the program can be easily extended.
  It is important to keep the main ideas of the algorithm reflected in the structure of the program, separated from the actual details of the computation. This implies a modular style of writing. Of course, a high level programming language enhances readability and makes changes and additions easier.

- How can the results be manipulated?
  It is of course not enough to be able to compute a certain OPE or normal ordered product. Different results have to be added, coefficients simplified or extracted, equations solved... It would lead the programmer of the package way to far if he/she had to implement all this functionality.

These remarks point towards an implementation in an interactive environment where it is possible to perform symbolic computations, and where a high level language is available, *i.e.* in a symbolic manipulation program. Indeed, these environments are developed to perform the dull – or algorithmic – calculations, once the necessary rules are implemented. The user can then use the programming power of the environment to solve his problem. It would be perfectly possible to write the proposed program in an "ordinary" programming language like $C^{++}$ or *lisp*. After all, the symbolic manipulation programs are written in one of these languages. However, the amount of code required would probably go beyond the scope of a physics PhD.

Many of the design principles we mentioned, are reminiscent of object oriented programming. As this is a fairly new concept, one immediately is directed towards the more recent environments. Of these, the best known and more powerful are *Axiom* [123], *Maple* [36] and *Mathematica* [210]. All three are suitable for implementation of the algorithm for working with OPEs.
From the viewpoint of the programmer, *Axiom* seems to be the most attractive choice. It is indeed the most recent environment, and its design reflects all points made above. The major disadvantages of this system are that it is not very popular yet, and the hardware requirements are very high (a powerful Unix workstation is needed).
*Maple* is being used more and more over the last few years. Its design is concentrated on run-time efficiency and small memory requirements. Presumably, a consequence of this is that the *Maple* programming language is more like *C* or *Pascal*, although with powerful extensions for symbolic manipulations. However, it does not seem to permit a high abstraction level and modularity compared to the other two environments.
Finally, the *Mathematica* environment is somewhere in between the two others. It is probably not as efficient as *Maple*, but it is a lot easier to program in. It runs on a lot of machines, from PC's to supercomputers and all versions are completely compatible. Many of the ideas of object oriented programming can be used in *Mathematica* and its pattern matching capabilities are very useful for any mathematical programming.

## 3.3 Implementation

In this section, we first discuss the algorithm which we used. In principle, it simply consists of repeated application of the rules given in subsection 2.3.2. In the next subsection, we give some parts of the actual code of *OPEdefs*. Finally, we comment on the performance of the package.

### 3.3.1 Algorithm

We will discuss the algorithm in two steps. First, we show how the singular part of the OPEs can be computed. Then we give the algorithm for simplifying composites. In *OPEdefs*, composites are simplified at each step of the calculation.

*OPEdefs* 3.1 can calculate OPEs and Poisson brackets. We concentrate on OPEs here and briefly comment which changes are needed for the other case.

#### Computing OPEs

The package should compute OPEs of arbitrarily complicated composites when a set of generators and their OPEs is given. Obviously, we will need the rules constructed in subsection 2.3.2. For easy reference, we list the rules we need, aside from linearity. First, there are the rules involving derivatives:

$$[\partial A\, B]_q \;=\; -(q-1)[AB]_{q-1} \tag{3.3.1}$$

$$[A\,\partial B]_q \;=\; (q-1)[AB]_{q-1} + \partial[AB]_q\,. \tag{3.3.2}$$

Next, we need the OPE of $B$ with $A$, given the OPE of $A$ with $B$:

$$[BA]_q = (-1)^{|A||B|}\sum_{l\geq q}\frac{(-1)^l}{(l-q)!}\partial^{(l-q)}[AB]_l\,. \tag{3.3.3}$$

OPEs with composites can be calculated using eq. (2.3.21):

$$[A\ [BC]_0]_q \;=\; (-1)^{|A||B|}[B\ [AC]_q]_0 + [[AB]_q\ C]_0 + \sum_{l=1}^{q-1} \binom{q-1}{l} [[AB]_{q-l}C]_l \tag{3.3.4}$$

where $q \geq 1$.

These rules are the only ones needed to compute every OPE in the OPA. Indeed, when computing the OPE of $A$ with $B$, we apply the following procedure:

- if $A$ and $B$ are generators whose OPE we know, return it as the result.
- apply linearity if necessary.
- if $A$ is an operator with derivatives, use eq. (3.3.1).
- if $B$ is an operator with derivatives, use eq. (3.3.2).
- if $B$ is a composite, use eq. (3.3.4).
- if $A$ is a composite, use eq. (3.3.3).
- if the OPE $B(z)\ A(w)$ is known, compute the OPE $A(z)\ B(w)$ using eq. (3.3.3).

This list should be used recursively until none of the rules applies, which means that the OPE has been calculated. The order in which we check the rules is in this case not important, but we will check them in a "top-down" order. Note that to compute an OPE of a composite with a generator, first eq. (3.3.3) is used, and eq. (3.3.4) in the next step. The question now is if this is a finite procedure.

First of all, all sums in the previous equations have a finite number of terms. This follows from the assumption that all OPEs between generators in the OPA have a finite number of poles. The only possible problem is that eq. (3.3.4) would require an infinite number of steps. Indeed, when computing the OPE $\underbracket{A\ [B}C]_0$, one needs all OPEs $[A\underbracket{B]_n\ C}$ for $n > 0$. It is now possible that $[AB]_n$ contains a composite again, leading to a new application of eq. (3.3.4). For graded OPAs (see def. 2.3.7), we can prove the following theorem.

**Theorem 3.3.1** *If the OPA is graded and there is a lower bound $D$ on the dimension of all the operators in the OPA, then the application of eqs. (3.3.1-3.3.4) requires only a finite number of steps.*

**Proof :**
The sum of the dimensions of the fields in the OPE $[A_1\ [B_1C_1]_0]$ is $\Delta = a_1+b_1+c_1$. The application of eq. (3.3.4) leads to the computation of the OPEs $[[AB]_n\ C]$ for $n > 0$. The sum of the dimensions is in this case $a_1 + b_1 + c_1 - n$. As this is strictly smaller than $\Delta$, we see that after (the integer part of) $\Delta - 2D$ steps, we would need the OPE of a field with dimension lower than $D$, which is not present in the OPA by assumption. ∎

Note that the assumption of the existence of a lower bound on the dimension is not very restrictive. First of all, all conformal OPAs (see def. 2.3.6) with generators of positive dimension trivially satisfy the criterion. Moreover, a finite number of fermionic generators can have negative dimension. Indeed, to form composites of negative dimension with these fermions, one needs extra derivatives – *e.g.* $[\partial A\ A]_0$ – imposed by eq. (3.3.8) below.

An important remark here is that the dimension used in the proof is not necessarily the conformal dimension of the operators. For instance, for a bosonic $bc$-system which commutes with all the other generators of the OPA, we can assign a dimension of 1/2 to both $b$ and $c$, such that the theorem can be used.

The algorithm used to compute Poisson brackets is essentially the same. Some small changes in the rule eq. (3.3.4), as discussed in subsection 2.3.5, are needed.

### Simplifying composites

We now discuss how we can reduce normal ordered products to a standard form. We define an order on the generators and their derivatives, *e.g.* lexicographic ordering. Given a composite, we apply the rules of subsection 2.3.2 until all composites are normal ordered from right to left and the operators are ordered, *i.e.* $[A[B[C\ldots]_0]_0]_0$, and $A \leq B \leq C$. The relevant formulas are:

$$\partial[AB]_0 = [A\ \partial B]_0 + [\partial A\ B]_0 \tag{3.3.5}$$

$$[AB]_{-q} = \frac{1}{q!}[(\partial^q A)\ B]_0\,, \qquad q \geq 1 \tag{3.3.6}$$

$$[BA]_0 = (-1)^{|A||B|}[AB]_0 + (-1)^{|A||B|}\sum_{l\geq 1}\frac{(-1)^l}{l!}\partial^l[AB]_l \tag{3.3.7}$$

$$[AA]_0 = -\sum_{l>0}\frac{(-1)^l}{2l!}\partial^l[AA]_l\ , \quad \text{for } A \text{ fermionic .} \tag{3.3.8}$$

$$[A\ [BC]_0]_0 = (-1)^{|A||B|}[B\ [AC]_0]_0 + \tag{3.3.9}$$

$$[([AB]_0 - (-1)^{|A||B|}[BA]_0)\ C]_0 \tag{3.3.10}$$

$$[A\ [AC]_0]_0 = [[AA]_0\ C]_0\ , \quad \text{for } A \text{ fermionic .} \tag{3.3.11}$$

The rule (3.3.10) follows from eq. (2.3.21).

We have to wonder if we again require only a finite number of steps to reach the end-result. We can not prove this along the same lines as theorem 3.3.1. Indeed, in eq. (3.3.10) all terms necessarily have the same dimension. We give an example of infinite recursion in the intermezzo.

**Intermezzo 3.3.1**
Consider an OPA with bosonic operators $A, B$ such that:

$$[AB]_1 = aB\,. \tag{3.3.12}$$

We do not specify the OPE of $B$ with itself. We want to reorder $[[AB]_0B]_0$ into standard order, *i.e.* $[A[BB]_0]_0$ + corrections. In the computation which follows, an ellipsis denotes terms containing more than one derivative or containing $[BB]_n$ with $n > 0$. (The equation numbers at the right correspond to the formula used in a certain step.)

$$\begin{aligned}
[[AB]_0B]_0 &= [B[AB]_0]_0 - \partial[B[AB]_0]_1 + \dots && \textit{eq.}\ (3.3.7)\\
&= [A[BB]_0]_0 + [\partial[BA]_1\ B]_0 - && \textit{eq.}\ (3.3.10)\\
&\quad \partial\left([A[BB]_1]_0 + [[BA]_1B]_0\right) + \dots && \textit{eq.}\ (3.3.4)\\
&= [A[BB]_0]_0 - a[\partial B\ B]_0 + a\partial[BB]_0 + \dots && \textit{eqs.}\ (3.3.5), (3.3.7)\\
&= [A[BB]_0]_0 + a[\partial B\, B]_0 + \dots
\end{aligned}$$

Now, suppose that there is a null operator:

$$\partial B - \frac{1}{a}[AB]_0 = 0\,. \tag{3.3.13}$$

in the OPA and we use this relation to eliminate $\partial B$. We find:

$$[[AB]_0B]_0 = [A[BB]_0]_0 + [[AB]_0B]_0 + \dots\,,$$

which means that we cannot determine $[[AB]_0B]_0$ in this way. Note that there is no problem if we use eq. (3.3.13) to eliminate $[AB]_0$.

This example occurs for a scalar field $X$ and the vertex operators defined in subsection 2.6.1. Take $A = \partial X$ and $B =: \exp(aX) :$. If we normalised $X$ such that $[\partial X\,\partial X]_2 = 1$, eq. (3.3.12) is satisfied. On the other hand, $\partial(: \exp(aX) :) = a[\partial X\quad : \exp(aX) :]_0$. Clearly, this corresponds to eq. (3.3.13) if $a = 1$.

---

The example in the intermezzo shows that infinite recursion can occur when null operators are present. If $N$ is a null operator, we call $N = 0$ a null-relation. As shown in the intermezzo, the algorithm does not work if we use a null-relation to eliminate operators with "less" derivatives in favour of those with "more" derivatives.

We make this more precise. For an operator $A$, we define $d(A)$ as the number of derivatives occuring in $A$. When we do not use any null-relations, we have:

$$\begin{aligned}
d(A) &= 0 && \text{if } A \text{ is a generator}\\
d(\partial A) &= d(A) + 1\\
d([AB]_0) &= d(A) + d(B)\\
d(A + B) &= \min(d(A), d(B))\,.
\end{aligned} \tag{3.3.14}$$

We say that we "use a null-relation in increasing derivative-number" if we use it to eliminate (one of) the operator(s) with the smallest derivative-number. In this case, two of the relations eq. (3.3.14) are converted to inequalities: $d(\partial^n A) \geq n$ and $d([AB]_0) \geq d(A) + d(B)$.

We are now in a position to prove the following theorem.

**Theorem 3.3.2** *In an OPA satisfying the conditions of theorem 3.3.1, and where any null-relations are used in increasing derivative-number, the application of eqs. (3.3.5-3.3.11) (together with eqs. (3.3.1-3.3.4)) to order the operators in a normal ordered product requires only a finite number of steps.*

**Proof :**
We define:

$$n \equiv d\left([A[BC]_0]_0\right)\,.$$

We wish to show that the derivative number of the correction terms in eq. (3.3.10) is always bigger than $n$:

$$d\left([A[BC]_0]_0 - (-1)^{|A||B|}[B[AC]_0]_0\right) \ > \ n\,.$$

When $n = 0$ this is immediately clear.
For $n = 1$ we have to consider the different orderings. Let us look first at the case where $A = \partial X$:

$$\begin{aligned}
[\partial X[BC]_0]_0 &= [B[\partial X\ C]_0]_0 - \sum_{l\geq 1}\frac{(-1)^l}{l!}[\partial^l[\partial X\ B]_l\ C]_0\\
&= [B[\partial X\ C]_0]_0 + \sum_{l\geq 2}\frac{(-1)^l(l-1)}{l!}[\partial^l[XB]_{l-1}\ C]_0\,,
\end{aligned} \tag{3.3.15}$$

where we used eq. (3.3.1) in the last step. We see that the derivative-number for the second term in the *rhs* is at least 2. The case where $B = \partial X$ follows from eq. (3.3.15) by moving the correction terms to the *lhs*. Finally, when $C = \partial X$, the correction terms will be of the form $[\partial^l[AB]_l\partial X]_0$, which again have derivative number greater than 2 because $l \geq 1$.

It is now easy to see that the same reasoning can be used for general $n$. For this we will need that:

$$d(\partial^l[\partial^m A\ \partial^n B]_l) \ \geq \ m + n + 1\,,$$

(or $[\partial^m A\ \partial^n B]_l = 0$) which follows from eqs. (3.3.1) and (3.3.2).

To conclude the proof, we note that with each application of eq. (3.3.10) the derivative-number of the correction terms increases. On the other hand, the dimension of the correction terms remains constant. However, for any operator $A$, one has that:

$$d(A) \leq \dim(A) - D\,,$$

where $D$ is the smallest dimension occuring in the OPA. This shows that after a finite number of applications of eq. (3.3.10), no correction terms can appear. ∎

Theorem 3.3.2 guarantees that the algorithm to reorder composites ends in a finite number of steps. However, we have not yet proven that the end-result of the algorithm gives a "standard form", *i.e.* is the result unique if two expressions are the same up to null-relations?

**Intermezzo 3.3.2**
Consider the following counterexample. Suppose there is a null-relation:

$$[AB]_0 - C = 0\,,$$

where $A, B, C$ are bosonic operators. If we use this relation to eliminate the composite $[AB]_0$, we find immediately

$$[[AB]_0 D]_0 = [CD]_0\,,.$$

On the other hand, two applications of eq. (3.3.7) and one of eq. (3.3.10) show that

$$[[AB]_0 D]_0 = [A[BD]_0]_0 + \ldots$$

where the ellipsis denotes terms with derivatives. Together, these relations show that $[A[BD]_0]_0 = [CD]_0 + \ldots$ while both *lhs* and *rhs* are not changed by the algorithm.

---

The intermezzo shows that null-relations should be used to eliminate the "simpler" fields in terms of the composites. We need the notion of "composite-number" $c(A)$. When we do not use any null-relations, we have:

$$\begin{aligned} c(A) &= 1 \qquad \text{if } A \text{ is a generator} \\ c(\partial A) &= c(A) \\ c([AB]_0) &= c(A) + c(B) \\ c(\sum_i A_i) &= \min_{\forall i \text{ such that } d(A_i) = d(\sum_j A_j)} c(A_i)\,. \end{aligned} \tag{3.3.16}$$

With these definitions, the composite-number is equal for both sides of all eqs. (3.3.5-3.3.10). We say that we "use a null-relation in increasing derivative- and composite-number" if we use it to eliminate (one of) the operator(s) with the smallest composite-number of those with the smallest derivative-number. Note that we have $c(\partial A) \geq c(A)$ in this case. We can now assert:

**Theorem 3.3.3** *In an OPA satisfying the conditions of theorem 3.3.1, and where any null-relations are used in increasing derivative- and composite-number, the application of eqs. (3.3.5-3.3.11) (together with eqs. (3.3.1-3.3.4)) to order the operators in a normal ordered product defines a "standard form" on the elements of the algebra.*

For the case of Poisson brackets, the rules (3.3.7) and (3.3.10) drastically simplify, only the first terms remain. Also, eq. (3.3.8) is changed to $[AA]_0 = 0$ when $A$ is fermionic.

**Improvements**

The rules given in this section up to now are sufficient to compute any OPE, and to reorder any composite into a standard form. However, some shortcuts exist. $[A\ [BC]_0]_q$ can be computed using eq. (3.3.4), but an alternative follows from eq. (2.3.22) using eq. (3.3.6):

$$\begin{aligned} [A\ [BC]_0]_q &= (-1)^{|A||B|}\Big([B\ [AC]_q]_0 + \sum_{l\geq 0} \frac{(-1)^{l+q}}{l!}[\partial^l [BA]_{l+q}\ C]_0 \\ &\qquad + \sum_{l=1}^{q-1} (-1)^l [[BA]_l\ C]_{q-l}\Big)\,. \end{aligned} \tag{3.3.17}$$

This rule is more convenient when we know the OPE $B(z)A(w)$ while $A(z)B(w)$ has to be computed using eq. (3.3.3).

Similarly, to compute an OPE where the first operator is a composite, the algorithm as presented above uses eqs. (3.3.3) and (3.3.4). Clearly, eq. (2.3.24) and eq. (3.3.6) implement this in one step:

$$\begin{aligned} [[AB]_0\ C]_q &= \sum_{l\geq 0} \frac{1}{l!}[\partial^l A\ [BC]_{l+q}]_0 + (-1)^{|A||B|} \sum_{l\geq 0} \frac{1}{l!}[\partial^l B\ [AC]_{l+q}]_0 \\ &\quad + (-1)^{|A||B|} \sum_{l=1}^{q-1} [B\ [AC]_{q-l}],, \end{aligned} \tag{3.3.18}$$

where $q \geq 1$ and:

$$\begin{aligned} [[AB]_0 C]_0 &= [A[BC]_0]_0 + \\ &\quad \sum_{l>0} \frac{1}{l!}[\partial^l A\ [BC]_l]_0 + (-1)^{|A||B|} \sum_{l>0} \frac{1}{l!}[\partial^l B\ [AC]_l]_0\,. \end{aligned} \tag{3.3.19}$$

We will see in subsection 3.3.2 where these rules are applied in *OPEdefs* 3.1.

### 3.3.2 The internals of *OPEdefs*

In this subsection, some crucial points in the implementation of the program are discussed. The rest of this chapter does not depend on the material presented here. Familiarity with the *Mathematica* programming language and packages is assumed. More information concerning these topics can be found in appendix C and [210]. For clarity, some of the rules in this subsection are slightly simplified versions of those appearing in the actual code of *OPEdefs*. Input for *Mathematica* is written in `typeset` font.

### 3.3.3 Operator handling

Before any calculations can be done, we have to distinguish between operators and scalars. The user of the package has to declare the bosonic and fermionic operators he/she want to use with the functions `Bosonic` and `Fermionic` *e.g.*

```
Bosonic[A,B]
```

This declaration simply sets the appropriate values for some internal functions: a function which distinguishes between operators and scalars (`OperatorQ`), and another for bosons and fermions (`BosonQ`).

```
BosonicHelp[A_] :=   (BosonQ[A]=True; OperatorQ[A]=True;
                      OPEposition[A] = OPEpositionCounter++)
Bosonic[A__] := Scan[BosonicHelp,{A}]
OPEpositionCounter = 0
```

and similar rules for `Fermionic`. The function `OPEposition` is used to record the order between the operators, which is determined by the order in which they were declared and the standard *Mathematica* order for operators declared with the same pattern:

```
OPEOrder[a_, b_] :=
    Block[{res = OPEposition[b] - OPEposition[a]},
        If[res == 0, Order[a, b], res]
    ]
```

We now list all rules that are necessary for testing if an expression is an operator or not.

```
OperatorQ[A_+B_] := OperatorQ[A]
OperatorQ[A_Times] :=
     Apply[Or, Map[OperatorQ, Apply[List,A]]
OperatorQ[Derivative[_][A_]] := OperatorQ[A]
OperatorQ[_NO] = True
OperatorQ[0] = True
OperatorQ[_] = False                                    (3.3.20)
```

The second rule is the most complicated one. It handles the testing of a product by testing if any of the factors is an operator. The fourth rule declares any composites (which we will give the head `NO`) to be operators. The last rule says that all expressions which were not handled by the previous rules (and the rules set by any declarations) are scalars.

Note that this simple way of defining `OperatorQ` is only possible because *Mathematica* reorders the rules such that more general rules are checked last. This means that when a user declares `A` to be bosonic, a `OperatorQ[A]=True` will be added on top of the list of rules given in (3.3.20).

The only rules we need for the function `BosonQ` are given below, together with the definition of `SwapSign` which gives the sign when interchanging two operators, *i.e.* $-1$ for two fermions and $+1$ otherwise[4]:

```
BosonQ[Derivative[_][A_]] := BosonQ[A]
Literal[BosonQ[NO[A_,B_]]]:= Not[Xor[BosonQ[A],BosonQ[B]]]
SwapSign[A_,B_] := If[Or[BosonQ[A], BosonQ[B]],1,-1]
```

The unit operator $\mathbb{1}$ is declared in *OPEdefs*, and is written as `One`. Introducing an explicit unit operator considerably simplifies many of the functions in *OPEdefs*.

#### Data representation

The representation we used in *OPEdefs* 2.0 was a Laurent series representation. This allowed us to use the built-in rules of *Mathematica* to add OPEs, take derivatives with respect to one of the arguments, .... However, storage space turned out to be a problem for complicated algebras. Therefore, we now use a representation for the singular part of an OPE which is a list of the operators at the different poles. The highest order pole occurs first in the list, the first order pole is the last. As an example, the OPE of a Virasoro operator with itself, eq. (2.3.2), is represented as:

```
OPEData[{c/2 One, 0, 2T, T'}]
```

This representation has the additional advantage that it is more general than a Laurent series representation. Poisson brackets form an obvious example, see subsection 2.3.5.

To hide the details of the representation to the user, the user-interface for constructing an OPE is the function `MakeOPE`. This enables us to allow a different syntax for the interface and the representation, see section 3.4.

With the representation we use, it is easy to determine the order of the highest pole:

```
MaxPole[OPEData[A_List]] := Length[A]
```

For the Virasoro OPE this obviously gives 4.

The actual definitions for the `OPEData` structure enable us to multiply an OPE with a scalar, and add two OPEs. Also higher order poles that are zero are removed. The addition turns out to be the most complicated to achieve. One is tempted to use the `Listable` attribute of `Plus`, *i.e.* adding two list of the same length gives a list with the sum of the corresponding elements. To be able to use this feature, we should make the lists of poles the same length by extending the shorter list with extra zeroes (corresponding to zero operators at those poles). As addition of OPEs will occur frequently, the corresponding rule should be as efficient as possible:

[4]`Literal` prevents *Mathematica* to simplify the left hand side of a definition.

```
OPEData /: n_ * OPEData[A_List] := OPEData[n*A]
OPEData /: A1_OPEData + A2__OPEData :=
        Block[{maxP = Max[Map[MaxPole, {A1,A2}]]},
            OPEData[ Plus @@
                Map[Join[ Table[0,{maxP-Length[#]}], # ]&,
                    Map[First, {A1,A2}]
                ]
            ]
        ]
OPEData[{0, A___}] := OPEData[{A}]
```

**Intermezzo 3.3.3**
In this intermezzo we explain the rule for the addition of `OPEData` structures. The *lhs* of the definition makes sure that `{A1,A2}` will be the list of all `OPEData`s which are added[5]. When adding a number of `OPEData` structures, `maxP` is set to the order of the highest pole occuring in the sum. Then `First` extracts the lists of poles, and we add the correct number of zeroes to the left. Finally, the sum of all these lists is made and `OPEData` wrapped around the result. As an example, consider the sum of `OPEData[{T,0}]` and `OPEData[{T'}]`, we get:

```
{A1,A2}         -> {OPEData[{T,0}], OPEData[{T'}]}
maxP            -> 2
First ...       -> {{T,0}, {T'}}
Join ...        -> {{T,0}, {0,T'}}
Plus @@ ...     -> {T,T'}
```

---

### OPE rules

We use the head `OPE` to represent an OPE. The bilinearity of the `OPE` function is defined by:

[5] A similar effect would be obtained with a *lhs* `Literal[Plus[A__OPEData]]`. However, this gives an infinite recursion when one adds erroneously a constant to an OPEData. Indeed, `Plus[1, OPEData[One]` matches `Literal[Plus[A__OPEData]]`, and *Mathematica* would keep applying the rule, changing nothing at all).

```
Literal[OPE[A___,s_ B_,C___]] :=
        s OPE[A,B,C] /; OperatorQ[B]

Literal[OPE[a___,b_Plus,c___]]:=
        Distribute[
                Lineartmp[a,b,c],
                Plus,Lineartmp,
                Plus,OPE
        ]
```
(3.3.21)

**Intermezzo 3.3.4**
The rule for an OPE of a sum uses the `Distribute[f[__],g,f,gn,fn]` built-in function, which implements distributivity of `f` with respect to `g`, replacing `g` with `gn` and `f` with `fn` in the result. A more obvious rule would be:

```
OPE[a___,b_+c_,d___] := OPE[a,b,d] + OPE[a,c,d]
```
(3.3.22)

However, this rule is much slower in handling sums with more than two terms. Indeed when there is a sum of $n_1$ ($n_2$) terms in the first (second) argument of `OPE`, this rule would have to be applied $n_1 \times n_2$ times, while `Distribute` handles all cases in one application. The difference in performance is shown in fig. 3.1. During the execution of the algorithm, OPEs of sums occur frequently, hence it is important to select the most efficient rule.

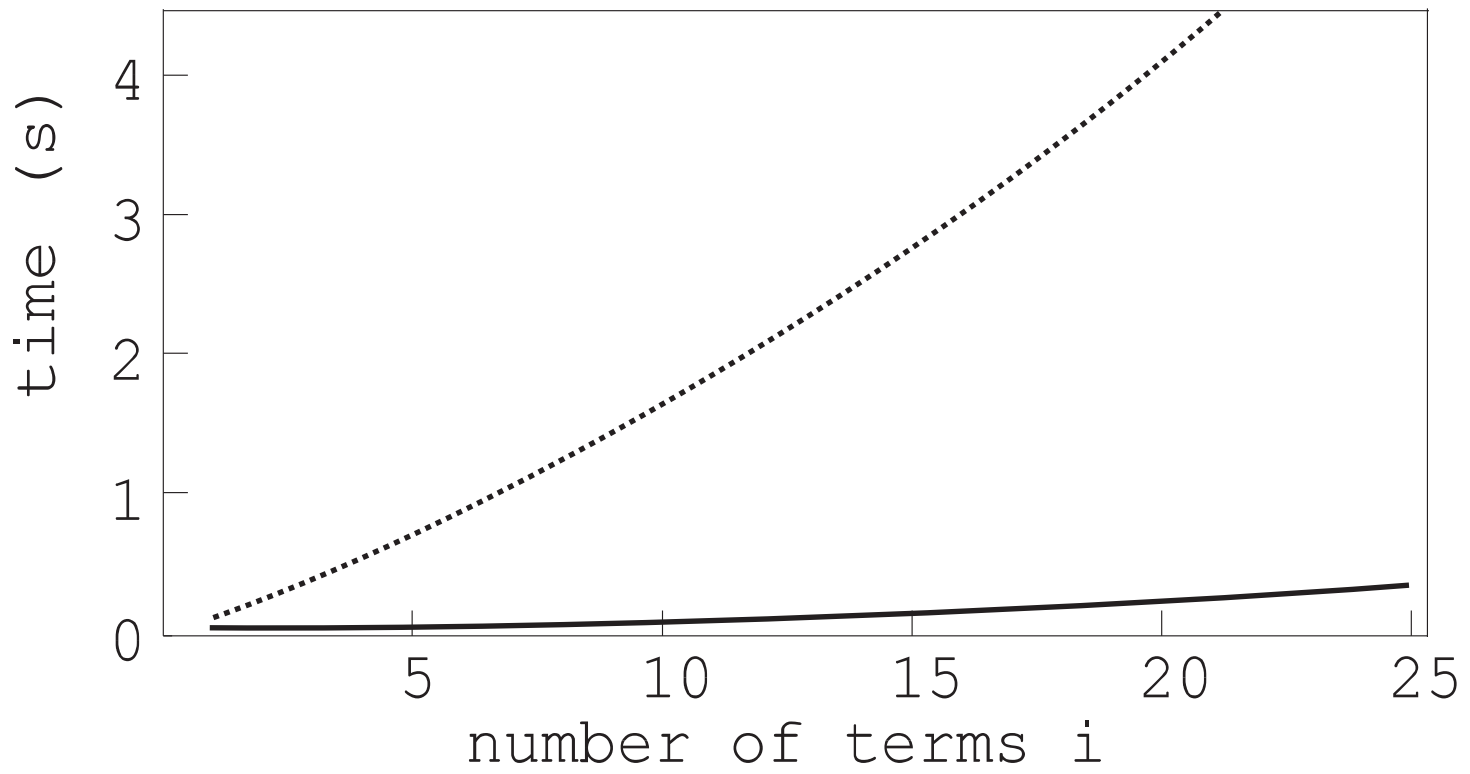

Figure 3.1: Timings for two different rules to implement bilinearity of OPEs. The timings are for `OPE[A,B]` with `A` of a sum of 10 terms and `B` a sum of $i$ terms. The plain line is for `Distribute` (3.3.21) and the dashed line is for the sum-rule (3.3.22).

---

The other rules that are needed for the `OPE` function are:

```
Literal[OPE[Derivative[i_][A_],B_]]:=
        OPEDerivativeHelpL[A,B,i]
Literal[OPE[A_,Derivative[i_][B_]]]:=
        OPEDerivativeHelpR[A,B,i]
Literal[OPE[A_,NO[B_,C_]]] :=
        CallAndSave[OPECompositeHelpR,A,B,C]
Literal[OPE[NO[A_,B_],C_]] :=
        CallAndSave[OPECompositeHelpL,A,B,C] /;
        Not[SameQ[Head[B],NO]]
Literal[OPE[B_,A_]] :=
        OPECommuteHelp[B,A] /;
        Or[SameQ[Head[B],NO], OPEOrder[A,B]>0]
Literal[OPE[_,_]]= OPEData[{}]                    (3.3.23)
```

Here, `OPEDerivativeHelpL`, `OPEDerivativeHelpR`, `OPECompositeHelpL`, `OPECommuteHelp` implement eqs. (3.3.1), (3.3.2), (3.3.18) and (3.3.3) respectively. `OPECompositeHelpR` differentiates between (3.3.4) and (3.3.17). `CallAndSave[f_,args__]` is defined in *OPEdefs* to call the function `f` with arguments `args`, saving the result when a global switch is set. The last rule declares all non-defined OPEs to be regular.

Using the `Help` functions has several advantages compared to inserting their definitions "in-place". First, the structure of the program becomes much clearer. Second, by simply redefining `OPECompositeHelpL` and `OPECompositeHelpR`, we can switch between OPEs and Poisson brackets.

The actual definition of the `Help` functions is of course rather technical, we will give one example in intermezzo 3.3.5. However, more important is to understand how the rules for `OPE` handle an OPE with arbitrarily complicated operators. First of all it is important to remember the evaluation sequence of *Mathematica*. The rules (3.3.23) are stored in the order they are defined, preceded by the definitions the user has given for the OPEs of the generators. Presented an OPE to evaluate, *Mathematica* starts on top of the list of rules and applies the first matching rule. Then the evaluation sequence is restarted. In this way, these rules obviously form an implementation of the algorithm presented in section 3.3.1.

Note that the rules (3.3.23) avoid making use of the more complicated function `OPECompositeHelpL` when the first argument of the OPE is a nested composite, instead `OPECommuteHelp` is used on the result of `OPECompositeHelpR`. This is more efficient in most cases.

Let us as an example follow *OPEdefs* in the computation of a fairly complicated OPE.

```
In[1] :=   Bosonic[A,B,C,D,E]
In[2] :=   TracePrint[OPE[NO[A,NO[C',E]],NO[B,D]],
                (OPEDerivativeHelpL|OPEDerivativeHelpR|
                 OPECompositeHelpL|OPECompositeHelpR|
                 OPECommuteHelp)[__]]
```

```
Out[2] =  OPECompositeHelpR[NO[A, NO[C', E]], B, D]
              OPECommuteHelp[NO[A, NO[C', E]], B]
               OPECompositeHelpR[B, A, NO[C', E]]
                  OPECommuteHelp[B, A]
                  OPECompositeHelpR[B, C', E]
                     OPEDerivativeHelpR[B, C, 1]
              OPECommuteHelp[NO[A, NO[C', E]], D]
               OPECompositeHelpR[D, A, NO[C', E]]
                  OPECommuteHelp[D, A]
                  OPECompositeHelpR[D, C', E]
                     OPEDerivativeHelpR[D, C, 1]
                      OPECommuteHelp[D, C]
```

**Intermezzo 3.3.5**
In this intermezzo, we give an example of the implementation of a `Help` function, `OPECompositeHelpRQ` which corresponds to eq. (3.3.4).

```
OPECompositeHelpRQ[A_,B_,C_] :=
    Block[{q,l,sign = SwapSign[A,B], ABC, AB, AC,
            maxAB, maxABC, maxq},
        AB = OPE[A,B];
        AC = If[ SameQ[B,C], AB, OPE[A,C]];
        maxAB = MaxPole[AB];
        ABC = Table[
                OPE[OPEPole[q][AB], C],
                {q,maxAB}
              ];
        maxABC = Map[MaxPole, ABC];
        maxq = Max[maxABC + Range[maxAB], MaxPole[AC]];
        maxABC = Max[maxABC,0];
        OPEData[
            Table[
                OPESimplify[
                    sign * NO[B,OPEPole[q][AC]] +
                    NO[OPEPole[q][AB],C] +
                    Sum[Binomial[q-1,l] *
                       OPEPole[l][ ABC[[q-l]] ] ,
                       {l,Max[1,q-maxAB], Min[q-1, maxABC]}
                    ]
                ],
                {q,maxq,1,-1}
            ]
        ]
    ]
```

Although this is rather lengthy, most of the lines are quite straightforward. The most important part of the routine is formed by the statements bracketed with `OPEData`. Due to the internal representation for OPEs that is used, the result of the routine is just a table of all the poles in the OPE – starting with the highest pole – with `OPEData` wrapped around it. Note that `ABC` is defined such that `OPEPole[l][ ABC[[q-l]] ]` is equal to $[[AB]_{q-l}C]_l$.

The determination of the order `maxq` of the highest pole in the OPE requires some explanation. Clearly, the first term in the *rhs* of eq. (3.3.4) gives `maxq` $\geq$ `maxAC`. Furthermore, the terms in the sum over `l` are zero unless

$$\mathtt{l} \leq \mathtt{MaxPole[ABC[[q-l]]]}$$

or also

$$\mathtt{q} - \mathtt{l} \leq \mathtt{MaxPole[ABC[[l]]]}$$

hence

$$\mathtt{q} \leq \mathtt{MaxPole[ABC[[l]]]} + \mathtt{l}$$

which is exactly the other boundary on `maxq` which is used.

This shows that the boundaries on `l` and `q` can be computed without any notion of the dimension of a field. The rule is thus suitable for any OPA, which is also true for all other `Help` functions.

---

### Rules for derivatives

The *Mathematica* symbol `Derivative` has almost no rules associated with it, except derivatives of the standard functions like `Sin`. However, we do not need these standard functions, but we do need linearity of derivatives – $\partial(A+B) = \partial A + \partial B$ – which is not included in the built-in definitions for `Derivative`. We chose not to use a new symbol for derivatives because of the convenient notation `A'` for `Derivative[1][A]`. The rules are similar to declaring linearity of OPEs:

```
Derivative[i_][a_Plus] := Map[Derivative[i], a]
Derivative[i_][a_ b_]  :=
        b Derivative[i][a] /; OperatorQ[a]
Derivative[_][0] = 0                                (3.3.24)
```

The first rule handles the $i$-th derivative of a sum. As `Derivative[i_]` expects only one argument, a call to `Distribute` is not necessary. Instead, we map the $i$-th derivative on all terms in the sum.

The second rule handles operators multiplied with scalars. We never need products of operators in our framework.

**Intermezzo 3.3.6**

We comment on the efficiency of the rule for a product given in (3.3.24). In *Mathematica* standard order, `b` follows `a`. Because `Times` has the attribute `Orderless`, this means that during pattern matching the `a_` pattern will be matched sequentially to every factor in the product, while `b_` will be the rest of the product. As we know that the argument of `Derivative[i]` consists of some scalars times only one operator, we see that for each scalar `OperatorQ` is called once, until the operator is found. Hence, it takes maximum $n$ evaluations of `OperatorQ` for a product of $n$ factors. The steps in evaluating `(a b c)'`, where only one of these is an operator, are:

```
if OperatorQ[a] then return(b c a')
elseif OperatorQ[b] then return(a c b')
elseif OperatorQ[c] then return(a b c')
endif
```

This has to be contrasted with the rule

```
Derivative[i__][b_ a_]  :=
        a Derivative[i][b] /; OperatorQ[b]          (3.3.25)
```

As `a` is ordered before `b` the steps in evaluating the same derivative `(a b c)'` are:

```
if OperatorQ[b c] then
   return( Times[a,
     if OperatorQ[c] then return(b c')
     elseif OperatorQ[b] then return(c b')
     endif
   ])
elseif OperatorQ[a c] then
   ...
elseif OperatorQ[a b] then
   ...
endif
```

This is clearly more complicated. Moreover, to test `OperatorQ[b c]` may require two evaluations of `OperatorQ`. In general, for $n$ factors the worst case would be that $n(n-1)/2$ tests are needed. The difference in performance is rather drastic, see fig. 3.2 where we include also two other possibilities using `Not[OperatorQ[a]]`.

---

### Other rules in *OPEdefs*

With the above rules we have already a working package to compute OPEs of arbitrary nested composites. We also need to bring composites in a standard form. This is done by `NO`. The rules attached to `NO` are entirely analogous to those for `OPE`.

One also needs some extra functions like `OPESimplify`. We do not give their implementation here. One can always consult the source of *OPEdefs* for further information.

One major ingredient of *OPEdefs* is not yet discussed: `OPEPole`. For the operation of the above rules to work, only a very simple definition for the `OPEPole` of an `OPEData` is needed:

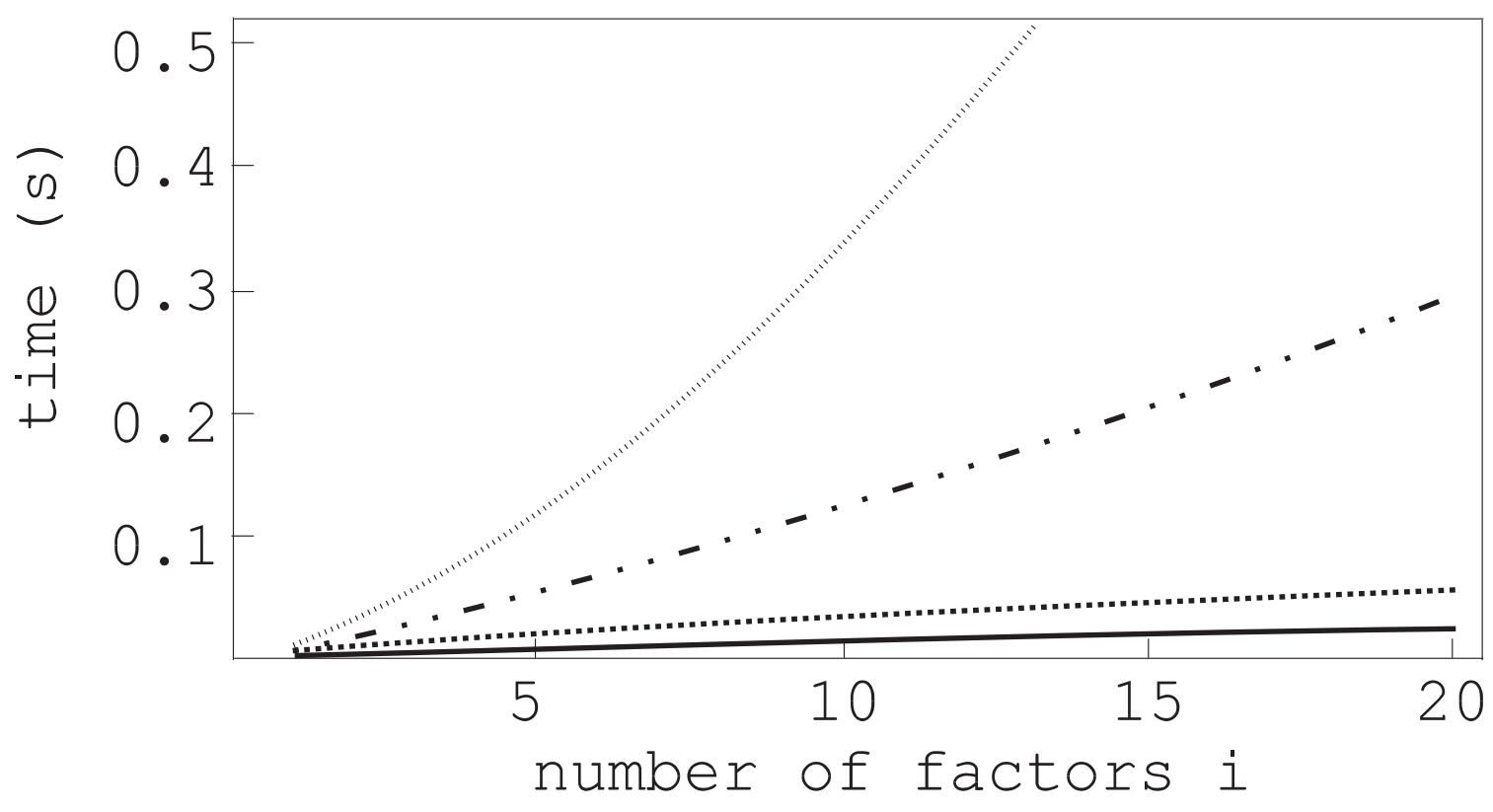

Figure 3.2: Timings for four different rules to implement extracting of scalars as a function of the number of scalars. The timings are for the evaluation of `d[A B]` where `A` is a product of $i$ scalars and `B` is an operator.
Plain line for `d[a_b_] := b d[a] /; Op[a]`,
dashed line for `d[a_b_] := b d[a] /; !Op[b]`,
dash-dot line for `d[a_b_] := a d[b] /; !Op[a]`,
dotted line for `d[a_b_] := a d[b] /; Op[b]`.

```
OPEPole[n_][OPEData[A_]] :=
        If [1 <= n <= Length[A], A[[-n]], 0]
```

*OPEdefs* allows also a syntax `OPEPole[i][A,B]`. This can be used to compute only one pole of the complete OPE, avoiding the computation of all the other poles. Currently, the rules for `OPEPole` are completely independent of those to compute a full OPE. This is because when computing *e.g.* `OPE[A, NO[B, C]]` (see `OPECompositeHelpR` above), it is more efficient to compute `OPE[A, B]` and `OPE[A, C]` once and save these results. It is not a good programming practice to keep two separate sets of rules computing essentially the same thing, but efficiency seemed to be more important at this point.

### 3.3.4 Performance

In [192], a Wakimoto [202] realisation for the Kač–Moody algebra $\widehat{B_2}$ with level $k$ using 2 free scalars and 4 (bosonic) $\beta$, $\gamma$ systems was constructed. Describing this realisation would lead us too far here, but to give an idea of the complexity of the calculation, the total number of composites in the realisation is rougly sixty, and composites of up to four free fields are used.

In table 3.1, we tabulate CPU times for computing an OPE of two of the currents, and the Sugawara tensor for this realisation. The first time given in the table is the time for evaluating the statement after loading the package and defining the realisation. The time between brackets is measured when the statement is repeated. The second execution is much faster because *OPEdefs* stores some of the OPEs with composites. Note that version 2.0 of *Mathematica* is roughly 1.4 times slower than version 1.2.

Table 3.1: CPU time for the computation of the OPE of the currents corresponding to the positive simple root of $\widehat{B_2}$ (statement 9) and the computation of the Sugawara tensor (statement 11) (see Ref. [192]) for *Mathematica* running on a PC 386 (25 Mhz).

| *Mathematica*-version<br>*OPEdefs*-version | 1.2<br>2.0 | 1.2<br>3.1 | 2.0<br>3.1 |
|---|---|---|---|
| `In[9]` | 23.5 (4.5) s | 14.9 (2.8) s | 19.3 (3.8) s |
| `In[11]` | 43.2 (11.6) s | 31.3 (9.4) s | 40.7 (12.1) s |

## 3.4 User's Guide

This section is intended as a user's guide to the package *OPEdefs* 3.1. Explicit examples are given for most operations. Note that *OPEdefs* 3.1 requires *Mathematica* 1.2 or later.

We introduce some special notations. Input for and output from *Mathematica* is written in `typeset` font. Input lines are preceded by "*In[n] :=*", and corresponding output statements by "*Out[n] =*", as in *Mathematica*.

As *OPEdefs* is implemented as a *Mathematica* package, it has to be loaded *before* any of its global symbols is used. Loading the package a second time will clear all previous definitions of operators and OPEs, as well as all stored intermediate results. Assuming that the package is located in the *Mathematica*-path, *e.g.* in your current directory, issue:

*In[1] :=* `<<OPEdefs.m`

After loading *OPEdefs* into *Mathematica*, help for all the global symbols is provided using the standard help-mechanism, *e.g.* `?OPE`.

Now, you need to declare the operators that will be used. If you want to define bosonic operators `T` and `J[i]` (any index could be used), and fermionic operators `psi[i]`, the corresponding statements are:

```
In[2] :=    Bosonic[T, J[i_]]
In[3] :=    Fermionic[psi[i_]]
```

The order of the declarations fixes also the ordering of operators used by the program:

```
T < J[1]' < J[1] < J[2] < J[i] < psi[1] < ...          (3.4.1)
```

By default, derivatives of an operator are considered "smaller" than the operator itself. This can be reversed using the global options `NOOrdering` (see below).

Finally, the non-regular OPEs between the basic operators have to be given. An OPE can be specified in two different ways.
The first way is by listing the operators that occur at the poles, the first operator in the list is the one at the highest non-zero pole, the last operator has to be the one at the first order pole, *e.g.* :

```
In[4] :=    OPE[T, T] = MakeOPE[{c/2 One, 0, 2T, T' }];
```

Note the operator `One` which specifies the unit-operator.
The second way is by giving the OPE as a Laurent series expansion, adding the symbol `Ord` which specifies the (implicit) arguments of the operators for which the OPE is defined[6]. The arguments for the operators can be any *Mathematica* expression.
**Warning**: it is important that the operators occuring as arguments of `OPE` in a definition should be given in standard order (3.4.1), otherwise wrong results will be generated.

The following statements define a $\widehat{SU(2)}_k$-Kač–Moody algebra:

```
In[5] :=    OPE[J[i_],J[i_]] :=
             MakeOPE[-k/2 (z-w)^-2+ Ord[z,w,0]]
In[6] :=    OPE[J[1],J[2]] =
             MakeOPE[J[3][w](z-w)^-1 +Ord[z,w,0]];
In[7] :=    OPE[J[2],J[3]] =
             MakeOPE[J[1][w](z-w)^-1 + Ord[z,w,0]];
In[8] :=    OPE[J[1],J[3]] =
             MakeOPE[-J[2][w](z-w)^-1 + Ord[z,w,0]];
```

In fact, with the above definitions, one has to use always the explicit indices $1, 2, 3$ for the currents $J$. If we would compute an OPE with current `J[i]` where the index $i$ is not $1, 2$ or $3$, wrong results will be given. One can circumvent this peculiarity by reformulating the definitions.

A normal ordered product $[AB]_0$ is entered in the form `NO[A,B]`. Multiple composites can be entered using only one `NO` head, *e.g.* `NO[A,B,C]`. This input is effectively translated into `NO[A, NO[B, C]]`. All output is normal ordered with the same convention, *i.e.* from right to left (input can be in any order). Also, the operators in composites will always be ordered according to the standard order (3.4.1).
As an example, we can define the Sugawara energy-momentum tensor for $\widehat{SU(2)}_k$. The *Mathematica* output of an OPE is a list of the operators at the poles.

```
In[9] :=    Ts = -1/(k+2)(NO[J[1],J[1]]+
                 NO[J[2],J[2]]+NO[J[3],J[3]]);
In[10] :=   OPESimplify[OPE[Ts, J[1]]]
Out[10] =   << 2|| J[1] ||1|| J[1]' >>
```

**Warning**: when computing OPEs with composites, or when reordering composites, *OPEdefs* remembers by default some intermediate results. Thus, it is dangerous to change the definition of the basic OPEs after some calculations have been performed. For example, consider a constant $a$ in an OPE. If calculations are performed after assigning a value to $a$, the intermediate results are stored with this value. Changing $a$ afterwards will give wrong results.

The other globally defined functions available from the package are:

- `OPEOperator[operator_, parity_]` provides a more general way to declare an operator than `Bosonic` and `Fermionic`. The second argument is the parity of the operator such that $(-1)^{\texttt{parity}}$ is $+1$ for a boson, and $-1$ for a fermion. It can be a symbolic constant. This is mainly useful for declaring a *bc*-system of unspecified parity, or a Kač–Moody algebra based on a super-Lie algebra. In such cases, the operator can contain a named pattern:
  ```
  In[11] :=   OPEOperator[J[i_],parity[i]]
  ```
  If one wants to declare more operators, one can group each operator and its parity in a list:
  ```
  In[12] :=   OPEOperator[{b[i_],parity[i]},{c[i_],parity[i]}]
  ```
  See also `SetOPEOptions[ParityMethod, _]`.

- `OPEPole[n_][ope_]` gets a single pole term of an OPE:
  ```
  In[13] :=   OPEPole[2][Out[10]]
  Out[13] =   J[1]
  ```
  `OPEPole[n_][A_,B_]` can also be used to compute only one pole term of an OPE:
  ```
  In[14] :=   Factor[OPEPole[4][Ts, Ts]]
  Out[14] =   (3 k One)/(2 (2 + k))
  ```
  `OPEPole` can also give terms in the regular part of the OPE:
  ```
  In[15] :=   OPEPole[-1][T, T]
  Out[15] =   NO[T', T]
  ```

- `MaxPole[ope_]` gives the order of the highest pole in the OPE.

- `OPEParity[A]` returns an even (odd) integer of $A$ is bosonic (fermionic).

[6]The first time you use this syntax, you may notice an unexpected delay. This is because *Mathematica* is loading the `Series` package.

- `OPESimplify[ope_, function_]` "collects" all terms in `ope` with the same operator and applies `function` on the coefficients. When no second argument is given, the coefficients are `Expand`ed.

```
In[16] :=  OPESimplify[OPE[J[1], NO[J[2], J[1]]]]
Out[16] =  << 2|| (1 - k/2) J[2] ||1||
              NO[J[1], J[3]] + J[2]' >>
```

  `OPESimplify[pole_, function_]` does the same simplifications on sums of operators.

- `OPEMap[function_, ope_]` maps `function` to all poles of `ope`.

- `GetCoefficients[expr_]` returns a list of all coefficients of operators in `expr` which can be (a list of) OPEs or poles.

- `OPEJacobi[op1_,op2_,op3_]` computes the Jacobi-identities (2.3.21) for the singular part of the OPEs of the three arguments. Due to the nature of eq. (2.3.21), the computing time will be smallest (in most cases) when `op1` $\leq$ `op2` $\leq$ `op3` in the order (3.4.1). In nonlinear algebras, the computation will use rules for OPEs of composites which assume that the Jacobi identities hold. This means that the result of `OPEJacobi` gives only necessary conditions. In some cases, different orderings of `op1-op3` have to be tried to find all conditions.
  The result of `OPEJacobi` is a double list of operators. It is generated by

```
Table[OPEPole[n][A,OPEPole[m][B,C]] +
        corrections, {m, maxm},{n,maxn}]
```

  All elements of the list should be zero up to null operators for the OPA to be associative.

- `Delta[i_,j_]` is the Kronecker delta symbol $\delta_{ij}$.

- `ClearOPESavedValues[]` clears all stored intermediate results, but not the definition of the operators and their OPEs. To clear everything, reload the package.

- `OPEToSeries[ope_]` converts an OPE to a Laurent series expansion in `z` and `w`. The arguments can be set to `x` and `y` with:

```
In[17] :=  SetOPEOptions[SeriesArguments, {x, y}]
```

- `TeXForm[ope_]` gives TeXoutput for an OPE. The same arguments are used as in `OPEToSeries`.

- `OPESave[filename_]` (with `filename` a string between double quotes) saves the intermediate results that *OPEdefs* remembers to file (see the option `OPESaving` below).

- `SetOPEOptions` is a function to set the global options of the package. The current options are:
  - `SetOPEOptions[SeriesArguments, {arg1_, arg2_}]` : sets arguments to be used by `TeXForm` and `OPEToSeries`. One can use any Mathematica expression for `arg1` and `arg2`.
  - `SetOPEOptions[NOOrdering, n_]` : if `n` is negative, order higher derivatives to the left (default), if `n` is positive, order them to the right.
  - `SetOPEOptions[ParityMethod, 0|1]` : makes it possible to use operators of an unspecified parity. When the second argument is 0 (default), all operators have to be declared to be bosonic or fermionic. When the argument is 1, `OPEOperator` can be used with a symbolic parity. Note that in this case, powers of $-1$ are used to compute signs, which is slightly slower than the boolean function which is used by the first method.
    This option is not normally needed as the use of `OPEOperator` with a non-integer second argument sets this option automatically.
  - `SetOPEOptions[OPESaving, boolean_]` : if `boolean` evaluates to `True` (default), *OPEdefs* stores the intermediate results when computing OPEs of composites and when reordering composites. This option is useful if *Mathematica* runs short of memory in a large calculation, or when computing with dummy indices[7].
  - `SetOPEOptions[OPEMethod, method_]` : with the parameter `method` equal to `QuantumOPEs` enables normal OPE computations (default setting), while `ClassicalOPEs` enables Poisson bracket computations. Using this option implicitly calls `ClearOPESavedValues[]`.

## 3.5 Example : The conformal anomaly in superstring theory

We consider only one free boson field $X$ and one free fermion field $\psi$ because additional free fields will have exactly the same OPEs and commute with each other. We denote $\partial X$ with `J` and $\psi$ with `psi` (we normalise them such that they have a $+1$ in their OPEs (2.6.7) and (2.6.26)). The ghosts are a fermionic $b$, $c$ system (operators `b,c`) and a bosonic $\beta$, $\gamma$ system (operators `B,G`) (normalised such that ${}_iA^j = (-1)^{i+1}{}_i\delta^j$ in eq. (2.6.31)). $b$ has conformal dimension 2 and $\beta$ has 3/2. It is now a trivial task to compute the conformal anomaly:

```
In[1] :=  <<OPEdefs.m
```

[7]No mechanism to use dummy indices is built-in in *OPEdefs*. I wrote a separate package *Dummies* to handle this.

```
In[2] :=  Bosonic[J,B,G]; Fermionic[b,c,psi];
          OPE[J,J] = MakeOPE[{One, 0}];
          OPE[psi,psi] = MakeOPE[{One}];
          OPE[b,c] = MakeOPE[{One}];
          OPE[B,G] = MakeOPE[{One}];
          Tb = 1/2 NO[J,J]; Tf = -1/2 NO[psi,psi'];
          Tbc = -2 NO[b,c'] - NO[b',c];
          TBG = 3/2 NO[B,G'] + 1/2 NO[B',G];
In[3] :=  OPESimplify[OPE[Tb,Tb]]
Out[3] =  << 4|| One/2 ||3|| 0 ||2|| NO[J, J] ||1||
             NO[J', J] >>
In[4] :=  OPESimplify[OPE[Tf,Tf]]
Out[4] =  << 4|| One/4 ||3|| 0 ||2|| NO[psi', psi] ||1||
             NO[psi'', psi]/2>>
In[5] :=  OPESimplify[OPE[Tbc,Tbc]]
Out[5] =  << 4|| -13 One ||3|| 0 ||2||
             -4 NO[b, c'] - 2 NO[b', c]||1||
             -2 NO[b, c''] - 3 NO[b', c'] - NO[b'', c] >>
In[6] :=  OPESimplify[OPE[TBG,TBG] - MakeOPE[{2 TBG,TBG'}]
Out[6] =  << 4|| 11 One/2 ||3|| 0 ||2|| 0 ||1|| 0 >>
```

We see that each bosonic (fermionic) field will contribute a central charge 1 (1/2) to the total central charge of the theory. The $b$, $c$ system contributes $-26$, and the $\beta$, $\gamma$ system 11. This gives the well known relation for the critical dimensions of the bosonic string $D_b - 26 = 0$ and the superstring $3/2D_s - 26 + 11 = 0$. Moreover, we can easily verify that the energy–momentum tensors obey the Virasoro algebra .

The reader without experience in CFT is invited at this point to take out some time and compute the OPE for $T_{BG}$, for instance, by hand. Although this computation is rather trivial with *OPEdefs*, the same calculation was attempted in [179] using the mode–algebra. There it proved not to be possible to compute the Virasoro algebra automatically due to difficulties with the infinite sums in the normal ordered products.

## 3.6 Future developments

A first extension would be to add the possibility to specify a range of poles one wants. This would unify the current implementations of `OPE` and `OPEPole`. The main difficulty in programming this, is that when saving intermediate results, one has to see which poles are already computed and which not.

The objective in writing *OPEdefs*, was to make a package available which is as general as possible. One could write specialised extensions which would outperform *OPEdefs*. For instance, the restriction to free fields would be very useful. Also $\mathcal{W}$–algebras, a conformal OPA generated by quasiprimaries, form a preferred subclass of the OPAs. In this case it would be advantageous to compute in a basis of quasiprimaries, see section 4.3. A package under development [194] collects some formulas for working with (quasi)primaries in a $\mathcal{W}$–algebra, but it relies on *OPEdefs* for computing OPEs.

The main restriction of *OPEdefs* is of course the requirement of poles of integer order. In particular, vertex operators are widely used in conformal field theory. Here the order of the powers in the generalised Laurent expansion remain integer spaced. This should make it possible to extend *OPEdefs* to handle this case. However, the notion of the singular and regular part of an OPE is not so important when using vertex operators as in other cases. Indeed, the normal ordered product of two vertex operators should not be defined as the zeroth order pole in their OPE. This makes it desirable to use a data representation which keeps the information to generate any “pole”, but stores already computed results[8]. This would make it possible to work with vertex operators $V_a$ with a symbolic weight $a$, and not only with fixed numbers. One could then also define an OPE between fields of (non-numeric) dimension $h_1, h_2$, *e.g.* with the unit operator at the pole of order $h_1 + h_2$. Finally, the extension to nonmeromorphic OPEs would allow to treat parafermions [67, 68, 95].

We are currently working at a version for computing with super OPEs in $N = 2$ superfields [137]. Extension to arbitrary $N$ will not give great difficulties.

Finally, it would be convenient to be able to use dummy indices in *OPEdefs*. I wrote a separate package *Dummies* to handle this. However, this package uses only a rudimentary algorithm to simplify expressions with dummy indices. There seems to exist no algorithm (except exhaustive enumeration) to do this simplification for the case at hand. The main difficulty is that correction terms are needed when interchanging operators in a normal ordered product.

## 3.7 Other packages

As shown in section 2.4, an other approach to the same problem would be to use modes. An attempt to compute commutators of the modes of normal ordered operators in REDUCE [179] was not completely successful due to difficulties in the reshuffling of the indices in infinite sums. It should be possible to avoid this by constructing the formulas for commutation with a mode of a composite operator by looking at the corresponding formula for OPEs. Apparently, this is done in a package by L. Romans, which is not published, but acknowledged by a few authors. The mode-approach has probably the advantage that contributions of the tower of derivatives of a quasiprimary operator (see section 4.3) are summed in advance. However, this definition (2.4.1) which makes the formulas for computing with modes convenient is restricted to a conformal OPA.

[8] This is a well known concept called “streams” in the symbolic manipulation program *Axiom*, where *e.g.* Taylor series are stored in this way.

Related to mode calculations is the approach taken by H. Kausch (using another symbolic manipulation program *Maple*). He uses the equivalence between conformal fields $\Phi(x)$ and states $\Phi(0)|0>$ discussed in [98]. As this reference restricts itself to bosonic fields of integer dimension, and fermionic fields of half-integer dimension, we suspect that Kausch's program will not be able to handle algebras outside this framework, like [160, 21, 49], while they present no problems for *OPEdefs*. However, Kausch did not publish his work.

Finally, A. Fujitsu recently developed a package, *ope.math*, in *Mathematica* for computing OPEs of free fields [93]. It is able to treat vertex operators at non-integer poles (this is not possible in *OPEdefs*). However, one cannot compute OPEs in a $\mathcal{W}$–algebra or a Kač–Moody algebra, except by working in a realisation with free fields. The current version is much slower than *OPEdefs* for calculations without vertex operators. *ope.math* uses a fixed notation for all fields, *e.g.* the $i$-th derivative of a complex boson (part of a bosonic $b, c$-system, see section 2.6) has to be called `be[i, j, z]` where $j$ is an index and $z$ the coordinate. The package is currently not able to work with an unspecified number of fields, *i.e.* `ope[be[0,i, z], ga[0,j, w]]` returns zero.

# Chapter 4

# $\mathcal{W}$-algebras

In this chapter, we focus our attention on $\mathcal{W}$–algebras. Recall that we defined a $\mathcal{W}$–algebra as an Operator Product Algebra (OPA) generated by a Virasoro operator $T$ and quasiprimary operators $W^i$. We first discuss the representations of $\mathcal{W}$–algebras, section 4.2. We then show how the representations of the global conformal group restrict the possible Operator Product Expansions (OPEs) of the generators, section 4.3. In the next section, the full conformal group is studied. We demonstrate by using the Jacobi identities how to reconstruct the complete OPEs from the knowledge of the coefficients of the primary operators in these OPEs. The ultimate goal of section 4.3 is to provide the formulas for an algorithmic computation of Virasoro descendants and the coefficients with which they appear in OPEs.

In section 4.5, we discuss a few of the more important methods to construct $\mathcal{W}$–algebras and comment on the classification of the $\mathcal{W}$–algebras. As an illustration of the ideas in this chapter, we conclude with an example in section 4.6: the $\mathcal{W}_c B_2$ algebra.

The presentation used in sections 4.3 and 4.4 is slightly more general than the literature, and in subsections 4.3.2, 4.4.2 and 4.4.3 new developments are given. The results on the $\mathcal{W}_c B_2$ algebra were published in [78].

## 4.1 Introduction

The chiral symmetry generators of a conformal field theory in general form a $\mathcal{W}$–algebra. The representation theory of $\mathcal{W}$–algebras will thus have its direct consequences for the correlation functions of the theory. For physical applications, highest weight representations are most useful. In certain special cases, the correlation functions are completely fixed by the symmetry algebra. These representations are the minimal models, pioneered for the Virasoro algebra in [13] and extended to other $\mathcal{W}$–algebras in [66, 63, 65, 6]. They consist of a finite number of highest weight operators which each give rise to degenerate a representation. Minimal models were first studied as a simple model for infinite dimensional highest weight representations. Almost all exactly solvable conformal field theories (CFT) are based on minimal models. They also appear in $\mathcal{W}$-string theory, see chapter 8.

The study of the $\mathcal{W}$–algebras is complicated by the fact that the OPEs can contain composite terms, or equivalently the mode algebra is nonlinear. The prototype of a nonlinear $\mathcal{W}$–algebra is the $\mathcal{W}_3$ algebra [211]. It is generated by the energy–momentum tensor $T$ and a dimension 3 current $W$ with OPEs given by:

$$\begin{aligned}
T(z)T(w) &= \frac{c}{2}(z-w)^{-4} + 2(z-w)^{-2}T(w) + (z-w)^{-1}\partial T(w) + \cdots \\
T(z)W(w) &= 3(z-w)^{-2}W(w) + (z-w)^{-1}\partial W(w) + \cdots \\
W(z)W(w) &= \frac{c}{3}(z-w)^{-6} + 2(z-w)^{-4}T(w) + (z-w)^{-3}\partial T(w) \\
&\quad +(z-w)^{-2}\left[2\beta\Lambda(w) + \frac{3}{10}\partial^2 T(w)\right] \\
&\quad +(z-w)^{-1}\left[\beta\partial\Lambda(w) + \frac{1}{15}\partial^3 T(w)\right] + \cdots, \qquad (4.1.1)
\end{aligned}$$

where:

$$\Lambda = [TT]_0 - \frac{3}{10}\partial^2 T \qquad (4.1.2)$$

and:

$$\beta = \frac{16}{22+5c}\,. \qquad (4.1.3)$$

The operator $\Lambda$ is defined such that it is quasiprimary, *i.e.* $[T\Lambda]_3 = 0$. Although this is the simplest nonlinear $\mathcal{W}$–algebra for generic $c$, checking the Jacobi identities by hand is already a nontrivial task.

We notice that the *rhs* of the OPE $[WW]$ can be written in terms of quasiprimaries or their derivatives. Using the global conformal transformations, generated by $L_{\pm 1}, L_0$, we will see in section 4.3 that the coefficients of the derivatives of the quasiprimaries, are numbers depending only on the dimensions of the operators involved. This result was already contained in the paper of Belavin, Polyakov and Zamolodchikov [13]. We present some new results on how the quasiprimaries in a given OPE can be found.

The full conformal group gives even more information. In case that the OPA is generated by primary operators and their descendants, the Jacobi identities with $\{T, \Psi_i, \Psi_j\}$ fix the form of the OPE of two primaries $\Psi_i, \Psi_j$. In fact, the coefficients of the descendants of the primaries occuring in the OPE now depend on the dimensions of the operators and the central charge. In the example of $\mathcal{W}_3$, the only primary in the singular part of the $W(z)W(w)$ OPE is the unit operator at the sixth order pole. The rest of this OPE is then completely determined by the Jacobi identity $\{T, W, W\}$.
It was shown in [13, 74] how the coefficients of the Virasoro descendants can be found. As they can be very complicated, automation of the computation of these coefficients is highly desirable[1]. However, in the basis for the descendants used in [13, 74] this computation is very CPU-intensive. We will therefore construct a basis of quasiprimaries, and give the necessary formulas to compute descendants and coefficients in this basis. A *Mathematica* package that implements these algorithms is now in testing phase [194].

The conclusion will be that we can reconstruct the OPEs of the primary generators of a $\mathcal{W}$–algebra from the list of the coefficients of all primaries. For primary operators $\Psi_i$, we will write symbolically:

$$\Psi_i \times \Psi_j \longrightarrow C_{ij}{}^{k}\ [\Psi_k]\,, \tag{4.1.4}$$

where $[\Psi_k]$ denotes the conformal family of the primary operator $\Psi_k$. We only have to keep track of the primaries in the singular part of the OPE, because any primaries in the regular part can be constructed from the information contained in the singular part of the OPE. The structure constants $C_{ij}{}^{k}$ are still restricted by the Jacobi identities discussed in 2.3, but now for triples of primaries. Unfortunately, the equations for the coefficients are too complicated to solve in the general case, preventing to classify the $\mathcal{W}$–algebras in this way. We will discuss briefly some other attempts towards the classification.

[1] The descendants themselves are easily computed using *OPEdefs*.

## 4.2 Highest weight representations and minimal models

In this section we consider representations of a $\mathcal{W}$–algebra which are constructed by acting repeatedly with the generators of the $\mathcal{W}$–algebra on a highest weight state $|\phi\rangle$.

**Definition 4.2.1** *For a mode algebra with generators $W^i{}_n$ (with $W^0 \equiv T$), a highest weight state (HWS) $|\phi\rangle$ with weights $w^i$ satisfies:*

$$W^i{}_0|\phi\rangle = \ w^i|\phi\rangle, \qquad W^i{}_n|\phi\rangle = \ 0 \quad n > 0\,,$$

In this definition it is understood that a generator with half-integral modes has no weight associated to it. We will use the correspondence between states and fields in meromorphic conformal field theory, as discussed in [98]. So, with every highest weight operator $\phi$ corresponds a highest weight state $|\phi\rangle$.

For the Virasoro algebra it is natural to take $\phi$ to be a primary operator of dimension $h$. This gives an extra condition $\widehat{L}_{-1}\phi = \partial\phi$. The notion of a primary operator, and in particular the requirement that the first order pole of the OPE of the energy–momentum tensor with a primary operator is the derivative of the operator, arises from the geometrical interpretation of a conformal transformation eq. (2.1.4). As no geometrical meaning is currently known for the transformations generated by currents with a nonlinear OPE, the concept of a $\mathcal{W}$–primary cannot be defined at present.

To define a highest weight representation, we first introduce some notation. We will denote a sequence $(W^i)_{-n_1}\ldots(W^i)_{-n_k}$ where $n_j \geq n_{j+1} > 0$, as $(W^i)_{-\{n\}}$. This notation is extended to sequences of modes of different generators $W^0{}_{-\{n_0\}}\ldots W^d{}_{-\{n_d\}}$, which we write as $W_{-\{n\}}$, where $\{n\}$ forms an ordered partition of $N$ with a different "colour" for every generator. We use the convention that for a sequence of positive modes, the order is the reverse as for negative modes, *i.e.* $W_{\{n\}}$ acts like $W^d{}_{\{n^d_{k_d}\}}\ldots W^d{}_{\{n^d_1\}}\ldots W^0{}_{\{n^0_1\}}$.

A Verma module is then defined as the space of all "descendants" $W_{-\{n\}}|\phi\rangle$. The level of the descendant is defined as its $L_0$ weight minus the $L_0$ weight of $|\phi\rangle$, *e.g.* the level of $L_{-n}(W^i)_{-m}\phi$ is $n+m$. Although the dimension of the Verma modula is infinite, the dimension of the space of descendants at a certain level is finite.

The representation can then be computed by using the (graded) commutators of the modes and the definition 4.2.1. The representation will (at least) depend on the weights $w^i$ and on the central charge $c$ of the $\mathcal{W}$–algebra, we will write $\mathcal{R}(w^i, c)$.

The representation is reducible, or degenerate, if at a certain level $N$ there occurs a new HWS, which will be the starting point for a new tower of states. One can then divide out this new representation $\mathcal{R}(\tilde{w}^i, c)$, where $\tilde{h} = \tilde{w}^0 = h+N$. If this process is repeated for all descendant HWSs, one finally obtains an irreducible representation.

The descendant state which is also a HWS is often called a singular vector of the representation.

We can define an inner product on the Verma module indirectly via the adjoint operation:

$$\left(W^i{}_n\right)^\dagger = W^i{}_{-n} \qquad <\!\phi|\phi\!>= 1\,. \tag{4.2.1}$$

For two states with weights $w^i, \tilde{w}^i$, eq. (4.2.1) implies that a nonzero inner product can only occur of $w^i = \tilde{w}^i$. In particular, two descendant states of different level have zero inner product. Singular vectors have inner product with all other states.

Singular vectors can be constructed using screening operators $S$. These are characterised by the fact that the singular part of the OPE of a generator $W^i$ with $S$ can be written as a total derivative:

$$W^i(z)S(w) = \sum_n \frac{[W^iS]_n(w)}{(z-w)^n} = \frac{d}{dw}[\text{something}] + O(z-w)^0\,. \tag{4.2.2}$$

Using eq. (2.3.16), this is seen to be equivalent to $[SW^i]_1 = 0$. For any screening current $S$, we can define an "intertwiner" $Q_S$ whose action on an operator $X$ is given by:

$$Q_S\, X(w) \;=\; \oint_{\mathcal{C}_w} \frac{dz}{2\pi i} S(z)\, X(w) \;= [SX]_1(w)\,. \tag{4.2.3}$$

Due to eq. (4.2.2), $Q_S$ commutes with the modes $W^i{}_m$. This means that $Q_S$ on a HWS $|\phi\!>$ is either zero, or another HWS. In general, $Q_S|\phi\!>$ will be a descendant of another HWS. Hence, the intertwiners can be used to construct singular vectors.

The matrix $S$ of the inproducts of all descendants at level $N$ is called the Šapovalov form. It depends only on the weights $w^i$ of the HWS and the central charge of the $\mathcal{W}$–algebra. The determinant of $S$ is the Kač–determinant [126]. Its zeroes are related to the singular vectors in the representation.

Because highest weight descendants are null operators, they generate partial differential equations on the correlation functions [13]. Completely degenerate representations have as much independent null vectors as possible [31]. It can then be proven [13] that the HWSs which give rise to a singular vector form a closed (possibly non-meromorphic) OPA. Minimal models are then defined as those cases where this OPA is finitely generated.

## 4.3 Consequences of the global conformal group

In this section, we will study the OPEs of the quasiprimaries in a $\mathcal{W}$–algebra. The global conformal group, generated by the modes $L_{-1}, L_0, L_{+1}$ of the energy–momentum tensor $T$, puts strong restrictions on these OPEs.

Recall the definition of a quasiprimary operator $\Phi_i$ of dimension $h_i$ (def. 2.3.6)[2]:

$$\begin{aligned} Ł_1\Phi_i &= 0 \\ Ł_0\Phi_i &= h_i\Phi_i \\ Ł_{-1}\Phi_i &= \partial\Phi_i\,, \end{aligned} \tag{4.3.1}$$

where the modes are defined, eq. (2.4.1), as:

$$L_n\Phi \equiv [T\Phi]_{n+2}\,. \tag{4.3.2}$$

From the definitions in section 4.2, we see that a quasiprimary operator furnishes a highest weight representation for the global conformal algebra.

We will use the following assumption:

**Assumption 4.3.1** *All elements of the OPA are linear combinations of quasiprimaries $\Phi_i$ and their global conformal descendants, namely their derivatives.*

This is a natural – and commonly used – assumption, but see intermezzo 4.3.1.

**Intermezzo 4.3.1**
Consider a free field theory with background charge, see subsection 2.6.1. Using the definition of the energy–momentum tensor, eq. (2.6.17), $T = 1/2[\partial X\partial X]_0 - q\,\partial^2 X$, we see that $\partial X$ is not quasiprimary when $q$ is not zero:

$$T\,\partial X \;= \ll 2q \mid \partial X \mid \partial^2 X \gg\;.$$

$X$ is not even a scaling operator:

$$T\;X \;= \ll q \mid \partial X \gg\;.$$

Hence, including $\partial X$ as a generator of the OPA, would make the assumption 4.3.1 invalid.

Consider now a vertex operator $V_a$, which is primary with conformal dimension $h_a = a(a+2q)/2$ with respect to $T$. Using eq. (2.6.14), we find for the OPE of $V_a$ with $V_{-a}$:

$$V_a(z)\,V_{-a}(w) = (z-w)^{-a^2}\,(1 + (z{-}w)a\partial X(w) + \ldots)$$

This OPE obeys all associativity conditions. In particular, this means that it satisfies Jacobi identities with $T$, given in eq. (4.3.6), for all $q$. Clearly, it provides an example of an OPE of two primary operators which cannot be written in a basis of quasiprimaries (if $q \neq 0$), even not in a basis of highest weight operators.

Of course, by taking a different $T$ as the Virasoro operator, the results of this chapter can be applied.

[2]For convenience, we will drop the hats on the $\hat{L}$ in the rest of this chapter.

### 4.3.1 Consequences for OPEs

We will now investigate how this assumption restricts the form of the OPEs. Consider the OPE of two quasiprimary operators. The assumption 4.3.1 implies that it can be written as:

$$\Phi_i(z)\Phi_j(w) \;=\; \sum_k \sum_{p\geq 0} a_{ij}^k(p)\; \partial^p \Phi_k(w)(z-w)^{p-h_{ijk}}\,, \qquad (4.3.3)$$

where we introduced the notation:

$$h_{ijk} = h_i + h_j - h_k\,. \qquad (4.3.4)$$

The coefficients $a_{ij}^k(p)$ are (partially) determined by the Jacobi identities, eq. (2.3.21), in the OPA. We now set out to find these coefficients. The method we use consists in acting with $L_1$ on eq. (4.3.3). For $L_1$ acting on the *lhs*, we use the Jacobi identities of $\{T, \Phi_i, \Phi_j\}$, while for the *rhs* of eq. (4.3.3) we use the identities of $\{T, T, \Phi_k\}$.

We write the *lhs* of eq. (4.3.3) with the notation introduced in eq. (2.3.3):

$$\Phi_i(z)\Phi_j(w) \;=\; \sum_n [\Phi_i\Phi_j]_n (z-w)^{-n}\,. \qquad (4.3.5)$$

From the Jacobi identities (2.3.21), or alternatively, using the commutation rules of the modes of $\Phi_i$ with $L_m$ given in eq. (2.4.7)[3], gives:

$${L_1}^m [\Phi_i\Phi_j]_n = (2h_i - m - n)_m [\Phi_i\Phi_j]_{m+n}\,, \qquad (4.3.6)$$

where the Pochhammer symbol $(a)_n$ is defined in eq. (2.A.1).

For the action of ${L_1}^m$ on the *rhs* of eq. (4.3.3), we use the following formula:

$${L_1}^m {L_{-1}}^p \Phi_k = (p-m+1)_m (2h_k + p - m)_m {L_{-1}}^{p-m}\Phi_k\,, \qquad (4.3.7)$$

which can be derived using (4.A.1). Note that we dropped the term $L_1\Phi_k$ as it is zero.

Comparing the powers of $(z-w)^{-n}$, we find:

$$\begin{aligned}(2h_i - m - n)_m \sum_k a_{ij}^k(h_{ijk} - m - n)\; \partial^{h_{ijk}-m-n}\Phi_k \;=\; \\ \sum_k a_{ij}^k(h_{ijk} - n)\; (h_{ijk} - n - m + 1)_m \\ (h_i + h_j + h_k - n - m)_m\; \partial^{h_{ijk}-m-n}\Phi_k\,. \end{aligned} \qquad (4.3.8)$$

[3] The commutation relations eq. (2.4.7) were derived for a primary field $\Phi$. Clearly, they are also valid for a quasiprimary if $L_m \in \{L_1, L_0, L_{-1}\}$.

We now temporarily assume that the $\{\Phi_i\}$ form an independent set, *i.e.* no linear combination of the $\Phi_i$ and their derivatives can be made which is a null operator. This means that the coefficients of all $\partial^p\Phi_k$ in the *lhs* and *rhs* have to be equal. Choosing $m = h_{ijk} - n$, gives:

$$a_{ij}^k(m)\; m!\; (2h_k)_m \;=\; (h_i - h_j + h_k)_m a_{ij}^k(0)\,. \qquad (4.3.9)$$

Introducing the notation:

$$a_{ij}^k(0) = \mathcal{C}_{ij}^k \qquad a_{ij}^k(n) = \mathcal{C}_{ij}^k \alpha(h_i, h_j, h_k, n)\,, \qquad (4.3.10)$$

we conclude that if all $h_k > 0$:

$$\Phi_i(z)\Phi_j(w) = \sum_k \sum_{n\geq 0} \mathcal{C}_{ij}^k \alpha(h_i, h_j, h_k, n)\; \partial^n\Phi_k(w)(z-w)^{n-h_{ijk}}\,, \qquad (4.3.11)$$

where:

$$\alpha(h_i, h_j, h_k, n) \;=\; \frac{(h_i - h_j + h_k)_n}{n!\; (2h_k)_n}\,. \qquad (4.3.12)$$

Eqs. (4.3.11,4.3.12) were derived in [13]. They enable us to reconstruct the complete OPE when the structure constants $\mathcal{C}_{ij}^k$ are given, *i.e.* the coefficients of the quasiprimaries at every pole. It is easily verified that the OPEs of the $\mathcal{W}_3$-algebra have the structure given in eqs. (4.3.11,4.3.12).

We now treat some special cases where eqs. (4.3.11,4.3.12) are not valid.
When $h_k$ is half-integer and not strictly positive, eq. (4.3.9) indicates that $a_{ij}^k(1-2h_k)$ is not determined by the global conformal transformations. This is because in this case $\partial^{1-2h_k}\Phi_k$ is a quasiprimary (with dimension $1-h_k$), as is easily checked using eq. (4.3.7). This situation will be mirrored in the next section when considering the full conformal group. It occurs because the highest weight representation generated by $\Phi_k$ for $h_k \leq 0$ is reducible.
When null operators occur in the OPA, extra free coefficients in eq. (4.3.8) can appear. However, we can continue to use eqs. (4.3.11,4.3.12) as indeed coefficients if null operators are arbitrary.

### 4.3.2 Finding the quasiprimaries in an OPE

For practical applications, the quasiprimaries at each pole have to be identified when the complete OPE is given. This is the reverse problem of the previous subsection.

We write the quasiprimary at the $m$-th order as $QP^m(\Phi_i, \Phi_j)$. It can be determined in a recursive way by using the results of the previous subsection. Indeed, the highest pole of the OPE is by assumption 4.3.1 quasiprimary. We can substract the complete "tower" of derivatives of this quasiprimary from the OPE to end up with a new Laurent series where the most singular term is again quasiprimary. We find:

$$QP^m(\Phi_i,\Phi_j) \;=\; [AB]_m - \sum_{n\geq 1}\alpha(h_i,h_j,h_i+h_j-m-n,n)\;\partial^n\big(QP^{m+n}(\Phi_i,\Phi_j)\big)\,. \tag{4.3.13}$$

In particular, for $m=0$ this formula provides a definition of a quasiprimary normal ordered product of two quasiprimaries, for example:

$$\mathcal{NO}(T\Phi_i) = [T\Phi_i]_0 - \frac{3}{2(2h_i+1)}\partial^2\Phi_i\,, \tag{4.3.14}$$

of which eq. (4.1.2) is a special case. More general, we see that $QP^m(\Phi_i,\Phi_j)$ for $m\leq 0$ are composite quasiprimaries. The formula (4.3.13) appears also in [26].

The recursive definition eq. (4.3.13) of the operators $QP^m$ is a very inefficient, and complicated, way of computing the quasiprimaries. To simplify this definition, we note that it can be rewritten as:

$$QP^m(\Phi_i,\Phi_j) = \sum_{n\geq 0} a^m_n(h_i,h_j)\;\partial^n[\Phi_i\Phi_j]_{n+m}\,, \tag{4.3.15}$$

where the coefficients $a^m_n(h_i,h_j)$ are determined by eq. (4.3.13). We find:

$$\sum_{k=0}^{n} a^{m+k}_{n-k}(h_i,h_j)\alpha(h_i,h_j,h_i+h_j-m-k,k) = 0\,. \tag{4.3.16}$$

We will provide a closed form for the solution of eq. (4.3.16) by requiring that eq. (4.3.15) defines a quasiprimary operator:

$$L_1\;QP^m(\Phi_i,\Phi_j) = 0\,. \tag{4.3.17}$$

To do this, we need the action of $L_1$ on all the terms in eq. (4.3.15):

$$L_1(\partial^n[\Phi_i\Phi_j]_p) \;=\; (2h_i-p-1)\partial^n[\Phi_i\Phi_j]_{p+1} + n(2h_i+2h_j-2p+n-1)\partial^{n-1}[\Phi_i\Phi_j]_p\,, \tag{4.3.18}$$

which can be derived by commuting $L_1$ to the right. The result contains a term similar to eq. (4.3.7), and a term coming from the action of the $L_1$ on $[\Phi_i\Phi_j]_p$, see eq. (4.3.6). We find for eq. (4.3.17):

$$\sum_{n\geq 1}\partial^{n-1}[\Phi_i\Phi_j]_{m+n}\Big(a^m_{n-1}(h_i,h_j)\,(2h_i-n-m) + a^m_n(h_i,h_j)\,n\,(2h_i+2h_j-2m-n-1)\Big) \;=\; 0\,. \tag{4.3.19}$$

In the general case, this amounts to a recursive relation[4]:

$$a^m_n(h_i,h_j) \;=\; -\frac{2h_i-n-m}{n(2h_i+2h_j-2m-n-1)}a^m_{n-1}(h_i,h_j)\,. \tag{4.3.20}$$

Choosing $a^m_0$ equal to 1, gives:

$$a^m_n(h_i,h_j) \;=\; (-1)^n\frac{(2h_i-n-m)_n}{n!(2h_i+2h_j-2m-n-1)_n}\,. \tag{4.3.21}$$

**Intermezzo 4.3.2**

We can directly check that the coefficients given in eq. (4.3.21) indeed satisfy eq. (4.3.16). This involves the summation of terms which are a product of factorials. This summation, and most of the other sums that will be used in this chapter, can be done using the `Algebra‘SymbolicSum‘` package of *Mathematica*, which provides an implementation of the Gosper algorithm. However, the current version of this package does not handle the `Pochhammer` function. This can be remedied by converting the Pochhammer functions to a quotient of Γ-functions, eq. (2.A.1). The results returned by `SymbolicSum` in the case at hand contain a cosecans. This can again be converted to Γ-functions using the identity:

$$\sin(\pi x) = \frac{\pi}{\Gamma(x)\Gamma(1-x)}\,.$$

The result then contains terms like $\Gamma(-n)$ which are infinite for positive $n$. However, as all sums in this chapter are finite, these infinities necessarily disappear against terms like $\Gamma(2-n)^{-1}$. It is possible to write a set of *Mathematica* rules which checks this cancellation automatically by converting quotients of Γ-functions back to Pochhammer symbols. All sums in this chapter are computed in this way.

In the case at hand, eq. (4.3.16), the final result contains a factor:

$$-2\,n\,\sqrt{\pi}\,\Gamma(2\,n) + 4^n\,\Gamma(\frac{1}{2}+n)\,\Gamma(1+n)\,,$$

which is zero (except at the singularities), proving that eq. (4.3.21) provides the solution of eq. (4.3.16).

---

As the operator $QP^m(\Phi_j,\Phi_i)$ extracts the quasiprimary at the $m$-th order pole, we expect that reversing $i$ and $j$ does not give a new operator. Indeed, an explicit calculation gives[5]:

$$QP^m(\Phi_j,\Phi_i) = (-1)^{ij+m}\;QP^m(\Phi_i,\Phi_j)\,, \tag{4.3.22}$$

[4] Of course, when $\partial^{n-1}[\Phi_i\Phi_j]_{m+n}$ is proportional (up to null operators) to $\partial^n[\Phi_i\Phi_j]_{m+n+1}$, eq. (4.3.19) fixes only a combination of $a^m_n$ and $a^m_{n+1}$, but it is this combination that will appear in eq. (4.3.15).

[5] Ref. [26] contains this formula for $m=0,-1$.

where $ij$ in the phase factor gives a sign depending on the parity of the quasiprimaries ($-1$ if both are fermionic, 1 otherwise). This equation can be proven using eq. (2.3.16) and the identity:

$$\sum_{m=0}^{n} \frac{(-1)^m}{(n-m)!} \mathrm{a}_m^q(h_j, h_i) = \mathrm{a}_n^q(h_i, h_j)\,. \tag{4.3.23}$$

To conclude this subsection, let us discuss a special case where eq. (4.3.21) is not valid, namely when $n \geq n_c$, where we define:

$$n_c = 2(h_i + h_j - m) - 1\,. \tag{4.3.24}$$

We observe that the coefficient $\mathrm{a}_{n_c}^m(h_i, h_j)$ is not determined by eq. (4.3.19). On the other hand, eq. (4.3.15) shows that for such $n \geq n_c$, the $\mathrm{a}_n^m$ are coefficients of the $n$-th derivative of a operator with dimension $h \leq 0$. As discussed in the previous subsection,for any quasiprimary of dimension $h_k \leq 0$, $\partial^{1-2h_k}\Phi_k$ is also quasiprimary. This is clearly the origin of the freedom in $\mathrm{a}_{n_c}^m(h_i, h_j)$.
Let us take as an example $n = n_c$, and assume that there are no poles of order higher than $m + n_c$. In this case $[\Phi_i\Phi_j]_{m+n_c}$ is a quasiprimary of negative or zero conformal dimension. In this case, $QP^m(\Phi_i, \Phi_j)$ is indeed quasiprimary for arbitrary $\mathrm{a}_{n_c}^m(h_i, h_j)$.

## 4.4 Consequences of the full conformal group

In the previous section, the global conformal transformations and their implications on OPEs were studied. Here, we extend the analysis to include all modes $L_n$ of the energy–momentum tensor $T$, acting on primary operators. A primary operator $\Psi_i$ of dimension $h_i$ satisfies (def. 2.3.6):

$$\begin{aligned} L_n\Psi_i &= 0 \qquad\qquad n > 0 \\ L_0\Psi_i &= h_i \\ L_{-1}\Psi_i &= \partial\Psi_i\,. \end{aligned} \tag{4.4.1}$$

Similarly to the previous section, we use the following assumption.

**Assumption 4.4.1** *All elements of the OPA are linear combinations of the primaries $\Psi_i$ and their Virasoro descendants.*

Note that we included the unit operator $\mathbb{1}$ in the list of primaries which generate the OPA. The energy–momentum tensor $T$ is then a Virasoro descendant of $\mathbb{1}$:

$$L_{-2}\mathbb{1} = T\,. \tag{4.4.2}$$

### 4.4.1 Restrictions of conformal covariance on the OPEs

As in the previous section, assumption 4.4.1 implies that the OPE of two primary operators can be written as:

$$\Psi_i(z)\Psi_j(w) = \sum_k \sum_{\{p\}} C_{ij}{}^k\ \beta(h_i, h_j, h_k, \{p\})\ L_{-\{p\}}\Psi_k(w)\ (z-w)^{P-h_{ijk}}\,, \tag{4.4.3}$$

where the sum is over all ordered sequences $\{p\}$ and $P$ is the level of $\{p\}$ (see section 4.2 for notations). To rewrite eq. (4.4.3) in terms of composites with $T$, we use eq. (2.3.14):

$$L_{-p}\Psi = [T\Psi]_{2-p} = \frac{1}{(p-2)!}[\partial^{p-2}T\ \Psi]_0 \qquad p \geq 2\,. \tag{4.4.4}$$

The $\beta$-coefficients defined in eq. (4.4.3) depend only on the dimensions of the primaries. They were introduced in [13], where the first two levels where computed explicitly. The equations which determine these coefficients where explicitly written down in [74] using the inner product given in section 4.2. We now show how these equations can be derived independently of the inproduct, in a way entirely analogous to the previous section.

We will act with positive modes $L_n$ on eq. (4.4.3). For the *lhs*, the Jacobi identities (2.3.21) with $T$ imply:

$$L_p[\Psi_i\Psi_j]_n \ = \ h_j(p-1)[\Psi_i\Psi_j]_n + [(L_{-1}\Psi_i)\Psi_j]_{n+1}\,. \tag{4.4.5}$$

Because $\Psi_i$ is a primary operator, we have that $L_{-1}\Psi_i = \partial\Psi_i$, eq. (4.4.1), which gives together with eq. (2.3.13):

$$L_p[\Psi_i\Psi_j]_{h_{ijk}-P} = (h_ip - h_j + h_k + P - p)[\Psi_i\Psi_j]_{h_{ijk}-P+p}\,. \tag{4.4.6}$$

Extending this result to a partition $\{p\}$ of $P$, we find:

$$L_{\{p\}}\left([\Psi_i\Psi_j]_{h_{ijk}-P}\right) = f(h_i, h_j, h_k, \{p\})[\Psi_i\Psi_j]_{h_{ijk}}\,, \tag{4.4.7}$$

where

$$f(h_i, h_j, h_k, \{p\}) \equiv \prod_l \left(h_ip_l - h_j + h_k + \sum_{l'>l} p_{l'}\right)\,. \tag{4.4.8}$$

On the other hand, the action of $L_{\{p\}}$ on the *rhs* of eq. (4.4.3) can be computed by commuting the positive modes to the right, where they annihilate $\Psi_k$. The result depends on the dimensions of the fields, but also on the central charge $c$ appearing in the commutators of the Virasoro algebra, eq. (2.4.5). No easy formula can be given for the resulting expression, we write:

$$L_{\{m\}}L_{-\{n\}}\Psi_k = S^{\{m\},\{n\}}(h_k, c)\ \Psi_k\,, \tag{4.4.9}$$

for partitions of the same level N. The matrix $S$ is the Šapovalov form defined in section 4.2. The dimension of $S$ is given by the number of (ordered) partitions of $N$, which we denote by $p(N)$.

**Intermezzo 4.4.1**
As an example we give the matrix $S$ at level 2:

$$\begin{array}{rcl|rcl} L_2 L_{-2}\Psi &=& (4h+c/2)\Psi & L_2 {L_{-1}}^2\Psi &=& 6h\Psi \\ {L_1}^2 L_{-2}\Psi &=& 6h\Psi & {L_1}^2 {L_{-1}}^2\Psi &=& 4h(2h+1)\Psi\,, \end{array}$$

where $\Psi$ is a primary with dimension $h$. $S$ is a symmetric matrix. This follows from the fact that the Virasoro commutators (2.4.5) are invariant under the substitution $L_n \to L_{-n}$.

---

Assuming independence of the Virasoro descendants, we get:

$$\sum_{\{m\}} S^{\{n\},\{m\}}\beta(h_i,h_j,h_k,\{n\}) \;=\; f(h_i,h_j,h_k,\{n\})\,, \qquad (4.4.10)$$

where the sum is over all partitions of $N$. This equation determines the $\beta$-coefficients at level $N$ completely, unless the matrix $S$ is singular. The singular vectors of $S$ correspond to primary Virasoro descendants. Clearly, their coefficients cannot be determined by the Jacobi identities with $T$. This corresponds to the poles in the $\beta$-coefficients.

**Intermezzo 4.4.2**
From the previous intermezzo, we can give the $\beta$-coefficients at level two:

$$\begin{aligned} \beta(h_i,h_j,h_k,\{1,1\}) \;=\;& \Big(c\,(h_i-h_j+h_k)\,(1+h_i-h_j+h_k)+ \\ & 4h_k\big(-4h_i+2{h_i}^2+h_j-4h_ih_j+2{h_j}^2- \\ & h_k+4h_ih_k-4h_jh_k+2{h_k}^2\big)\Big)\Big/ \\ & \big(4h_k\,(c\,(1+2h_k)+2h_k\,(-5+8h_k))\big) \\ \beta(h_i,h_j,h_k,\{2\}) \;=\;& \big(h_i-3{h_i}^2+h_j+6h_ih_j-3{h_j}^2-h_k \\ & +2h_ih_k+2h_jh_k+{h_k}^2\big)\Big/ \\ & \big(c\,(1+2h_k)+2h_k\,(-5+8h_k)\big). \end{aligned}$$

These expressions are already complicated. The coefficient of ${L_{-1}}^2\Psi_k$ is undetermined when $h_k=0$ because $\partial\Psi_k$ is a primary of dimension 1 in this case. Also, when

$$c=\frac{2\,(5-8\,h_k)\,h_k}{1+2\,h_k}$$

the quasiprimary combination:

$$[T\Psi_k]_0-\frac{3\partial^2\Psi_k}{2(1+2h_k)}$$

turns out to be primary, which explains the second pole in the $\beta$-coefficients.

---

When trying to work out the $\beta$-coefficients at higher level, the main problem lies in solving eq. (4.4.10). Standard numerical algorithms cannot be used, as the matrix $S$ contains the constant $c$, which we want to keep as a free parameter. Also, the dimension $p(N)$ of the matrix increases rapidly with the required level, which makes solving eq. (4.4.10) for high levels unfeasible. In the next subsection, we will work in a basis of quasiprimaries. This will increase the complexity of computing $S$, but brings it to block-diagonal form, making the solution of equation eq. (4.4.10) much easier. Also, the primary descendants are of course necessarily quasiprimaries, so they are much easier to find in this basis.

We wish to stress that this analysis for Virasoro-primary operators cannot be extended to highest weight operators. Indeed, to determine the coefficients of the Virasoro descendants in an OPE, we explicitly used extra information about $L_{-1}$ on a primary operator, namely that $L_{-1}\Psi=\partial\Psi$ (see eq. (4.4.5)). As we do not have such information at present for nonlinear algebras, it is not yet possible to write down $\mathcal{W}$-covariant OPEs. See also section 4.1.1 in [177].

### 4.4.2 Virasoro descendants in a quasiprimary basis

Our aim in this subsection is to use the information of section 4.3 to bring the system of equations in eq. (4.4.10) in a block-diagonal form. This will be done by changing to a basis of quasiprimary Virasoro descendants and their derivatives. This basis is defined in terms of operators $\tilde{L}_m$ and $L_{\pm1}$, where:

$$\tilde{L}_m\Psi \equiv QP^{m+2}(T,\Psi)\,, \qquad (4.4.11)$$

for $\Psi$ quasiprimary (see eqs. (4.3.15) and (4.3.21))[6] The definition of $\tilde{L}_m$ depends on the dimension of the operator on which it is acting. This can be avoided by using $L_0$. However, in the normalisation we chose, one needs to introduce factors $(2L_0+i)^{-1}$. To avoid this, we will write $\tilde{L}_m(h)$ when confusion can arise.

The action of $\tilde{L}_m$ is well-defined for any quasiprimary $\Phi$. When $m<-1$, the sum in eq. (4.3.15) contains only a finite number of terms because $a_n^{m+2}(2,h_j)=0$ for $n\geq 2-m$. When $m>1$, $L_{n+m}\Phi=[T\Phi]_{n+m+2}$ is zero for $n$ large enough. In particular, when $\Phi$ is a level $N$ descendant of a primary operator, $L_{n+m}\Phi=0$ if $n+m>N$.

---

[6] These operators were also defined in [131], but apparently not used.

We will denote partitions which do not contain the number 1 as $\{\tilde{n}\}$, and write $p_1(N)$ for the number of such partitions. Note that all partitions of level $N$ which do contain a 1 can be obtained by adding a 1 to the normal partitions of level $N-1$. This implies $p_1(N) = p(N) - p(N-1)$.

The descendants

$$L_{-1}{}^{N-N_k}\tilde{L}_{-\{\tilde{n}_k\}}\Psi_j \tag{4.4.12}$$

(where the partition $\{\tilde{n}_k\}$ has level $N_k$) for $N_k = 0, 2, 3, \dots N$ span the space of all Virasoro descendants at level $N$. When no primary descendants exist, they are independent.

Once we know the coefficients of the quasiprimaries at all levels $M \le N$, we can use the results of the previous section to find the complete OPE:

$$\Psi_i(z)\Psi_j(w) = \sum_k \sum_{\{\tilde{n}\}} C_{ij}{}^k\ \tilde{\beta}(h_i, h_j, h_k, \{\tilde{n}\})$$

$$\sum_{m\ge 0} \alpha(h_i, h_j, h_k + N, m)\ \partial^m \tilde{L}_{-\{n\}}\Psi_k(w)\ (z-w)^{N+m-h_{ijk}}\,. \tag{4.4.13}$$

To determine the $\tilde{\beta}$-coefficients, we proceed as before, after a projection on the quasiprimaries in eq. (4.4.13). We act with a sequence of $\tilde{L}_m$ operators, $m \ge 2$, on eq. (4.4.13). For the *rhs*, we need the matrix $\tilde{S}$ which appears in:

$$\tilde{L}_{\{\tilde{m}\}}\tilde{L}_{-\{\tilde{n}\}}\Psi_k = \tilde{S}^{\{\tilde{m}\},\{\tilde{n}\}}(h_k, c)\ \Psi_k\,, \tag{4.4.14}$$

To compute $\tilde{S}$, we have to know the commutation rules for the $\tilde{L}_n$ operators. We will need only the case where we commute a positive mode through a negative mode. We find for $m, n > 1$:

$$\tilde{L}_m(h+n)\tilde{L}_{-n}(h) \;=\; \delta_{m-n} f_1(h,m) + \tilde{L}_{m-n} f_2(h, m, -n)$$

$$+ \sum_{p \ge 2-\min(n,m)} \tilde{L}_{-n-p}(h-m-p)\tilde{L}_{m+p}(h)\ f_3(h, m, -n, p)\,, \tag{4.4.15}$$

where we take $\tilde{L}_{\pm 1} = 0$ and $\tilde{L}_0 = L_0$. The coefficients $f_i$ appearing in eq. (4.4.15) are quite involved and are given in appendix 4.A, see eqs. (4.A.10, 4.A.16). The sum in eq. (4.4.15) contains an infinite number of terms, but we have to keep only the terms with $p + m \le N$ when acting on a Virasoro descendant of level $N$.

Applying $\tilde{L}_{\{\tilde{m}\}}$ to the *lhs* of eq. (4.4.13), after projecting on the quasiprimaries, we get the analogue of eq. (4.4.7):

$$\tilde{L}_{\{\tilde{n}\}}\left(QP^{h_{ijk}-\tilde{N}}(\Psi_i, \Psi_j)\right) = \tilde{f}(h_i, h_j, h_k, \{\tilde{n}\}) QP^{h_{ijk}}(\Psi_i, \Psi_j)\,, \tag{4.4.16}$$

where the $\tilde{f}(h_i, h_j, h_k, \{\tilde{n}\}$ are given in the appendix, eq. (4.A.25).

We have the final equation:

$$\sum_{\{\tilde{m}\}} \tilde{S}^{\{\tilde{n}\},\{\tilde{m}\}}\tilde{\beta}(h_i, h_j, h_k, \{\tilde{n}\}) \;=\; \tilde{f}(h_i, h_j, h_k, \{\tilde{n}\})\,, \tag{4.4.17}$$

where the sum is over all partitions of $N$, not containing the number 1.

**Intermezzo 4.4.3**

As an example, at level 2 we find:

$$\tilde{S}^{\{2\},\{2\}}(h,c) \;=\; \frac{c - 10h + 2ch + 16h^2}{2\,(1+2h)}\,. \tag{4.4.18}$$

This gives for $\tilde{\beta}(h_i, h_j, h_k, \{2\})$ immediately the result of intermezzo 4.4.2. In contrast to when using the $L_{-\{n\}}$ basis, no problem of inverting $\tilde{S}$ occurs when we consider a dimension 0 operator[7]. We find:

$$\tilde{\beta}(h_i, h_j, 0, \{2\}) = \frac{h_i - 3h_i^2 + h_j + 6h_ih_j - 3h_j^2}{c}\,, \tag{4.4.19}$$

which reduces to $2h_i/c$ for $h_j = h_i$.

A particular example of a dimension zero operator is the unit operator $\mathbb{1}$. We see that the conventional normalisation of a primary operator of dimension $h$ with OPE $\Psi(z)\,\Psi(w) = c/h(z-w)^{-2h} + \dots$ gives as level two descendant simply $2T$, *i.e.* with a $c$-independent coefficient.

---

The complexity of the coefficients in eqs. (4.4.15) and (4.4.16) makes these formulas unsuited for pen and paper calculations, but is no problem when using a symbolic manipulation program. As discussed in intermezzo 4.4.4, it turns out that the calculation of $S$ takes even more time than needed for $\tilde{S}$. Actually, we are more interested in the quasiprimary basis, so we do not need $S$ at all. Even more important for computer implementation [194] is that we reduced the dimension of the system of equations eq. (4.4.17) from $p(N)$ to $p_1(N) = p(N) - p(N-1)$. That this is a significant simplification is easily seen on a few examples. For levels 2 through 8, $p(N)$ is $\{2,3,5,7,11,15,22\}$ while $p_1(N)$ is $\{1,1,2,2,4,4,7\}$.

**Intermezzo 4.4.4**

As an illustration of the above arguments, we will present some timings when using an implementation in *Mathematica* of the formulas in this and the previous section. Timings were done in *Mathematica* 2.2 running on a 486 (50 MHz).

To compare the computation of the $S$ and $\tilde{S}$ matrices, we compute the case of an OPE

[7] $\tilde{f}(h_i, h_j, h_j - h_k - 2, \{2\})$ as given in eq. (4.A.25) does have an apparent factor $h_k$ in the denominator, but this is a consequence of writing $\tilde{f}$ in a closed form. No factor $h_k^{-1}$ appears in the definition of $\tilde{f}(h_i, h_j, h_j - h_k - 2, 2)$, see eq. (4.A.23). Indeed, the factor $h_k^{-1}$ disappears after factorisation.

where descendants of a primary occur up to level 7, *i.e.* we need the matrices for level 1 to 7. As the matrices are computed recursively, we use an algorithm that stores and reuses its results. For $S$, the total time needed is $20s$, while for $\tilde{S}$ it is only $4s$. For level 9, the timings become $110s$ and $40s$ respectively. When going to even higher level, the difference in timing decreases. However, for the most complicated algebras explicitly constructed up to now (*e.g.* $\mathcal{W} A_5$ in [113]) the maximum level of a descendant of a primary (not equal to the unit operator) is 8.

The advantage of using the quasiprimary basis shows up even more when solving the equations for the $\beta, \tilde{\beta}$-coefficients, eqs. (4.4.10) and (4.4.17). We computed the coefficients for $h_i = 6, h_j = 7, h_k = 10$ and $c$ arbitrary. Although the equations we need to solve are linear equations, serious problems occured for moderately high level.
The *Mathematica* (version 2.2 or lower) function `Solve` has a serious deficiency in that it does not simplify intermediate results. This means that when arbitrary constants are present in the equations, the solution is usually a quite big expression which simplifies drastically after factorisation. Unfortunately, this factorisation can take a very long time. For even more complicated equations `Solve` runs into problems because the intermediate results grow too fast. K. Hornfeck and I wrote a separate package to solve linear equations using Gaussian elimination, simplifying the coefficients of the variables at each step.
We find the following CPU-times for the solution of the equations (4.4.10) and (4.4.17).

| level | $\beta$ with `Solve` | $\beta$ with `LinSolve` | $\tilde{\beta}$ with `Solve` | $\tilde{\beta}$ with `LinSolve` |
|---|---|---|---|---|
| 4 | 6.3s | 4.7s | 1.6s | 1.9s |
| 5 | * 18.4s | 14.5s | 1.6s | 1.9s |
| 6 | * > 400s | 69.6s | * 15.3s | 54.6s |

In this table, a star means that factorisation of the result of `Solve` did not succeed in a reasonable amount of time.

---

As before, eq. (4.4.17) only determines the $\tilde{\beta}$-coefficients when $\tilde{S}$ is nonsingular. Clearly, any singular vectors of $\tilde{S}$ correspond to primary descendants, whose coefficients remain unfixed by Jacobi identities with $T$. These free coefficient will appear automatically when solving the linear equations.

### 4.4.3 Virasoro descendants of the unit operator

We now consider the case of the Virasoro descendants of the unit operator, which is a primary of dimension zero. In this case, it turns out that the matrix $\tilde{S}$ has a large amount of singular vectors, even for general $c$. We usually need the descendants of $\mathbb{1}$ to a much higher level, because most $\mathcal{W}$–algebras studied up to now have primary generators with dimension larger than 2. This means we should reduce the number of (linear) equations in eq. (4.4.17) as much as possible when computing $\alpha(h_i, h_j, 0, m)$.

The singular vectors of $\tilde{S}$ for a dimension zero field are in general primary fields. For the unit operator however, the singular vectors are exactly zero. There is thus no use in keeping them in the calculation. This provides an additional reason for studying the singular vectors of $\tilde{S}$ for dimension zero.

Some examples of these singular vectors are easily constructed. Obviously we have that $L_{-1}\mathbb{1} = \partial\mathbb{1} = 0$. $\tilde{L}_{-n}\mathbb{1}$ is zero for $n > 2$ as well. Indeed, the definition (4.3.2) shows that $\tilde{L}_{-n}\mathbb{1}$ is proportional to $\partial^{n-2}T$, which is clearly not quasiprimary unless $n = 2$, hence it is zero for $n \neq 2$. One can also prove that $\tilde{L}_{-n}T$ is zero for odd $n$. More complicated relations at higher level exist.

Before discussing how to find a basis for the quasiprimary descendants of $\mathbb{1}$, let us determine the number of independent quasiprimaries at level $N$, which we will denote as $p_2(N)$. First, the number of independent descendants is simply $p_1(N)$. Indeed, when ordering the $L_{-\{n\}}$ as before, the $L_{-1}$ will act on $\mathbb{1}$ first, giving zero. Hence, all partitions containing 1 should be dropped. For general central charge, no further relations between the remaining descendants exist. To compute the number of quasiprimary descendants of $\mathbb{1}$ at level $N$, we take all descendants, and "substract" those which are derivatives of the descendants at the previous level, *i.e.* $p_2(N) = p_1(N) - p_1(N-1)$.

We can use this information to find other singular vectors than the ones given already. Because $p_2(7) = 0$, there are no descendants at level 7, which means that $\tilde{L}_{-3}\tilde{L}_{-2}\tilde{L}_{-2}\mathbb{1}$ is zero, which can be checked explicitly. At level 9, there is only one independent quasiprimary. Eliminating the zeroes we found at previous levels, two candidates remain. We find:

$$\tilde{L}_{-\{3,2,2,2\}}\mathbb{1} = -\frac{8}{5}\tilde{L}_{-\{5,2,2\}}\mathbb{1}\,. \tag{4.4.20}$$

Similar relations exist at higher (not necessarily odd) levels.

Remembering the rule for the number $p_2(N)$ of independent quasiprimary descendants of $\mathbb{1}$, we propose[8]:

**Conjecture 4.4.1** *The partitions which give the independent descendants of the unit operator at level $N$ can be found as follows:*

- *Take the partitions (not containing the number 1) at level $N-1$ and order them in increasing lexicographic order ($\{32\} < \{33\}$).*
- *"Increment" these partitions one by one, such that no partition of level $N$ is obtained twice.*
- *The Virasoro descendants of $\mathbb{1}$ at level $N$ correspond to partitions at level $N$ which are not in the list constructed in the previous step.*

---

[8] This conjecture was established together with K. Hornfeck after suggestions of M. Flohr. In [83], the same conjecture is made, but not in detail. Actually, [83] claims to have proved the conjecture, but his argument seems to be solely based on counting the number of independent descendants.

With "incrementing" a partition $\{n_1, n_2, \ldots n_k\}$, we mean adding a 1 at a position $i$ where $n_{i-1} \geq n_i + 1$. As an example, incrementing $\{3, 2, 2\}$ gives $\{3, 3, 2\}$ or $\{4, 2, 2\}$.

The first nontrivial example of this procedure is at level 9. At level 8 the partitions are $\{2222\ 332\ 44\ 53\ 62\ 8\}$. Incrementing the first three partitions gives $\{3222\ 333\ 54\}$. We now have to increment the partition $\{53\}$. The partition $\{54\}$ is already in our list, so we should take $\{63\}$. In the next steps $\{72\}$ and $\{9\}$ are found. The only partition of level 9 that remains is $\{522\}$, in agreement with eq. (4.4.20).

This conjecture was checked up to level 16.

### 4.4.4 Finding the primaries in an OPE

Entirely analogously to eq. (4.3.13), we can define:

$$P^m(\Psi_i, \Psi_j) = QP^m(A, B) - \sum_{\{\tilde{n}\}} \tilde{\beta}(h_i, h_j, h_i + h_j - m - N, \{n\})\ \tilde{L}_{-\{\tilde{n}\}} P^{m+N}(\Phi_i, \Phi_j)\,. \tag{4.4.21}$$

This can be rewritten as:

$$P^m(\Psi_i, \Psi_j) = \sum_{\{\tilde{n}\}} a^m_{\{\tilde{n}\}}(h_i, h_j)\ \tilde{L}_{-\{\tilde{n}\}} QP^{m+N}(\Psi_i, \Psi_j)\,, \tag{4.4.22}$$

where the sum is over all partitions of level not equal to one. We have not found a closed form for this expression. Even determining the recursion relations for the coefficients $a^m_{\{\tilde{n}\}}(h_i, h_j)$ seems unfeasible in general.

In many cases, the action of $P^m$ will give zero, as there simply is no primary at a certain pole.

The operators $P^m$ are particularly convenient to construct primary composites. Indeed, for $m \geq 2$, the operator $P^{-m}(\Psi_i, \Psi_j)$ contains $[\Psi_i \Psi_j]_{-m} \sim [\partial^m \Psi_i\ \Psi_j]_0$. In [201], a formula is given to count the number of primaries that can be constructed.

## 4.5 An overview of $\mathcal{W}$–algebras

In this section, we present a overview on the $\mathcal{W}$–algebras which are known. For a more in-depth review, see [31], which we also follow for the nomenclature of the $\mathcal{W}$–algebras.

We first distinguish different classes of $\mathcal{W}$–algebras. "Deformable" algebras are algebras that exist for generic values of the central charge. $\mathcal{W}$–algebras which are only associative for specific values of $c$ are called non-deformable. In a "freely generated" $\mathcal{W}$–algebra, for generic $c$ no relations exist between the generators, *i.e.* no null operators appear.

Up to now, all efforts have been concentrated on algebras in the following class:

**Assumption 4.5.1**

- *The OPA is generated by a set of primary operators and the Virasoro operator.*
- *All generators have strictly positive conformal dimension.*
- *The unit operator occurs only in OPEs between primaries of the same conformal dimension.*
- *Generators which are null operators are discarded.*

Not all $\mathcal{W}$-algebras of interest fall within this class. For instance, it was recently shown [136] that the (nonlinear) $\mathcal{W}_3$ algebra and the Bershadsky algebra [160, 21] with bosonic generators of dimensions $(2, 3/2, 3/2, 2)$ can be viewed as subalgebras of *linear* algebras with a null operator as generator, and a non-primary dimension 1 operator. Yet even with the restrictions 4.5.1, no complete classification has been found yet, see [62].

The current status in our knowledge of $\mathcal{W}$–algebras can be compared to the study of Lie algebras before Cartan presented his classification. Different construction methods of $\mathcal{W}$–algebras are known, giving rise to series of $\mathcal{W}$–algebras with common features. In a sense, they are analogous to the "classical Lie algebras", which were also known via a realisation. A few isolated cases have also been constructed, although they currently seem to fit in the general pattern.

We will now comment on some of the more important construction methods and end with the (incomplete) classification proposed in [34].

### 4.5.1 Direct construction

One can start with a list of primaries of a certain dimension and attempt to construct a $\mathcal{W}$–algebra. One has then to determine the structure constants of the primaries. Several methods are in use to check that the OPA is associative.

A first method is based on the claim in [13] that it is sufficient to check whether all three- and four-point functions are crossing symmetric. In the perturbative conformal bootstrap method [13, 29, 76], one analyses:

$$G^{lk}_{nm}(x) \equiv \lim_{\substack{z\to\infty \\ \varepsilon\to 0}} z^{2h_k} \langle \Psi_j(z)\Psi_l(1)\Psi_n(x)\Psi_m(\varepsilon)\rangle\,. \tag{4.5.1}$$

The structure constants are then constrained by requiring crossing symmetry of the $G^{lk}_{nm}(x)$. This can be checked perturbatively around $x = 0$. The crossing symmetry constraints can be implemented using a group theoretical method due to Bouwknegt [29], that was generalised in [76].

The equations resulting from this method simplify drastically when considering the $c \to \infty$ limit. This is used in [201] to check in a very efficient way if an algebra exists in this limit, which can be considered to be a necessary condition for the algebra to exist at generic values of the central charge $c$ (see also subsection 4.5.5). Of course, this method gives the structure constants only in the large $c$ limit.

Another approach is to check the Jacobi identities with the help of the mode algebra. In [33], the necessary formulas are given to compute the commutator of two quasiprimaries in a basis of modes of quasiprimaries. This was used in [26, 131, 129] to explicitly construct the $\mathcal{W}$–algebras with a small number of primary operators of low dimensions.

Finally, one can use OPEs to construct the OPA, and use the formulas of subsection 2.3.2 (in particular eq. (2.3.21) for $q, p > 0$) to check the associativity. In [113] *OPEdefs* was used to construct all algebras with primary generators with dimensions $3, 4, 5$ and $3, 4, 5, 6$. These are the most complicated $\mathcal{W}$–algebras explicitly constructed up to now.

It is clear that none of these methods can give a classification. Direct construction is however the only known way that can give an exhaustive list of algebras with a certain operator-content.

### 4.5.2 Subalgebras of known $\mathcal{W}$–algebras

Any subalgebra of an OPA trivially satisfies the associativity conditions. Two main methods exist to construct subalgebras. First, we can factor out any free fermions and bosons, as is discussed in chapter 5. This is a powerful result, as the classification can now be restricted to $\mathcal{W}$–algebras where such operators are not present.

A second method to construct a subalgebra is by "orbifolding". If a certain discrete symmetry exists in the $\mathcal{W}$–algebra, the elements which are invariant under this symmetry necessarily form a subalgebra. As an example, the $N = 1$ superconformal algebra is invariant under a sign change of the supersymmetry generator $G$. The corresponding subalgebra is purely bosonic and is generated by $T$ and two more operators of dimension 4 and 6 respectively [111, 130].

Deformable $\mathcal{W}$–algebras will contain null operators at a certain value of $c$. This gives rise to a quotient algebra. In many cases, this algebra is only associative for this value of $c$. It is a common belief that all non-deformable algebras can be obtained in this way. An example of this mechanism will be given in section 4.6. Recently, it has been found that in some cases, a new deformable algebra results [27]. In the quantum case, these new algebras are finitely but non-freely generated, *i.e.* contain null operators for all values of $c$. In the classical case, an infinite number of generators is needed to construct the complete OPA. A particular feature is that the $\mathcal{W}$–algebras based on Lie-algebras in the same Cartan series (see next subsection) seem to contain the same subalgebra of this new type (for different values of $c$). The study of these "unifying" $\mathcal{W}$–algebras is based on a formula for the structure constants of the operators of the lowest dimensions in the $\mathcal{W}$–algebra [114].

### 4.5.3 Constructing a $\mathcal{W}$–algebra via a realisation

If an algebra is found that is realised in terms of the generators of an associative algebra, the Jacobi identities are automatically satisfied. One has to prove that the algebra closes on a finite number of operators.
The main advantage of using a representation is that the representation theory of the underlying OPA can be used to study the representations of the $\mathcal{W}$–algebra itself. As we will discuss below, a whole series of $\mathcal{W}$–algebras has been found via realisations in a Kač–Moody algebra. Using the representation theory of the Kač–Moody algebras, the minimal models for the corresponding $\mathcal{W}$–algebras have been completely characterised [31].

A first example is given by the coset construction [99, 6]. For a Kač–Moody algebra $\hat{g}$ of level $k$ which has a subalgebra $\hat{g}'$, one defines the coset algebra $\mathcal{W}_c\,[\hat{g}/\hat{g}', k]$ as the set of elements in $\hat{g}$ which commute with the elements of $\hat{g}'$. The structure of the algebra generically depends on $k$, due to the appearance of null operators at specific levels. For certain cases, a representation of the same deformable algebra is found for all $k$, *e.g.* for a diagonal coset $\mathcal{W}_c\,[\hat{g} \oplus \hat{g}/\hat{g}_{\text{diag}}, k]$.

One can define a second type of coset algebras by considering all elements in $\hat{g}$ which commute with the originating Lie algebra $g$. These algebras can be obtained in a certain limit from the diagonal coset algebras [6]. For a simply-laced Lie algebra $\bar{g}$ and at level $k = 1$, the resulting $\mathcal{W}$–algebras are the same as the Casimir algebras defined in [5]. This is proven in [31] using character techniques. For non-simply laced algebras, the situation is more complicated. Nevertheless, one still refers to $\mathcal{W}_c\,[\hat{g}/g, k]$ as the Casimir algebra of $\hat{g}$ at level $k$.
As an example, the Casimir algebra for $B_n$ at level 1 turns out to be the bosonic projection of an algebra generated by the Casimir operators of $\hat{B}_n$ and a fermionic operator of dimension $n + 1/2$. For convenience, we call the underlying $\mathcal{W}$–algebra, the Casimir algebra of $\hat{B}_n$, $\mathcal{W}_c B_n$. Section 4.6 contains an explicit construction of $\mathcal{W}_c B_2$.

A third method to construct realisations of $\mathcal{W}$–algebras is by using Drinfeld-Sokolov reduction. This method was developed for classical $\mathcal{W}$–algebras in [60, 7], quantisation is discussed in chapter 6. Given a (super) Kač–Moody algebra, one chooses an $sl(2)$ embedding. One then puts certain constraints on the currents of the Kač–Moody algebra. These constraints are characterised by the $sl(2)$. As such, one finds a realisation of a finitely generated $\mathcal{W}$–algebra for any $sl(2)$-embedding of a (super) Lie algebra. The dimensions of the operators are given by $j_\alpha + 1$, where the adjoint representation of the Lie algebra contains the $sl(2)$ -irreducible representations $\underline{j_\alpha}$. Clearly, no dimension 1/2 operators can be present in a $\mathcal{W}$–algebra arising from Drinfeld-Sokolov reduction. This is no severe restriction, as it is shown in chapter 5 that any $\mathcal{W}$–algebra can be written as the direct product of some free operators with a $\mathcal{W}$–algebra without dimension 1/2 operators.

All known finitely generated $\mathcal{W}$–algebras in the class 4.5.1 can be obtained from Drinfeld-Sokolov reduction, either directly or by applying the methods of the previous subsection, see [71]. As an example, it is argued in [31] that, for simply laced algebras, the diagonal coset algebras $\mathcal{W}_c\ [\hat{g} \oplus \hat{g}/\hat{g}_{\rm diag}, k]$ are the same as the Drinfeld-Sokolov reductions $\mathcal{W}_{DS}[\hat{g}, k']$ if one takes the principal $sl(2)$ embedding of $g$[9]. For non-simply laced algebras, the relation between coset and DS-algebras is more complicated. As an example, one finds at least classically, that $\mathcal{W}_c\, B_n = \mathcal{W}_{DS}[B(0,n), k]$ [121, 205], for generic $n$, see also section 4.6.

Before the Drinfeld-Sokolov reduction was quantised, several other attempts have been made to find a method to construct realisations of quantum $\mathcal{W}$–algebras. We mention only the work of Fateev and Lukyanov [63, 64, 65], which is based on a quantisation of the Miura transformation. This transformation relates different gauge choices in the constrained phase space, and gives a realisation in terms of "simpler" operators [45], in particular in terms of free operators for the simply laced algebras. In section 4.6, a free operator realisation of $\mathcal{W}_c\, B_2$ is constructed which is shown to correspond to the Fateev-Lukyanov construction.

### 4.5.4 Superconformal algebras

For superconformal algebras, the extra structure given by the supersymmetry transformations puts strong restrictions on the number of primary operators in the theory. By studying the Jacobi identities for the linear superconformal algebras with nonzero central extension, we were able to classify all possible *linear* algebras with generators of positive dimension [109]. The only such algebras that exist have a number of supersymmetry generators $N \leq 4$. The $N = 3$ and the "large" $N = 4$ algebras appear in section 5.3. The $N = 1, 2, 3$ and the "small" $N = 4$ algebra are all subalgebras of the large $N = 4$. The superconformal algebras with quadratic nonlinearity were classified in [85].

### 4.5.5 Attempts towards a classification

The methods which rely on a construction of a $\mathcal{W}$–algebra, either directly or via a realisation, can give no clue if all $\mathcal{W}$–algebras are obtained. In [34], a first attempt is made to classify all classical positive-definite $\mathcal{W}$–algebras, which are algebras in the class 4.5.1 with the additional condition that the central extensions define a positive-definite metric. It is proven that the finite-dimensional algebra defined by the linearised commutators of the "vacuum preserving modes" $(\Psi_i)_m$ for $|m| < h_i$, is the direct sum of a semisimple Lie algebra with an abelian algebra[10]. Moreover, this Lie algebra necessarily contains an $sl(2)$ formed by the $-1, 0, +1$ modes of $T$. The vacuum preserving modes of $\Psi_i$ form a spin $h_i - 1$ representation of the $sl(2)$. This means that by classifying all possible $sl(2)$-embeddings, the dimensions which can occur in a classical positive-definite $\mathcal{W}$–algebra are obtained. Ref. [34] then proceeds by investigating under which conditions a quantum $\mathcal{W}$–algebra has a positive-definite classical limit ($c \to \infty$).

The algebras arising from Drinfeld-Sokolov reduction (both quantum and classical) satisfy the criteria of [34], proving that at least one algebra exists for every $sl(2)$-embedding. It is not proven that this algebra is unique. Moreover, for many algebras constructed by orbifolding no classical limit can be found, and hence they fall outside this attempt towards classification. For more details, see [70, 43].

[9] The principal $sl(2)$ of $g$ is defined by taking for $e_\pm$ the sum of all positive (resp. negative) simple roots.

[10] Free fermions have dimension 1/2 and hence do not contribute to this Lie algebra.

## 4.6 An example : $\mathcal{W}_c\, B_2$

In this section we will construct $\mathcal{W}_c\, B_2$, the Casimir algebra of $\hat{B}_2$. We refer to subsection 4.5.3 for the terminology. This algebra contains, besides $T$, an extra dimension 4 operator and a fermionic dimension 5/2 operator. The algebra will be written down in terms of quasiprimary families. We will construct the most general realisation of this algebra with two free bosons and one free fermion and show that it is equivalent to the realisation proposed by Fateev and Lukyanov [65]. Finally, using this realisation, we check that the screening operators are related to the long and short root of $B_2$ leading to the degenerate representations.

### 4.6.1 The $\mathcal{W}_c B_2$- algebra

The Lie algebra $B_2$ has two independent Casimir operators of order two and four. The energy–momentum tensor $T$ of $\mathcal{W}_c B_2$ corresponds to the second order Casimir. The $\mathcal{W}$–algebra contains also a Virasoro primary operator $W$ of dimension 4, which corresponds to the fourth order Casimir. Although there exists a deformable OPA with only this operator content [29, 108], this is not yet the Casimir algebra of $B_2$, see subsection 4.5.3. To get $\mathcal{W}_c\ B_2$, one has to introduce a primary weight 5/2 operator $Q$. This is reminescent of the fact that the Frenkel-Kač level 1 realisation of the non-simply laced affine Lie algebras $\widehat{B}_n$, requires, due to the short root, the introduction of an extra fermion [100].

The OPA of $\mathcal{W}_c B_2$ can be written schematically as:

$$\begin{aligned} Q \times Q &\longrightarrow \frac{2c}{5}\ [\mathbb{1}] + C_{\frac{5}{2}\frac{5}{2}}{}^{4}\ [W], \\ W \times W &\longrightarrow \frac{c}{4}[0] + C_{44}{}^{4}\ [W] + C_{44}{}^{6}\ [\Psi], \\ Q \times W &\longrightarrow C_{\frac{5}{2}4}{}^{\frac{5}{2}}\ [Q], \end{aligned} \tag{4.6.1}$$

with $[\mathbb{1}]$ the conformal family of the identity. From the Jacobi identities we find the symmetry property $C_{\frac{5}{2}\frac{5}{2}}{}^{4} = \frac{8}{5} C_{\frac{5}{2}4}{}^{\frac{5}{2}} = \frac{8}{5} C_{4\frac{5}{2}}{}^{\frac{5}{2}}$. The algebra generated by $\{T, W\}$, for ${C_{44}}^6 = 0$, is well known to be associative for all values of $c$ [29]. Here we discuss the solution where $\Psi$ is the dimension 6 Virasoro primary $\Psi \propto P^{-1}(Q,Q) = \partial QQ +$ *corrections*, see eq. (4.4.22). Using Jacobi identities (2.3.21) or the conformal bootstrap, we find that the algebra is associative for all values of the central charge[11] provided the couplings are given by:

$$\begin{aligned}
{C_{\frac52\frac52}}^4 &= \sqrt{\frac{6(14c+13)}{5c+22}}\epsilon_1,\\
{C_{44}}^4 &= \frac{3\sqrt{6}(2c^2+83c-490)}{\sqrt{(14c+13)(5c+22)}(2c+25)}\epsilon_1,\\
{C_{44}}^6 &= \frac{12\sqrt{5(6c+49)(4c+115)(c-1)}(5c+22)}{\sqrt{(14c+13)(7c+68)(2c-1)(c+24)}(2c+25)}\epsilon_2,
\end{aligned} \tag{4.6.2}$$

where $\epsilon_1$ and $\epsilon_2$ are arbitrary signs.

For $c = -13/14$ the subalgebra generated by $\{T, Q\}$ corresponds to the spin 5/2 algebra of Zamolodchikov [211]. This is an example of a non-deformable $\mathcal{W}$–algebra.

As an application of sections 4.3 and 4.4, we now present the more complicated OPEs:

$$Q \times Q = \ll \frac{2c}{5} \mid 0 \mid 2T \mid \partial T \mid \frac{3}{10}\partial^2 T + \frac{27}{5c+22}\Lambda + {C_{\frac52\frac52}}^4 W \gg \tag{4.6.3}$$

with $\Lambda$ defined in eq. (4.1.2). For the OPE $W(z)W(w)$, we only list the quasiprimaries appearing in its singular part. The OPE can then be reconstructed using eqs. (4.3.11,4.3.12). Denoting $QP^m(W,W)$ as $\phi_{8-m}$, we have:

$$\begin{aligned}
\phi_0 &= \frac{c}{4}\mathbb{1}, \qquad \phi_2 \;=\; 2T,\\
\phi_4 &= \frac{42}{5c+22}\Lambda \;+\; {C_{44}}^4 W,\\
\phi_6 &= -\frac{95c^2+1254c-10904}{6(7c+68)(5c+22)(2c-1)}\left(\partial T\partial T - \frac{4}{5}\partial^2 TT - \frac{1}{42}\partial^4 T\right)\\
&\quad + \frac{24(72c+13)}{(7c+68)(5c+22)(2c-1)}\left((T(TT)) - \frac{9}{10}\partial^2 TT - \frac{1}{28}\partial^4 T\right)\\
&\quad - \frac{14}{9(c+24)}{C_{44}}^4\left(\partial^2 W - 6TW\right) \;+\; {C_{44}}^6\Psi\,.
\end{aligned} \tag{4.6.4}$$

[11] Except, of course, for these $c$-values for which ${C_{44}}^4$ has poles; the extra poles in ${C_{44}}^6$ are cancelled by the zeros in the normalisation constant $\mathcal{N}$ (4.6.6). Notice also that the coupling constants are imaginary for $-22/5 < c < -13/14$ [153].

Finally, $\Psi$ is the unique dimension 6 Virasoro primary given by:

$$\begin{aligned}
\Psi &= \mathcal{N}\Big(Q\partial Q + \alpha_1\partial^2 W + \alpha_2 TW + \alpha_3\partial T\partial T +\\
&\qquad \alpha_4 T\partial^2 T + \alpha_5\partial^4 T + \alpha_6(T(TT))\Big)\,,
\end{aligned} \tag{4.6.5}$$

where the normalisation factor[12]:

$$\mathcal{N} = -\sqrt{\frac{5(2c-1)(c+24)(7c+68)}{(c-1)(4c+115)(6c+49)(14c+13)}} \tag{4.6.6}$$

is fixed such that $\Psi(z)\Psi(w) = (c/6)(z-w)^{-12} + \ldots$ and the coefficients $\alpha_1, \ldots, \alpha_6$ are:

$$\begin{aligned}
\alpha_1 &= -\tfrac{13c+350}{36(c+24)}{C_{\frac52\frac52}}^4, & \alpha_2 &= \tfrac{19}{3(c+24)}{C_{\frac52\frac52}}^4,\\
\alpha_3 &= -6(566c^2+5295c-3998)/N, & \alpha_4 &= -6(530c^2+6141c-1487)/N,\\
\alpha_5 &= -(46c^3-125c^2-6235c+1778)/N, & \alpha_6 &= 12(734c+49)/N,
\end{aligned}$$

with $N = 12(7c+68)(5c+22)(2c-1)$.

The fermionic primary $Q$ can be viewed as some kind of "generalised supersymmetry" generator. The appearance of the dimension 4 operator is then similar to the appearance of the affine $U(1)$ current in the $N = 2$ super Virasoro algebra. Similarly, one can view the $\mathcal{W}_c B_n$ algebras as some "generalised supersymmetry" algebras, the role of the higher spin operators being to ensure associativity of the OPA for generic values of the central charge.

### 4.6.2 Coulomb Gas Realisation

The next step in the analysis of the $\mathcal{W}_c B_2$ algebra consists in constructing a Coulomb gas realisation, *i.e.* a realisation of the primary operators of the OPA in terms of free operators. In order to find this realisation, one needs two free bosons and one free fermion. The OPEs for the free operators are defined to be:

$$\begin{aligned}
\partial\varphi_i \times \partial\varphi_j &= \ll \delta_{ij} \mid 0 \gg\\
\psi \times \psi &= \ll 1 \gg\ .
\end{aligned} \tag{4.6.7}$$

The energy–momentum tensor is simply the free energy–momentum tensor of the two free bosons, with a background charge, and the free fermion. Due to rotational invariance, one can always transform the background charge into one direction. Without limiting the generality, we take :

$$T = \frac{1}{2}\partial\varphi_1\partial\varphi_1 + \frac{1}{2}\partial\varphi_2\partial\varphi_2 + \alpha_0\partial^2\varphi_1 + \frac{1}{2}\partial\psi\psi, \tag{4.6.8}$$

[12] This formula corrects a misprint in [78]. $\mathcal{N}$ is $-5$ times the factor given there.

which satisfies a Virasoro algebra with central charge:

$$c = \frac{5}{2} - 12\alpha_0^2. \tag{4.6.9}$$

It is a long and boring task to find the explicit form of the dimension 4 and dimension 5/2 primaries. Using *OPEdefs*, one can try to construct the most general primary dimension four operator and require the $W(z)W(w)$ OPE to be satisfied. This, however, leads to a system of quadratic equations which is very difficult to solve. A somewhat easier way is to construct the most general primary dimension 5/2 operator. This leads to a four parameter family of such primaries. The $Q(z)Q(w)$ OPE gives (up to some discrete automorphisms) three possible solutions for these parameters, and hence also three candidate dimension four primary operators. Finally, matching the $W(z)W(w)$ OPE eliminates two of the solutions, and yields a unique construction. We wish to stress that we have checked all these statements explicitly using *OPEdefs*, including the appearance of the dimension 6 primary mentioned earlier.

Let us now present the explicit solution[13]. The dimension 5/2 operator is given by:

$$\begin{aligned} Q &= \xi\Big(\frac{3}{2}\partial\varphi_1\partial\varphi_1\psi - \frac{3}{2}\partial\varphi_2\partial\varphi_2\psi + 4\partial\varphi_1\partial\varphi_2\psi + \alpha_0\partial^2\varphi_1\psi + \\ &\quad 3\alpha_0\partial^2\varphi_2\psi + 4\alpha_0\partial\varphi_1\partial\psi + 2\alpha_0\partial\varphi_2\partial\psi + 2\alpha_0^2\partial^2\psi\Big), \end{aligned} \tag{4.6.10}$$

where $\xi = 1/\sqrt{5(5-4\alpha_0^2)}$. The dimension 4 operator is somewhat more complicated:

$$\begin{aligned} W &= \sigma\Big(N^{ijkl}\partial\varphi_i\partial\varphi_j\partial\varphi_k\partial\varphi_l + N^{ijk}\partial^2\varphi_i\partial\varphi_j\partial\varphi_k + N^{ij}\partial^3\varphi_i\partial\varphi_j \\ &\quad +\tilde{N}^{ij}\partial^2\varphi_i\partial^2\varphi_j + N^i\partial^4\varphi_i + M^{ij}\partial\varphi_i\partial\varphi_j\partial\psi\psi \\ &\quad +M^i\partial^2\varphi_i\partial\psi\psi + \tilde{M}^i\partial\varphi_i\partial^2\psi\psi + K_1\partial^3\psi\psi + K_2\partial^2\psi\partial\psi\Big), \end{aligned} \tag{4.6.11}$$

with $N^{ijkl}$, $\tilde{N^{ij}}$ and $M^{ij}$ completely symmetric and $N^{ijk}$ symmetric in the last two indices. The coefficients appearing in (4.6.11) are given explicitly by:

[13] By convention, normal ordering is always from the right to the left, *e.g.* $\partial\varphi_i\partial\varphi_j\partial\varphi_k = (\partial\varphi_i(\partial\varphi_j\partial\varphi_k))$.

$$\begin{array}{rclrcl}
N^{1111} &=& N^{2222} = 81/80, & N^{1112} &=& -N^{1222} = -\mu/20,\\
N^{1122} &=& (560\alpha_0^2-79)/720, &&&\\
N^{111} &=& 81\alpha_0/20, & N^{112} &=& -7\alpha_0\mu/60,\\
N^{122} &=& \alpha_0(80\alpha_0^2+197)/60, & N^{211} &=& -11\alpha_0\mu/30,\\
N^{212} &=& -\alpha_0\mu/10, & N^{222} &=& \alpha_0\mu/5,\\
N^{11} &=& (128\alpha_0^2-25)/60, & N^{12} &=& -\alpha_0^2\mu/30,\\
N^{21} &=& -\alpha_0^2\mu/5, & N^{22} &=& (80\alpha_0^4+82\alpha_0^2-25)/60,\\
\tilde{N}^{11} &=& (34\alpha_0^2+25)/40, & \tilde{N}^{12} &=& -\alpha_0^2\mu/15,\\
\tilde{N}^{22} &=& (80\alpha_0^4-174\alpha_0^2+25)/40, &&&\\
N^{1} &=& \alpha_0(128\alpha_0^2-25)/360, & N^{2} &=& -\alpha_0^3\mu/30,\\
M^{11} &=& M^{22} = (82\alpha_0^2-35)/6, & M^{12} &=& 0,\\
M^{1} &=& \alpha_0(2\alpha_0^2+11)/3, & M^{2} &=& \alpha_0\mu/3,\\
\tilde{M}^{1} &=& -\alpha_0\mu/3, & \tilde{M}^{2} &=& -\alpha_0\mu/6,\\
K_1 &=& (36\alpha_0^4-22\alpha_0^2+5)/18, & K_2 &=& (52\alpha_0^4-26\alpha_0^2-15)/6,
\end{array}$$

where

$$\sigma = \sqrt{\frac{3}{2(2-7\alpha_0^2)(23-40\alpha_0^2)}\frac{1}{4\alpha_0^2-5}}, \quad \mu = 23 - 40\alpha_0^2. \tag{4.6.12}$$

This solution corresponds to the sign choices $\epsilon_1 = \epsilon_2 = +1$ in (4.6.2).

Some remarks are in order.

The primary dimension 5/2 operator $Q$ can be rewritten in a more suggestive form. Indeed, rotating the free scalars:

$$\begin{aligned} \bar{\varphi}_1 &= \frac{1}{\sqrt{10}}(3\varphi_1 - \varphi_2), \\ \bar{\varphi}_2 &= \frac{1}{\sqrt{10}}(\varphi_1 + 3\varphi_2), \end{aligned} \tag{4.6.13}$$

one can rewrite:

$$Q = 5\left(\sqrt{\frac{2}{5}}\alpha_0\partial + \partial\bar{\varphi}_1\right)\left(\sqrt{\frac{2}{5}}\alpha_0\partial + \partial\bar{\varphi}_2\right)\psi, \tag{4.6.14}$$

which is exactly the starting point of the analysis of Fateev and Lukyanov [65]. Our construction hence proves that the algebra generated by (4.6.14) is indeed finitely generated for all values of the central charge, as was conjectured in [65].

A second remark concerns the fact that, (up to some discrete automorphisms), there is (up to some discrete automorphisms) only one free field realisation of $\mathcal{W}_c B_2$ with two free bosons and one free fermion. In the case of $\mathcal{W}_3$, Fateev and Zamolodchikov found two inequivalent free field realisations with two free bosons [66]. The simplest one was related to $su(3)$ and led Fateev and Lukyanov to generalise this to the $\mathcal{W}A_n$ algebras using $n$ free bosons [63]. The more complicated one has been [155] shown to be related to parafermions (at least for a specific value of the central charge). In fact, it was argued in [155] that such a realisation exists (for fixed $c$) for all $\mathcal{W}A_n$ algebras, and that in the limit $n \to \infty$ it corresponds to the $c = 2$ free field realisation of $\mathcal{W}_\infty$ by Bakas and Kiritsis [8]. For $\mathcal{W}_c B_2$, there does not seem to be a similar construction[14].

### 4.6.3 Highest weight representations

Examples of $\mathcal{W}_c B_2$-highest weight operators are easily constructed in the Coulomb gas realisation. They are given by the vertex operators defined in subsection 2.6.1:

$$V_{\vec{\beta}}(z) = e^{\vec{\beta}.\vec{\varphi}}(z)\,, \tag{4.6.15}$$

with $\vec{\beta} \equiv (\beta_1, \beta_2)$ and $\vec{\varphi}(z) \equiv (\varphi_1(z), \varphi_2(z))$. $V_{\vec{\beta}}(z)$ has Virasoro dimension:

$$\Delta_{\vec{\beta}} = \frac{1}{2}\beta_1^2 + \frac{1}{2}\beta_2^2 - \beta_1\alpha_0, \tag{4.6.16}$$

To write down the $W$-weight, we introduce the following notation, connected to the root system of $B_2$. $\vec{e}_L \equiv \sqrt{\frac{2}{5}}(1,-2)$ and $\vec{e}_S \equiv \sqrt{\frac{1}{10}}(1,3)$ are the positive simple roots. $\vec{\rho} = \frac{1}{2}(3\vec{e}_L + 4\vec{e}_S)$ is half the sum of the positive roots; note that $\vec{\alpha_0} \equiv (\alpha_0, 0)$ is parallel to $\vec{\rho}$. Finally, $W$ denotes the Weyl group of $B_2$. With this notation, the $W$-weight of the vertex operator (4.6.15) can be written as:

$$\begin{aligned} w_{\vec{\beta}} = \frac{\sigma}{480}\Big((40\alpha_0^2 - 23)\left(\prod_{w\in W} w(\vec{\beta} - \vec{\alpha_0}).\vec{\rho}\right)^{1/2} \\ +8(128\alpha_0^2 - 25)\Delta_{\vec{\beta}} + 1944\Delta_{\vec{\beta}}^2\Big)\,, \end{aligned} \tag{4.6.17}$$

where $\sigma$ was defined in eq. (4.6.12). From this formula, it follows immediately that the weights $\Delta_{\vec{\beta}}$ and $w_{\vec{\beta}}$ are invariant under $\vec{\beta} \to w(\vec{\beta} - \vec{\alpha_0}) + \vec{\alpha_0}$, with $w \in \mathcal{W}$ [65].

In the case at hand, one finds at the third order pole $[W\ V_{\vec{\beta}}]_3$ a new Virasoro primary operator, proportional to $(\beta_2\partial\varphi_1 + (2\alpha_0 - \beta_1)\partial\varphi_2))V_{\vec{\beta}}$. As mentioned in section 4.2, we have no intrinsic geometric way to express this new Virasoro primary as well as the ones appearing in the first and second order poles.

Using the realisation in terms of free fields of the previous section, one can construct two different kinds of screening operators. They can be written in such a way that their relation with the root system of $B_2$ is manifest:

$$\begin{aligned} V_\pm^L &= \exp(\beta_\pm \vec{e}_L.\vec{\varphi}) \\ V_\pm^S &= \psi \exp(\beta_\pm \vec{e}_S.\vec{\varphi})\,, \end{aligned} \tag{4.6.18}$$

with $\beta_+ + \beta_- = \sqrt{\frac{2}{5}}\alpha_0$ and $\beta_+\beta_- = -1$. They are thus seen to be equal to the screening charges presented in [65].

Given the screening operators, it is a standard construction to derive the degenerate representations, see [128, 66, 65, 203] to which we refer for details.

[14] In [78] it was conjectured that no realisation of $\mathcal{W}_{DS}B_2$ [29] exists for generic $c$ using two free scalars. This was based on an explicit calculation with no background charge $\alpha_0$. However, it was proven in [132] that such a realisation does exist, except at isolated values of $\alpha_0$, including $\alpha_0 = 0$.

## 4.7 Discussion

In this chapter we analysed in detail what the consequences are of the global and local conformal group on the OPEs in a $\mathcal{W}$–algebras. We provided explicit formulas for working with primaries and quasiprimaries. Although some of the formulas are quite complicated, computer implementation presents no problem [194]. A further step would be to implement the formulas of working with OPEs of (quasi)primaries in a future version of *OPEdefs*, *i.e.* without expanding the (quasi)primaries in order to calculate an OPE.

As an example of the power of using the techniques of primary quasiprimary operators, combined with automated OPEs, we have proven the existence, for generic $c$, of the Casimir algebra of $B_2$ by explicitly constructing it. Using a Coulomb gas realisation in terms of two free bosons and one free fermion, we have been able to show the equivalence with the results conjectured by Fateev and Lukyanov [65].

## 4.A Appendix

This appendix collects some of the more technical details of this chapter.

### A few formulas with the modes of the energy–momentum tensor

First we give a number of identities which follow from the Virasoro algebra of the modes of the energy–momentum tensor (2.4.5).

For all $m \in \mathbb{Z}$, $n \in \mathbb{N}$ we have:

$$L_m\, L_{-1}^n = \sum_{k=0}^{m+1} \binom{n}{k} (m-k+2)_k\, {L_{-1}}^{n-k}\, L_{m-k} \tag{4.A.1}$$

$$L_{-1}^n\, L_m = \sum_{k=0}^{m+1} \binom{n}{k} (-m-1)_k\, L_{m-k}\, {L_{-1}}^{n-k}\,. \tag{4.A.2}$$

## Derivation of eq. (4.4.15)

We repeat eq. (4.4.15) here for convenience:

$$\tilde{L}_m(h+n)\tilde{L}_{-n}(h) \;=\; \delta_{m-n} f_1(h,m) + \tilde{L}_{m-n} f_2(h,m,-n) + \sum_{p\geq 2-\min(n,m)} \tilde{L}_{-n-p}(h-m-p)\tilde{L}_{m+p}(h)\; f_3(h,m,-n,p)\,, \tag{4.A.3}$$

where $n,m>1$ for the remainder of this appendix.

We also introduce a shorter notation for the coefficients in eq. (4.4.11):

$$\tilde{L}_m(\Phi) = \sum_{n\geq 0} b_n^m(h) {L_{-1}}^n\, L_{n+m}\,\Phi\,, \qquad m\in\mathbb{Z}\setminus\{1,0,-1\}\,, \tag{4.A.4}$$

where we defined:

$$b_n^m(h) \;\equiv\; a_n^{m+2}(2,h) \;=\; (-1)^n \frac{(2-n-m)_n}{n!(2h-2m-n-1)_n}\,. \tag{4.A.5}$$

We will need the following lemma.

**Lemma 4.A.1** *On a quasiprimary $\Phi$ of dimension $h$, the action of $\tilde{L}_m$ can be written as:*

$$\tilde{L}_{-m}(\Phi) = \sum_{n\geq 0} \tilde{b}_n^m(h) L_{n-m}\, {L_{-1}}^n\,\Phi\,, \qquad m>1\,, \tag{4.A.6}$$

*where:*

$$\tilde{b}_n^m(h) \;=\; (-1)^n \binom{1+m}{n} \frac{(2h+n)_{m-n-2}}{(2h+m+1)_{m-2}}\,. \tag{4.A.7}$$

The proof follows immediately from the definitions eq. (4.A.4) and eq. (4.A.5) of $\tilde{L}$ and eq. (4.A.2).

We now set out to prove eq. (4.A.3). Our strategy consists of ordering all $L_k$ modes in (4.A.3) such that higher modes are moved to the right. After reordering, we look only at terms which do not contain $L_{-1}$. This is sufficient as the other terms are fixed by requiring that both *lhs* and *rhs* of eq. (4.A.3) are quasiprimary.

For the terms in the *rhs* of eq. (4.A.3), the lemma immediately gives:

$$\tilde{L}_{-n}(h-m)\tilde{L}_m(h) \;\;\rightarrow\;\; \frac{(2h-2m)_{n-2}}{(2h-2m+n+1)_{n-2}} L_{-n}L_m\,, \tag{4.A.8}$$

where the rightarrow means that we drop terms containing $L_{-1}$.

For the *lhs* of eq. (4.A.3), we find:

$$\begin{aligned}
&\tilde{L}_m(h+n)\tilde{L}_{-n}(h)\\
&\quad= \sum_{i,j\geq 0} b_i^m(h+n)\,\tilde{b}_j^n(h)\,{L_{-1}}^i\Big(L_{-n+j}L_{m+i} + (m+i+n-j)L_{m+i-n+j} + \frac{c}{2}\binom{m+i}{3}\delta_{m+i-n+j}\Big){L_{-1}}^j\\
&\quad\rightarrow \sum_{i,j\geq 0} b_i^m(h+n)\,\tilde{b}_j^n(h)\\
&\qquad\Big((n-j-1)_i\,(m+i-j+2)_j\; L_{-n+j-i}L_{m+i-j} + \frac{c}{2}\binom{m}{3}\delta_{m-n}\delta_i\delta_j\\
&\qquad+(m+i+n-j)L_{m-n}\times\\
&\qquad\big((m-n-2)_j\,\delta_i\delta_{m-n+j\geq 0} + (n-m-i-1)_i\,\delta_j\delta_{m-n+i<0}\big)\Big).
\end{aligned} \tag{4.A.9}$$

Let us look at the term with $L_{-n+j-i}L_{m+i-j}$. An additional reordering of the modes is necessary when $-n+j-i>m+i-j$, which can only happen if $n>m$. This reordering will give an additional contribution to the term proportional to $L_{m-n}$ in eq. (4.A.9).

We will now find the lower boundary for $p$ in the sum $\tilde{L}_{-n-p}\tilde{L}_{m+p}$ in eq. (4.A.3). We notice that the coefficient of $L_{-n+j-i}L_{m+i-j}$ is zero unless $m+i-j+2>0$ or $j=0$. This means that the lowest $p$ that occurs is certainly larger than $-2-m$. Furthermore, $p=-1-m$ gives a $L_{-1}$ contribution which we dropped, $p=1-m$ gives $L_1$ which is zero on a quasiprimary and so does not contribute either. Finally, the term $p=-m$ simply gives $L_0$, and hence $h$, and should be added to the linear term in eq. (4.A.3). We can conclude that the lowest $p$ which gives a quadratic contribution is larger than or equal to $2-m$, such that the rightmost $\tilde{L}$ mode of the quadratic term is always a positive mode. On the other hand, from the definition of $\tilde{b}_j^n$ eq. (4.A.7), we have that $j\leq n+1$. Together with the factor $(n-j-1)_i$, this means that for the leftmost $\tilde{L}$ the highest possible mode has $p=-n$. For the same reasons as before, we can conclude that the minimal $p$ should also be larger than or equal to $2-n$ (making the leftmost $\tilde{L}$ always a negative mode). In this way, we see that eq. (4.A.3) has the correct form.

To determine the coefficients in eq. (4.A.3), we compare eq. (4.A.9) to eq. (4.A.8). Some of the sums in these coefficients can be found using *Mathematica*. We find for $f_1$ and $f_2$:

$$f_1(h,m) \;=\; \frac{c}{12}(m-1)m(m+1)\frac{(2h)_{m-2}}{(2h+m+1)_{m-2}}$$

$$f_2(h,m,-n) \;=\; \Big((-1)^m(-2-2h+2h^2+m-3hm+2h^2m+m^2-hm^2-n$$

$$+2hn-2mn+2hmn+n^2)\,(2-m+n)_m +$$

$$(2-m+2hm-m^2-n+2hn+2mn+n^2)\,(-2+2h-m+n)_m$$

$$\Big)\Big/\,(-2+2h-m+2n)_{1+m}\,, \qquad \text{for } n \geq m \qquad (4.A.10)$$

In the case $n < m$, $f_2$ can be determined from the following lemma.

**Lemma 4.A.2**

$$f_2(h, m, -n) = f_2(h - m + n, n, -m)\,. \qquad \textit{for } n < m \qquad (4.A.11)$$

**Proof :**
This relation is most easily proven by considering the inproduct (for $n > m$):

$$< \Psi|\ \tilde{L}_{n-m}(h)\tilde{L}_m(h+n)\tilde{L}_{-n}(h)\Psi >=$$

$$f_2(h, m, -n) < \Psi|\ \tilde{L}_{n-m}(h - m + n)\tilde{L}_{m-n}(h)\Psi >, \qquad (4.A.12)$$

where $\Psi$ is a primary of dimension $h$ (with $< \Psi|\Psi >$ nonzero) and we used eq. (4.A.3) on the two last operators of the *lhs*. The inproduct is defined in eq. (4.2.1). We can compute the inproduct in the *lhs* of eq. (4.A.12) as:

$$< \Psi|\ \left(\tilde{L}_{-n}(h)\right)^+ \left(\tilde{L}_m(h+n)\right)^+ \left(\tilde{L}_{n-m}(h)\right)^+ \Psi >\ . \qquad (4.A.13)$$

Now, substituting the definition eq. (4.4.11) for the rightmost operator, only the term $L_{m-n}$ of the sum remains, as $L_1\Psi$ is zero. Also, the first two operators acting on the left state create a quasiprimary state, which is annihilated (from the right) by $L_{-1}$. Hence, we can effectively replace $\left(\tilde{L}_{n-m}(h)\right)^+$ by $\tilde{L}_{m-n}(h)$ in eq. (4.A.13). The same reasoning can be followed for the other operators, but we have to shift the dimensions[15]. We get:

$$< \Psi|\ \tilde{L}_n(h+n)\tilde{L}_{-m}(h-m+n)\tilde{L}_{m-n}(h)\Psi >\ . \qquad (4.A.14)$$

However, using eq. (4.A.3) on the two first operators, this is also equal to:

$$f_2(h - m + n, n, -m) < \Psi|\ \tilde{L}_{n-m}(h - m + n)\tilde{L}_{m-n}(h)\Psi >, \qquad (4.A.15)$$

which proves eq. (4.A.11). ■

To conclude the computation of eq. (4.A.3), we give the expression for $f_3$:

$$f_3(h, n, m, p) = f_4(h, n, m, p)\frac{(1+2h-2m+n-p)_{-2+n+p}}{(2h-2m-2p)_{-2+n+p}}\,, \qquad (4.A.16)$$

[15] When computing correlation functions of quasiprimaries, a similar reasoning shows that $(\tilde{L}_n(h))^+ = \tilde{L}_{-n}(h-n)$.

where

$$f_4(h, m, n, p) =$$

$$\begin{cases} (-1)^n(1+n)\,(2+m-n)_n\,/\,(-1+2h+n)_n & \text{if } p = -n \\ (-2+m+m^2-n-hn+2h^2n-mn-2hm \\ \quad +3hn^2-mn^2+n^3)\frac{(-1)^n(1+n)(3+m-n)_{-1+n}}{2(1-h+m-n)(-1+2h+n)_n} & \text{if } p = 1-n \\ (-1)^p\ {}_4\mathcal{F}_3(-1+m, 2+m, -1-n-p, 2-n-p; \\ \quad 2-2h+2m-2n, 1-p, 2h-p; 1) \\ \quad \frac{(2h-p)_{1+n+p}(2+m+p)_{-p}(2+n+p)_{-p}}{(-p)!(-2+2h+n)_{1+n}} & \text{if } 1-n < p < 0 \\ {}_4\mathcal{F}_3(-1-n, 2-n, -1+m+p, 2+m+p; \\ \quad 2h, 1+p, 2-2h+2m-2n+p; 1) \\ \quad \frac{(2h)_{-2+n}(-1+m)_p(-1+n)_p}{p!(1+2h+n)_{-2+n}(-1+2h-2m+2n-p)_p} & \text{if } p \geq 0 \end{cases} \qquad (4.A.17)$$

We did not find a simple expression for the sums that are involved. We rewrote them in terms of generalised hypergeometric functions:

$${}_p\mathcal{F}_q(n_i; m_j; z) = \sum_{k\geq 0} \frac{\prod_i (n_i)_k}{k! \prod_j (m_j)_k} z^k\,, \qquad (4.A.18)$$

where $i$ runs from 1 to $p$ and $j$ from 1 to $q$. In $f_4$, the infinite sum always reduces to a finite number of terms because one of the $n_i$ is negative. One can now use identities for the generalised hypergeometric functions [188] to prove that:

$$f_3(h, m, -n, p) = f_3(h + n - m, n, -m, p)\,. \qquad (4.A.19)$$

Alternatively, this can be checked using an inproduct with four $\tilde{L}$ operators.

## Determination of the coefficient in eq. (4.4.16)

We first prove:

$$\tilde{L}_n(h_j - m)QP^{h_i+m}(\Psi_i, \Psi_j) = f_5(h_i, h_j, m, n)\ QP^{h_i+m+n}(\Psi_i, \Psi_j)\,, \quad n \geq 2\,. \qquad (4.A.20)$$

along the same lines as eq. (4.A.3), *i.e.* we will order the modes, and drop $L_{-1}$ contributions. We write down the definition of the *lhs* using modes:

$$\sum_{k,l\geq 0} a_k^{n+2}(2, h_j - m)L_{-1}{}^k L_{n+k}\ a_l^{m+h_i}(h_i, h_j)L_{-1}{}^l\widehat{(\Psi_i)}_{m+l}\Psi_j\,, \qquad (4.A.21)$$

where the $a_n^m$ are given in eq. (4.3.21). We move the $L_{-1}$ to the left using eq. (4.A.1), and drop terms containing $L_{-1}$:

$$\sum_{l\geq 0} a_l^{m+h_i}(h_i,h_j)\,(n-l+2)_l\;\; L_{n-l}\widehat{\Psi}_{im+l}\Psi_j\,. \tag{4.A.22}$$

Now, the factor $(n-l+2)_l$ restricts $l$ to be smaller than $n+2$. In fact, for $l=n+1$ we get a $L_{-1}$ term which we should drop. For $l$ less than $n$, we can commute $L_{n-l}$ through the $\Psi_i$ mode. Only the commutator eq. (2.4.7) remains, as $L_{n-l}$ annihilates the primary operator $\Psi_j$. We get for the *lhs* of eq. (4.A.20), dropping $L_{-1}$ terms:

$$\Big(\sum_{l=0}^{n-1} a_l^{m+h_i}(h_i,h_j)\,(n-l+2)_l\,(n(h_i-1)-m-l)$$
$$+a_n^{m+h_i}(h_i,h_j)\,(2)_n\,(h_j-m-n)\Big)\widehat{\Psi}_{im+n}\Psi_j\,. \tag{4.A.23}$$

The coefficient of $\widehat{\Psi}_{im+n}\Psi_j$ in this equation is $f_5$, as the only term in the *rhs* of eq. (4.A.20) without $L_{-1}$ is simply $\widehat{\Psi}_{im+n}\Psi_j$. After summation, we find:

$$\begin{aligned} f_5(h_i,h_j,m,n) = &\\ &\Big((-1)^n\big(-h_i(h_i-1)+h_j(h_j-1)+M(M-1)+\\ &\quad h_j n(2M+n-1)\big)\,(h_i-h_j+M)_n\\ &\quad +(h_i(h_i-1)-h_j(h_j-1)+M(M-1)+\\ &\quad h_i n(2M+n-1))\,(-h_i+h_j+M)_n\\ &\Big)\Big/\,(2M+n-2)_{n+1} \end{aligned} \tag{4.A.24}$$

where $M=h_j-m-n$.
Although we only looked at the term free of $L_{-1}$, the others have to be such that the *rhs* of eq. (4.A.20) is quasiprimary. Moreover, we see that the *lhs* of eq. (4.A.20) is of the form $\sum x_l {L_{-1}}^l(\widehat{\Psi_i})_{m+n+l}\Psi_j$. We found in subsection 4.3.2 that requiring this form to be quasiprimary fixed all $x_j$ in terms of $x_0$.

It is now clear that the proportionality constant in eq. (4.4.16) is given by:

$$\tilde{f}(h_i,h_j,h_k,\{\tilde{n}\}) \equiv \prod_l f_5(h_i,h_j,h_j-h_k-\sum_{k\geq l} n_k,\; n_l)\,. \tag{4.A.25}$$

# Chapter 5

# Factoring out Free Fields

An Operator Product Algebra (OPA) is factored in two parts if we can write it as a direct product structure $\mathcal{A} \otimes \mathcal{A}'$. All operators of $\mathcal{A}$ have nonsingular OPEs with operators of $\mathcal{A}'$, or equivalently, their modes commute. Some years ago, Goddard and Schwimmer [103] proved that every OPA can be factorised into a part with only free fermions (of dimension $\frac{1}{2}$) and a part containing no free fermions. As a consequence, in the classification of $\mathcal{W}$–algebras, spin $\frac{1}{2}$ fermions need never be considered. This is very fortunate, since the main method of constructing a large number of $\mathcal{W}$–algebras, Drinfeld-Sokolov reduction (see chapter 7), does not yield dimension $\frac{1}{2}$ fields. (Supersymmetric reduction, see [86], does give weight $\frac{1}{2}$ fields.)

The first section of this chapter extends the result of Goddard and Schwimmer for other free fields. We present an algorithmic procedure for the factorisation. We start with a derivation of the result of [103] in our formalism. We then treat bosonic fields of weight $\frac{1}{2}$, which were not treated in [103]. It was already noticed in [103] that in some cases (*e.g.* the $N = 4$ linear superconformal algebra) dimension 1 bosons can also be decoupled from a conformal theory. This is certainly not a general property, and the factorisation-algorithm presented at the end of the first section gives an easy criterium to decide when free bosons can be decoupled.

In the second section of this chapter, we show how the generating functionals of the algebra obtained by factoring out free fields can be found. Also, the criterium for factorisable dimension 1 bosons is rederived from Ward identities.

Finally, the linear and nonlinear $N = 3, 4$ superconformal algebras are discussed as an example.

The first and second section of this chapter contains material published in [47], see also [45]. However, the factorisation algorithm is considerably simplified. Section 5.3 is based on [182].

## 5.1 Algorithms for factorisation

In the following subsections, we will show how various free fields can be decoupled by introducing certain projection operators on the vectorspace of fields in the OPA. These operators were found in [47], but we will show some additional properties which make the formulation of the algorithm simpler.

The method explained in this section is valid in any OPA. In fact, we do not require the presence of a Virasoro operator. Therefore, we define the modes $\hat{A}_m$ in this chapter by:

$$\hat{A}_m B \equiv [AB]_m \,, \qquad m \in \mathbb{Z}\,. \tag{5.1.1}$$

This is a shift with respect to the usual definition (2.4.1).

### 5.1.1 Free fermions

For completeness, we first rederive the result of [103] in our formalism and give an explicit algorithm for the decoupling. Consider a theory containing a free fermion $\psi$, see section 2.6.2. From the OPE (2.6.26) we find the following anticommutation relations for the modes:

$$\hat{\psi}_m \hat{\psi}_n = -\hat{\psi}_n \hat{\psi}_m + \lambda \delta_{m+n-1}\,, \qquad m, n \in \mathbb{N}\,, \tag{5.1.2}$$

where $\lambda$ is a normalisation constant.

Our method consists of defining a set of projection operators $\mathcal{P}_n$ in the OPA. $\mathcal{P}_n$ projects on the kernel of the mode $\hat{\psi}_n$. Together, they project the OPA to a subalgebra which commutes with $\psi$. It is then easy to show that the OPA is the direct product of this subalgebra with the OPA generated by $\psi$.

The projection operators are in the case of free fermions defined by:

$$\mathcal{P}_n \equiv 1 - \frac{1}{\lambda} \hat{\psi}_{1-n} \hat{\psi}_n \,, \qquad n > 0\,. \tag{5.1.3}$$

From eq. (5.1.2) we see that:

$$\begin{aligned}\hat{\psi}_n \mathcal{P}_n &= 0\\ \mathcal{P}_n \hat{\psi}_{1-n} &= 0\\ \hat{\psi}_n \mathcal{P}_m &= \mathcal{P}_m \hat{\psi}_n, \qquad m \neq n, 1-n\,.\end{aligned} \tag{5.1.4}$$

These relations lead to:

$$\begin{aligned}\mathcal{P}_n \mathcal{P}_n &= \mathcal{P}_n\\ \mathcal{P}_n \mathcal{P}_m &= \mathcal{P}_m \mathcal{P}_n\,.\end{aligned} \tag{5.1.5}$$

Together, eqs. (5.1.4) and (5.1.5) show that $\mathcal{P}_n$ is a projection operator in the kernel of $\hat{\psi}_n$. Moreover, the different projection operators commute. Clearly, the projection operator:

$$\mathcal{P} \equiv \prod_{n>0} \mathcal{P}_n \tag{5.1.6}$$

is such that for any field $A$ of the OPA, the OPE $(\mathcal{P}A)(z)\ \psi(w)$ is nonsingular. By using the relation (2.3.14) for the regular part of an OPE, we see that $\mathcal{P}A$ is equal to $A$ plus composites containing $\psi$.

As an example, it is easy to check that an energy–momentum tensor $T$ for which $\psi$ is a primary field with dimension 1/2 gets the expected correction:

$$\mathcal{P}T \;=\; T - \frac{1}{2\lambda}\partial\psi\psi\,, \tag{5.1.7}$$

*i.e.* the energy–momentum tensor of a free fermion is substracted. This means that the central charge of $\mathcal{P}T$ is equal to $-1/2$ the central charge of $T$.

When the OPA is generated by operators $T^i$, we see that $\mathcal{P}T^i$ generate a subalgebra where all fields commute with $\psi$. Finally, because $\mathcal{P}A = A + \ldots$, the complete OPA is generated by $\mathcal{P}T^i$ and $\psi$. This proves the factorisation.

### 5.1.2 Symplectic bosons

Suppose we have a pair of symplectic bosons $\xi^+, \xi^-$ with OPEs given in section 2.6.3. The modes (5.1.1) satisfy the commutation relations:

$$\begin{aligned}\hat{\xi}^\pm_m \hat{\xi}^\pm_n &= \hat{\xi}^\pm_n \hat{\xi}^\pm_m,\\ \hat{\xi}^+_m \hat{\xi}^-_n &= \hat{\xi}^-_n \hat{\xi}^+_m + \lambda\delta_{m+n-1}\,.\end{aligned} \tag{5.1.8}$$

The method we apply is completely similar to the previous case. We define the operators:

$$\mathcal{P}^\pm_n \equiv \sum_{i\geq 0} \frac{(\mp 1)^i}{i!\lambda^i} \left(\hat{\xi}^\mp_{1-n}\right)^i \left(\hat{\xi}^\pm_n\right)^i, \qquad n>0\,. \tag{5.1.9}$$

The action of these operators on a field $\Phi$ of the OPA is well-defined when only a finite number of terms in the sum is non-zero when $\mathcal{P}^\pm_n$ acts on $\Phi$. For a graded OPA (see definition 2.3.7), a sufficient condition is that $\dim(\xi^\pm) = 1/2$, and that there is a lower bound on the dimension of the fields in the algebra. In the case that $h^+ = \dim(\xi^+) \geq 1$, a similar argument does not exist. Indeed, if $n \leq h^+$, it cannot be argued on dimensional grounds that $\left(\hat{\xi}^+_n\right)^i$ on a field has to be zero for $i$ large enough, because $\dim\left(\hat{\xi}^+_n\Phi\right) = \dim(\Phi) + h^+ - n \geq \dim(\Phi)$. However, we expect that the projection operators (5.1.9) can be used in most cases.

Assuming that the infinite sums give no problems, we proceed as in the previous subsection. Using (5.1.8), we find:

$$\begin{aligned}\hat{\xi}^\pm_n \mathcal{P}^\pm_m &= (1-\delta_{m-n})\mathcal{P}^\pm_m \hat{\xi}^\pm_n\\ \mathcal{P}^\mp_m \hat{\xi}^\pm_n &= (1-\delta_{m+n-1})\hat{\xi}^\pm_n \mathcal{P}^\mp_m\,.\end{aligned} \tag{5.1.10}$$

This allows us to prove that $\mathcal{P}^\pm_n$ is a projection operator on the kernel of $\xi^\pm_n$, and all these projection operators commute. Note that we can rewrite the definition (5.1.9) as:

$$\mathcal{P}^\pm_n \;= : \exp\left(\mp\frac{1}{\lambda}\hat{\xi}^\mp_{1-n}\hat{\xi}^\pm_n\right) :, \tag{5.1.11}$$

where normal ordering with respect to the modes is used. The complete projection operator is:

$$\mathcal{P} \;\equiv\; \prod_{n>0} \mathcal{P}^+_n \mathcal{P}^-_n\,. \tag{5.1.12}$$

This proves that symplectic bosons can all be decoupled when the action of the operators (5.1.9) is well-defined in the OPA. This is always the case for dimension 1/2 symplectic bosons. Note that when the dimension $h^+ \neq 1/2$, fields with zero or negative dimension are present in the OPA. For applications in conformal field theory this is undesirable. Of course, $\beta\gamma$–systems used in BRST quantisation are already factored from the rest of the OPA.

As an example, consider a Virasoro operator $T$. $\xi^+$ and $\xi^-$ are primaries of dimension $h^+$ and $1-h^+$ respectively with respect to $T$. We find:

$$\mathcal{P}^+_1\mathcal{P}^-_1 T = T - \frac{h^+}{\lambda}\xi^+\partial\xi^- + \frac{h^-}{\lambda}\partial\xi^+\xi^-\,. \tag{5.1.13}$$

$\mathcal{P}^+_1\mathcal{P}^-_1 T$ already commutes with $\xi^\pm$, so no further projections are necessary. The central extension of the new Virasoro operator is given by $c - 2(6(h^+)^2 - 6h^+ + 1)$.

Finally, we consider the case of a fermionic $bc$-system. It can be decoupled using a formula similar to (5.1.11). Here, all sums are reduce to only two terms, proving that fermionic $b, c$ can always be factored out. Alternatively, one can define two free fermions $b \pm c$ for which the results of the previous subsection can be used. This can be done even if $b$ and $c$ have different dimensions, as subsection 5.1.1 does not rely on the dimension at all.

### 5.1.3 $U(1)$ currents

A (bosonic) $U(1)$ current $J$ has the OPE (the notation for OPEs is introduced in subsection 2.3.4):

$$J \times J = \ll \lambda \mid 0 \gg \ . \tag{5.1.14}$$

The derivative of a free scalar treated in section 2.6.1 provides a realisation of this OPE. The commutation rules for the modes are:

$$\hat{J}_m \hat{J}_n = \hat{J}_n \hat{J}_m + \lambda(m-1)\delta_{m+n-2} \, . \tag{5.1.15}$$

From eq. (5.1.15), we can see that the desired projection operators are:

$$\mathcal{P}_n \ \equiv : \exp\left(\frac{1}{(1-n)\lambda}\hat{J}_{2-n}\hat{J}_n\right) : , \tag{5.1.16}$$

except for $n = 1$. Notice that in a graded OPA (see def. 2.3.7) $\dim(J) = 1$, such that $\dim(\hat{J}_1 A) = \dim(A)$. Hence, we cannot argue on dimensional grounds that $(\hat{J}_1)^i A$ is zero even for very large $i$. However, the situation is disctinctly different from previous subsection, where we could still hope that for most OPAs the projection operators (5.1.9) are well-defined. In the case of a $U(1)$-scalar, $\mathcal{P}_1$ simply does not exist.

**Theorem 5.1.1** *A $U(1)$ current $J$ can be decoupled if and only if:*

$$\hat{J}_1 A = [JA]_1 = 0 \tag{5.1.17}$$

*for all fields $A$ of the OPA. Or in words, all fields have zero $[U(1)]$ charge with respect to the current $J$.*

**Proof :**
The fact that eq. (5.1.17) is a sufficient condition follows because we can define the projection operators (5.1.16).

We now show that (5.1.17) is also a necessary condition using a special case of the Jacobi identity (2.3.21):

$$[J[BC]_n]_1 = [[JB]_1 C]_n + [B[JC]_1]_n \tag{5.1.18}$$

and

$$[J\,\partial A]_1 = \partial([JA]_1)\, . \tag{5.1.19}$$

Suppose the currents $J$ and $T^k$ generate the algebra and $[JT^k]_1$ is nonzero for some $k$. We look for an alternative set $\tilde{T}^k$ which still generate the algebra (together with $J$) and for which $[J\tilde{T}^k]_1 = 0$. We can always choose

$$\tilde{T}^k = T^k + X^k\, , \tag{5.1.20}$$

where $X^k$ has strictly positive derivative-number (see eq. (3.3.14)), or a composite-number (see eq. (3.3.16)) larger than 1. We define:

$$D = \min_k d([JT^k]_1) \qquad C = \min_k c([JT^k]_1)\, . \tag{5.1.21}$$

We will now prove that $d([JX^k]_1) > D$ or $c([JX^k]_1) > C$. If $d(X^k) > 0$, eq. (5.1.19) proves the first inequality. If $d(X^k) = 0$, we necessarily have that $c(X^k) > 1$. In this case the second inequality follows from eq. (5.1.18) with $n = 0$. This means that it is not possible to find $X^k$ which cancel the contributions of $T^k$ completely. ■

The condition (5.1.17) is not equivalent to $[AJ]_1 = 0$ because, see eq. (2.3.16):

$$[AJ]_1 = [JA]_1 + \sum_{i\geq 2} \frac{(-1)^i}{i!} [AJ]_i \ . \tag{5.1.22}$$

In a $\mathcal{W}$–algebra generated by primary fields (except $T$ itself), the criterion becomes that for any primary generator $A$, $[AJ]_1$ may not contain any primary fields. Indeed, if $J$ and $C$ are primaries with respect to $T$, we see from eq. (5.1.18) with $B = T$ that $[JC]_1$ is primary. Because the primary at $[JC]_1$ is the same as the one in $[CJ]_1$, eq. (5.1.17) translates in the requirement that there is no primary field in $[AJ]_1$.

Finally, we remark that the condition eq. (5.1.17) is in fact quite natural, as $U(1)$-scalars can be viewed as the derivative of a dimension zero field, and for any fields $A$ and $B$, eq. (2.3.12) implies that $[\partial A\ B]_1 = 0$.

## 5.2 Generating functionals

In this section, we study the relation between the generating functionals of the algebras related by factoring out a free field. The results are especially important as these functionals define the induced actions of the corresponding $\mathcal{W}$-gravities (see chapter 7).

Recall the definition of the generating functions (2.5.1) and (2.5.4) for an OPA with generators $T^k$:

$$Z[\mu] = \exp\left(-\Gamma[\mu]\right) = \left\langle \exp\left(-\frac{1}{\pi}\int \mu_k T^k\right)\right\rangle\ . \tag{5.2.1}$$

Suppose the OPA contains a free field $F$ that can be factored out. We will denote by $\tilde{T}^k$ the redefined generators (anti-) commuting with $F$. By inverting the algorithms of the previous section, we can write:

$$T^k = \tilde{T}^k + P^k[\tilde{T}, F]\, , \tag{5.2.2}$$

where the $P^k[\tilde{T}, F]$ are some differential polynomials with all terms at least of order 1 in $F$. We now provide some heuristic arguments – *i.e.* based on path integrals – that the generating functional $\tilde{Z}[\tilde{\mu}]$ of the reduced gravity theory, which is defined similarly to eq. (5.2.1), can be obtained from $Z[\mu]$ by integrating over the source of the free field $\mu_F$:

$$\tilde{Z}[\tilde{\mu}] = \int [d\mu_F]\, Z\left[\tilde{\mu}, \mu_F\right] . \tag{5.2.3}$$

We can compute $Z$ as follows

$$Z[\tilde{\mu}, \mu_F] = \left\langle \exp\left(-\frac{1}{\pi}\int \tilde{\mu}_k(\tilde{T}^k + P^k[\tilde{T}, F]) + \mu_F F\right)\right\rangle_{\rm OPE} . \tag{5.2.4}$$

We assume that there exists a path integral formulation for this expression, *i.e.* the $\tilde{T}^k$ are expressed in terms of some matter fields $\varphi$. In this case, the polynomials $P^k$ in the path integral would be the classical limit of those in (5.2.2), and some regularisation procedure has to be applied to find (5.2.4). In particular, short distance singularities should be resolved for example by point splitting. We have:

$$Z[\tilde{\mu}, \mu_F] = \int [d\varphi][dF]$$
$$\exp -\left(S[\varphi] + S_F[F] + \frac{1}{\pi}\int \tilde{\mu}_k\left(\tilde{T}^k[\varphi] + P^k[\tilde{T}[\varphi], F]\right) + \mu_F F\right) . \tag{5.2.5}$$

Here $S_F$ is the free field action which gives the correct OPE for $F$. For $\mu_k = 0$ the matter fields and $F$ are not coupled. This implies that $\tilde{T}^k$ and $F$ have a non-singular OPE as is required.
We now integrate eq. (5.2.5) over $\mu_F$ and interchange the order of integration. The last term in the exponential gives us $\delta(F)$, such that all terms containing $F$ can be dropped. The remaining expression is exactly $\tilde{Z}$.

Going to the effective theory (see chapter 7), we define:

$$\exp\left(-W[\check{T}]\right) = \int [d\mu]\, Z[\mu] \exp\left(\frac{1}{\pi}\int \mu_k \check{T}^k\right) . \tag{5.2.6}$$

From relation (5.2.3), we immediately see:

$$\tilde{W}[\check{T}] = W[\check{T}, \check{T}_F = 0] . \tag{5.2.7}$$

Therefore, the two theories are related by a *quantum* Hamiltonian reduction.

Finally, let us see how the classical limits of the Ward identities for the generating functionals (2.5.18) are related. The first step is to note that for both theories, the same Ward identities are satisfied by the Legendre transform[1] $W^{(0)}$ of the classical limit of $\Gamma$ by replacing $\{\mu, \delta\Gamma^{(0)}/\delta\mu\}$ with $\{\delta W^{(0)}/\delta\check{T}, \check{T}\}$. Furthermore, $W^{(0)}$ is the classical limit of $W$ (upto some factors), as it is the saddle-point value in (5.2.6). From eq. (5.2.7), we see that the classical limit of the Ward identities of the reduced theory can be obtained by putting $\delta\Gamma^{(0)}/\delta\mu_F = 0$ in the original identities.

[1]See the introduction of chapter 7 for the definition of these functionals.

Indeed, when we factor out a fermion, the Ward identity corresponding to $\mu_i = \mu_\psi$ is, see (2.5.18):

$$\bar{\partial} u_\psi = -\frac{\lambda}{\pi}\mu_\psi + F\left[\mu, u, u_\psi\right] , \tag{5.2.8}$$

where:

$$u^i \equiv \frac{\delta\Gamma}{\delta\mu_i} . \tag{5.2.9}$$

Setting $u_\psi = 0$, we can solve for $\mu_\psi$ and substitute the solution in the other Ward identities. In this way, the fermion $\psi$ completely disappears from the theory. The same can be done for a couple $(\xi^+, \xi^-)$ of symplectic bosons, by looking at the equations with $\mu_i = \xi^\pm$.

We now treat the decoupling of a $U(1)$-scalar $J$. The Ward identity (2.5.18) of $\mu_J$ has an anomalous term proportional to $\partial\mu_J$. This means that we will only be able to remove the scalar field if $\mu_J$ never appears underived. So our criterium for the factoring out of a scalar field $J$ should be that in all Ward identities of the theory, the coefficient of $\mu_J$, without derivative, vanishes. If we look at eq. (2.5.18) for some source $\mu_i$, this term is given by:

$$-\frac{1}{\pi}\mu_J\left\langle [JT_i]_1 \exp\left(-\frac{1}{\pi}\int \mu_k T^k\right)\right\rangle . \tag{5.2.10}$$

We see that requiring this term to vanish, yields precisely the classical limit of the condition (5.1.17).

## 5.3 Examples

In this section we discuss the $N = 3, 4$ linear [1, 171, 185] and nonlinear [20, 133, 103] superconformal algebras. For both cases, the factorisations were performed in [103]. The relation (5.2.3) between the induced actions will be derived here in a more explicit way without relying on an underlying path integral formalism for the factorised algebra. In this section, we do *not* follow the conventions of appendix B for summation indices. No signs are implied in the summations and indices are raised and lowered with the Kronecker delta.

### 5.3.1 $N = 3$ superconformal algebras

Both $N = 3$ superconformal algebras contain the energy–momentum tensor $T$, supercharges $G^a$, $a \in \{1, 2, 3\}$ and an $so(3)$ affine Lie algebra with generators $U^a$, $a \in \{1,2,3\}$. The linear algebra [1] contains in addition a dimension 1/2 fermion $Q$. The OPEs of the generators are (we use tildes for the nonlinear algebra and omit OPEs with $T$ and $\widetilde{T}$):

$$\begin{array}{rcl|rcl}
G^a\,G^b &=& \delta^{ab}\frac{2c}{3}[\mathbb{1}] - \varepsilon^{abc}2[U^c] & \widetilde{G}^a\,\widetilde{G}^b &=& \delta^{ab}\frac{2(\tilde{c}-1)}{3}[\mathbb{1}] - \frac{2(\tilde{c}-1)}{\tilde{c}+1/2}\varepsilon^{abc}[\widetilde{U}^c] \\
&&&&& +\frac{3}{\tilde{c}+1/2}[\widetilde{U}^{(a}\widetilde{U}^{b)} - \frac{2\tilde{c}+1}{3\tilde{c}}\delta^{ab}\widetilde{T}] \\
U^a\,U^b &=& -\frac{c}{3}\delta^{ab}[\mathbb{1}] + \varepsilon^{abc}[U^c] & \widetilde{U}^a\,\widetilde{U}^b &=& -\frac{\tilde{c}+1/2}{3}\delta^{ab}[\mathbb{1}] + \varepsilon^{abc}[\widetilde{U}^c] \\
U^a\,G^b &=& \delta^{ab}[Q] + \varepsilon^{abc}[G^c] & \widetilde{U}^a\,\widetilde{G}^b &=& \varepsilon^{abc}[\widetilde{G}^c] \\
Q\,G^a &=& [U^a] &&& \\
Q\,Q &=& -\frac{c}{3}[\mathbb{1}], &&&
\end{array} \tag{5.3.1}$$

where we list only the primaries in the OPEs (see section 4.4).

The relation [103] between the linear and nonlinear algebras is that $Q$ commutes with the combinations that constitute the nonlinear algebra:

$$\begin{aligned}
\widetilde{T} &\equiv T + \frac{3}{2c}\partial Q Q, \\
\widetilde{G}^a &\equiv G^a + \frac{3}{c}U^a Q, \\
\widetilde{U}^a &\equiv U^a,
\end{aligned} \tag{5.3.2}$$

while the central charges are related by $\tilde{c} = c - 1/2$. These relations are easily found by applying the algorithms of section 5.1.

We derive the Ward identities for the induced actions (2.5.4) $\Gamma$ and $\tilde{\Gamma}$ by considering their transformation properties under $N = 3$ supergravity transformations. We will use the notations $h \equiv \mu_T, \psi_a \equiv \mu_{G^a}, A_a \equiv \mu_{U^a}, \eta \equiv \mu_Q$. The transformations read, for the linear case:

$$\begin{aligned}
\delta h &= \bar{\partial}\varepsilon + \varepsilon\partial h - \partial\varepsilon h + 2\theta^a\psi_a, \\
\delta\psi^a &= \bar{\partial}\theta^a + \varepsilon\partial\psi^a - \frac{1}{2}\partial\varepsilon\psi^a + \frac{1}{2}\theta^a\partial h - \partial\theta^a h - \varepsilon^{abc}(\theta_b A_c + \omega_b\psi_c) \\
\delta A^a &= \bar{\partial}\omega^a + \varepsilon\partial A^a - \varepsilon^{abc}(\partial\theta_b\psi_c - \theta_b\partial\psi_c) + \theta^a\eta - \varepsilon^{abc}\omega_b A_c - \partial\omega^a h + \tau\psi^a \\
\delta\eta &= \bar{\partial}\tau + \varepsilon\partial\eta + \frac{1}{2}\partial\varepsilon\eta + \theta^a\partial A_a - \partial\omega^a\psi_a - \frac{1}{2}\tau\partial h - \partial\tau h\,.
\end{aligned} \tag{5.3.3}$$

They are the same for the nonlinear case, except that there is no field $\eta$ and no parameter $\tau$, and $\delta A_a$ contains a $\tilde{c}$ dependent extra term:

$$\delta_{\rm extra}A^a = \frac{3}{2\tilde{c}}\varepsilon^{abc}(\partial\theta_b\psi_c - \theta_b\partial\psi_c)\,. \tag{5.3.4}$$

The anomaly for the linear theory is:

$$\delta\Gamma[h,\psi,A,\eta] = -\frac{c}{12\pi}\int\varepsilon\partial^3 h - \frac{c}{3\pi}\int\theta^a\partial^2\psi_a + \frac{c}{3\pi}\int\omega^a\partial A_a + \frac{c}{3\pi}\int\tau\eta\,. \tag{5.3.5}$$

Defining[2]:

$$t = \frac{12\pi}{c}\frac{\delta\Gamma}{\delta h} \qquad g^a = \frac{3\pi}{c}\frac{\delta\Gamma}{\delta\psi^a} \qquad u^a = -\frac{3\pi}{c}\frac{\delta\Gamma}{\delta A^a} \qquad q = -\frac{3\pi}{c}\frac{\delta\Gamma}{\delta\eta} \tag{5.3.6}$$

we obtain the Ward identities for the linear theory by combining eqs. (5.3.3) and (5.3.5):

$$\begin{aligned}
\partial^3 h &= \overline{\nabla}t - (2\psi_a\partial + 6\partial\psi_a)\,g^a + 4\partial A_a u^a - (2\eta\partial - 2\partial\eta)\,q \\
\partial^2\psi^a &= \overline{\nabla}g^a - \frac{1}{2}\psi^a t + \varepsilon^{abc}A_b g^c + \eta u^a + \varepsilon^{abc}\left(2\partial\psi_b + \psi_b\partial\right)u_c + \partial A^a q \\
\partial A^a &= \overline{\nabla}u^a - \varepsilon^{abc}\psi_b g_c + \varepsilon^{abc}A_b u_c - (\psi^a\partial + \partial\psi^a)\,q \\
\eta &= \overline{\nabla}q - \psi_a u^a\,,
\end{aligned} \tag{5.3.7}$$

where:

$$\overline{\nabla}\Phi = \left(\bar{\partial} - h\partial - h_\Phi\partial h\right)\Phi, \tag{5.3.8}$$

with $h_\Phi = 2,\ \frac{3}{2},\ 1,\ \frac{1}{2}$ for $\Phi = t,\ g^a,\ u^a,\ q$.

Because these functional differential equations have no explicit dependence on $c$, the induced action can be written as:

$$\Gamma[h,\psi,A,\eta] = c\ \Gamma^{(0)}[h,\psi,A,\eta]\,, \tag{5.3.9}$$

where $\Gamma^{(0)}$ is $c$-independent.

The nonlinear theory can be treated in a parallel way. The anomaly is now:

$$\begin{aligned}
\delta\widetilde{\Gamma}[h,\psi,A] &= -\frac{\tilde{c}}{12\pi}\int\varepsilon\partial^3 h - \frac{\tilde{c}-1}{3\pi}\int\theta^a\partial^2\psi_a + \frac{\tilde{c}+1/2}{3\pi}\int\omega^a\partial A_a \\
&\quad -\frac{3}{\pi(\tilde{c}+1/2)}\int\theta_a\psi_b\left(U^{(a}U^{b)}\right)_{\rm eff}\,.
\end{aligned} \tag{5.3.10}$$

The last term, which is due to the nonlinear term in the algebra eq. (5.3.1), can be rewritten as:

$$\left(U^{(a}U^{b)}\right)_{\rm eff}(x) = \left\langle \widetilde{U}^{(a}\widetilde{U}^{b)}(x)\exp\left(-\frac{1}{\pi}\int\left(h\widetilde{T} + \psi_a\widetilde{G}^a + A_a\widetilde{U}^a\right)\right)\right\rangle\Big/$$

[2] All functional derivatives are left derivatives.

$$\begin{aligned}
&\exp\Big(-\widetilde{\Gamma}\Big) \\
&= \left(\frac{\tilde c+1/2}{3}\right)^2 u^a(x)\,u^b(x) + \\
&\quad \frac{(\tilde c+1/2)\pi}{6}\lim_{y\to x}\left(\frac{\partial u^a(x)}{\partial A_b(y)} - \frac{\partial}{\bar\partial}\delta^{(2)}(x-y)\delta^{ab} + a \rightleftharpoons b\right).
\end{aligned} \tag{5.3.11}$$

The limit in the last term of eq. (5.3.11) reflects the point-splitting regularisation of the composite terms in the $\widetilde{G}\widetilde{G}$ OPE (5.3.1). One notices that in the limit $\tilde c \to \infty$, $u$ becomes $\tilde c$ independent and one has simply:

$$\lim_{\tilde c\to\infty}\left(\frac{3}{\tilde c+1/2}\right)^2\Big(U^{(a}U^{b)}\Big)_{\rm eff}(x) = u^a(x)\,u^b(x)\,. \tag{5.3.12}$$

Using eq. (5.3.11), we find that eq. (5.3.10) can be rewritten as:

$$\begin{aligned}
\delta\widetilde{\Gamma}[h,\psi,A] &= -\frac{\tilde c}{12\pi}\int \varepsilon\partial^3 h - \frac{\tilde c-1}{3\pi}\int\theta^a\partial^2\psi_a + \frac{\tilde c+1/2}{3\pi}\int\omega^a\partial A_a \\
&\quad -\frac{\tilde c+1/2}{3\pi}\int\theta_a\psi_b u^a u^b \\
&\quad -\lim_{y\to x}\int\theta^{\,(a}\psi^{b)}\left(\frac{\partial u^a(x)}{\partial A_b(y)} - \frac{\partial}{\bar\partial}\delta^{(2)}(x-y)\delta_{ab}\right),
\end{aligned} \tag{5.3.13}$$

where the last term disappears in the large $\tilde c$ limit. The term proportional to $\int\theta_a\psi_b u^a u^b$ in eq. (5.3.13) can be absorbed by adding a field dependent term in the transformation rule for $A$:

$$\delta^{\rm nl}_{\rm extra}A_a = -\theta_a\psi_b u^b\,. \tag{5.3.14}$$

Doing this, we find that in the large $\tilde c$ limit, the anomaly reduces to the minimal one.

Combining the nonlinear transformations with eq. (5.3.13), and defining:

$$\tilde t = \frac{12\pi}{\tilde c}\frac{\delta\widetilde\Gamma}{\delta h}\qquad \tilde g^a = \frac{3\pi}{\tilde c-1}\frac{\delta\widetilde\Gamma}{\delta\psi_a}\qquad \tilde u^a = -\frac{3\pi}{\tilde c+1/2}\frac{\delta\widetilde\Gamma}{\delta A_a}, \tag{5.3.15}$$

we find the Ward identities for $\widetilde\Gamma[h,\psi,A]$ (they can also be found in [54]):

$$\begin{aligned}
\partial^3 h &= \overline{\nabla}\tilde t - \left(1-\frac{1}{\tilde c}\right)(2\psi_a\partial + 6\partial\psi^a)\,\tilde g^a + 4\left(1+\frac{1}{2\tilde c}\right)\partial A_a\tilde u^a \\
\partial^2\psi_a &= \overline{\nabla}\tilde g^a - \left(\frac12 + \frac{1}{2\tilde c-2}\right)\psi_a\tilde t + \varepsilon^{abc}A_b\tilde g^c + \varepsilon^{abc}\left(2\partial\psi_b + \psi_b\partial\right)\tilde u^c \\
&\quad -\left(1+\frac{3}{2\tilde c-2}\right)\left(\frac{3}{\tilde c+1/2}\right)^2\psi_b\left(U^{(a}U^{b)}\right)_{\rm eff} \\
\partial A_a &= \overline{\nabla}\tilde u^a - \left(1-\frac{3}{2\tilde c+1}\right)\varepsilon^{abc}\psi_b\tilde g^c + \varepsilon^{abc}A_b\tilde u^c\,.
\end{aligned} \tag{5.3.16}$$

The normalisation of the currents has been chosen such that the anomalous terms on the *lhs* have coefficient one. The explicit $\tilde c$ dependence of the Ward identities arises from several sources: some couplings in the nonlinear algebra eq. (5.3.1) are explicitly $\tilde c$-dependent, the transformation rules eq. (5.3.4) are $\tilde c$-dependent, and the field-nonlinearity. The dependence implies that the induced action is given by a $1/\tilde c$ expansion:

$$\widetilde\Gamma[h,\psi,A] = \sum_{i\geq 0}\tilde c^{1-i}\widetilde\Gamma^{(i)}[h,\psi,A]\,. \tag{5.3.17}$$

Returning to the Ward identities for the linear theory eq. (5.3.7), we observe that when we take $\tilde c = c+1/2$ and put $q=0$, we find from the last identity in eq. (5.3.7) that $\eta = -\psi_a u^a$. Substituting this into the first three identities of eq. (5.3.7) yields precisely the Ward identities for the nonlinear theory eq. (5.3.16) in the $c\to\infty$ limit. Also, the extra term in the nonlinear $\delta A_a$ (eq. (5.3.4)), that was added to bring the anomaly to a minimal form, now effectively reinserts the $\theta^a\eta$ term that disappeared from the linear transformation, eq. (5.3.3). This is in accordance with the observations at the end of the previous section.

We will now prove relation (5.2.3) between $Z$ and $\tilde Z$. First we rewrite the definition of $Z$ (2.5.1) using eq. (5.3.2), the crucial ingredient being that $Q$ commutes with the nonlinear algebra, thus factorising the averages:

$$\begin{aligned}
Z[h,\psi,A,\eta] &= \Bigg\langle\exp\Big(-\frac1\pi\int(h\widetilde T + \psi_a\widetilde G^a + A_a\widetilde U^a)\Big) \\
&\qquad \Big\langle\exp\Big(-\frac1\pi\int(hT_Q + \hat\eta Q)\Big)\Big\rangle_Q\Bigg\rangle \\
&= \Big\langle\exp\Big(-\frac1\pi\int(h\widetilde T + \psi_a\widetilde G^a + A_a\widetilde U^a) - \Gamma[h,\hat\eta]\Big)\Big\rangle
\end{aligned} \tag{5.3.18}$$

where:

$$T_Q = \tfrac{3}{2c}Q\partial Q,\quad \hat\eta = \eta - \frac{1}{3c}\psi_a\widetilde U^a\,. \tag{5.3.19}$$

The $Q$ integral can be expressed in terms of the Polyakov action (2.5.15):

$$\Gamma[h,\hat\eta] = \frac{1}{48\pi}\Gamma_{\rm Pol}[h] - \frac{c}{6\pi}\int\hat\eta\frac{1}{\overline{\overline{\nabla}}}\hat\eta\,, \tag{5.3.20}$$

where $\overline{\nabla} = \bar\partial - h\partial - \frac{1}{2}\partial h$ and

$$\Gamma_{\rm Pol}[h] = \int \partial^2 h \frac{1}{\bar\partial}\frac{1}{1-h\partial\bar\partial^{-1}}\frac{1}{\partial}\partial^2 h\,. \tag{5.3.21}$$

Using eqs. (5.3.18) and (5.3.20), we find:

$$\begin{aligned}\exp\Big(-\widetilde\Gamma[h,\psi,A]\Big) \;&=\; \\ &\exp\Big(\Gamma[h,\hat\eta = \eta + \frac{\pi}{3c}\psi_b\frac{\delta}{\delta A_b}]\Big)\;\exp\Big(-\Gamma[h,\psi,A,\eta]\Big)\,.\end{aligned} \tag{5.3.22}$$

We checked this formula explicitly on the lowest order correlation functions using *OPEdefs*. Introducing the Fourier transform of $\Gamma$ with respect to $A$:

$$\exp\Big(-\Gamma[h,\psi,A,\eta]\Big) = \int [du]\exp\Big(-\Gamma[h,\psi,u,\eta] + \frac{c}{3\pi}\int u^a A_a\Big)\,, \tag{5.3.23}$$

eq. (5.3.23) further reduces to:

$$\begin{aligned}\exp\Big(-\widetilde\Gamma[h,\psi,A]\Big) \;=\; &\exp\Big(\frac{1}{48\pi}\Gamma_{\rm Pol}[h]\Big)\int[du]\exp\Big(-\Gamma[h,\psi,u,\eta] \\ &-\frac{c}{6\pi}\int\Big(\eta+\psi_a u^a\Big)\frac{1}{\overline{\overline{\nabla}}}\Big(\eta+\psi_b u^b\Big) + \frac{c}{3\pi}\int u^a A_a\Big)\,.\end{aligned} \tag{5.3.24}$$

As the *lhs* of eq. (5.3.25) is $\eta$-independent, the *rhs* should also be. We can integrate both sides over $\eta$ with a measure chosen such that the integral is equal to one:

$$\exp\Big(-\frac{1}{48\pi}\Gamma_{\rm Pol}[h]\Big)\int[d\eta]\exp\Big(\frac{c}{6\pi}\int\Big(\eta+\psi_a u^a\Big)\frac{1}{\overline{\overline{\nabla}}}\Big(\eta+\psi_b u^b\Big)\Big) = 1\,. \tag{5.3.25}$$

Combining this with eq. (5.3.25), we finally obtain eq. (5.2.3).

### 5.3.2 $N = 4$ superconformal algebras

Now we extend the method applied for $N = 3$ to the case of $N = 4$. Again, there is a linear $N = 4$ algebra and a nonlinear one, obtained [103] by decoupling four free fermions and a $U(1)$ current. In the previous case we made use in the derivation of the explicit form of the action induced by integrating out the fermions. In the present case no explicit expression is available for the corresponding quantity, but we will see that it is not needed.

The $N = 4$ superconformal algebra [171, 185] is generated by the energy–momentum tensor $T$, four supercharges $G^a$, $a \in \{1,2,3,4\}$, an $so(4)$ affine Lie algebra with generators $U^{ab} = -U^{ba}$, $a,b \in \{1,2,3,4\}$, 4 free fermions $Q^a$ and a $U(1)$ current $P$. The two $su(2)$- algebras have levels $k_+$ and $k_-$. The supercharges $G^a$ and the dimension 1/2 fields $Q^a$ form two $(2,2)$ representations of $su(2)\otimes su(2)$. The central charge is given by:

$$c = \frac{6\,k_+\,k_-}{k_+ + k_-}\,. \tag{5.3.26}$$

The OPEs are (we omit the OPEs of $T$):

$$\begin{aligned}
G^a\,G^b &= \frac{3c}{2}\delta^{ab}[\mathbb{1}] + [-2\,U^{ab} + \zeta\,\varepsilon^{abcd}U^{cd}] \\
U^{ab}\,U^{cd} &= \frac{k}{2}\left(\delta^{ad}\,\delta^{bc} - \delta^{ac}\,\delta^{bd} - \zeta\,\varepsilon^{abcd}\right)[\mathbb{1}] \\
&\quad +[\delta^{bd}\,U^{ac} - \delta^{bc}\,U^{ad} - \delta^{ad}\,U^{bc} + \delta^{ac}\,U^{bd}] \\
U^{ab}\,G^c &= -\zeta\,\left(\delta^{bc}\,[Q^a] - \delta^{ac}\,[Q^b]\right) + \varepsilon^{abcd}\,[Q^d] - (\delta^{bc}\,[G^a] - \delta^{ac}\,[G^b]) \\
Q^a\,G^b &= \delta^{ab}\,[P] - \frac{1}{2}\varepsilon^{abcd}\,[U^{cd}] \\
Q^a\,U^{bc} &= \delta^{ac}\,[Q^b] - \delta^{ab}\,[Q^c] \\
P\,G^a &= [Q^a] \\
P\,P &= -\frac{k}{2}[\mathbb{1}] \\
Q^a\,Q^b &= -\frac{k}{2}\delta^{ab}[\mathbb{1}]
\end{aligned} \tag{5.3.27}$$

where $k = k_+ + k_-$ and $\zeta = (k_+ - k_-)/k$. Note that $[PA]_1 = 0$ for all primaries, and hence for all elements of the OPA.

We will write the induced action (2.5.4) as $\Gamma[h,\psi,A,b,\eta]$. All the structure constants of the linear algebra (5.3.27) depend only on the ratio $k_+/k_-$. Apart from this ratio, $k$ enters as a proportionality constant for *all* two-point functions. As a consequence, $\Gamma$ depends on that ratio in a nontrivial way, but its $k$-dependence is simply an overall factor $k$.

Using the following definitions:

$$t = \frac{12\pi}{c}\frac{\delta\Gamma}{\delta h},\quad g^a = \frac{3\pi}{c}\frac{\delta\Gamma}{\delta\psi_a},\quad u^{ab} = -\frac{\pi}{k}\frac{\delta\Gamma}{\delta A_{ab}}, \tag{5.3.28}$$

$$q^a = -\frac{2\pi}{k}\frac{\delta\Gamma}{\delta\eta_a},\quad p = -\frac{2\pi}{k}\frac{\delta\Gamma}{\delta b} \tag{5.3.29}$$

and $\gamma = 6k/c$, the Ward identities are:

$$\partial^3 h \;=\; \overline{\nabla}t - 2\,(\psi_a\partial + 3\partial\psi_a)\,g^a + 2\,\gamma\,\partial A_{ab}u^{ab} + \frac{\gamma}{2}\,(\partial\eta_a - \eta_a\partial)\,q^a + \gamma\,\partial b\,p$$

$$\begin{aligned}
\partial^2\psi_a &= \overline{\nabla} g^a - 2\,A_{ab}g^b - \frac{1}{2}\psi_a t + \frac{\gamma}{4}\partial b\, q^a + \frac{\gamma}{2}\zeta\,\partial A_{ab}q^b + \frac{\gamma}{4}\varepsilon_{abcd}\,\partial A_{bc}q^d \\
&\quad + \frac{\gamma}{4}\varepsilon_{abcd}\,\eta_b u^{cd} + \frac{\gamma}{4}\left(\psi_b\partial + 2\partial\psi_b\right)\left(2u^{ab} - \zeta\varepsilon_{abcd}\,u^{cd}\right) + \frac{\gamma}{4}\,\eta_a p \\
\partial A_{ab} &+ \frac{\zeta}{2}\varepsilon_{abcd}\,\partial A_{cd} = \overline{\nabla} u^{ab} - 4\,A_{c[a}u^{b]c} - \frac{4}{\gamma}\psi_{[a}g^{b]} + \eta_{[a}q^{b]} \\
&\quad -\zeta\left(\psi_{[a}\partial + \partial\psi_{[a}\right)q^{b]} - \frac{1}{2}\varepsilon_{abcd}\left(\psi_c\partial + \partial\psi_c\right)q^d \\
\eta_a &= \overline{\nabla} q^a - 2\,A_{ab}q^b - \psi_a p + \varepsilon_{abcd}\,\psi_b u^{cd} \\
\partial b &= \overline{\nabla} p - \left(\psi_a\partial + \partial\psi_a\right)q^a\,. \qquad (5.3.30)
\end{aligned}$$

Note that $b$ appears only with at least one derivative. The brackets denote antisymmetrisation in the indices.

The nonlinear $N=4$ superconformal algebra has the same structure as eq. (5.3.27) but there is no $P$ and $Q^a$. The central charge is related to the $su(2)$-levels by $\tilde c = \frac{3(\tilde k + 2\tilde k_+\tilde k_-)}{2+\tilde k}$. We only give the $\widetilde{G}\widetilde{G}$ OPE explicitly:

$$\begin{aligned}
\widetilde{G}^a\,\widetilde{G}^b &= \frac{4\tilde k_+\tilde k_-}{\tilde k+2}\,\delta^{ab}\,[\mathbb{1}] - \frac{2\tilde k}{\tilde k+2}\,[\widetilde{U}^{ab}] + \frac{\tilde k_+ - \tilde k_-}{\tilde k+2}\varepsilon_{abcd}\,[\widetilde{U}^{cd}] \qquad (5.3.31)\\
&\quad + [\frac{2\tilde k}{\tilde k + 2\tilde k_+\tilde k_-}\,\delta^{ab}\,\widetilde{T} + \frac{1}{4(\tilde k+2)}\,\varepsilon_{acdg}\,\varepsilon_{befg}(\widetilde{U}^{cd}\widetilde{U}^{ef} + \widetilde{U}^{ef}\widetilde{U}^{cd})]\,.
\end{aligned}$$

To write down the Ward identities in this case, we define:

$$\tilde t = \frac{12\pi}{\tilde c}\frac{\delta\widetilde\Gamma}{\delta h}, \qquad \tilde g^a = \frac{(\tilde k+2)\pi}{2\tilde k_+\tilde k_-}\frac{\delta\widetilde\Gamma}{\delta\psi_a}, \qquad \tilde u^{ab} = -\frac{\pi}{\tilde k}\frac{\delta\widetilde\Gamma}{\delta A_{ab}}, \qquad (5.3.32)$$

and

$$\tilde\gamma = \frac{\tilde k(\tilde k+2)}{\tilde k_+\tilde k_-}, \qquad \tilde\kappa = \frac{6\tilde k}{\tilde c}, \qquad \tilde\zeta = \frac{\tilde k_+ - \tilde k_-}{\tilde k}\,. \qquad (5.3.33)$$

$$\begin{aligned}
\partial^3 h &= \overline{\nabla}\tilde t - \frac{2\tilde\kappa}{\tilde\gamma}\left(\psi_a\partial + 3\partial\psi_a\right)\tilde g^a + 2\tilde\kappa\,\partial A_{ab}\tilde u^{ab} \\
\partial^2\psi_a &= \overline{\nabla}\tilde g^a - 2\,A_{ab}\tilde g^b - \frac{\tilde\gamma}{2\tilde\kappa}\,\psi_a\tilde t \\
&\quad - \frac{\tilde\gamma}{4\tilde k(\tilde k+2)}\,\varepsilon_{acdg}\,\varepsilon_{befg}\,\psi_b\left(\left(\widetilde{U}^{cd}\widetilde{U}^{ef}\right)_{\mathit{eff}} + \left(\widetilde{U}^{ef}\widetilde{U}^{cd}\right)_{\mathit{eff}}\right) \\
&\quad + \frac{\tilde\gamma\tilde k}{4(\tilde k+2)}\left(\psi_b\partial + 2\partial\psi_b\right)\left(2\tilde u^{ab} - \tilde\zeta\,\varepsilon_{abcd}\tilde u^{cd}\right) \\
\partial A_{ab} &+ \frac{\tilde\zeta}{2}\,\varepsilon_{abcd}\,\partial A_{cd} = \overline{\nabla}\tilde u^{ab} - 4A_{c[a}\tilde u^{b]c} - \frac{4}{\tilde\gamma}\psi_{[a}\tilde g^{b]}\,. \qquad (5.3.34)
\end{aligned}$$

As in the previous subsection, we will use the results of section 5.1 or [103] to eliminate the free fermion fields and the $U(1)$-field $P$. The new currents are:

$$\begin{aligned}
\widetilde{T} &= T + \frac{1}{k}PP + \frac{1}{k}\partial Q^c Q^c \\
\widetilde{G}^a &= G^a + \frac{2}{k}PQ^a + \varepsilon_{abcd}\left(\frac{2}{3k^2}\,Q^bQ^cQ^d + \frac{1}{k}Q^b\widetilde{U}^{cd}\right) \\
\widetilde{U}^{ab} &= U^{ab} - \frac{2}{k}\,Q^aQ^b \qquad (5.3.35)
\end{aligned}$$

and the constants in the algebras are related by $\tilde k_\pm = k_\pm - 1$, and thus $\tilde c = c - 3$. Again, we find agreement between the large $k$-limit of the Ward identities putting $q^a$ and $p$ to zero, and solving $\eta^a$ and $\partial b$ from the two last identities of (5.3.30).

We now find the nonlinear effective action in terms of the effective action of the linear theory. In analogy with the $N=3$ case, it seems that the operators of the nonlinear theory can be written as the difference of the operators of the linear theory, and a realisation of the linear theory given by the free fermions and $P$. In the present case, this simple linear combination is valid for the integer spin currents $\widetilde{T}$ and $\widetilde{U}$, but not for $\widetilde{G}$. A second complication is that, due to the presence of a trilinear term (in $Q$) in the relation between $\widetilde{G}$ and $G$, integrating out the $Q$-fields is more involved. Nevertheless, we can still obtain (5.2.3).

There is a variety of ways to derive this relation, starting by rewriting the decompositions of (5.3.35) in different ways. We will use the following form:

$$G^a + \frac{1}{k}\varepsilon_{abcd}Q^bU^{cd} = \widetilde{G}^a - \frac{2}{k}PQ^a + \frac{4}{3k^2}\varepsilon_{abcd}Q^bQ^cQ^d\,. \qquad (5.3.36)$$

This leads immediately to[3]:

$$\begin{aligned}
&\left\langle \exp\left(-\frac{1}{\pi}\int\left(hT + \psi_a G^a + A_{ab}U^{ab} + bP + \eta_a Q^a + \frac{1}{k}\varepsilon_{abcd}\psi_a Q^b U^{cd}\right)\right)\right\rangle = \\
&\left\langle \exp\left(-\frac{1}{\pi}\int\left(\left(h\widetilde{T} + \psi_a\widetilde{G}^a + A_{ab}\widetilde{U}^{ab}\right)\right.\right.\right. \qquad (5.3.37)
\end{aligned}$$

[3]There are no normal ordering problems as the OPEs of the relevant operators turn out to be nonsingular (e.g. the term cubic in the $Q^a$ is an antisymmetric combination).

$$+\frac{1}{k}\left(-hP^2 - h\partial Q^a Q_a - 2\psi_a P Q^a + 2A_{ab}Q^aQ^b + bP + \eta_a Q^a\right)$$
$$+\frac{4}{3k^2}\varepsilon_{abcd}\psi_a Q^bQ^cQ^d\Big)\Big)\Big\rangle .$$

Again the crucial step is that in the *rhs*, the expectation value factorises: the average over $Q^a$ and $P$ can be computed separately, since these fields commute with the nonlinear SUSY-algebra. This average is in fact closely related to the partition function for the linear $N = 4$ algebra with $k_+ = k_- = 1$ and $c = 3$, up to the renormalisation of some coefficients. We have:

$$Z^{c=3}[h,\psi,A,b,\eta] = \Big\langle \exp\Big( -\frac{1}{\pi}\int\Big(-\frac{h}{2}\left(\hat{P}\hat{P} + \partial\hat{Q}_a\hat{Q}^a\right) \qquad (5.3.38)$$
$$-\psi_a\left(\hat{P}\hat{Q}^a + \frac{1}{6}\varepsilon_{abcd}\hat{Q}^b\hat{Q}^c\hat{Q}^d\right) + A_{ab}\hat{Q}^a\hat{Q}^b + b\hat{P} + \eta_a\hat{Q}^a\Big)\Big)\Big\rangle$$

where the average value is over free fermions $\hat{Q}^a$ and a free $U(1)$-current $\hat{P}$. These are normalised in a $k$-independent fashion:

$$\hat{P}\times\hat{P} \to -[\mathbb{1}] \qquad \hat{Q}^a\times\hat{Q}^b \to -\delta^{ab}[\mathbb{1}] \qquad (5.3.39)$$

and the explicit form [170, 171, 185] of the currents making up the $c = 3$ algebra has been used. The average can be represented as a functional integral with measure:

$$[d\hat{Q}][d\hat{P}]\exp\left( -\frac{1}{2\pi}(\hat{P}\frac{\bar\partial}{\partial}\hat{P} + \hat{Q}^a\bar\partial\hat{Q}_a)\right) . \qquad (5.3.40)$$

The (nonlocal) form of the free action for $\hat{P}$ follows from its two-point function: it is the usual (local) free scalar field action if one writes $\hat{P} = \partial X$. The connection between the linear theory, the nonlinear theory, and the $c = 3$ realisation is then:

$$\exp\left( -\frac{\pi}{k}\varepsilon^{abcd}\psi_a\frac{\delta}{\delta\eta_b}\frac{\delta}{\delta A_{cd}}\right) Z[\psi,A,\eta,b] = \widetilde{Z}[\psi,A] \qquad (5.3.41)$$
$$\exp\left(\frac{\pi^2}{3k^2}(4+\sqrt{2k})\varepsilon^{abcd}\psi_a\frac{\delta}{\delta\eta_b}\frac{\delta}{\delta\eta_c}\frac{\delta}{\delta\eta_d}\right) Z^{c=3}[h,\psi,A,\eta\sqrt{k/2},b\sqrt{k/2}].$$

Contrary to the $N = 3$ case, where the Polyakov partition function was obtained very explicitly, this connection is not particularly useful, but the representation (5.3.40) of $Z^{c=3}$ as a functional integral can be used effectively. Indeed, when we take the Fourier transform of eq. (5.3.41), *i.e.* we integrate (5.3.41) with:

$$\int [d\,h][d\,\psi][d\,A][d\,b][d\,\eta]\exp\left(\frac{1}{\pi}\int\left(h\,t + \psi_a\,g^a + A_{ab}\,u^{ab} + b\,p + \eta_a\,q^a\right)\right), \quad (5.3.42)$$

we obtain using eqs. (5.3.38) and (5.3.40)[4]:

$$\exp\left( - W[t,\, g^a - \frac{1}{k}\varepsilon_{abcd}q^bu^{cd},\, u,\, p,\, q]\right) = \exp\left( -\frac{1}{\pi k}(p\frac{\bar\partial}{\partial}p + q^a\bar\partial q_a)\right)$$
$$\exp\left( - \widetilde{W}[t + \frac{1}{k}\left(p^2 + \partial q\, q\right),\, g^a + \frac{2}{k}p\,q^a - \frac{4}{3k^2}\varepsilon_{abcd}q^bq^cq^d, u^{ab} - \frac{2}{k}q^aq^b]\right)$$
$$(5.3.43)$$

giving the concise relation:

$$\widetilde{W}[t,\, g^a,\, u^{ab}] \;+\; \frac{1}{\pi k}(p\frac{\bar\partial}{\partial}p + q^a\bar\partial q_a)$$
$$= W[t - \frac{1}{k}\left(p^2 + \partial q^a\, q_a\right),$$
$$g^a - \frac{2}{k}pq^a - \frac{1}{k}\varepsilon_{abcd}q^bu^{cd} - \frac{2}{3k^2}\varepsilon_{abcd}q^bq^cq^d,$$
$$u^{ab} + \frac{2}{k}q^aq^b,\, p,\, q^a] .$$
$$(5.3.44)$$

Putting the free $p$ and $q^a$-currents equal to zero, we obtain (5.2.7).

## 5.4 Discussion

We have given explicit algorithms for the factoring out of free fields, including a simple criterion for the factorisation of $U(1)$–scalars. We have worked purely at the quantummechanical level, but the algorithms do not need any modification for the classical case if we use the definition eq. (2.3.32) for the negative modes. In fact, recently [71, 167] a number of classical $\mathcal{W}$–algebras were constructed by hamiltonian reduction, containing bosons of dimensions 1 and $\frac{1}{2}$ that could be decoupled.

Recently [52], it was shown that factoring out all dimension 1 fields of a classical $\mathcal{W}$–algebra gives rise to algebras which are finitely generated by rational polynomials, *i.e.* fractions of generators are allowed. When regarding the fractions as new generators, a nonlinear $\mathcal{W}$–algebra with an infinite number of generators results. In the quantum case only a finite number of generators are needed, due to the appearance of null fields.

The examples in section 5.3 have shown that factorising a linear algebra gives in general rise to nonlinear algebras. Conversely, one could attempt to linearise a

[4] The effective action $W$ is defined by the Fourier transform of $Z$, and similarly for $\widetilde{W}$.

$\mathcal{W}$–algebra by adding free fields and performing a basis transformation. This was achieved for $W_3$ and the bosonic $N = 2$ superconformal algebra in ref. [136]. Some important features of the construction are that the free fields are not quasiprimary with respect to the original energy–momentum tensor, and that the linear algebra contains a "null" field $G$ in the sense that it has no OPE where $\mathbb{1}$ occurs. In fact, for $W_3$ the field $G$ is equal to $W - W_J$, where $W_J$ is the expression of the one scalar realisation of $W_3$. Clearly, this linear algebra is only of interest if one does not put $G$ to zero.

The results of this chapter can be used to see if the effective action of a nonlinear theory is related to its classical limit by introducing some renormalisations, see section 7.4. If we take for granted that this is true for the linear theory, eq. (5.2.7) immediately transfers this property to the nonlinear theory. Moreover, since the 'classical' parts are equal also (as implied by the $c \to \infty$ limit of the Ward identities) the renormalisation factors for both theories are the same (for couplings as well as for fields) if one takes into account the shifts in the values for the central extensions $c$, $k_+$ and $k_-$. This fact can be confirmed by looking at explicit calculations of these renormalisation factors, see chapter 7.

# Chapter 6

# Extensions of the Virasoro algebra and gauged WZNW–models

One of the most general methods to obtain a realisation of an extended Virasoro algebra is Drinfeld-Sokolov reduction [60, 7]. This method consists of imposing certain constraints on the currents of a Kač–Moody algebra $\bar{g}$. The constraints are associated to an embedding of $sl(2)$ in the Lie algebra $g$. In the classical case, the reduced phase space then forms a realisation of a $\mathcal{W}$–algebra [60, 7], as reviewed in section 6.1.

In [7], it was shown that the constraints reduce the classical WZNW Ward identities to those of the extended Virasoro algebra. This points towards a connection between the induced actions of the Kač–Moody algebra and the $\mathcal{W}$–algebra. This connection was made explicit in the case of $\mathcal{W}_3$ in [39]. In section 6.2 this is generalised to arbitrary (super) Kač–Moody algebras [181]. The main idea is to implement the constraints on a gauged WZNW–model. This leads to a path integral formulation in the Batalin-Vilkovisky formalism of the induced action of the $\mathcal{W}$–algebra in terms of the WZNW action.

To complete the construction, we show in section 6.4 that the quantum currents of the constrained WZNW–model form a realisation of a quantum $\mathcal{W}$–algebra. These results are used in section 6.6 to find the quantum corrections to the Batalin-Vilkovisky action.

The $osp(N|2)$ affine Lie algebra will be used as an example in sections 6.3 and 6.5. The $sl(2)$ embedding we use, gives a realisation of the $N$-extended $so(N)$ superconformal algebras [133, 20].

This chapter contains results published in [181, 183]. Several results presented here are new. For conventions on WZNW–models, we refer to section 2.6.4. Appendix B fixes our conventions for (super) Lie algebras and summation indices.

## 6.1 Classical Drinfeld-Sokolov reduction

In this section we will briefly explain the classical Hamiltonian reduction that gives a realisation of a (classical) $\mathcal{W}$–algebra in terms of a Kač–Moody algebra. References [60, 7, 69, 86, 45] can be consulted for further details.

Consider a Kač–Moody algebra $\hat{g}$ with level $\kappa$, with a (super) Lie algebra $\bar{g}$ valued field $J_z(z) = J_z^a(z)\ {}_at$, where $t_a$ are matrices representing $\bar{g}$. For a given $sl(2)$-embedding, we impose a set of constraints on the currents $J_z^a$:

$$\Pi_-\left(J_z - \frac{\kappa}{2}e_-\right) = 0\,. \tag{6.1.1}$$

These constraints are all first class in the terminology of Dirac [57], except $\Pi_{-1/2}J_z = 0$. It is more convenient to have all constraints first class. For a (bosonic) Lie algebra, this can be achieved by taking only half of the constraints at grading $-1/2$. It is possible [69] to do this by choosing an alternative integral grading. However, this procedure does not work for most super Lie algebras. Therefore we will introduce auxiliary fields instead. We postpone the discussion of this case with half-integral $sl(2)$ grading to the next section. So we ignore in this section any sign issues for the super case.

In the reduced phase space, there is a gauge freedom because elements are considered equivalent modulo the constraints. The infinitesimal gauge transformations (2.6.43):

$$\begin{aligned}\delta_\eta J_z(x) &= -\int \left\{ str\left(\eta(y)J_z(y)\right), J_z(x)\right\}_{\rm PB}\\ &= \frac{\kappa}{2}\partial\eta(x) + [\eta(x), J_z(x)], \qquad\qquad \eta\in\Pi_+\bar{g}\end{aligned} \tag{6.1.2}$$

can be used to impose suitable gauge conditions. Several choices are possible, but we will mainly use the highest weight gauge:

$$\Big(1-\Pi_{hw}\Big)\Big(J_z-\frac{\kappa}{2}e_-\Big)=0\,. \tag{6.1.3}$$

We will now compute the reduced Poisson algebra in the highest weight gauge. This can be done in several ways. First, one can view the gauge conditions as an extra set of constraints. Because the constraints are then second class, Dirac brackets should be used. However, this approach requires inverting the matrix of the Poisson brackets of the constraints, which is in general quite difficult.

A second way is by using gauge invariant polynomials. These are polynomials in the currents which are gauge invariant under eq. (6.1.2) up to the constraints (6.1.1). It can be shown [9] that for each highest weight current, there is a gauge invariant polynomial [1]:

$$\tilde J_z^{(j,\alpha_j)} = J_z^{(jj,\alpha_j)}\Big|_{\text{constraints eq. (6.1.3)}}\,. \tag{6.1.4}$$

The $\tilde J_z^{(j,\alpha_j)}$ are unique modulo the constraints (6.1.1). They generate all gauge invariant polynomials by using addition and multiplication. The polynomials (6.1.4) can be computed by finding the unique (finite) gauge transformation which brings the constrained current eq. (6.1.1) to the gauge fixed form eq. (6.1.3). This is used in [45, 46, 187] to construct an algorithm to obtain the polynomials $\tilde J_z^{(j,\alpha_j)}$. Because the gauge transformations are compatible with the Poisson brackets, the gauge invariant polynomials form a (nonlinear) subalgebra with respect to the original Poisson bracket. Moreover, we can see that this algebra is isomorphic to the reduced Poisson algebra.

**Proof :**
To show that the Poisson algebra of the gauge invariant polynomials is isomorphic to the reduced Poisson algebra, we observe that their Dirac and Poisson brackets are the same (modulo (6.1.1)). Recall the definition of the Dirac bracket for some second class constraints $\chi^\alpha$:

$$\{F(x),G(y)\}_D = \{F(x),G(y)\}_{\rm PB} - \int d^2x_0 d^2y_0 \{F(x),\chi^\alpha(x_0)\}_{\rm PB}\Delta_{\alpha\beta}(x_0,y_0)\{\chi^\beta(y_0),G(y)\}_{\rm PB}\,,$$

with $\Delta_{\alpha\beta}$ the inverse matrix of $\{\chi^\alpha,\chi^\beta\}_{\rm PB}$. When $F(x)$ is gauge invariant, *i.e.* $\{F,\chi^\alpha\}_{\rm PB}$ is zero modulo constraints, we see that $\{F,G\}_D=\{F,G\}_{\rm PB}$ modulo constraints. ■

A third way to compute the reduced phase space is to use cohomological techniques [138]. For every constraint $\chi^\alpha$ in (6.1.1), we introduce a ghost $c_\alpha$ and antighost $b^\alpha$ pair, to which we assign ghost number +1 and −1 respectively. We define the Poisson brackets:

$$\{b^\alpha(x),c_\beta(y)\}_{\rm PB} = \delta^\alpha{}_\beta\,\delta^2(x-y)\,. \tag{6.1.5}$$

The BRST operator $Q$ acting in the complex $\mathcal{A}$ generated by $\{J_z^a, b^\alpha\}$ is defined by:

$$Q(A)(x)=\int d^2y\,\{\mathcal{J}(y),A(x)\}_{\rm PB}\,,\qquad A\in\widehat{\mathcal{A}}\,, \tag{6.1.6}$$

where the ghost number +1 BRST current:

$$\mathcal{J}=c_\alpha(x)\ {}^\alpha\chi(x)+\text{higher order in ghostfields}\,, \tag{6.1.7}$$

is determined by requiring that $Q^2=0$. Because the constraints (6.1.1) form a linear Poisson algebra, we know that this extra part contains only ghosts. We now study the cohomology of $Q$:

$$\mathcal{H}^*(Q;\widehat{\mathcal{A}})=\frac{\ker(Q)}{\mathrm{im}(Q)}\,, \tag{6.1.8}$$

*i.e.* we look for fields annihilated by $Q$, but as $Q^2=0$, we use the equivalence relation:

$$A\sim A+Q(B)\,. \tag{6.1.9}$$

Let us see how this is related to the constraints. $Q$ on a field $f$, which does not contain $b$ or $c$, gives the gauge transformation (6.1.2) with parameter $c$ of $f$. Obviously, we are only interested in fields modulo the gauge transformations, *i.e.* modulo im$(Q)$. Now consider a ghost number zero functional of the currents and (anti)ghosts $F[J,b,c]$. We will call $f[J]\equiv F[J,b=0,c=0]$. Because $Q$ acts as a derivation, the BRST transformation of $F$ is schematically given by:

$$Q(F)=Q(f)+G_\alpha Q(b^\alpha)+H^\alpha Q(c_\alpha)+Q(J)I\,, \tag{6.1.10}$$

where $G,H,I$ are other functionals with $I$ depending on $b,c$. Because of the Poisson brackets (6.1.5) this gives:

$$Q(F)=\delta_c(f)+G_\alpha\chi^\alpha+\text{ terms with antighosts}\,, \tag{6.1.11}$$

where $\delta_c$ denotes the gauge transformation (6.1.2) with parameter $c$. This means that if $Q(F)=0$, its ghost–free part $f$ is gauge invariant modulo constraints. Moreover, it can be shown that the "dressing" of a gauge invariant polynomial to an element of the cohomology is unique, see *e.g.* [110]. Hence, computing the cohomology of $Q$ corresponds to finding the gauge invariant polynomials, and the gauge choice (6.1.3) corresponds to a choice of representative for an equivalence class (6.1.9). In [43], an iterative method was provided to construct the generators of the cohomology for the case of bosonic Lie algebras. We will discuss the generalisation of this method in section 6.4.

[1] See appendix B for the basis for the generators of the super Lie algebra.

Quantisation of the Drinfeld-Sokolov reduction using a BRST approach was initiated in [24]. The general case for bosonic Kač–Moody algebras, was treated in [43]. We will treat superalgebras in section 6.4.

Having characterised the reduced phase space, we now give some general comments about its structure. The current corresponding to $e_+$ (which has $j=1$) is common to all reductions (as we are treating $sl(2)$ embeddings). It turns out that it satisfies a Virasoro Poisson bracket. This is still the case if we add the Sugawara tensor $T_S^{(0)}$ for the Kač–Moody algebra formed by the highest weight $j=0$ currents. We define:

$$T_s = 2y\tilde{J}_z^{(1,0)} + sT_S^{(0)}\,, \tag{6.1.12}$$

which is a Virasoro current for $s=0,1$. With respect to $T_1$, all the other currents $\tilde{J}_z^{(j,\alpha_j)}$ are primary and have spin $j+1$ [69], see also section 6.4. So, we have found a representation of a classical $\mathcal{W}$–algebra for any $sl(2)$ embedding in a (super) Lie algebra. See [87] for a classification of all embeddings and the dimensions of the generators of the resulting $\mathcal{W}$–algebra.

## 6.2 Quantum reduction

In this section, the previous scheme for constraining the currents of a Kač–Moody algebra is implemented by gauging a WZNW–model. This will enable us to quantise the reduction and to give a path integral expression for the induced action of the $\mathcal{W}$–algebra which results from the construction.

### 6.2.1 Gauged WZNW Model

The affine Lie algebra $\hat{g}$ is realised by a WZNW–model with action $\kappa S^-[g]$. This action has the global symmetry (2.6.42). To make part of the symmetries local, we introduce gauge fields $A_z \in \Pi_-\bar{g}$ and $A_{\bar{z}} \in \Pi_+\bar{g}$ and the action:

$$\begin{aligned}\mathcal{S}_0 &= \kappa S^-[g] - \frac{\kappa}{2\pi x}\int str\left(A_z\, g^{-1}\bar{\partial}g\right) + \frac{\kappa}{2\pi x}\int str\left(A_{\bar{z}}\partial g g^{-1}\right) + \\ &\quad \frac{\kappa}{2\pi x}\int str\left(A_{\bar{z}} g A_z g^{-1}\right)\,. \end{aligned}\tag{6.2.1}$$

Parametrising the gauge fields as $A_z \equiv \partial h_-\, h_-^{-1}$ and $A_{\bar{z}} \equiv \bar{\partial}h_+\, h_+^{-1}$ where $h_\pm \in \Pi_\pm\bar{g}$, we see that $\mathcal{S}_0$ is invariant under:

$$\begin{aligned} h_\pm &\to \gamma_\pm h_\pm \\ g &\to \gamma_+ g \gamma_-^{-1}, \end{aligned}\tag{6.2.2}$$

where

$$\gamma_\pm \in \Pi_\pm\bar{g}. \tag{6.2.3}$$

This can be shown by using the Polyakov-Wiegman formula (2.6.36) to bring eq. (6.2.1) in the form $\mathcal{S}_0 = \kappa S^-[h_+^{-1}gh_-]$ and noting that $S^-[h_\pm]=0$. We will always work in the gauge $A_z=0$ and hence drop all contributions of this gauge field.

In order to impose the constraints (6.1.1), we would like to add the following term to the action:

$$\mathcal{S}_{\rm extra} = -\frac{\kappa}{2\pi x}\int str\left(A_{\bar{z}}e_-\right), \tag{6.2.4}$$

using the gauge field as a Lagrange multiplier. The variation of this term under the gauge transformations eq. (6.2.2) is:

$$\delta\mathcal{S}_{\rm extra} = -\frac{\kappa}{2\pi x}\int str\left([\eta, A_{\bar{z}}]e_-\right). \tag{6.2.5}$$

The variation is nonzero when $\Pi_{1/2}\bar{g}$ is not empty. We would like to preserve the gauge invariance. As already mentioned, this can be done for bosonic algebras by choosing an integral grading. This is used in [43] to quantise the Drinfeld-Sokolov reduction for bosonic algebras. However, for a super Lie algebra, another method has to be used. We restore gauge invariance by introducing the "auxiliary" field $\tau \in \Pi_{1/2}\bar{g}$ and define:

$$\begin{aligned}\mathcal{S}_1 &= \kappa S^-[g] + \frac{1}{\pi x}\int str\left(A_{\bar{z}}\left(J_z - \frac{\kappa}{2}e_- - \frac{\kappa}{2}[\tau, e_-]\right)\right) \\ &\quad + \frac{\kappa}{4\pi x}\int str\left([\tau, e_-]\bar{\partial}\tau\right), \end{aligned}\tag{6.2.6}$$

with the affine currents (2.6.40) $J_z = \frac{\kappa}{2}\partial g g^{-1}$. If $\tau$ transforms under (6.2.2) as:

$$\tau \to \tau + \Pi_{1/2}\eta\,, \tag{6.2.7}$$

where $\exp\eta \equiv \gamma^+$, the action $\mathcal{S}_1$ turns out to be invariant. To prove this we need:

$$str\left([\eta, A_{\bar{z}}][\tau, e_-]\right) = 0\,, \tag{6.2.8}$$

which follows from $\Pi_{-1/2}[\eta, A_{\bar{z}}]=0$.

The Lagrange multipliers $A_{\bar{z}}$ impose now the constraints:

$$\Pi_- J_z - \frac{\kappa}{2}e_- - \frac{\kappa}{2}[\tau, e_-] \;=\; 0\,. \tag{6.2.9}$$

Using the gauge symmetry we can further reduce the current by choosing $\Pi_{\geq 0}J_z \in \mathcal{K}_+$ and $\tau=0$, thus indeed reproducing the highest weight gauge eq. (6.1.3).

As discussed in the previous section, the polynomials in the affine currents and $\tau$ which are gauge invariant modulo the constraints, *i.e.* the equations of motion of $A_{\bar z}$, form a realisation of a classical extended Virasoro algebra. The $\mathcal{W}$–algebra is generated by the polynomials $\tilde J_z^{(j,\alpha_j)}$ (6.1.4). Of course, we can take a different set of generators. For example, it is customary to use the Virasoro operator $T_1$ (6.1.12). We will denote the new generators as $\mathcal{T}^{(j,\alpha_j)}$, and assemble them for convenience in a matrix:

$$\mathcal{T} \equiv \mathcal{T}^{(j,\alpha_j)}\ {}_{(jj,\alpha_j)}t \in \mathcal{K}_+ \,. \tag{6.2.10}$$

### 6.2.2 The induced action

We couple the generators $\mathcal{T}$ of the $\mathcal{W}$–algebra to sources $\mu \in \mathcal{K}_-$ and add this term to the action $\mathcal{S}_1$ (6.2.6):

$$\mathcal{S}_2 = \mathcal{S}_1 + \frac{1}{4\pi xy}\int str(\mu\mathcal{T}) \,. \tag{6.2.11}$$

However, $\mathcal{T}$ is only invariant up to the constraints (6.2.9), *i.e.* equations of motions of the gauge fields $A_{\bar z}$. To make $\mathcal{S}_2$ gauge invariant, we modify the transformation rules of the $A_{\bar z}$ with $\mu$-dependent terms. As the gauge fields appear linearly in the action, no further modifications are needed. The action is then invariant under $\Pi_+\bar g$ gauge transformations. Due to the new transformation rules of $A_{\bar z}$, the gauge algebra now closes only on-shell.

Provided that at quantum level the currents $\mathcal{T}^{(j,\alpha_j)}$, up to multiplicative renormalisations and normal ordering, satisfy a quantum version of the $\mathcal{W}$–algebra, we obtain a realisation of the induced action for this $\mathcal{W}$–algebra:

$$\exp\left(-\Gamma[\mu]\right) = \int [\delta g\, g^{-1}][d\tau][dA_{\bar z}]\left(\mathrm{Vol}\left(\Pi_+\bar g\right)\right)^{-1}\exp\left(-\mathcal{S}_2[g,\tau,A_{\bar z},\mu]\right) \,. \tag{6.2.12}$$

Of course, we should properly define the above expression. The gauge fixing procedure is most easily performed using the Batalin–Vilkovisky (BV) method [11]. We skip the details here, a readable account of the BV method can be found in *e.g.* [51, 196, 197, 199]. For every generator of the gauge algebra, we introduce a ghost field $c^a$ with statistic $(-1)^{1+a}$, $c \in \Pi_+\bar g$. Furthermore, for every field we introduce an antifield of opposite statistics, $J_z^* \in \bar g$, $A_{\bar z}^* \in \Pi_-\bar g$, $\tau^* \in \Pi_{-1/2}\bar g$ and $c^* \in \Pi_-\bar g$. We denote all fields $\{J_z, \tau, A_{\bar z}, c\}$ collectively with $\Phi^\alpha$ with corresponding antifields $\Phi^*_\alpha$. The first step of the BV method consists of extending the action $\mathcal{S}_2$:

$$\mathcal{S}_{\mathrm{BV}} = \mathcal{S}_2 + \frac{1}{\pi}\int \Phi^*_\alpha \delta_c \Phi^\alpha + \ldots \,, \tag{6.2.13}$$

where $\delta_c$ denotes the gauge transformations with the parameter $\eta$ replaced by the ghost $c$. The ellipsis denotes extra terms at least quadratic in the ghosts and antifields such that the classical BV master equation is satisfied:

$$(\mathcal{S}_{\mathrm{BV}}, \mathcal{S}_{\mathrm{BV}}) = 0 \,, \tag{6.2.14}$$

where the bracket denotes the BV antibracket:

$$(A,B) \equiv \int d^2x \left( \frac{\overleftarrow{\delta}\, A}{\delta\Phi^\alpha(x)} \frac{\overrightarrow{\delta}\, B}{\delta_\alpha\Phi^*(x)} - \frac{\overleftarrow{\delta}\, A}{\delta\Phi^*_\alpha(x)} \frac{\overrightarrow{\delta}\, B}{\delta^\alpha\Phi(x)} \right) . \tag{6.2.15}$$

The master equation is guaranteed to have a solution by closure of the gauge algebra [10, 82, 110, 198]. Note that the master equation (6.2.14), together with Jacobi identities for the antibracket, implies that the operator $Q_{\mathrm{BV}}$ defined by:

$$Q_{\mathrm{BV}}(A) \;=\; (\mathcal{S}_{\mathrm{BV}}, A) \tag{6.2.16}$$

is nilpotent. This is provides a link to the BRST operator. In particular, for gauge algebras which close without using the equations of motion, this definition coincides with eq. (6.1.6) for $Q$ acting on fields.

We find that the solution to the master equation is given by:

$$\begin{aligned} \mathcal{S}_{\mathrm{BV}} \;=\; & \mathcal{S}_1 + \frac{1}{2\pi x}\int str\Big( - c^*cc + J_z^*\left(\frac{\kappa}{2}\partial c + [c, J_z]\right) + \tau^* c \\ & + A^*_{\bar z}\left(\bar\partial c + [c, A_{\bar z}]\right) + \frac{1}{2y}\mu\hat{\mathcal{T}}\Big) \,, \end{aligned} \tag{6.2.17}$$

where $\mu_{(j,\alpha_j)}$ appears linearly in the extended action. $\hat{\mathcal{T}}^{(j,\alpha_j)}$ reduces to the gauge invariant polynomials $\mathcal{T}^{(j,\alpha_j)}$, when antifields and ghosts are set to zero, see [46, 187]. We will determine the exact form of $\hat{\mathcal{T}}^{(j,\alpha_j)}$ at the end of this section.

The gauge is fixed in the BV method by performing a canonical transformation and by putting the new antifields to zero. A canonical transformation between fields and antifields leaves the antibracket invariant. One should use a canonical transformation such that the new fields have a well-defined propagator.
When going to the quantum theory, a quantum term has to be added to the BV master equation (6.2.14) in order to ensure that the results of the theory are gauge-independent. This implies that quantum corrections have to be added to the extended action. We will not make a fully regularised quantum field theory computation, as this does not seem feasible for the general case. Instead, we use BRST invariance as a guide. We will use OPE-techniques without specifying a regularisation underlying this method in renormalised perturbation theory. The OPEs will enable us to present results to all orders in the coupling constant $\kappa$.

To be able to use OPEs, we should choose a gauge where we can assign definite OPEs to the fields. We will put $A_{\bar z} = 0$, which is different from the highest weight gauge. In the BV formalism, this amounts to using a canonical transformation which simply interchanges $A^*_{\bar z}$ and $A_{\bar z}$. Moreover, we rename the old antigauge fields $A^*_{\bar z}$ to BRST-antighosts $b$. The gauge-fixed action reads:

$$\begin{aligned} \mathcal{S}_{\rm gf} &= \kappa S^-[g] + \frac{\kappa}{4\pi x}\int str\left([\tau, e_-]\bar\partial\tau\right) + \frac{1}{2\pi x}\int str\left(b\bar\partial c\right) \\ &\quad + \frac{1}{4\pi x y}\int str\left(\mu\hat{\mathcal{T}}\right) . \end{aligned} \tag{6.2.18}$$

**Intermezzo 6.2.1**

Using the results of section 2.6, we see that the fields satisfy the OPEs:

$$\begin{aligned} \tau^a\tau^b &= \ll \frac{2}{\kappa}{}^a h^b \gg \\ b^a c^b &= \ll {}^a g^b \gg , \end{aligned} \tag{6.2.19}$$

where ${}^a h^b$ is the inverse of:

$$ {}_a h_b = (e_-)^c\ {}_{ca}f^d\ {}_d g_b\,. \tag{6.2.20}$$

The currents $J^a_z$ satisfy the Kač–Moody algebra $\hat g$ with level $\kappa$ (in the classical and quantum case). The tensors $g$ and $f$ are defined in appendix B.

---

Corresponding to the nilpotent operator eq. (6.2.16), we have the BRST charge[2]:

$$Q = \frac{1}{4\pi i x}\oint str\left(c\left(J_z - \frac{\kappa}{2}e_- - \frac{\kappa}{2}[\tau, e_-] + \frac{1}{4}\{b,c\}\right)\right) . \tag{6.2.21}$$

It is easy to see that indeed $Q(\Phi^\alpha) = \delta_c\Phi^\alpha$.

We now comment on the explicit form of the currents $\hat{\mathcal{T}}^{(j,\alpha_j)}$. BRST invariance of the action requires $\hat{\mathcal{T}}$ to be BRST invariant. This determines $\hat{\mathcal{T}}^{(j,\alpha_j)}$ up to BRST exact pieces. This will lead us to the study of the BRST cohomology in section 6.4. In particular, we will prove in subsection 6.4.2 that the classical cohomology is generated by the gauge invariant polynomials $\tilde J_z^{(j,\alpha_j)}$ after replacing the current $J_z$ with the "total current" $\hat J_z^{(j,\alpha_j)}$:

$$\hat J_z \equiv J_z + \frac{1}{2}\{b,c\}\,. \tag{6.2.22}$$

This means that in the gauge $A_{\bar z} = 0$, $\hat{\mathcal{T}}^{(j,\alpha_j)}$ is given by performing the same substitution in the invariant polynomials $\mathcal{T}^{(j,\alpha_j)}$. This result determines $\hat{\mathcal{T}}^j$ in the extended action (6.2.17), *i.e.* before putting antifields to zero:

$$\hat{\mathcal{T}}[J_z, \tau, A^*_{\bar z}, c] = \mathcal{T}[J_z + \frac{1}{2}\{A^*_{\bar z}, c\}, \tau]\,, \tag{6.2.23}$$

where we indicated the functional dependence of the gauge invariant polynomials as $\mathcal{T}[J_z, \tau]$. An elegant argument for this formula purely relying on BV methods can be found in [187].

We can now check, using the classical OPEs given in intermezzo 6.2.1, that the currents $\Pi_+\hat J_z$ defined in (6.2.22) satisfy the same Poisson brackets as $\Pi_+ J_z$. Because $\mathcal{T}$ is a functional of only the positively graded currents (and $\tau$), the polynomials $\hat{\mathcal{T}}^{(j,\alpha_j)}$ still form a representation of the $\mathcal{W}$–algebra.

We are now in a position to go to the quantum theory. BRST invariance will be our guideline. We notice that the expression eq. (6.2.21) for the BRST charge has no normal ordering ambiguities. We will take it as the definition for the quantum BRST operator, and we will determine any quantum corrections to extended action by requiring invariance under the quantum BRST transformations. For $\mu = 0$, the gauge fixed action (6.2.18) is quantum BRST invariant. Hence, any corrections for the gauge fixed action reside only in the $\hat{\mathcal{T}}^{(j,\alpha_j)}$. The quantum corrections to the generators of the classical $\mathcal{W}$–algebra will be determined in section 6.4 by studying the quantum cohomology.

## 6.3 An example : $osp(N|2)$

In this section we present an explicit example to make our treatment of the classical case more concrete.

The Lie algebra $osp(N|2)$ is generated by a set of bosonic generators:

$$t_{\mp}, t_0, t_{=}, t_{ab}, \qquad t_{ab} = -t_{ba} \text{ and } a, b \in \{1, \cdots, N\}\,, \tag{6.3.1}$$

which form an $sl(2)\times so(N)$ Lie algebra and a set of fermionic generators:

$$t_{+a}, t_{-a}, \qquad a \in \{1, \cdots, N\}\,. \tag{6.3.2}$$

We will represent a Lie algebra valued field $X$ by $X^{\mp}\ {}_{\mp}t + X^0\ {}_0 t + X^{=}\ {}_{=}t + \frac{1}{2}X^{ab}\ {}_{ab}t + X^{+a}\ {}_{+a}t + X^{-a}\ {}_{-a}t$, where the matrices ${}_a t$ are in the fundamental representation (which has index $x = 1/2$):

$$X \equiv \begin{pmatrix} X^0 & X^{\mp} & X^{+b} \\ X^{=} & -X^0 & X^{-b} \\ X^{-a} & -X^{+a} & X^{ab} \end{pmatrix} , \tag{6.3.3}$$

and $X^{ab} = -X^{ba}$. From this, one reads the generators of $osp(N|2)$ in the fundamental representation and one can easily compute the (anti)commutation relations. The dual Coxeter number for this algebra is $\tilde h = \frac{1}{2}(4-N)$. In the fundamental representation, the supertrace (B.8) of two fields is:

$$\begin{aligned} str(X\,Y) &= 2\,X^0\,Y^0 + X^{\mp}Y^{=} + X^{=}\,Y^{\mp} - X^{ab}\,Y^{ba} + \\ &\quad (-1)^Y\,2\,X^{-a}Y^{+a} - (-1)^Y\,2\,X^{+a}Y^{-a}\,, \end{aligned} \tag{6.3.4}$$

[2]Note that $\{b,c\}$ denotes an anticommutator, not a Poisson bracket.

with $(-1)^Y$ a phase factor depending on the parity of $Y$, which is $(-1)$ for (anti)ghosts and $(+1)$ for all other fields. In this equation we inserted the phase factor explicitly. In this section, and in section 6.5, the summation convention is applied without introducing extra signs. Also, indices are raised and lowered using the Kronecker delta.

**Intermezzo 6.3.1**
We list the $osp(N|2)$ classical OPEs for convenience:

$$
\begin{array}{llll}
J^0_zJ^0_z & = \ll \frac{\kappa}{8} \mid 0 \gg & J^i_zJ^j_z & = \ll \frac{\kappa}{8}\delta^{ij} \mid -\frac{\sqrt{2}}{4}f^{ij}{}_kJ^k_z \gg \\
J^{\ddagger}_zJ^{=}_z & = \ll \frac{\kappa}{4} \mid J^0_z \gg & J^i_zJ^{\pm a}_z & = \ll \frac{\sqrt{2}}{4}\lambda_{ab}{}^iJ^{\pm b}_z \gg \\
J^0_zJ^{\overset{\ddagger}{=}}_z & = \ll \pm\frac{1}{2}J^{\overset{\ddagger}{=}}_z \gg & J^{\pm a}_zJ^{\pm b}_z & = \ll \mp\frac{1}{4}\delta^{ab}J^{\overset{\ddagger}{=}}_z \gg \\
J^0_zJ^{\pm a}_z & = \ll \pm\frac{1}{4}J^{\pm a}_z \gg & J^{\overset{\ddagger}{=}}_zJ^{\mp a}_z & = \ll \frac{1}{2}J^{\pm a}_z \gg \\
J^{+a}_zJ^{-b}_z & = \ll \frac{\kappa}{8}\delta^{ab} \mid \frac{1}{4}\left(\delta^{ab}J^0_z + \sqrt{2}\lambda^i_{ab}J^i_z\right) \gg , & &
\end{array}
$$

where the index $i$ stands for a pair of indices $(pq)$ with $1 \leq p < q \leq N$, and $\delta^{(pq)(rs)} = \delta^{pr}\delta^{qs}$. The second order "poles" are given by $-\kappa/2\ {}^ag^b$ where the metric agrees with eq. (6.3.4). The normalisations are such that $\lambda_{ab}{}^{(pq)} \equiv 1/\sqrt{2}(\delta^p_a\delta^q_b - \delta^q_a\delta^p_b)$, $[\lambda^i, \lambda^j] = f^{ij}{}_k\lambda^k$, $tr(\lambda^i\lambda^j) = -\delta^{ij}$, $f^{ik}{}_lf^{jl}{}_k = -(N-2)\delta^{ij}$ where $f^{ij}{}_k$ are the structure constants of the $so(N)$ subalgebra. Note that the classical and quantum OPEs are the same.

---

We consider the embedding corresponding to the $sl(2)$ subalgebra formed by $\{t_{\ddagger}, t_0, t_{=}\}$. It has index of embedding (B.12) $y = 1$. The constraints eq. (6.2.9) are in our example simply[3]:

$$
J^{=}_z = \frac{\kappa}{2} \qquad J^{-a}_z = -\frac{\kappa}{2}\tau^{+a}\,. \tag{6.3.5}
$$

The elements of the classical extended Virasoro algebra are polynomials which are invariant (up to equations of motion of the gauge fields) under the $\Pi_+OSp(N|2)$ gauge transformations. The generators are found by carrying out the unique (finite) $\Pi_+OSp(N|2)$ gauge transformation which brings the currents $J_z$ and $\tau$ in the highest weight gauge, *i.e.* $\tilde{J}^0_z = \tilde{\tau}^{+a} = 0$ for the transformed currents. We find:

$$
\begin{aligned}
\tilde{J}^{\ddagger}_z &= J^{\ddagger}_z + \frac{2}{\kappa}J^0_zJ^0_z + 2J^{+a}_z\tau^{+a} - \sqrt{2}\lambda_{ab}{}^iJ^i_z\tau^{+a}\tau^{+b} - \partial J^0_z - \frac{\kappa}{2}\partial\tau^{+a}\tau^{+a}, \\
\tilde{J}^{+a}_z &= J^{+a}_z + \sqrt{2}\lambda_{ab}{}^iJ^i_z\tau^{+b} + J^0_z\tau^{+a} - \frac{\kappa}{2}\partial\tau^{+a} \\
\tilde{J}^i_z &= J^i_z + \frac{\kappa}{2\sqrt{2}}\lambda_{ab}{}^i\tau^{+a}\tau^{+b}\,.
\end{aligned} \tag{6.3.6}
$$

[3] $\tau^{+a}$ changed sign with respect to [181] because $[\tau, e_-] = \tau_{[181]}$.

We refer to [181] for the gauge transformations of these polynomials, and give only one example:

$$
\delta\tilde{J}^i_z = -\frac{\pi}{2\sqrt{2}}\lambda_{ab}{}^i\eta^{+a}\frac{\delta\mathcal{S}_1}{\delta A^{+b}_{\bar{z}}}\,. \tag{6.3.7}
$$

Computing the Poisson brackets of the generators, we find that the operator $T_s$ (6.1.12) is given by:

$$
T_s = 2\left(\tilde{J}^{\ddagger}_z + s\frac{2}{\kappa}\tilde{J}^i_z\tilde{J}^i_z\right)\,. \tag{6.3.8}
$$

It satisfies the Virasoro Poisson brackets for $s = 0$ and 1 (with central charge $-6\kappa$). The latter choice is to be preferred as the other generators $\tilde{J}^j_z$ are then primary fields [69]. We can now readily identify:

$$
T \equiv T_1, \qquad G^a \equiv 4i\,\tilde{J}^{+a}_z, \qquad U^i = -2\sqrt{2}\tilde{J}^i_z\,, \tag{6.3.9}
$$

as the generators of the classical $N$-extended $so(N)$ superconformal algebra, with the level $k$ of the $so(N)$ currents given by $k = -2\kappa$, see intermezzo 6.3.2.

**Intermezzo 6.3.2**
The $N$-extended $so(N)$ superconformal algebras [133, 20] are generated by the energy–momentum tensor $T$, $N$ dimension $3/2$ supersymmetry currents $G^a$ and affine $so(N)$ currents $U^i$. For $N = 1$ and $N = 2$ these are just the standard $N = 1$ and $N = 2$ superconformal algebras. For $N \geq 3$, the algebra is nonlinearly generated. The subalgebra of transformations globally defined on the sphere, form an $osp(N|2)$ algebra. The nontrivial (classical and quantum) OPEs are given by:

$$
\begin{aligned}
G^aG^b &= \ll \delta^{ab}\beta \mid \frac{\beta}{k}\lambda_{ab}{}^iU^i \mid 2\delta^{ab}T + \frac{\beta}{2k}\lambda_{ab}{}^i\partial U^i + \gamma\Pi^{ij}_{ab}(U^iU^j) \gg \\
U^iU^j &= \ll -\frac{k}{2}\delta^{ij} \mid f^{ij}{}_kU^k \gg \\
U^iG^a &= \ll \lambda_{ba}{}^iG^b \gg,
\end{aligned} \tag{6.3.10}
$$

where $k$ is the level of the $so(N)$ Kač–Moody algebra. $\lambda_{ab}{}^i$ and $f^{ij}{}_k$ are defined in intermezzo 6.3.1, and $\Pi^{ij}_{ab} = \Pi^{ij}_{ba} = \Pi^{ji}_{ab} = \lambda_{ac}{}^i\lambda_{cb}{}^j + \lambda_{ac}{}^j\lambda_{cb}{}^i + \delta_{ab}\delta^{ij}$. The only difference between the classical and quantum OPEs of the generators is in the constants $c, \beta$ and $\gamma$. Although we need only the classical OPEs here, we give the expressions for the quantum case to avoid repetition:

$$
c = \frac{k}{2}\frac{6k + N^2 - 10}{k + N - 3} \qquad \beta = k\frac{2k + N - 4}{k + N - 3} \qquad \gamma = \frac{2}{k + N - 3}\,.
$$

The values of $c, \beta$ and $\gamma$ for classical OPEs are given by the large $k$ limit of these expressions.

---

We choose for the generators $T^{(j,\alpha_j)}$ of the $so(N)$ superconformal algebra the fields appearing in eq. (6.3.9) $T_s, G^a, U^i$, keeping $s$ arbitrary. We couple these generators to sources $h, \psi_a, A_i$ and add this term to the action $\mathcal{S}_1$ as in eq. (6.2.11):

$$\mathcal{S}_2 = \mathcal{S}_1 + \frac{1}{\pi}\int hT_s + \psi_a G^a + A_i U^i\,, \tag{6.3.11}$$

To preserve the gauge invariance of the resulting action, we have to modify the transformation of the gauge fields $A_{\bar z}$ such that the equation of motion terms in (6.3.7) are canceled. This reflects itself in the terms proportional to the antifields $A^*_{\bar z}$ in the Batalin-Vilkovisky action. Because the new gauge algebra closes only on-shell, we need terms quadratic in the antifields to find an extended action satisfying the BV master equation. The result is (see eq. (6.2.17), and eq. (4.20) of [181]) [4]:

$$\begin{aligned}
\mathcal{S}_{\rm BV} \;=\;& \text{(terms independent of } h,\psi_a,A_i)\\
&+\frac{1}{\pi}\int A^{\ddagger *}_{\bar z}\left(-c^{\ddagger}\partial h - \frac{4}{\kappa}c^{\ddagger}hJ^0_z - 2ic^{\ddagger}\psi_a\tau^{+a}\right)\\
&-\frac{2}{\pi}\int A^{+b*}_{\bar z}\left(-\frac{1}{2}c^{+b}\partial h + c^{\ddagger}h\tau^{+b} - \frac{2}{\kappa}c^{+b}hJ^0_z + s\frac{2\sqrt{2}}{\kappa}\lambda_{ab}{}^i c^{+a}J^i_z h\right.\\
&\qquad\left. -ic^{\ddagger}\psi_b + ic^{+a}\psi_a\tau^{+b} - ic^{+b}\psi_a\tau^{+a} - ic^{+a}\psi_b\tau^{+a} - \lambda_{ab}{}^i c^{+a}A_i\right)\\
&-\frac{2}{\pi\kappa}\int A^{\ddagger *}_{\bar z}A^{+a*}_{\bar z}c^{+a}c^{\ddagger}h + s\frac{1}{\pi\kappa}\int A^{+a*}_{\bar z}A^{+a*}_{\bar z}c^{+b}c^{+b}h\,.
\end{aligned} \tag{6.3.12}$$

We now fix the gauge with the condition $A_{\bar z}=0$. Renaming the anti-gauge fields into antighosts:

$$A^{+a*}_{\bar z} = b^{-a}, \qquad A^{\ddagger *}_{\bar z} = b^{=}\,, \tag{6.3.13}$$

we end up with the gauge-fixed action eq. (6.2.18):

$$\begin{aligned}
\mathcal{S}_{\rm gf} \;=\;& \kappa S^{-}[g] - \frac{\kappa}{\pi}\int \tau^{+a}\bar\partial\tau^{+a} + \frac{1}{\pi}\int b^{=}\bar\partial c^{\ddagger} - \frac{2}{\pi}\int b^{-a}\bar\partial\gamma^{+a} +\\
&\frac{1}{\pi}\int\left(h\hat T_s + \psi_a\hat G^a + A_i\hat U^i\right)\,,
\end{aligned} \tag{6.3.14}$$

[4] We use a more convenient normalisation for the antifields here compared to [181]. We have $(A^{+a*}_{\bar z}) = -\frac{\pi}{2}(A^{+a*}_{\bar z})_{[181]}$ and $(A^{\ddagger *}_{\bar z}) = \pi(A^{\ddagger *})_{\bar z})_{[181]}$.

where $\hat T_s$, $\hat G_a$ and $\hat U_i$ have precisely the same form as in eqs. (6.3.8) and (6.3.9), but the currents $J$ are replaced by their hatted counterparts (6.2.22):

$$\begin{aligned}
\hat J^{\ddagger}_z &= J^{\ddagger}_z\\
\hat J^{+a}_z &= J^{+a}_z - \frac{1}{2}c^{\ddagger}b^{-a}\\
\hat J^{0}_z &= J^{0}_z - \frac{1}{2}b^{=}c^{\ddagger} + \frac{1}{2}b^{-a}\gamma^{+a}\\
\hat J^{i}_z &= J^{i}_z + \frac{1}{\sqrt{2}}\lambda_{ab}{}^i b^{-a}\gamma^{+b}\,.
\end{aligned} \tag{6.3.15}$$

This substitution rule is valid independent of the value of $s$ one takes. The hatted generators are classically invariant under the action of the BRST-charge eq. (6.2.21).

## 6.4 Cohomology

In [181], the quantum cohomology of the BRST operator (6.2.21) was studied. Ref. [184] completely characterised the cohomology. This section summarises the results and extends the study to the classical case. In addition we prove that the construction method as given in (6.2.21) is not unique.

When treating the quantum case, all products of fields are considered regularised using point splitting, and normal ordering is performed from right to left. Where algebraic expressions are involved – like when taking the supertrace or the commutator of two fields – it is always understood that the fields are not reordered:

$$[X,Y] \equiv -(-1)^{aY}[X^aY^b]_0\;{}_{ba}f^c{}_c t\,. \tag{6.4.1}$$

### 6.4.1 Computing the quantum cohomology

Consider the OPA $\mathcal{A}$ generated by the basic fields $\{b, \hat J_z, \tau, c\}$. In the quantum case the currents $\Pi_{\geq 0}\hat J_z$ satisfy the same OPEs as $\Pi_{\geq 0}J_z$, except for the central extension. For two currents of zero $sl(2)$ grading we find [5] :

$$\hat J^a_z \hat J^b_z = \ll -\frac{\kappa}{2}\;{}^a g^b + (-1)^{\underline{c}}\;{}_{\underline{c}}f^{ad}\;{}_d f^{b\underline{c}} \;\mid\; \hat J^c_z\;{}_c f^{ab} \gg\,, \tag{6.4.2}$$

where an index $\underline{c}$ is limited to generators of strictly negative grading. Similarly, we will use $\overline{c}$ for an index restrained to strictly positive grading.

[5] In [184] this formula was claimed to be true also when $\hat J^a_z$ and $\hat J^b_z$ do not have zero $sl(2)$ grading. This is not correct as such an OPE involves explicit (anti)ghosts. However, these OPEs are never needed in the computation of the cohomology, so the results of [184] are not influenced.

To every field $\Phi$, we assign a double grading $[\Phi] = (k, l)$, where $k \in \frac{1}{2}\mathbb{Z}$ is the $sl(2)$ grading and $k+l \in \mathbb{Z}$ is the ghost number. The "auxiliary" fields $\tau$ are assigned the grading $(0, 0)$. The algebra $\mathcal{A}$ acquires a double grading:

$$\mathcal{A} = \bigoplus_{\substack{m,n \in \frac{1}{2}\mathbf{Z} \\ m+n \in \mathbf{Z}}} \mathcal{A}_{(m,n)} \tag{6.4.3}$$

and OPEs preserve the grading. The BRST charge (6.2.21) decomposes into three parts of definite grading, $Q = Q_0 + Q_1 + Q_2$, with $[Q_0] = (1, 0)$, $[Q_1] = (\frac{1}{2}, \frac{1}{2})$ and $[Q_2] = (0, 1)$:

$$\begin{aligned} Q_0 &= -\frac{\kappa}{8\pi i x} \oint str(ce_-) \\ Q_1 &= -\frac{\kappa}{8\pi i x} \oint str(c\,[\tau, e_-]) \ . \end{aligned} \tag{6.4.4}$$

As illustrated in fig. 6.1, the operators $Q_0$, $Q_1$ and $Q_2$, map $\mathcal{A}_{(m,n)}$ to $\mathcal{A}_{(m+1,n)}$, $\mathcal{A}_{(m+\frac{1}{2},n+\frac{1}{2})}$ and $\mathcal{A}_{(m,n+1)}$ respectively.

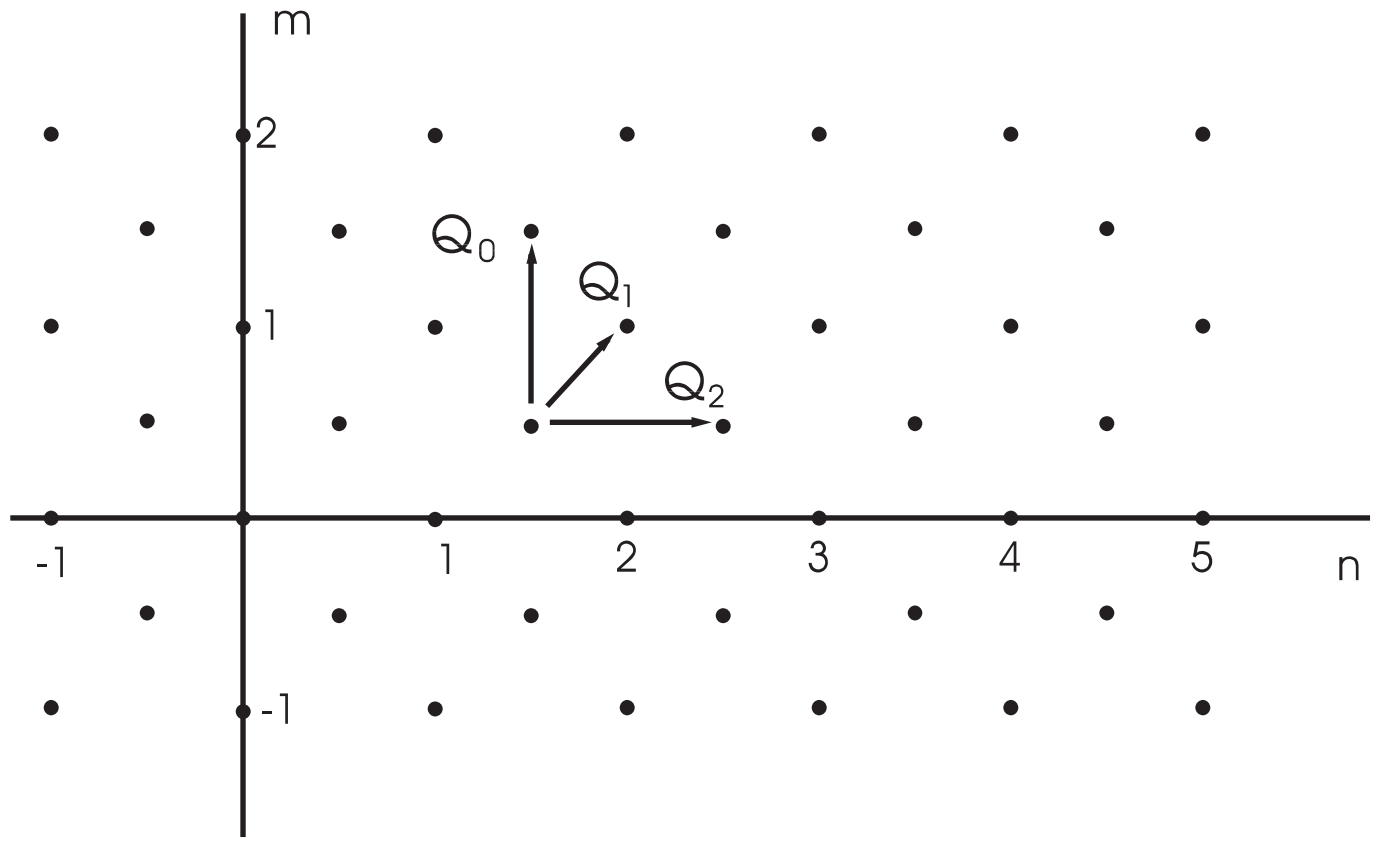

Figure 6.1: $Q_0$, $Q_1$ and $Q_2$ acting on $\mathcal{A}_{(\frac{1}{2},\frac{3}{2})}$.

It follows from $Q^2 = 0$ that $Q_0^2 = Q_2^2 = \{Q_0, Q_1\} = \{Q_1, Q_2\} = Q_1^2 + \{Q_0, Q_2\} = 0$, but:

$$Q_1^2 = -\{Q_0, Q_2\} = \frac{\kappa}{32\pi i x} \oint str\left(c\,[\Pi_{1/2}c, e_-]\right) \tag{6.4.5}$$

does not vanish. The presence of $Q_1$ is the main difference with the case of bosonic Kač–Moody algebras treated in [43], as the auxiliary field $\tau$ was not introduced there.

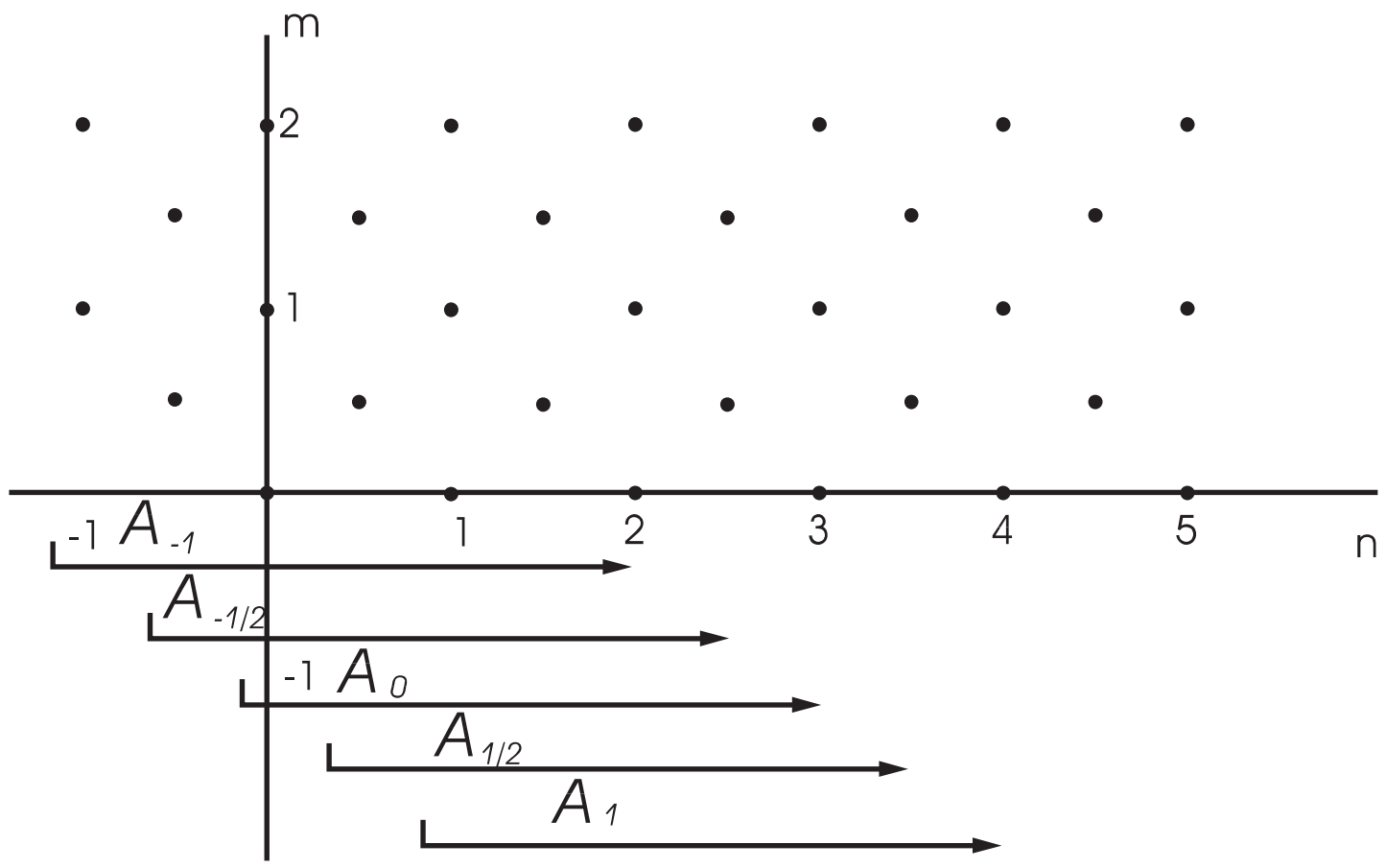

Figure 6.2: $\widehat{\mathcal{A}}$ and its filtration $\widehat{\mathcal{A}}^n$.

The action of $Q_0$, $Q_1$ and $Q_2$ on the basic fields is given in table 6.1. Note that due to eq. (2.3.21), the BRST charges $Q_i$ act as derivations on a normal ordered product of fields.

| $Q_0$ | $Q_1$ | $Q_2$ |
|---|---|---|
| $b \to -\frac{\kappa}{2}e_-$ | $b \to -\frac{\kappa}{2}[\tau, e_-]$ | $b \to \Pi_-\hat{J}_z$ |
| $c \to 0$ | $c \to 0$ | $c \to \frac{1}{2}cc$ |
| $\hat{J}_z \to -\frac{\kappa}{4}[e_-, c]$ | $\hat{J}_z \to -\frac{\kappa}{4}[[\tau, e_-], c]$ | $\hat{J}_z \to \frac{1}{2}[c, \Pi_{\geq 0}\hat{J}_z] + \frac{\kappa}{4}\partial c$ $-\frac{1}{2}\partial c^a{}_a f^{\underline{b}\overline{c}}{}_{\overline{c}\underline{b}} f^d{}_d t$ |
| $\tau \to 0$ | $\tau \to \frac{1}{2}\Pi_{+1/2}c$ | $\tau \to 0$ |

Table 6.1: The action of $Q_i$

The subcomplex $\mathcal{A}^{(1)}$, generated by $\{b,\ \Pi_-\hat{J}_z - \frac{\kappa}{2}[\tau, e_-]\}$ has a trivial cohomology $\mathcal{H}^*(\mathcal{A}^{(1)}; Q) = \mathbb{C}$. Moreover, its elements do not appear in the action of the charges $Q_i$ on the other fields. One can show that the cohomology of $\mathcal{A}$ is then equal to the cohomology of the reduced complex $\widehat{\mathcal{A}} = \mathcal{A}/\mathcal{A}^{(1)}$, generated by $\{\Pi_{\geq 0}\hat{J}_z, \tau, c\}$: $\mathcal{H}^*(\mathcal{A}; Q) = \mathcal{H}^*(\widehat{\mathcal{A}}; Q)$. Because $\widehat{\mathcal{A}}$ contains only positively graded operators, the OPEs close on the reduced complex. Obviously the double grading on $\mathcal{A}$ is inherited by $\widehat{\mathcal{A}}$. At this point, the theory of spectral sequences [32] is applied to compute the cohomology on $\widehat{\mathcal{A}}$. We summarise the results. Details can be found in [184]. The filtration $\widehat{\mathcal{A}}^m$, $m \in \frac{1}{2}\mathbf{Z}$ of $\widehat{\mathcal{A}}$ (see fig. 6.2):

$$\widehat{\mathcal{A}}^m \equiv \bigoplus_{k\in\frac{1}{2}\mathbf{Z}} \bigoplus_{l\geq m} \widehat{\mathcal{A}}_{(k,l)}\,, \tag{6.4.6}$$

leads to a spectral sequence $E_r = \mathcal{H}^*(E_{r-1}; d_{r-1})$, $r \geq 1$ which converges to $\mathcal{H}^*(\widehat{\mathcal{A}}; Q)$. The sequence starts with $E_0 = \widehat{\mathcal{A}}, d_0 = Q_0$. The derivation $d_r$ represents the action of $Q$ at that level. One shows that $E_1 = \mathcal{H}^*(\widehat{\mathcal{A}}; Q_0) \simeq \widehat{\mathcal{A}}\left[\Pi_{hw}\hat{J}_z\right] \otimes \widehat{\mathcal{A}}[\tau] \otimes \widehat{\mathcal{A}}\left[\Pi_{\frac{1}{2}}c\right]$, where $\widehat{\mathcal{A}}[\Phi]$ denotes the subalgebra of $\widehat{\mathcal{A}}$ generated by $\Phi$. The spectral sequence collapses after the next step and we find:

$$\mathcal{H}^*(\mathcal{A}; Q) \simeq E_2 = \mathcal{H}^*(E_1; Q_1) = \widehat{\mathcal{A}}\left[\Pi_{hw}\left(\hat{J}_z + \frac{\kappa}{4}[\tau, [e_-, \tau]]\right)\right]\,. \tag{6.4.7}$$

This result means that the dimension of the cohomology equals the number $n_j$ of $sl(2)$ irreducible representations in the branching of the adjoint representation of $\bar{g}$. Furthermore, the generators of the cohomology have ghost number zero. Because the reduced complex has no antighosts, the generators consist only of the currents $\Pi_{\geq 0}\hat{J}_z$ and $\tau$.

We now outline a recursive procedure to obtain explicitly the generators of the cohomology: the tic-tac-toe construction [32] (see fig. 6.3). The generators of the cohomology are split up as:

$$W^{(j,\alpha_j)} = \sum_{r=0}^{2j} W_r^{(j,\alpha_j)}\,, \tag{6.4.8}$$

where $W_r^{(j,\alpha_j)}$ has grading $(j - \frac{r}{2}, -j + \frac{r}{2})$. The leading term $W_0^{(j,\alpha_j)}$ is given by:

$$W_0^{(j,\alpha_j)} = \hat{J}_z^{(jj,\alpha_j)} + \delta_{j,0}\frac{\kappa}{4}[\tau, [e_-, \tau]]^{(00,\alpha_j)} \tag{6.4.9}$$

and the remaining terms are recursively determined by:

$$Q_0 W_r^{(j,\alpha_j)} = -Q_1 W_{r-1}^{(j,\alpha_j)} - Q_2 W_{r-2}^{(j,\alpha_j)}\,, \tag{6.4.10}$$

where $W_r^{(j,\alpha_j)} = 0$ for $r < 0$ or $r > 2j$. As an example, one can check that for $j > 1/2$:

$$W_1^{(j,\alpha_j)} = -[\tau, \Pi_{\geq 0}\hat{J}]^{(jj,\alpha_j)}\,. \tag{6.4.11}$$

For $j = 1/2$ this equation gets a correction term that depends only on $\tau$ such that eq. (6.4.10) is satisfied, see [184], and also eq. (6.5.1).

There is a certain ambiguity in this construction because any combination of the generators $W^{(j,\alpha_j)}$ is still BRST invariant (and non-trivial). This corresponds to the fact that $Q_0$ annihilates $\hat{J}_z^{(jj,\alpha_j)}$, and thus one can freely add at each step in the iteration (6.4.10) any combination of correct grading of these currents to $W_r^{(j,\alpha_j)}$. We will use the notation $W^{(j,\alpha_j)}$ for the elements of the cohomology which reduce to a single highest weight current when discarding non-highest weight currents, more precisely:

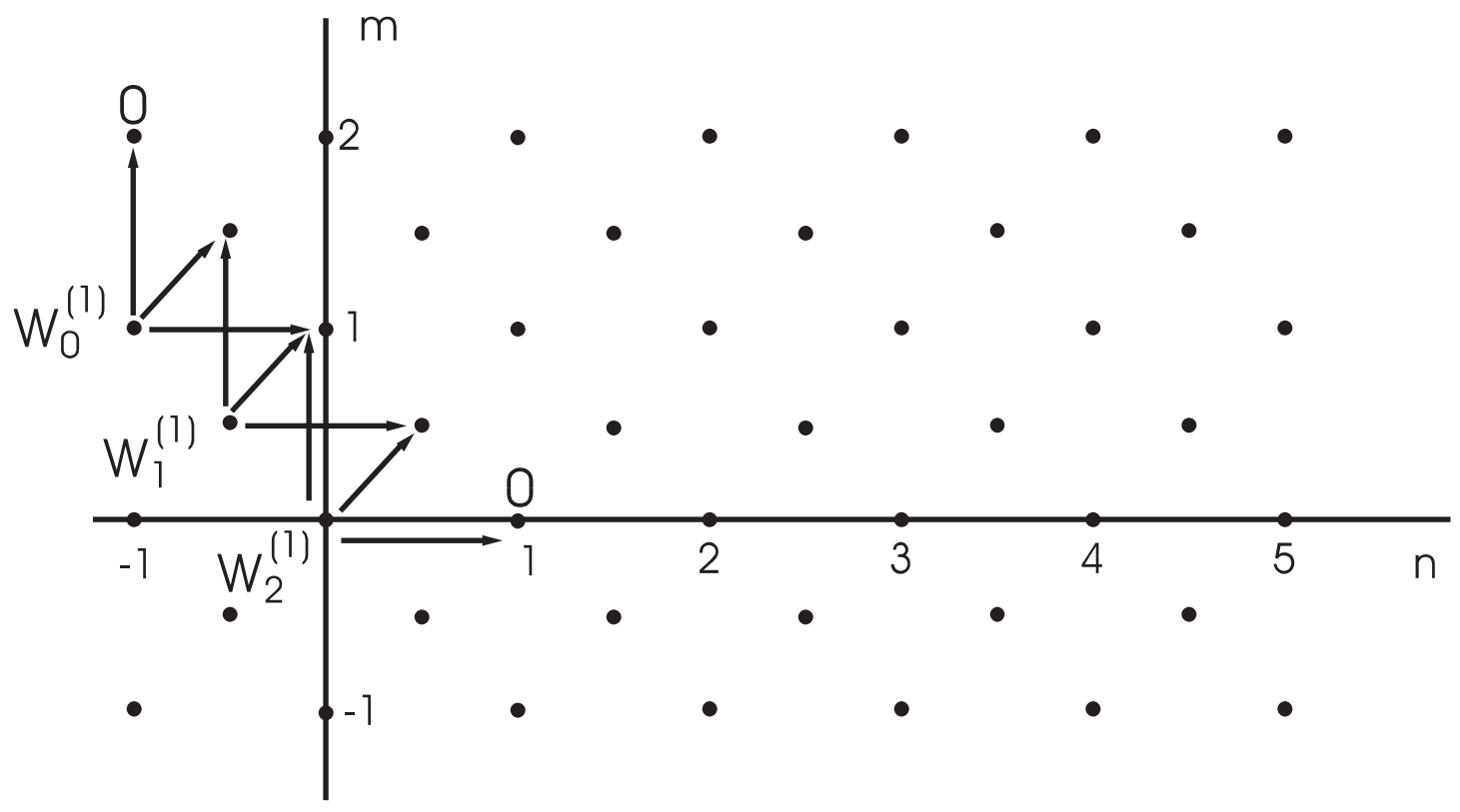

Figure 6.3: The tic-tac-toe construction for a conformal dimension 2 current.

$$W^{(j,\alpha_j)} = \hat{J}_z^{(j,\alpha_j)}\Big|_{(\Pi_+ - \Pi_{hw})\hat{J} \;=\; \tau \;=\; 0}\,. \tag{6.4.12}$$

An important step for practical applications of this construction, is the "inversion" of $Q_0$ in (6.4.10). Table 6.1 shows that it is necessary to find a field $X$ such that $Q_0(X) = \Pi_{\geq 1}c$. To find this field, we introduce an operator $L$ [40] in $\bar{g}$, which is defined by:

$$\begin{cases} L({}_at) = 0, & \text{if } {}_at \in \Pi_{hw}\bar{g} \\ [e_-, L({}_at)] \;=\; {}_at, & \text{otherwise}\,. \end{cases} \tag{6.4.13}$$

Using table 6.1 we see that:

$$Q_0(L\hat{J}_z) = \frac{\kappa}{4}\Pi_{\geq 1}c\,. \tag{6.4.14}$$

We will need $L$ in subsection 6.4.3 to construct $W^{(1,0)}$.

### 6.4.2 Classical cohomology

Before discussing the extended Virasoro algebra in this cohomology, let us see how these results are modified when computing the classical cohomology.

The action of $Q_i$ on the fields, table 6.1, remains the same except for $Q_2(\hat{J}_z)$. Here the last and most complicated term disappears as it originates from double contractions. This means that exactly the same reasoning can be followed as in the quantum case. The cohomology can be computed in the reduced complex $\widehat{\mathcal{A}}$ generated by $\{\Pi_{\geq 0}\hat{J}_z, \tau, c\}$. Because the generators $W^{(j,\alpha_j)}$ have ghost number zero, we see that any element of the cohomology can be written in terms of $\{\Pi_{\geq 0}\hat{J}_z, \tau\}$.

On the other hand, we know from general arguments that the elements from the cohomology (at ghostnumber zero) reduce to gauge invariant polynomials when $b$ and $c$ are set to zero (see the discussion at the end of section 6.1). We conclude that the elements of the classical cohomology can be found by substituting $J_z$ with $\hat{J}_z$ in the gauge invariant polynomials. In particular, it follows from eqs. (6.1.4) and (6.4.12) that :

$$W^{(j,\alpha_j)}_{\text{classical}} = \tilde{J}_z^{(j,\alpha_j)}\Big|_{J_z \to \hat{J}_z} . \tag{6.4.15}$$

This result was used in subsection 6.2.2 to determine the terms that couple to the sources $\mu$ in the classical extended action (6.2.17).

### 6.4.3 The quantum $\mathcal{W}$–algebra

The generators $W^{(j,\alpha_j)}$ of the quantum cohomology form (a realisation of) a quantum OPA. We still have to prove that we obtained a quantum version of the classical $\mathcal{W}$–algebra. One can check that:

$$2y\,\mathcal{C}\,W^{(1,0)} \tag{6.4.16}$$

satisfies a Virasoro OPE, where we extracted a factor:

$$\mathcal{C} = \frac{\kappa}{\kappa + \tilde{h}}\,, \tag{6.4.17}$$

which goes to 1 in the classical limit and $y$ is the index of embedding defined in (B.12). However, we choose to add the “diagonal” quadratic combination of the $W^{(0,\alpha_0)}$, which corresponds to the Sugawara tensor of the affine algebra of the $j=0$ highest weight currents. In other words, we choose $s=1$ in the quantum analogue of eq. (6.1.12). As in the classical case [69], this choice for the energy–momentum tensor is expected to give that the other generator s $W^{(j,\alpha_j)}$ are primary. We define:

$$\begin{aligned} \hat{T}^{\text{EM}} \;=\;& \frac{\kappa}{x(\kappa+\tilde{h})}\left( str\left(\hat{J}_z e_-\right) + str\left([\tau, e_-]\hat{J}_z\right)\right) \\ & + \frac{1}{x(\kappa+\tilde{h})} str\left(\Pi_0(\hat{J}_z)\Pi_0(\hat{J}_z)\right) + \frac{1}{x} str\left(e_-\partial L \hat{J}_z\right) \\ & + \frac{1}{x(\kappa+\tilde{h})} str\left(\left[\Pi_0(t^A), \left[\Pi_0(t_A), \partial L\hat{J}_z\right]\right] e_-\right) \\ & - \frac{\kappa}{4x} str\left([\tau, e_-]\partial\tau\right)\,, \end{aligned} \tag{6.4.18}$$

where $L$ is defined in eq. (6.4.13). The first term is $2y\mathcal{C}\left(W_0^{(1,0)} + W_1^{(1,0)}\right)$, the remainder forms $2y\,\mathcal{C}\,\left(W_2^{(1,0)} + \frac{2}{xy\kappa} str(\Pi_0 W)(\Pi_0 W)\right)$. To bring this in a more recognisable form, we add a BRST exact term to $\hat{T}^{\text{EM}}$:[6]

$$\begin{aligned} \hat{T}^{\text{IMP}} \;\equiv\;& \hat{T}^{\text{EM}} + Q\left(\frac{2}{x(\kappa+\tilde{h})} str\, b(J_z + \frac{1}{4}\{b, c\})\right) \\ =\;& \frac{1}{x(\kappa+\tilde{h})} str J_z J_z - \frac{1}{2x} str e_0 \partial J_z - \frac{\kappa}{4x} str\left([\tau, e_-]\partial\tau\right) \\ & + \frac{1}{4x} str b[e_0, \partial c] - \frac{1}{2x} str b\partial c + \frac{1}{4x} str \partial b[e_0, c]\,. \end{aligned} \tag{6.4.19}$$

We now analyse the various parts of the improved energy–momentum tensor, eq. (6.4.19). The fact that it differs by a BRST exact form from the $T^{\text{EM}}$ does not change the value of the central charge. The first term is the Sugawara tensor (2.6.47) for the Kač–Moody algebra $\hat{g}$ with central charge:

$$c_{\text{Sug}} = \frac{\kappa(d_B - d_F)}{\kappa + \tilde{h}}\,. \tag{6.4.20}$$

Here $d_B$ and $d_F$ are the number of bosonic and fermionic generators of $\bar{g}$ .
The second term of eq. (6.4.19) is the so-called improvement term. Given an affine current $J \in \Pi_m\bar{g}$, it changes the conformal dimension of $J$ from $m$ to $m+1$. It was added by hand in [69] to ensure that all terms in the constraint $str\left((J_z - \kappa/2e_-)e_+\right)$ have conformal dimension zero. The contribution of the second term to the central charge is:

$$c_{\text{imp}} = -6y\kappa\,, \tag{6.4.21}$$

In the case where $\tilde{h}$ is different from zero, we can evaluate eq. (B.12) in the adjoint representation by using eq. (B.7):

$$y = \frac{1}{3\tilde{h}} \sum_{j\alpha_j} (-1)^{\alpha_j} j(j+1)(2j+1)\,, \tag{6.4.22}$$

and $(-1)^{\alpha_j} = +1$ $(-1)$ if $t_{(jm,\alpha_j)}$ is bosonic (fermionic).

The next term of eq. (6.4.19) is the energy–momentum tensor for the auxiliary $\tau$ fields with central charge:

$$c_\tau = \frac{1}{2}\left(\dim\left(\Pi^F_{1/2}\bar{g}\right) - \dim\left(\Pi^B_{1/2}\bar{g}\right)\right)\,, \tag{6.4.23}$$

[6] This formula corrects a misprint in [181] in the coefficient of the second term of the second line.

where the superscript $B\,(F)$ denotes a projection on the bosonic (fermionic) part of the algebra.

Finally, the last terms of eq. (6.4.19) form the energy–momentum tensor for the ghost–antighost system $T_{\rm gh}$. To each generator $t$ of the gauge group, $t \in \Pi_m\bar{g}$ where $m > 0$, we associated a ghost $c \in \Pi_m\bar{g}$ and an antighost $b \in \Pi_{-m}\bar{g}$. They have respectively conformal dimension $m$ and $1-m$ with respect to $T_{\rm gh}$. Such a pair contributes $\mp 2(6m^2-6m+1)$, where we have the minus sign if $b$ and $c$ are fermionic, and a plus when they are bosonic, see eq. (2.6.34). As such, the total contribution to the central charge coming from the ghosts is given by:

$$c_{\rm gh} = -\sum_{j,\alpha_j}(-1)^{\alpha_j}2j(2j^2-1) + \frac{1}{2}\left(\dim\left(\Pi^B_{1/2}\bar{g}\right) - \dim\left(\Pi^F_{1/2}\bar{g}\right)\right). \tag{6.4.24}$$

Adding all this together, we obtain the full expression for the total central charge $c$ as a function of the level $\kappa$:

$$c = c_{\rm Sug} + c_{\rm imp} + c_\tau + c_{\rm ghost}\,. \tag{6.4.25}$$

Using the explicit form for the index of embedding $y$ given in eq. (6.4.22) and some elementary combinatorics, we can rewrite eq. (6.4.25) in the following pretty form:

$$c = \frac{1}{2}c_{\rm crit} - \frac{(d_B-d_F)\tilde{h}}{\kappa+\tilde{h}} - 6y(\kappa+\tilde{h}), \tag{6.4.26}$$

where $c_{\rm crit}$ is the critical value of the central charge for the extension of the Virasoro algebra under consideration:

$$c_{\rm crit} = \sum_{j,\alpha_j}(-1)^{\alpha_j}2(6j^2+6j+1)\,. \tag{6.4.27}$$

Knowing the leading term of the generators $W^{(j,\alpha_j)}$, eq. (6.4.9), we find that their conformal dimension is $j+1$. We are convinced that they are primary with respect to $T^{\rm EM}$, but leave the verification of this to the industrious reader.

We now come to the main result:

**Theorem 6.4.1** *The generators $W^{(j,\alpha_j)}$ form an extension of the Virasoro algebra with $n_j$*(adjoint) *generators of conformal dimension $j+1$ and with the value of the central charge given by eq. (6.4.26).*

To finish the proof of this statement, it only needs to be shown that the fields $W^{(j,\alpha_j)}$ close under OPEs. From the fact that the OPEs close on $\widehat{\mathcal{A}}$ and preserve the grading, one deduces that the OPEs of the generators $W$ close modulo BRST exact terms. However, there are no fields of negative ghostnumber in the reduced complex $\widehat{\mathcal{A}}$. This implies that the OPEs of the generators $W$ close.

## 6.5 An example : $osp(N|2)$, continued

We now continue our example of section 6.3 for the quantum case using the results of the previous section. For this we need the OPEs of $\tau$ and the (anti)ghosts. They follow from the action $\mathcal{S}_{\rm gf}$ (6.3.14), see intermezzo 6.2.1 and [181]. We can use the results for the cohomology of the BRST operator of the previous section. In the $osp(N|2)$ case, $L\hat{J}_z$ is given by $-\hat{J}^0_z t_{\ddagger}$. We get the following representants of the cohomology of $Q$:

$$\begin{aligned}
W^{(0,i)} &= \hat{J}^i_z + \frac{\kappa}{2\sqrt{2}}\lambda_{ab}{}^i\tau^{+a}\tau^{+b}\\
W^{(1/2,a)} &= \hat{J}^{+a}_z + \sqrt{2}\lambda_{ab}{}^i\hat{J}^i_z\tau^{+b} + \hat{J}^0_z\tau^{+a} - \frac{\kappa+1}{2}\partial\tau^{+a}\\
W^{(1,0)}_s &= \hat{J}^{\ddagger}_z + \frac{2}{\kappa}\hat{J}^0_z\hat{J}^0_z + 2\hat{J}^{+a}_z\tau^{+a} - \sqrt{2}\lambda_{ab}{}^i\hat{J}^i_z\tau^{+a}\tau^{+b} - \frac{\kappa+1}{\kappa}\partial\hat{J}^0_z\\
&\quad -\frac{2\kappa+3}{4}\partial\tau^{+a}\tau^{+a} + s\frac{2}{\kappa}W^{(0,i)}W^{(0,i)}\,.
\end{aligned} \tag{6.5.1}$$

These expressions are the same as the classical ones (eqs. (6.3.6) and (6.3.8) with $J_z \to \hat{J}_z$), up to finite renormalisation factors related to normal ordering. The operator $2\kappa/(\kappa+\tilde{h})\,\hat{T}^{(1,0)}_s$ satisfies the Virasoro algebra for $s=1$ or $(-2\kappa+N-3)^{-1}$, which corresponds to the classical values (6.3.8) in the large $\kappa$ limit.

We checked using *OPEdefs* (with an extension to use dummy arguments), that these currents (with $s=1$) form a representation of the quantum $so(N)$ superconformal algebra (6.3.10) with the normalisation factors:

$$G^a = 4i\sqrt{\frac{\kappa}{\kappa+\tilde{h}}}W^{(1/2,a)} \quad \text{and} \quad U^i = -2\sqrt{2}W^{(0,i)}\,, \tag{6.5.2}$$

and the value for the central charge as given by eq. (6.4.26). Alternatively, we can express the level of the $so(N)$ Kač–Moody algebra in terms of $\kappa$:

$$k = -2\kappa - 1\,. \tag{6.5.3}$$

Comparing eq. (6.5.2) to eq. (6.3.9) we see that the quantum currents have an overall renormalisation factor $\mathcal{C}^j$, where $\mathcal{C}$ is defined in eq. (6.4.17).

## 6.6 Quantum corrections to the extended action

Using the results on the cohomology of the BRST operator (6.2.21) in section 6.4, we will now determine the quantum corrections to the gauge-fixed action (6.2.18).

As already explained, we find the corrections by requiring BRST invariance of the gauge-fixed action in a point-splitting regularisation scheme.

No modifications to the terms which are independent of $\mu$ are needed. The fields $\hat{\mathcal{T}}^{(j,\alpha_j)}$ proportional to $\mu_{(j,\alpha_j)}$ in eq. (6.2.18) have to be elements of the quantum cohomology. The classical $\hat{\mathcal{T}}^{(j,\alpha_j)}$ can be viewed as functionals of the generators of the classical cohomology $W^{(j,\alpha_j)}_{\rm classical}$. To find the quantum corrections to $\hat{\mathcal{T}}$, we cannot simply substitute the generators of the classical cohomology $W$ by their quantum counterparts. Indeed, for the Virasoro operator we proved in subsection 6.4.3 that an additional renormalisation factor $\mathcal{C}$, given in eq. (6.4.17), is necessary to preserve the Virasoro OPE. The example of section 6.5 suggests that a convenient renormalisation for the other generators is given by[7]:

$$\hat{W}^{(j,\alpha_j)} \;=\; \mathcal{C}^j W^{(j,\alpha_j)}\,, \tag{6.6.1}$$

where the normalisation of the fields $W^{(j,\alpha_j)}$ is fixed in eq. (6.4.12). In summary, the quantum corrections to the gauge-fixed action (6.2.18) in the gauge $A_{\bar z}=0$ are obtained by replacing the classical $W$ currents with the quantum $\hat{W}$ in $\hat{\mathcal{T}}^{(j,\alpha_j)}$, which we write as $\hat{\mathcal{T}}[\hat{W}]$.

Because $\hat{\mathcal{T}}$ does not contain $J^*_z, c^*, \tau^*$ or $A_{\bar z}$, we can extend these results to the extended action. Also, no modifications to the terms which arise from the quantum transformation laws are needed, see table 6.1. In this way, we have determined the quantum BV action completely in an OPE regularisation. Giving the antighosts $b$ their original name $A^*_{\bar z}$, we find:

$$\begin{aligned}
\mathcal{S}^q_{\rm BV} \;=\;& \kappa S^-[g] + \frac{1}{\pi x}\int str\, A_{\bar z}\left(J_z + \frac{1}{2}\{A^*_{\bar z}, c\} - \frac{\kappa}{2}e_- - \frac{\kappa}{2}[\tau, e_-]\right)\\
&+\frac{\kappa}{2\pi x}\int str\left(\frac{1}{2}[\tau,e_-]\bar\partial\tau - c^*cc + \tau^*c + A^*_{\bar z}\bar\partial c\right)\\
&+\frac{\kappa}{2\pi x}\int str\left(J^*_z\left(\frac{\kappa}{2}\partial c + [c, J_z]\right)\right)\\
&+\frac{1}{4\pi xy}\int str\left(\mu\,\hat{\mathcal{T}}[\hat{W}]\right)\,.
\end{aligned} \tag{6.6.2}$$

We feel confident that $Q^2=0$ guarantees the gauge invariance of the quantum theory. Therefore we will use eq. (6.6.2) as it stands, also for a different gauge in chapter 7.

## 6.7 Discussion

We have proven that any (super) Lie algebra with an $sl(2)$ embedding gives rise to a realisation of a (classical and quantum) $\mathcal{W}$–algebra, which is generated by $n_j$ (adjoint) generators of conformal dimension $j+1$. $\mathcal{W}$–algebras with dimension $1/2$ fields cannot be realised by a Drinfeld-Sokolov reduction. However, this is no shortcoming of the present approach since such fields can always be factored out, as we have explained in chapter 5.

For a classical bosonic $\mathcal{W}$–algebra, it was shown in [34, 70] that one can identify a finite Lie algebra with an embedded $sl(2)$ in the mode algebra, called the "vacuum preserving algebra", see also subsection 4.5.5. Furthermore, for a $\mathcal{W}$–algebra constructed by a Drinfeld-Sokolov reduction of a Lie algebra $\bar g$, this Lie algebra is isomorphic to $\bar g$. We expect that the results of section 6.4 on the quantum Drinfeld-Sokolov reduction can be used to prove that the vacuum preserving algebra of the quantum $\mathcal{W}$–algebra and the classical $\mathcal{W}$–algebra are the same.

For any quantum $\mathcal{W}$–algebra arising in a Drinfeld-Sokolov reduction, we have constructed a path integral formulation of the induced action in terms of a gauged WZNW–model. In a point-splitting regularisation, the results on the quantum cohomology give the quantum corrections to the gauge fixed action to all orders. This will be used in the next chapter to discuss the effective action of the corresponding induced $\mathcal{W}$–gravity theory.

[7] This is a slightly different choice as used in [184].

# Chapter 7

# Renormalisation factors in $\mathcal{W}$–Gravity

In this chapter, we will study induced gravity theories in the light-cone gauge. For $\mathcal{W}$–algebras which can be realised in terms of a constrained WZNW–model, we prove in section 7.2 that the effective action can be computed by simply inserting renormalisation factors in the classical result. We give explicit expressions for these factors to all orders in the coupling constant.

As an example, the supergravity theories based on the $N$-extended $so(N)$ superconformal algebras will be presented in section 7.3.

In section 7.4 the renormalisation factors for the linear superconformal algebras are computed in a semiclassical approximation. These results are compared with the all–order expressions of 7.3.

The discussion relies on the results of chapter 6. Conventions on WZNW–models are given in section 2.6.4.

This chapter is based on [181]. However, several topics have been expanded, especially in section 7.2.

## 7.1 Introduction

In two dimensions, the Einstein-Hilbert action for the metric $g^{ij}$:

$$\int d^2x \, \sqrt{g}R \,, \tag{7.1.1}$$

where $R$ is the Riemann curvature scalar, is a topological invariant and hence trivial. This means that any action where the matter fields are covariantly coupled to the metric can be regarded as an action for the matter coupled to gravity. Consider as an example for the matter system a scalar boson:

$$S[X,g] = -\frac{1}{4\pi}\int dx^2 \, \sqrt{g} g^{ij}\nabla_i X(x)\nabla_j X(x) \,, \tag{7.1.2}$$

where $\nabla_i$ denotes a covariant derivative. In the light-cone gauge $ds^2 = 2dzd\bar{z} + 2\mu d\bar{z}d\bar{z}$, this action reduces to the free field action with a flat metric (2.6.1) $S_s(X,\delta^{ij})$ plus a coupling of the metric to the holomorphic component of the energy–momentum tensor:

$$S[X,g] = S_s(X,\delta^{ij}) + \frac{1}{\pi}\int \mu T \,. \tag{7.1.3}$$

By integrating out the matter field $X$, we obtain an induced action for gravity in two dimensions. We see that the induced action is equal to the generating functional $\Gamma$ (2.5.4) of the Virasoro algebra. Similarly, when the matter system specifies a conformal invariant system, the only remnant of the matter system in the induced action is the central charge of $T$.

Consider now a matter system, where we denote matter fields collectively by $\varphi$, with a set of holomorphic global symmetries generated by $T^i[\varphi]$. We will assume that the matter system is conformally invariant, and the $T^i[\varphi]$ generate a $\mathcal{W}$–algebra. The symmetries can be gauged to make them local. The gauge fields $\mu_i$ correspond to the remnant of the metric and its generalisations for the higher dimension fields in the light-cone gauge. The induced action for $\mathcal{W}$–gravity coupled to the matter $\varphi$ in the light-cone gauge is defined by:

$$\exp\left(-\Gamma_{\rm ind}[\mu]\right) = \int [d\varphi] \exp\left(-S[\varphi] - \frac{1}{\pi}\int \mu_i T^i[\varphi]\right). \tag{7.1.4}$$

The gauge fields $\mu_i$ are called generalised Beltrami differentials. When after quantisation, the $T^i$ form a quantum $\mathcal{W}$–algebra, the induced action (7.1.4) is equal to the generating functional $\Gamma$ (2.5.4) of the $\mathcal{W}$–algebra.

For $\mathcal{W}_3$–gravity, it was shown in [172] that the induced action (7.1.4) $\Gamma_{\rm ind}$ can be expanded in a power series of the inverse central charge $1/c$:

$$\Gamma[\mu] = \sum_{i\geq 0} c^{1-i}\Gamma^{(i)}[\mu] \,. \tag{7.1.5}$$

A similar expression will be true for other nonlinear $\mathcal{W}$–algebras for which a classical limit $c \to \infty$ exists, while for most linear algebras only $\Gamma^{(0)}$ is non-zero. The subleading terms in $1/c$ in eq. (7.1.5) arise from a proper treatment of the composite terms in the OPEs. In [157], an explicit form for the classical term $\Gamma^{(0)}$ of $\mathcal{W}_3$ was obtained through the classical reduction of an $Sl(3, \mathbf{R})$ WZNW–model. The higher order terms are more difficult to compute and do not bear any straightforward relation to $\Gamma^{(0)}$.

This complicated behaviour seems to simplify drastically when going to the effective theory – *i.e.* quantising the metric and other gauge fields of the $\mathcal{W}$–gravity theory. The generating functional $W$ of the connected Green's functions, which upon a Legendre transform becomes the effective action[1] is defined by:

$$\exp\left(-W[\check{T}]\right) = \int [d\mu]\, \exp\left(-\Gamma[\mu] + \frac{1}{\pi}\int \mu_i \check{T}^i\right) . \tag{7.1.6}$$

On the other hand, the Legendre transform $W^{(0)}[\check{T}]$ of $\Gamma^{(0)}[\mu]$ is defined by:

$$W^{(0)}[\check{T}] = \min_{\mu}\left(\Gamma^{(0)}[\mu] - \frac{1}{\pi}\int \mu\check{T}\right) . \tag{7.1.7}$$

$W^{(0)}[\check{T}]$ satisfies a functional equation which is given by the classical (*i.e.* large $c$) limit of the Ward identity for $\Gamma[\mu]$ by interchanging $\{\check{T}, \delta\Gamma^{(0)}/\delta\mu\}$ and $\{\delta W^{(0)}/\delta\check{T}, \mu\}$.

We will prove in this chapter, for $\mathcal{W}$–algebras arising from a Drinfeld-Sokolov reduction, that the relation between these functionals is:

$$W[\check{T}] = Z_W\, W^{(0)}\left[Z^{(\check{T})}\check{T}\right] , \tag{7.1.8}$$

where the renormalisation factors are functions of the central charge $c$. This was a long-existing conjecture in the literature. It was proven in [159, 134] for the Virasoro algebra, and conjectured for $\mathcal{W}_3$ in [157, 173]. The conjecture was based on a computation of the first order quantum corrections to $W^{(0)}$ which showed that the corrections split into two parts: one part contributes to the multiplicative renormalisations of $W^{(0)}$ while the other cancels $\Gamma^{(1)}$.

Eq. (7.1.8) implies that the one particle irreducible, or effective action is simply given by:

$$\Gamma_{\text{eff}}[\mu] = Z_W\, \Gamma^{(0)}\left[\frac{1}{k_c Z_{(\check{T})}}\mu\right] . \tag{7.1.9}$$

**Intermezzo 7.1.1**

The resemblance of the definition (7.1.6) to a Fourier transform, leads to the idea to prove eq. (7.1.8) by considering the functional equations arising from Ward identities. Unfortunately, this does not work. We will as an example consider the Virasoro algebra, for which the Ward identity (2.5.13) is:

$$-\frac{c}{12\pi}\partial^3\mu\; Z[\mu] = \bar{\partial}\frac{\delta Z}{\delta\mu} - \mu\partial\frac{\delta Z}{\delta\mu} - 2\partial\mu\frac{\delta Z}{\delta\mu} . \tag{7.1.10}$$

We multiply eq. (7.1.10) with $\exp\left(\frac{1}{\pi}\mu\check{T}\right)$ and integrate over $\mu$. Assuming we can use functional partial integration and discard boundary terms, we find:

$$\begin{aligned} -\frac{c}{12}\partial^3\frac{\delta F}{\delta\check{T}} &= -\bar{\partial}\check{T}\; F[\check{T}] - \pi\left(\frac{\delta}{\delta\check{T}}\partial + 2\partial\frac{\delta}{\delta\check{T}}\right)\int [d\mu]\; \exp\left(\int \frac{1}{\pi}\mu\check{T}\right)\frac{\delta Z}{\delta\mu} \\ &= -\bar{\partial}\check{T}\; F[\check{T}] + \left(\frac{\delta}{\delta\check{T}}\partial + 2\partial\frac{\delta}{\delta\check{T}}\right)\left(\check{T}F[\check{T}]\right) , \end{aligned} \tag{7.1.11}$$

where we defined:

$$F[\check{T}] \equiv \exp\left(-W[\check{T}]\right) . \tag{7.1.12}$$

Obviously, eq. (7.1.11) contains infinite $\delta^{(2)}(0)$ terms. This is because the Ward identity (7.1.10) suffers short-distance singularities when $\mu$ is a quantum field.

---

The conjecture (7.1.8) was elegantly proven for the case of $\mathcal{W}_3$ in [39] through the use of a quantum Drinfeld-Sokolov reduction. This chapter presents the generalisation for Drinfeld-Sokolov reductions of any Kač–Moody algebra [181]. The method is based on observations in [2, 24]. Using the definition of the induced action eq. (7.1.4), the generating functional of its connected Green's functions (7.1.6) becomes:

$$\exp\left(-W[\check{T}]\right) = \int [d\varphi]\; \delta(T[\varphi] - \check{T})\exp\left(-S[\varphi]\right) . \tag{7.1.13}$$

Evaluating this functional integral is in general impossible. Even when the number of degrees of freedom in the matter fields is the same as the number of generators $T$, it involves the computation of a usually very complicated Jacobian. However, if one takes for the matter system a gauged WZNW–model, we can actually compute the Jacobian and as a result obtain an all–order expression for the effective action.

## 7.2 All–order results

Consider a WZNW–model with action $\kappa S^-[g]$. The holomorphic symmetry currents form a Kač–Moody algebra $\hat{g}$. In the previous chapter, we proved that by imposing certain constraints on the currents of $\hat{g}$, we find a realisation of a $\mathcal{W}$–algebra. Moreover, we obtained an all-order expression for the induced action of the corresponding $\mathcal{W}$–gravity theory. To make the transition to the effective theory, we first calculate the induced action in the highest weight gauge.

[1] We will use the term 'effective action' also for the generating functional of the connected Green's functions. We trust the symbol $W$ instead of $\Gamma$ is enough to avoid confusion.

The induced action $\Gamma$ (7.1.4) for the $\mathcal{W}$–gravity is given by:

$$\exp(-\Gamma[\mu]) = \int [\delta g\, g^{-1}][d\tau][dA_{\bar z}]\left(\mathrm{Vol}\left(\Pi_+\bar g\right)\right)^{-1} \exp\left(-\mathcal{S}_1 - \frac{1}{4\pi xy}\int str(\mu\mathcal{T})\right), \quad (7.2.1)$$

where $\mathcal{S}_1$ is defined in (6.2.6). The generators of the $\mathcal{W}$–algebra $\mathcal{T}^{(j,\alpha_j)}$ are assembled in a matrix:

$$\mathcal{T} \equiv \mathcal{T}^{(j,\alpha_j)}\ {}_{(jj,\alpha_j)}t\,, \quad (7.2.2)$$

where the matrices ${}_{(jj,\alpha_j)}t$ form a representation of the (super) Lie algebra $\bar g$ (see appendix B). A similar matrix $\in \mathcal{K}_-$, exists for the sources $\mu^{(j,\alpha_j)}$.

We will define the path integral in eq. (7.2.1) by performing a gauge fixing in the BV formalism, where we use the quantum extended action found in the previous chapter (6.6.2). Going to the highest weight gauge $\Pi_{\geq 0}J_z \in \mathcal{K}_+$ and $\tau = 0$ is accomplished by a canonical transformation which interchanges fields and antifields for $\tau$ and $(\Pi_+ - \Pi_{hw})J_z$ and dropping all (new) antifields. To obtain the induced action, we integrate the resulting gauge-fixed action over the new fields. For clarity, we do not rename the fields involved in the transformation and we insert explicit delta functions in eq. (6.6.2):

$$\begin{aligned}\exp(-\Gamma[\mu]) = &\int [\delta g\, g^{-1}][d\tau^*][dA_{\bar z}][dJ_z^*] \\ &\delta\left((\Pi_{\geq 0} - \Pi_{hw})J_z\right)\ \delta\left(\Pi_{lw}J_z^*\right)\ \delta\left(\Pi_+ J_z^*\right)\ \delta(\tau)\ \delta\left(A_{\bar z}^*\right) \\ &\exp\Bigg(-\kappa S^-[g] - \frac{1}{\pi x}\int str\left(A_{\bar z}\left(J_z - \frac{\kappa}{2}e_-\right)\right) - \frac{\kappa}{2\pi x}\int str(\tau^* c) \\ &- \frac{\kappa}{2\pi x}\int str\left(J_z^*\left(\frac{\kappa}{2}\partial c + [c, J_z]\right)\right) - \frac{1}{4\pi xy}\int str\left(\mu\,\hat{\mathcal{T}}\right)\Bigg). \quad (7.2.3)\end{aligned}$$

The matrix $\hat{\mathcal{T}}$ in eq. (7.2.3) was determined in the previous chapter, see section 6.6. Important at this moment is that $\hat{\mathcal{T}}$ does not contain the fields $A_{\bar z}, \tau^*$ and $J_z^*$. This enables us to prove the following lemma.

**Lemma 7.2.1** *The functional integration over $A_{\bar z}$, $\tau^*$ and $J_z^*$ in eq. (7.2.3) gives two additional delta functions:*

$$\delta\left(\Pi_-\left(J_z - \frac{\kappa}{2}e_-\right)\right)\delta(c)\,.$$

**Proof :**
It is clear that the functional integration over $A_{\bar z}$ and $\tau^*$ gives:

$$\delta\left(\Pi_-\left(J_z - \frac{\kappa}{2}e_-\right)\right)\ \delta\left(\Pi_{1/2}c\right)\,.$$

We now integrate over $\Pi_{-m}J_z^*$ in increasing order of $m$. We show by recursion that each step in $m$ gives an additional $\delta(\Pi_{m+1}c)$, relying on the fact that $\Pi_{<(m+1)}c$ can be set to zero due to the previous steps. The recursion assumption is valid for $m = 0$.

Consider the part of the exponent in eq. (7.2.3) which contains the anticurrents:

$$\int str\left(\Pi_{-m}J_z^*\left(\frac{\kappa}{2}\partial\Pi_m c + \sum_{m'}[\Pi_{m'}c, \Pi_{m-m'}J_z]\right)\right).$$

The recursion assumption implies that the $\partial\Pi_m c$ term can be set to zero, and restricts the sum to $m' \geq m+1$. However, because of the delta functions of the currents, the terms in $\Pi_{m-m'}J_z$ are either highest weight currents (hence $m \geq m'$), or proportional to $e_-$ (where $m - m' = -1$). Clearly, only the latter gives a nonzero contribution. We end up with:

$$\int str\left(\frac{\kappa}{2}\left(\left(\Pi_{-m}J_z^*\right)[\Pi_{m+1}c, e_-]\right)\right).$$

As the supertrace gives a nondegenerate metric on the super Lie algebra, integration over $\Pi_{-m}J_z^*$ gives a delta function for all components of $\Pi_{m+1}c$. Note that due to the delta function $\delta\left(\Pi_{lw}J_z^*\right)$ in eq. (7.2.3), no lowest weight components are present in $\Pi_{-m}J_z^*$. This agrees with the fact that there are no highest weights in the commutator with $e_-$. This concludes the proof. ∎

Applying the previous lemma, we see that we can set $A_{\bar z} = \tau = c = 0$ in eq. (7.2.3). Then the matrix $\hat{\mathcal{T}}$ reduces to a very simple form. It was discussed in section 6.6 that the elements of this matrix can be written in terms of the generators of the quantum cohomology: $\hat{\mathcal{T}}[\hat W^{(j,\alpha_j)}]$. The equations (6.6.1), (6.4.12) and (6.2.22) show that[2]:

$$\hat W^{(j,\alpha_j)}[\Pi_{hw}J_z, \tau{=}0, A_{\bar z}^*{=}0, c{=}0] = \mathcal{C}^j J_z^{(jj,\alpha_j)}\,, \quad (7.2.4)$$

where the renormalisation constant $\mathcal{C}$ is defined in eq. (6.4.17). We will indicate this result with the slightly misleading notation $\hat{\mathcal{T}}[\mathcal{C}\Pi_{hw}J_z]$.

We can now perform the change of variables alluded to in the introduction. Passing from the Haar measure $[\delta g\, g^{-1}]$ to the measure $[dJ_z]$, we pick up a Jacobian:

$$[\delta g g^{-1}] = [dJ_z]\exp\left(-2\tilde h S^-[g]\right)\,. \quad (7.2.5)$$

[2] Any quantum corrections in $\hat W$ arise from the reordering of nonhighest weight currents, which cannot produce a highest weight current.

Combining the eqs. (7.2.3), (7.2.4) and (7.2.5), we obtain the induced action in a path integral formulation for the WZNW–model in the highest weight gauge:

$$\exp\left(-\Gamma[\mu]\right) = \int [dJ_z]\, \delta\left((1-\Pi_{hw})(J_z - \frac{\kappa}{2}e_-)\right)$$
$$\exp\Big(-\kappa_c S^-[g] - \frac{1}{4\pi xy}\int str\left(\mu \hat{\mathcal{T}}[\mathcal{C}J_z]\right)\Big)\,, \qquad (7.2.6)$$

where:

$$\kappa_c = \kappa + 2\tilde{h}\,. \qquad (7.2.7)$$

We are now in a position to study the effective theory of the (super)gravity theory. The effective action $W[\check{T}]$ (7.1.6) is, for the particular choice of matter sector, given by:

$$\exp\left(-W[\check{T}]\right) = \int [dJ_z][d\mu]\, \delta\left((1-\Pi_{hw})(J_z - \frac{\kappa}{2}e_-)\right)$$
$$\exp\Big(-\kappa_c S^-[g] - \frac{1}{4\pi xy}\int str\left(\mu(\hat{\mathcal{T}}[\mathcal{C}J_z] - \check{T})\right)\Big)\,, \qquad (7.2.8)$$

where $\check{T} \in \mathcal{K}_+$. The integration over $\mu$ gives an additional delta function:

$$\delta\left(\hat{\mathcal{T}}[\mathcal{C}J_z] - \check{T}\right)\,. \qquad (7.2.9)$$

This means that we find for the effective action $W[\check{T}]$ simply $\kappa_c S^-[g]$ where the groupelements $g$ are determined by the constraints imposed via the delta-functions.

To make this more explicit, we have to specify which set of generators $\mathcal{T}$ we choose for the $\mathcal{W}$–algebra. As discussed in subsection 6.4.3, it is convenient to choose $T^{\rm EM}$, defined in (6.4.18), for the Virasoro operator at $(j,\alpha_j) = (1,0)$. We take all other generators $\hat{\mathcal{T}}^{(j,\alpha_j)}$ equal to the generators of the cohomology $\hat{W}^{(j,\alpha_j)}$. We can write this choice of generators as[3]:

$$\hat{\mathcal{T}}[\mathcal{C}J_z] = \exp\left(\sqrt{\mathcal{C}}e_0\right)\Pi_{hw}J_z \exp\left(-\sqrt{\mathcal{C}}e_0\right) +$$
$$\frac{\mathcal{C}}{2xy\kappa} str\big((\Pi_{hw}\Pi_0 J_z)\ (\Pi_{hw}\Pi_0 J_z)\big)e_+\,. \qquad (7.2.10)$$

To express the constrains imposed on the WZNW–model by eq. (7.2.9) in terms of the groupelements of the WZNW–model, we should take into account that the quantum currents $J_z$ are renormalised, see eqs. (2.6.40) and (2.6.45):

$$J_z = \frac{\alpha_\kappa}{2}\partial g\, g^{-1}\,. \qquad (7.2.11)$$

[3]In fact, we take $\hat{\mathcal{T}}^{(1,0)} = \frac{1}{2y}T^{\rm EM}$ to avoid notational difficulties. We will show in intermezzo 7.2.1 that normalisation factors of the generators which do not depend on $\kappa$, do not influence the coupling constant or wave function renormalisation.

Combining this last equation with eqs. (7.2.8), (7.2.9), (7.2.10), we find the final result:

$$W[\check{T}] = \kappa_c S^-[g]\,, \qquad (7.2.12)$$

where $\kappa_c$ is defined in eq. (7.2.7). The WZNW–model in eq. (7.2.12) is constrained by:

$$\frac{\alpha_\kappa}{2}\partial g g^{-1} + \frac{{\alpha_\kappa}^2}{4xy\kappa} str\Big(\Pi_{\rm NA}\left(\partial g g^{-1}\right)\Pi_{\rm NA}\left(\partial g g^{-1}\right)\Big)e_+ =$$
$$\frac{\kappa}{2}e_- + \exp\left(-\ln\sqrt{\mathcal{C}}e_0\right)\check{T}\exp\left(\ln\sqrt{\mathcal{C}}e_0\right)\,. \qquad (7.2.13)$$

$\Pi_{\rm NA}\bar{g}$ are those elements of $\Pi_0\bar{g}$ which are highest weight, *i.e.* the centraliser of $sl(2)$ in $\bar{g}$. Using a global group transformation:

$$g \to \exp\left(\ln\left(\sqrt{\frac{\alpha_\kappa}{\kappa}}\right)e_0\right)g\,, \qquad (7.2.14)$$

which leaves $S^-[g]$ invariant, we bring the constraints in the standard form used in [7]:

$$\partial g g^{-1} + \frac{1}{4xy} str\Big(\Pi_{\rm NA}\left(\partial g g^{-1}\right)\ \Pi_{\rm NA}\left(\partial g g^{-1}\right)\Big)e_+$$
$$= e_- + \sum_{j,\alpha_j}\frac{2\kappa^j}{\mathcal{C}^j {\alpha_\kappa}^{j+1}}\check{T}^{(j,\alpha_j)}{}_{(jj,\alpha_j)}t\,. \qquad (7.2.15)$$

From eq. (6.4.26), we get the level $\kappa$ as a function of the central charge:

$$12y\kappa = -12y\tilde{h} - \left(c - \frac{1}{2}c_{\rm crit}\right) - \sqrt{\left(c - \frac{1}{2}c_{\rm crit}\right)^2 - 24(d_B - d_F)\tilde{h}y} \qquad (7.2.16)$$

We will now reformulate these results in terms of the Legendre transform $W^{(0)}$ of the classical limit of the induced action, eq. (7.1.7), and prove the conjecture eq. (7.1.8):

$$W[\check{T}] = Z_W\, W^{(0)}\left[Z^{(\check{T})}\check{T}\right]\,.$$

It is possible, using the results of [7], to find $W^{(0)}$ by comparing the Ward identities of the constrained WZNW–model to those of the classical extended Virasoro algebra. We will follow a different road. Because:

$$W_{cl}[\check{T}] = cW^{(0)}[\frac{1}{c}\check{T}]\,, \qquad (7.2.17)$$

we can obtain $W^{(0)}$ by taking the large $c$ limit of eqs. (7.2.12) and (7.2.15). We begin by observing that:

$$\frac{\kappa^j}{\mathcal{C}^j {\alpha_\kappa}^{j+1}} \approx \kappa^{-1} \approx -\frac{c}{6y}\,, \qquad \text{for large } c\,, \tag{7.2.18}$$

where we used eq. (7.2.16) in the second step. Eq. (7.2.18) is valid whatever the values of $\mathcal{C}$ and $\alpha_\kappa$ turn out to be, as their classical limit is fixed to 1 and $\kappa$ respectively. Together with (7.2.12) and (7.2.15), this gives[4]:

$$W[\check{T}] = -\frac{6y(\kappa+2\tilde{h})}{c} W_{cl}[-\frac{c}{6y}\frac{\kappa^j}{\mathcal{C}^j {\alpha_\kappa}^{j+1}} \check{T}^{(j,\alpha_j)}{}_{(j,\alpha_j)} t]\,. \tag{7.2.19}$$

Combining this result with eq. (7.2.17), we find:

$$Z_W = -6y\left(\kappa + 2\tilde{h}\right), \tag{7.2.20}$$

$$Z^{(\check{T}^{(j,\alpha_j)})} = -\frac{\kappa^j}{6y\,\mathcal{C}^j {\alpha_\kappa}^{j+1}}\,. \tag{7.2.21}$$

Together with eq. (7.2.16), this provides an all-order expression for renormalisation factors for the chosen normalisation eq. (7.2.10) of the generators of the $\mathcal{W}$–algebra.

**Intermezzo 7.2.1**
We now discuss the case where a different normalisation from eq. (7.2.10) for the generators of the $\mathcal{W}$–algebra is used. Suppose that there is an additional factor $\hat{n}^{(j,\alpha_j)}$:

$$\hat{T}^{(j,\alpha_j)} = \hat{n}^{(j,\alpha_j)} \hat{W}^{(j,\alpha_j)}\,.$$

This extra factor can depend on $\kappa$. If we denote its classical limit as $n^{(j,\alpha_j)}$, we see that the wavefunction renormalisation is changed to:

$$Z^{(\check{T}^{(j,\alpha_j)})} = -\frac{n^{(j,\alpha_j)}}{\hat{n}^{(j,\alpha_j)}} \frac{\kappa^j}{6y\,\mathcal{C}^j {\alpha_\kappa}^{j+1}}\,.$$

As discussed in section 6.6, we expect that all $\kappa$–dependence of the normalisation factor is absorbed in $\mathcal{C}$, eq. (6.6.1), such that the wavefunction renormalisation factor is not changed with respect to eq. (7.2.21).

---

We want to stress here that while the value of the coupling constant renormalisation is unambiguously determined, the computation of the value of the wavefunction renormalisation is very delicate. Indeed, when the gauged WZW model serves as a guideline [180], we expect that the precise value of the wavefunction renormalisation depends on the chosen regularisation scheme. As mentioned before, the computations leading to the quantum effective action are performed in the operator formalism using point-splitting regularisation. Within this framework, we should use $\alpha_\kappa = \kappa + \tilde{h}$ (2.6.46), and $\mathcal{C} = \kappa/(\kappa+\tilde{h})$ (6.4.17). The wavefunction renormalisation factor eq. (7.2.21) simplifies to:

$$Z^{(\check{T}^{(j,\alpha_j)})} = -\frac{1}{6y(\kappa+\tilde{h})}\,. \tag{7.2.22}$$

We expect that this result is fully consistent when using the operator formalism. To provide further support for this claim, we will compare the results for $so(N)$ supergravity of section 7.3 to perturbative computations in section 7.4. The perturbative calculations also rely on operator methods and give results which are fully consistent with both eqs. (7.2.20) and (7.2.22). See [187] for a further discussion of the wavefunction renormalisation factors.

[4] We hope that there can be no confusion between $\alpha_\kappa$ and $\alpha_j$.

## 7.3 An example : $so(N)$–supergravity

We will use the supergravity theories based on the $N$-extended $so(N)$ superconformal algebra [133, 20] as an example. A realisation of the matter sector, referred to above, is constructed from gauged $OSp(N|2)$ WZNW–model in sections 6.3 and 6.5. Applying the results of the previous section, we can give all–loop results for the effective theory for arbitrary $N$. Aspects of $N=1$ and $N=2$ supergravity were studied in [107, 162, 25]. Features of the $N=3$ theory were examined in [54], where one–loop results for the effective action were given.

The induced action $\Gamma[h,\psi,A]$ for the $N$-extended $so(N)$ supergravity is defined as:

$$\exp\left(-\Gamma[h,\psi,A]\right) = \left\langle \exp\left(-\frac{1}{\pi}\int \left(hT + \psi^a G_a + A^i U_i\right)\right)\right\rangle. \tag{7.3.1}$$

The OPEs of the $so(N)$- superconformal algebra are given in intermezzo 6.3.2. The Ward identities can be obtained by constructing the chiral supergravity transformations which give a minimal anomaly for the transformation of $\Gamma$. Alternatively, one can directly use the OPEs, eq. (2.5.18). We only list the result. Introducing:

$$t = \frac{12\pi}{c}\frac{\delta\Gamma[h,\psi,A]}{\delta h} \quad \text{and} \quad g^a = \frac{2\pi}{\beta}\frac{\delta\Gamma[h,\psi,A]}{\delta\psi_a}\,, \tag{7.3.2}$$

the Ward identities for $\Gamma$ are:

$$\partial^3 h = \overline{\nabla} t - \frac{3\beta}{c}\left(\psi^a\partial + 3\partial\psi^a\right) g_a + \frac{6k}{c}\partial A^i u_i$$

$$\partial^2\psi^a = \overline{\nabla} g^a - \frac{c}{3\beta}\psi^a t + \lambda_{ab}{}^i A_i g^b - \lambda_{ab}{}^i\left(2\partial\psi^b + \psi^b\partial\right) u_i$$

$$-\frac{k^2\gamma}{2\beta}\Pi^{ij}_{ab}u_iu_j - \frac{k\gamma\pi}{\beta}\Pi^{ij}_{ab}\lim_{y\to x}\left(\frac{\partial u_i(x)}{\partial A^j(y)} - \frac{\partial}{\bar\partial}\delta^{(2)}(x-y)\delta_{ij}\right)$$

$$\partial A^i = \overline{\nabla} u^i + \frac{\beta}{k}\lambda_{ab}{}^i\psi^a g^b + f^{ij}{}_k A^j u^k\,, \qquad (7.3.3)$$

where:

$$\overline{\nabla}\Phi = \left(\bar\partial - h\partial - h_\Phi(\partial h)\right)\Phi, \qquad (7.3.4)$$

with:

$$h_\Phi = 2, \frac{3}{2},\ 1 \quad \text{for} \quad \Phi = t, g^a,\ u^a\,. \qquad (7.3.5)$$

The Ward identities provide us with a set of functional differential equations for the induced action. Because of the explicit dependence of the Ward identities on $k$, we immediately see that the induced action is given in a $1/k$ expansion as:

$$\Gamma[h,\psi,A] = \sum_{i\geq 0} k^{1-i}\Gamma^{(i)}[h,\psi,A]\,. \qquad (7.3.6)$$

This definition of $\Gamma^{(i)}$ is different from eq. (7.1.5), where an expansion in the central charge $c$ was used. The relation between these two parameters is (see intermezzo 6.3.2):

$$c = \frac{k}{2}\frac{6k+N^2-10}{k+N-3} \qquad (7.3.7)$$

Inverting this formula would give rise to square roots which we wish to avoid. Therefore we will work in this section in the large $k$ limit, which coincides with the large $c$ limit upto some numerical factors. In this limit, the Ward identities become local and they are solved by $\Gamma[h,\psi,A] = k\Gamma^{(0)}[h,\psi,A]$.

If one defines the Legendre transform of $\Gamma^{(0)}[h,\psi,A]$ as:

$$W^{(0)}[t,g,u] = \min_{\{h,\psi,A\}}\left(\Gamma^{(0)}[h,\psi,A] - \frac{1}{4\pi}\int\left(h\,t + 4\psi^a\,g_a - 2A^i\,u_i\right)\right), \qquad (7.3.8)$$

we can view the Ward identities (7.3.3) in the large $k$ limit as a set of functional equations for $W^{(0)}$ (7.1.7).

Consider now the $OSp(N|2)$ WZNW–model with action $\kappa S^-[g]$. We will then show that by imposing the highest weight gauge constraints on its Ward identities (2.6.52), we recover the Ward identities of the $so(N)$ supergravity eq. (7.3.3) in the classical limit [7].

The Ward identities of a WZNW–model eq. (2.6.52) correspond to the zero curvature condition on the connections $A_{\bar z}$ and $u_z$:

$$R_{z\bar z} = \partial A_{\bar z} - \bar\partial u_z - [u_z, A_{\bar z}] = 0\,, \qquad (7.3.9)$$

where $u_z$ is defined in eq. (2.6.53). We impose the following constraint on $u_z$:

$$u_z \equiv \begin{pmatrix} 0 & u_z^{\ddagger} & u_z^{+b} \\ 1 & 0 & 0 \\ 0 & -u_z^{+a} & u_z^{ab}\end{pmatrix}. \qquad (7.3.10)$$

Using eq. (7.3.10), we find that some of the components of $R_{z\bar z}=0$ become algebraic equations. Indeed, $R^{=}_{z\bar z} = R^{0}_{z\bar z} = R^{-a}_{z\bar z} = 0$ for $0\leq a\leq N$ can be solved for $A^0_{\bar z}$, $A^{\ddagger}_{\bar z}$ and $A^{+a}_{\bar z}$, giving:

$$\begin{aligned} A^0_{\bar z} &= \frac{1}{2}\partial A^{=}_{\bar z} \\ A^{\ddagger}_{\bar z} &= -\frac{1}{2}\partial^2 A^{=}_{\bar z} + A^{=}_{\bar z}u^{\ddagger}_z + A^{-a}_{\bar z}u^{+a}_z \\ A^{+a}_{\bar z} &= \partial A^{-a}_{\bar z} + A^{=}_{\bar z}u^{+a}_z - \sqrt{2}\lambda_{ab}{}^i A^{-b}_{\bar z}u^i_z, \end{aligned} \qquad (7.3.11)$$

where:

$$u^i_z \equiv \frac{1}{\sqrt2}\lambda_{ab}{}^i u^{ab}_z. \qquad (7.3.12)$$

The remaining curvature conditions $R^{\ddagger}_{z\bar z} = R^{+a}_{z\bar z} = R^{ab}_{z\bar z} = 0$ reduce to the Ward identities eq. (7.3.3) in the limit $k\to\infty$ upon identifying:

$$\begin{aligned} h &\equiv A^{=}_{\bar z} \\ \psi^a &\equiv iA^{-a}_{\bar z} \\ A^i &\equiv -\sqrt2\left(A^i_{\bar z} - A^{=}_{\bar z}u^i_z\right) \\ t &\equiv -2\left(u^{\ddagger}_z + u^i_z u^i_z\right) \\ g^a &\equiv iu^{+a}_z \\ u^i &\equiv -\sqrt2 u^i_z, \end{aligned} \qquad (7.3.13)$$

where:

$$A^i_{\bar z} \equiv \frac{1}{\sqrt2}\lambda_{ab}{}^i A^{ab}_{\bar z}. \qquad (7.3.14)$$

This gives for the classical limit of the effective action eq. (7.1.7) of the $so(N)$–supergravity:

$$\begin{aligned} W^{(0)}\Big[t = -2(\partial gg^{-1})^{\ddagger} - 2(\partial gg^{-1})^i(\partial gg^{-1})^i,& \\ g^a = i(\partial gg^{-1})^{+a}, u^i = -\sqrt2(\partial gg^{-1})^i\Big] &= -\frac{1}{2}S^-[g]\,, \end{aligned} \qquad (7.3.15)$$

where $S^-[g]$ is the action for the WZNW–model with the constraints given in eq. (7.3.10).

The effective action $W[t,g,u]$ is defined by:

$$\exp\Big(-W[t,g,u]\Big) = \int [dh][d\psi][dA]$$
$$\exp\bigg(-\Gamma[h,\psi,A] + \frac{1}{4\pi}\int\Big(h\,t + 4\psi^a\,g_a - 2A^i\,u_i\Big)\bigg). \qquad (7.3.16)$$

One finds to leading order in $k$ (compare with eq. (7.2.17)):

$$W[t,g,u] = kW^{(0)}[t/k, g/k, u/k] = \kappa S^-[g] \qquad (7.3.17)$$

and the level $\kappa$ of the $osp(N|2)$ Kač–Moody algebra is related to the $so(N)$ level $k$ as $k \approx -2\kappa$ in the large $k$ limit.

We can now perform the analysis given in section 7.2 to find the all–order result for the effective action. The classical and quantum expressions for the realisation of the $so(N)$ $\mathcal{W}$–algebra in terms of the $osp(N|2)$ currents was given in sections 6.3 and 6.5 respectively. In particular, the normalisation constants can be found in eqs. (6.3.6), (6.3.8) and (6.5.2).
We do not give the explicit derivation of eqs. (7.2.8) and (7.2.9) here, as it is completely analogous to the general method, see [181] for a detailed calculation. We find that $W[t,g,u]$ is given by:

$$W[t,g,u] = \kappa_c S^-[g]\,, \qquad (7.3.18)$$

where

$$\kappa_c = \kappa + 2\tilde h = -\frac{1}{2}\left(k - 7 + 2N\right)\,, \qquad (7.3.19)$$

where we used $k = -2\kappa - 1$ (6.5.3). The WZNW functional in eq. (7.3.18) is constrained by:

$$\begin{aligned}
\left(\partial g g^{-1}\right)^{=} &= \frac{\kappa}{\alpha_\kappa}\\
\left(\partial g g^{-1}\right)^{-a} &= \left(\partial g g^{-1}\right)^{0} = 0\\
\left(\partial g g^{-1}\right)^{\ddagger} + \frac{\alpha_\kappa}{\kappa}\left(\partial g g^{-1}\right)^{i}\left(\partial g g^{-1}\right)^{i} &= \frac{t}{4\mathcal{C}\alpha_\kappa}\\
\left(\partial g g^{-1}\right)^{+a} &= \frac{1}{2i\sqrt{\mathcal{C}}\alpha_\kappa}g^a\\
\left(\partial g g^{-1}\right)^{i} &= \frac{1}{2\sqrt{2}\alpha_\kappa}u^i\,,
\end{aligned} \qquad (7.3.20)$$

which is the analogue of eq. (7.2.13). Performing the global group transformation eq. (7.2.14), substituting the value $\alpha_\kappa = \kappa + \tilde h$ and comparing with (7.3.15), we get:

$$W[t,g,u] = Z_W W^{(0)}\left[Z^{(t)}t, Z^{(g)}g, Z^{(u)}u\right], \qquad (7.3.21)$$

where:

$$Z_W = -2\kappa_c = (k - 7 + 2N) \qquad (7.3.22)$$

$$Z^{(t)} = Z^{(g)} = Z^{(u)} = \frac{1}{k + N - 3}\,. \qquad (7.3.23)$$

This agrees with the results of section 7.2 if we remember that $W^{(0)}$ is in this section defined via a large $k$ expansion. Using $k = -2\kappa - 1$ (6.5.3), we can convert eqs. (7.2.22) and (7.2.20) to this convention by simply substituting the factor $6y$ by 2, corresponding to the difference between the large $c$ and $k$ limit of $\alpha_\kappa = \kappa + \tilde h$.

## 7.4 Semiclassical evaluation for the linear superconformal algebras

In the previous section we computed the renormalisation factors for the nonlinear $so(N)$ superconformal algebras by realising them as WZNW–models. On the other hand, we showed in chapter 5 that for $N = 3$ and 4 these algebras can be obtained from linear ones by eliminating the dimension 1/2 fields and for $N = 4$ an additional $U(1)$ factor. Moreover, we showed in section 5.3 that the effective actions $W$ of the linear theories can be obtained from the linear effective actions simply by putting the currents corresponding to the free fields to zero, eq. (5.2.7). In this section we compute the effective actions for the linear theories in the semiclassical approximation. To facilitate comparison with the results for the $so(N)$ algebras, we first write the analogues of eqs. (7.3.22) and (7.3.23) in the large $c$ expansion:

$$Z_W = -3\left(k - 7 + 2N\right) \qquad (7.4.1)$$

$$Z^{(t)} = Z^{(g)} = Z^{(u)} = \frac{1}{3(k + N - 3)}\,, \qquad (7.4.2)$$

where the relation between $c$ and $k$ is given in intermezzo 6.3.2:

$$c = \frac{k}{2}\frac{6k + N^2 - 10}{k + N - 3}\,. \qquad (7.4.3)$$

Let us first explain the method which we shall use for the semiclassical evaluation [212, 176]. We restrict ourselves to linear algebras. In the semiclassical approxima-tion, the effective action is computed by a steepest descent method:

$$\begin{aligned} \exp\left(-W[\check{T}]\right) &= \int[d\mu]\exp\left(-\Gamma[\mu]-\frac{1}{\pi}\int\check{T}\mu\right) \\ &\simeq \exp\left(-\Gamma[\mu_{\rm cl}]-\frac{1}{\pi}\check{T}\mu_{\rm cl}\right)\int[d\tilde{\mu}]\exp\left(-\frac{1}{2}\tilde{\mu}\frac{\delta^2\Gamma[\mu_{\rm cl}]}{\delta\mu_{\rm cl}\delta\mu_{\rm cl}}\tilde{\mu}\right), \end{aligned} \tag{7.4.4}$$

where $\mu_{\rm cl}[\check{T}]$ is the saddle point value that solves

$$-\frac{\delta\Gamma}{\delta\mu}[\mu_{\rm cl}] = \frac{1}{\pi}\check{T}\,, \tag{7.4.5}$$

and $\tilde{\mu}$ is the fluctuation around this point. Therefore, all that has to be done is to compute a determinant:

$$W[\check{T}] \simeq W_{\rm cl}[\check{T}] + \frac{1}{2}\log sdet\frac{\delta^2\Gamma[\mu_{\rm cl}]}{\delta\mu_{\rm cl}\delta\mu_{\rm cl}}\,. \tag{7.4.6}$$

To evaluate this determinant, one may use the Ward identities. Schematically, they have the form:

$$\overline{D_1}[\mu]\frac{\delta\Gamma}{\delta\mu} \sim \partial_2\mu$$

where on the *lhs* there is a covariant differential operator, and on the *rhs* the term resulting from the anomaly. The symbol $\partial_2$ is standing for a differential operator of possibly higher order (see for example eq. (7.3.3) without the nonlinear term). Taking the derivative with respect to $\mu$, and transferring some terms to the *rhs* one obtains:

$$\overline{D_1}[\mu]\frac{\delta^2\Gamma}{\delta\mu\delta\mu} \sim D_2[\check{T}]\,, \tag{7.4.7}$$

where there now also appears a covariant operator on the *rhs* with $\check{T}$ and $\mu$ again related by eq. (7.4.5). The sought-after determinant is then formally the quotient of the determinants of the two covariant operators in eq. (7.4.7).

For the induced and effective actions of fields coupled to affine currents, the covariant operators are both simply covariant derivatives, and their determinants are known. Both induce a WZNW–model action. For the 2-D gravity action $\mu \to h$ and $\check{T} \to t$, the operator on the *rhs* is $\partial^3+t\partial+\partial t$ and the determinant is given in [212]. A similar computation for the semiclassical approximation to $W_3$ can be found in [173]. From these cases, one may infer the general structure of these determinants. In the gauge where the fields $\mu$ are fixed, the operator $\overline{D_1}[\mu]$ corresponds to the ghost Lagrangian. In BV language, the relevant piece of the extended action is $\mu^*\overline{D_1}[\mu]c + b^*\lambda$ and as in section 6.2 the $\mu^*$-field is identified with the Faddeev-Popov antighost. The determinant is then given by the induced action resulting from the ghost currents. Since these form the same algebra as the original currents (at least for a linear algebra), with a value of the central extension that can be computed, one has:

$$\log sdet\overline{D_1}[\mu] = c_{ghost}\Gamma^{(0)}[\mu]\,. \tag{7.4.8}$$

The second determinant can similarly be expressed as a functional integral over some auxiliary $bc$ and/or $\beta\gamma$ system. Let us, to be concrete, take $D_2[h] = \frac{1}{4}(\partial^3 + t\partial + \partial t)$ as an illustration. Then we have:

$$(sdetD_2[t])^{1/2} = \int[d\sigma]\exp\frac{1}{8\pi}\left(\sigma(\partial^3\sigma + t\partial\sigma + \partial(t\sigma))\right) \tag{7.4.9}$$

where it is sufficient to use a single fermionic integral since $D_2$ is antisymmetric. We can rewrite this as:

$$(sdetD_2[t])^{1/2} = \exp\left(-\tilde{W}[t]\right) = \langle\exp\left(\frac{1}{\pi}\int tH\right)\rangle_\sigma$$

where $H = \frac{1}{4}\sigma\partial\sigma$.

The propagator of the $\sigma$-field fluctuations is given by:

$$\sigma(z,\bar{z})\sigma(w,\bar{w}) = 2\frac{(z-w)^2}{\bar{z}-\bar{w}} + \text{regular terms} \tag{7.4.10}$$

and the induced action has been called $\tilde{W}$ instead of $\Gamma$ for reasons that will be clear soon.

The Lagrangian eq. (7.4.9) has an invariance:

$$\begin{aligned} \delta t &= \partial^3\omega + 2(\partial\omega)t + \omega\partial t \\ \delta\sigma &= \omega\partial\sigma - (\partial\omega)\sigma \end{aligned} \tag{7.4.11}$$

and, correspondingly, $\tilde{W}[t]$ obeys a Ward identity. We find:

$$\partial^3\frac{\delta\tilde{W}[t]}{\delta t} + 2t\partial\frac{\delta\tilde{W}[t]}{\delta t} + (\partial t)\frac{\partial\tilde{W}[t]}{\partial t} = -\frac{1}{\pi}\overline{\partial}t\,. \tag{7.4.12}$$

Another way to obtain this Ward identity (7.4.12) is by evaluating the OPE of the $H$-operator. Due to the propagator eq. (7.4.10) this OPE is not holomorphic. One finds:

$$H(z,\bar{z})H(0) = -\frac{c'}{4}\left(\frac{z}{\bar{z}}\right)^2 - \frac{z}{\bar{z}}H(0) - \frac{z^2}{2\bar{z}}\partial H(0) + \cdots\,, \tag{7.4.13}$$

where an ellipsis denotes higher order terms in $\bar{z}$. This OPE also appears in [212].

Eq. (7.4.12) is nothing but the usual chiral gauge conformal Ward identity 'read backwards', *i.e.* $t \leftrightarrow \frac{\delta\Gamma[h]}{\delta h}$ and $h \leftrightarrow \frac{\partial\tilde{W}[t]}{\partial t}$. We conclude that $\tilde{W}[t]$ is proportional to the Legendre transform of $\Gamma^{(0)}$:

$$\tilde{W}[t] = -6c'W^{(0)}[t] \tag{7.4.14}$$

with $c' = 2$.

Note that in [212] different bosonic realisations of the algebra (7.4.13) were used. Starting from the action $\frac{1}{2}\varphi(\partial^2 + \frac{t}{2})\varphi$ one finds that $H_\varphi = \frac{1}{4}\varphi^2$ satisfies (7.4.13) with $c' = -1/2$. This realisation will appear naturally when we discuss $N = 1$. Another realisation, starting from $\varphi_1[\partial^3 + t\partial + \partial t]\varphi_2$, has $c' = -4$ and is a bosonic twin of the one we used. The same algebra also realises a connection with $sl_2$, through ([212]) $H(z,\bar z) = -\frac{z^2}{2}j^+(\bar z) + zj^0(\bar z) + \frac{1}{2}j^-(\bar z)$: The antiholomorphic components $j^a$ of $H(z,\bar z)$ generate an affine $sl_2$ algebra.

The upshot is that whereas the first determinant is proportional to the classical induced action $\Gamma$, the second one is proportional to the classical effective action $W$. The proportionality constants are pure numbers independent of the central extension of the original action. From these numbers the renormalisation factors for the quantum effective action in the semiclassical approximation follow:

$$\begin{aligned}
W[\check T] &\simeq cW^{(0)}\left[\frac{\check T}{c}\right] - 6c'W^{(0)}\left[\frac{\check T}{c}\right] - \frac{c_{ghost}}{2}\Gamma^{(0)}[\mu_{\rm cl}] \\
&\simeq \left(c - 6c' - \frac{c_{ghost}}{2}\right)W^{(0)}\left[\frac{\check T}{c}\right] + \frac{c_{ghost}}{2}\check T\frac{\partial W^{(0)}}{\partial \check T}\left[\frac{\check T}{c}\right] \\
&\simeq \left(c - 6c' - \frac{c_{ghost}}{2}\right)W^{(0)}\left[\frac{\check T}{c}(1 + \frac{c_{ghost}}{2c})\right].
\end{aligned} \tag{7.4.15}$$

For $\mathcal{W}_2$-gravity, the results of [212] follow. We now turn to $N = 1\cdots 4$ linear supergravities.

### 7.4.1 $N = 1$

This case has been treated also in [212, 149]. The induced action is

$$\exp\left(-\Gamma[h,\psi]\right) = \langle \exp\left(-\frac{1}{\pi}(hT + \psi G)\right)\rangle\,, \tag{7.4.16}$$

where $T$ and $G$ generate the $N = 1$ superconformal algebra with central charge $c$. The Ward identities read[5], with $\Gamma = c\Gamma^{(0)}$:

$$\begin{aligned}
\big(\bar\partial - h\partial - 2(\partial h)\big)\frac{\delta\Gamma^{(0)}}{\delta h} - \frac{1}{2}\big(\psi\partial - 3(\partial\psi)\big)\frac{\delta\Gamma^{(0)}}{\delta\psi} &= \frac{1}{12\pi}\partial^3 h \\
\big(\bar\partial - h\partial - \frac{3}{2}(\partial h)\big)\frac{\delta\Gamma^{(0)}}{\delta\psi} - \frac{1}{2}\psi\frac{\delta\Gamma^{(0)}}{\delta h} &= \frac{1}{3\pi}\partial^2\psi\,.
\end{aligned} \tag{7.4.17}$$

[5] All functional derivatives are left derivatives.

From this we read off $\overline{D_1}$ and $D_2$ of eq. (7.4.7):

$$\begin{aligned}
\overline{D_1} &= \begin{pmatrix} \bar\partial - h\partial - 2(\partial h) & -\frac{1}{2}\psi\partial + \frac{3}{2}(\partial\psi) \\ -\frac{1}{2}\psi & \bar\partial - h\partial - \frac{3}{2}(\partial h) \end{pmatrix} \\
D_2 &= \frac{1}{3\pi}\begin{pmatrix} \frac{1}{4}(\partial^3 + (\partial\hat t) + 2\hat t\partial) & -\frac{1}{2}(\partial\hat y) - \frac{3}{2}\hat g\partial \\ (\partial\hat g) + \frac{3}{2}\hat g\partial & \partial^2 + \hat t/2 \end{pmatrix}
\end{aligned} \tag{7.4.18}$$

We abbreviated $\hat t = -12\pi\frac{\partial\Gamma^{(0)}[h]}{\partial h} = t/c$ and $\hat g = g/c$. For ease of notation, we will drop the hats in the computation of $sdetD_2$.

$\overline{D_1}$ gives rise to the ghost-realisation for $N = 1$, so we have:

$$sdet\overline{D_1} = 15\,\Gamma^{(0)}[h,\psi]\,. \tag{7.4.19}$$

$D_2$ is a (super)antisymmetric operator, as can be seen by rewriting:

$$\begin{aligned}
\partial^3 + (\partial t) + 2t\partial &= \partial^3 + \partial t + t\partial\,, \\
\frac{1}{2}(\partial g) + \frac{3}{2}g\partial &= \frac{1}{2}\partial g + g\partial\,, \\
(\partial g) + \frac{3}{2}g\partial &= \frac{1}{2}g\partial + \partial g\,.
\end{aligned} \tag{7.4.20}$$

The relevant action is then:

$$\frac{1}{\pi}\int\left(\frac{1}{8}\sigma(\partial + \partial t + t\partial)\sigma - \frac{1}{2}\sigma(\partial g)\varphi - \frac{3}{2}\sigma g\partial\varphi + \frac{1}{2}\varphi\left(\partial^2 + \frac{t}{2}\right)\varphi\right). \tag{7.4.21}$$

The determinant is:

$$(sdetD_2)^{1/2} = \langle\exp\left(-\frac{1}{\pi}(tH + g\Psi)\right)\rangle = \exp\left(-\tilde W(t,g)\right)\,, \tag{7.4.22}$$

where $H = \frac{1}{4}\sigma\partial\sigma + \frac{1}{4}\varphi^2$ and $\Psi = -\frac{1}{2}(\partial\sigma)\varphi + \sigma\partial\varphi$ and the average is taken in a free field sense with propagators eq. (7.4.10) and:

$$\langle\varphi(z,\bar z)\varphi(w,\bar w)\rangle = \frac{z - w}{\bar z - \bar w}\,. \tag{7.4.23}$$

This leads to the OPEs:

$$\begin{aligned}
H(z,\bar z)H(0) &= -\frac{c'}{4}\frac{z^2}{\bar z^2} + \frac{z}{\bar z}H(0) + \frac{1}{2}\frac{z^2}{\bar z}\partial H(0) + \cdots \\
H(z,\bar z)\Psi(0) &= \frac{1}{2}\frac{z}{\bar z}\Psi(0) + \frac{1}{2}\frac{z^2}{\bar z}\partial\Psi(0) + \cdots \\
\Psi(z,\bar z)\Psi(0) &= 2c'\frac{z}{\bar z^2} - \frac{4}{\bar z}H(0) - \frac{2z}{\bar z}\partial H(0) + \ldots\,,
\end{aligned} \tag{7.4.24}$$

with the value for the central extension $c' = 2 - \frac{1}{2}$.
The resulting Ward identities for $\tilde{W}$, eq. (7.4.22), are:

$$\begin{aligned}(\partial^3 + (\partial t) + 2t\partial)\frac{\partial\tilde{W}}{\partial t} - (2(\partial g) + 6g\partial)\frac{\partial\tilde{W}}{\partial g} &= -\frac{c'}{2\pi}\overline{\partial}t \\ (\partial^2 + \frac{t}{2})\frac{\partial\tilde{W}}{\partial g} + (\ (\partial g) + \frac{3}{2}g\partial)\frac{\partial\tilde{W}}{\partial t} &= -\frac{2c'}{\pi}\overline{\partial}g\,. \end{aligned} \tag{7.4.25}$$

Comparing with eq. (7.4.17), and reverting to the proper normalisation of $t$ and $g$, we have:

$$\tilde{W}[\hat{t},\hat{g}] = 6c'\ W^{(0)}[\hat{t},\hat{g}] = 6c'\ W^{(0)}[t/c, g/c]\,, \tag{7.4.26}$$

where we used:

$$W^{(0)}[\hat{t},\hat{g}] = \min_{\{h,\psi\}}\left(\Gamma^{(0)}[h,\psi] - \frac{1}{12\pi}\int h\hat{t} - \frac{1}{3\pi}\int \hat{g}\psi\right)\,. \tag{7.4.27}$$

Putting everything together in eq. (7.4.15) we find, for $N=1$, $c' = 3/2$:

$$W[t,g] \simeq cW^{(0)}\left[\frac{t}{c},\frac{g}{c}\right] - \frac{15}{2}\Gamma^{(0)}[h,\psi] - 9W^{(0)}\left[\frac{t}{c},\frac{g}{c}\right]\,. \tag{7.4.28}$$

Writing these results as:

$$W^{(N)}[\Phi] \simeq Z_W^{(N)} W^{(0)}[Z_\Phi^{(N)}\Phi]$$

we have:

$$\begin{aligned} Z_W^{(1)} &\simeq c - \frac{33}{2} \\ Z_t^{(1)} = Z_g^{(1)} &\simeq \frac{1}{c}(1 + \frac{15}{2c})\,. \end{aligned} \tag{7.4.29}$$

For reference, the corresponding equations for $N=0$ are ($c' = 2$),

$$\begin{aligned} W[t] &= cW^{(0)}[t/c] - \frac{26}{2}\Gamma^{(0)}[h] - 12W^{(0)}[t/c] \\ Z_W^{(0)} &= c - 25 \\ Z_t^{(0)} &= \frac{1}{c}(1 + \frac{13}{c})\,. \end{aligned} \tag{7.4.30}$$

These values are in complete agreement with [212, 134, 162, 149] and with our eqs. (7.4.2), (7.4.2) and eq. (7.4.3).

Before going to $N=2$, we comment on the technique we used to obtain eqs. (7.4.13) and (7.4.24). The easiest way is to expand the fields $\sigma, \varphi$ in solutions of the free field equations:

$$\begin{aligned} \sigma(z,\bar{z}) &= \sigma^{(0)}(\bar{z}) + z\sigma^{(1)}(\bar{z}) + \frac{z^2}{2}\sigma^{(2)}(\bar{z}) \\ \varphi(z,\bar{z}) &= \varphi^{(0)}(\bar{z}) + z\varphi^{(1)}(\bar{z}) \end{aligned} \tag{7.4.31}$$

and read off the OPEs for the antiholomorphic coefficients from eqs. (7.4.10), (7.4.23). Then all singular terms are given in eq. (7.4.24). An alternative would be to use Wick's method, with the contractions given by the propagators. The resulting bilocals then give, upon Taylor-expanding, the same algebra as in eq. (7.4.24), up to terms proportional to equations of motion. This ambiguity was already present in [212], see also [173]. We have simply used an antiholomorphic mode expansion like in eq. (7.4.31) in the following calculation. A disadvantage is, that in this way one loses control over equation of motion terms.

Let us close the $N=1$ case by noting that, just as the antiholomorphic modes corresponding to eq. (7.4.13) generate an $sl(2)$ Kač–Moody algebra, we get an affine $osp(1|2)$ from the modes of $H$ and $\Psi$ of eq. (7.4.24).

### 7.4.2 $N = 2$

For $N=2$ the extension of the scheme above has two $\Psi$-fields and a free fermion $\tau$, with $<\tau(x)\tau(0)>= -\frac{1}{x}$. This last field does not contribute to $H$:

$$\begin{aligned} H &= \frac{1}{4}\sigma\partial\sigma + \frac{1}{4}\sum_{a=1}^{2}\phi_a^2 \\ \Psi_a &= -\frac{1}{2}(\partial\sigma)\phi_a + \sigma\partial\phi_a - \varepsilon_{ab}\phi_b\tau \\ A &= \varepsilon_{ab}\partial\phi_a\phi_b + \sigma\partial\tau\,. \end{aligned} \tag{7.4.32}$$

Note that $\partial A$ is proportional to the equations of motion for $\phi_a$ and $\tau$. Neglecting terms proportional to equations of motion, we find that in the algebra of eq. (7.4.24) the first two equations are supplemented with:

$$\begin{aligned} H(z,\bar{z})A(0) &= 0 + \cdots \\ \Psi_a(z,\bar{z})\Psi_b(0) &= \delta_{ab}\left(\frac{2c'z}{\bar{z}^2} - \frac{4H(0)}{\bar{z}} - \frac{2z}{\bar{z}}\partial H(0)\right) + \varepsilon_{ab}\frac{z}{\bar{z}}A(0) + \cdots \\ A(z,\bar{z})\Psi_a(0) &= \frac{\varepsilon_{ab}}{\bar{z}}\Psi_b(0) + \cdots \\ A(z,\bar{z})A(0) &= \frac{c'}{\bar{z}^2} + \cdots\,, \end{aligned} \tag{7.4.33}$$

and $c' = 2 - 2 \cdot \frac{1}{2} = 1$. With the central charge of the ghosts being $c_{ghost} = +6$, we can immediately write down the $Z$-factors for $N = 2$:

$$Z_W^{(2)} = c - 9 \qquad Z_t^{(2)} = Z_g^{(2)} = Z_a \simeq \frac{1}{c}(1 + 3/c)\,, \tag{7.4.34}$$

which agrees with eqs. (7.4.1) and (7.4.2) using eq. (7.4.3). The algebra of antiholomorphic coefficients of $H, \Psi_a$ and $A$ is now $osp(2|2)$.

It should be remarked that for $N = 2$ (and higher) the algebra eq. (7.4.33) does not quite reproduce the Ward identities for the induced action. Here also, the equations of motion are involved. The difference is in the Ward identity:

$$\frac{1}{4}(\partial^3 + 2t\partial + (\partial t))h - \frac{1}{2}((\partial g_a) + 3g_a\partial)\psi^a - (\partial A)\cdot u = \overline{\partial} t\,. \tag{7.4.35}$$

The last term, as noted below eq. (7.4.32), is proportional to equations of motion of the free part of the action of the auxiliary system, and is not recovered from the procedure outlined above. We surmise that, as for $N = 0$ and 1, these terms do not change the result.

### 7.4.3 $N = 3, 4$

For $N = 3$ and 4, we refrain from writing out the action and transformation laws, but the same procedure as before is valid (see [181] for $N = 3$). The algebra of eq. (7.4.33) only changes in that more $\Psi_a$ and $A$ fields are present. The value of the central charge $c'^{(N)}$ in that algebra can most simply be obtained from $H(z)H(0)$, since only $\sigma$ and $\varphi_a$ fields contribute to it:

$$c'^{(N)} = 2 - \frac{N}{2}\,. \tag{7.4.36}$$

The resulting antiholomorphic coefficients constitute the $osp(N|2)$ Kač–Moody algebra: the dimension 1/2 field contributes no antiholomorphic modes.

The ghost system central charges vanish for $N = 3, 4$. As a result, for $N = 3$:

$$Z_W^{(3)} = c - 3, \qquad Z_t^{(3)} \simeq \frac{1}{c}\,, \tag{7.4.37}$$

and for $N = 4$, all $Z$-factors are equal to their classical values.

Now we compare these results for the renormalisation factors with the results of eqs. (7.4.1) and (7.4.2) for the *nonlinear* algebras, using the result of section 5.3. According to section 5.3, the respective effective actions are equal upon putting the appropriate currents to zero. This means that the other renormalisation factors are the same for the linear and the nonlinear theory.

Recall that the linear algebras reduce to the nonlinear ones when eliminating one spin 1/2 field for $N = 3$, and four spin 1/2 fields and one spin 1 field for $N = 4$. In this process, the central charge is modified:

$$\begin{aligned} c^{(3)}_{nonlinear} &= c^{(3)}_{linear} - 1/2 \\ c^{(4)}_{nonlinear} &= c^{(4)}_{linear} - 3\,. \end{aligned} \tag{7.4.38}$$

Furthermore, the $so(4)$ superconformal algebra is a special case of the nonlinear $N = 4$ algebra discussed in subsection 5.3.2. We find that they coincide for $k^{so(4)} = \tilde{k}_+ = \tilde{k}_-$, where $\tilde{k}_\pm$ are the $su(2)$-levels used in subsection 5.3.2. With these substitutions, the agreement is complete, both for the overall renormalisation factor and for the field renormalisations.

For $N = 3$ a similar computation was made [54] directly for the nonlinear supergravity, using Feynman diagrams to compute the determinants. In that case the classical approximation is not linear in $c$, but can be written as a power series. The determinant replacing our $sdet\overline{D_1}$ is not directly proportional to the induced action $\Gamma^{(0)}$ as in eq. (7.4.8). In fact, this part vanishes since $c_{ghost} = 0$ for $N = 3$. Instead, the determinant contains extra terms. These terms are computed in [54]. They cancel the non-leading terms of the classical induced action, at least to the extent they are relevant here (next-to-leading order). A similar cancellation was also observed in the computation of the $\mathcal{W}_3$ effective action [157]. The non-leading contribution and its cancellation with some of the loop contributions seems to have been overlooked in [54]. We have recomputed the renormalisation factors for the nonlinear algebra with the method of [54], taking into account the non-leading terms also. We again find agreement with the results obtained above. Note in particular that all field renormalisation factors are equal. This alternative computation of the determinants, using Feynmann diagrams, implicitely confirms our treatment of equation of motion terms in the Ward identities.

## 7.5 Discussion

From eq. (7.2.16), one deduces that for generic values of $\kappa$, no renormalisation of the coupling constant beyond one loop occurs if and only if either $d_B = d_F$ or $\tilde{h} = 0$ (or both). We get $d_B = d_F$ for $su(m \pm 1|m)$, $osp(m|m)$ and $osp(m+1|m)$ and $\tilde{h} = 0$, for $su(m|m)$, $osp(m+2|m)$ and $D(2,1,\alpha)$. Note that $P(m)$ and $Q(m)$ have not been considered, since we need an an invariant metric on $\bar{g}$. The non-renormalisation of the couplings is reminiscent of the $N = 2$ non-renormalisation theorems [106] and [94] (p.358) for extended supersymmetry. These imply that under suitable circumstances at most one loop corrections to the coupling constants are present (the wave function renormalisation may have higher order contributions). Comparing our list with the tabulation [86] of super $\mathcal{W}$–algebras obtained from a (classical) reduction of superalgebras, we find that many of them, though not all, have $N=2$ supersymmetry.
We first give an example of a theory where no renormalisations occur, although there is no $N=2$ subalgebra. There is an $sl(2)$ embedding in $osp(3|2)$ which gives the $N=1$ super-$\mathcal{W}_2$ algebra of [77], which contains four fields (dimensions $5/2$, $2$, $2$, $3/2$). Although $osp(3|2)$ is in our list, the $N=1$ super-$\mathcal{W}_2$ algebra does not contain an $N=2$ subalgebra. Still, the dimensions seem to fit in an $N=2$ multiplet. However, we checked using a *Mathematica* package for super OPEs in $N=2$ superconformal theory [137] that it is not possible to find an associative algebra in $N=2$ superspace with only a dimension $3/2$ superfield with SOPE closing on itself and a central extension.
As an example of the opposite case (renormalisation but $N=2$), it seems that all superalgebras based on the reduction of the unitary superalgebras $su(m|n)$ contain an $N=2$ subalgebra, however our list contains only the series $|m-n|\leq 1$. Clearly, the structural reason behind the lack of renormalisation beyond one loop remains to be clarified.

# Chapter 8

# Critical $\mathcal{W}$–strings

In this chapter some aspects of critical $\mathcal{W}$–strings are studied. In an introduction we review some basic knowledge about the bosonic string which serves as an example for the generalisation in section 8.2 to strings based on a nonlinear gauge algebra. We discuss the BRST-quantisation of $\mathcal{W}$–strings. In section 8.3 the simplest possible classical $\mathcal{W}$–algebras $\mathcal{W}_{2,s}$ are discussed. These are generated by a Virasoro operator and a bosonic primary field with dimension $s$. We construct realisations for all the classical $\mathcal{W}_{2,s}$ algebras, and obtain the corresponding classical BRST operators. We show by example that a graded structure can be given to these BRST operators by performing canonical field redefinitions involving the ghost and the matter fields. We find that these graded classical BRST operators can be promoted to fully quantum-nilpotent operators by the addition of $\hbar$-dependent terms. For $s \leq 7$, the spectra of these theories have been studied in [143, 147]. Section 8.4 explores the relation of the resulting effective Virasoro string theories to certain $\mathcal{W}$ minimal models. Explicit result are given for $s = 4, 5, 6$. In particular, we show how the highest weight states of the $\mathcal{W}$ minimal models decompose into Virasoro primaries.

The cohomologies of $\mathcal{W}$–string theories, and their connection to minimal models, are indicative of a kind of hierarchical structure of string theories, which was first articulated in the case of supersymmetric extensions of string theories by Berkovits and Vafa [14]. We examine the possibility of fermionic higher-dimension extensions of the hierarchical structure in section 8.5.

The results of section 8.4 are published in [144]. Sections 8.3 and 8.5 are based on [148].

## 8.1 The bosonic string

Before studying critical $\mathcal{W}$–strings, we briefly discuss the bosonic string, concentrating on the BRST approach to string theory. The reader may consult the general references [105, 127] for further details. Our discussion will be restricted to strings with a worldsheet which is a Riemann surface of zero genus.

The Polyakov action [158] for a bosonic string in $D$ dimensions is given by:

$$S[X^\mu, g^{ij}] = -\frac{1}{4\pi}\int dx^2\, \sqrt{g(x)} g^{ij}(x)\partial_i X^\mu(x)\partial_j X^\nu(x)\eta_{\mu\nu}\,, \tag{8.1.1}$$

where $g$ is the absolute value of the determinant of $g_{ij}$ and $\eta_{\mu\nu}$ is the $D$-dimensional Minkowski metric. $x$ are coordinates on the two-dimensional worldsheet and the values of the fields $X^\mu$ are coordinates in a $D$-dimensional flat space. The metric $g_{ij}$ is regarded as a fluctuating field, although it is classically not propagating because there are no derivatives of it in the action. In the case of noncritical strings, considering $g_{ij}$ is essential for a consistent theory. Due to the definition of the energy–momentum tensor $T^{ij}$ (2.2.8), the equations of motion of the worldsheet metric precisely constrain $T^{ij}$ to zero. One can use these equations of motion to show that the action (8.1.1) is classically equivalent to the Nambu-Goto action [152, 104], which is proportional to the surface area of the worldsheet of the string.

The action (8.1.1) clearly has general (worldsheet) coordinate invariance. It is also invariant under local Weyl rescalings of the metric if $X^\mu$ is assigned a zero scaling dimension and spin. In the conformal gauge, these gauge invariances are fixed by[1]:

$$g^{ij} = \exp(\phi)\delta^{ij}\,, \tag{8.1.2}$$

with $\phi$ an arbitrary field. The gauge fixing procedure is conveniently carried out in the BRST formalism. We introduce two ghost pairs $(b, c)$ and $(\bar{b}, \bar{c})$, one for each gauge fixing condition in eq. (8.1.2). The total action becomes (in the complex basis on the worldsheet):

$$S_{gf}[X^\mu, \phi, b, c, \bar{b}, \bar{c}] = S[X^\mu, \exp(\phi)\delta^{ij}] + \frac{1}{\pi}\int c\bar{\partial}b + \bar{c}\partial\bar{b}\,. \tag{8.1.3}$$

[1] After a Wick rotation to the euclidean plane.

The field $\phi$ formally drops out of the action. However, on the quantum level this is only true when the gauge symmetries, and in particular the Weyl symmetry, survives at the quantum level. Independence of the quantum theory on the metric implies that the total energy–momentum tensor $T^{ij}_{\rm tot}$ has a zero expectation value. The holomorphic component $T^{ij}_{\rm tot}$ is given by:

$$T_{\rm tot} = T_{\rm mat} + T_{bc}\,, \tag{8.1.4}$$

where $T_{\rm mat} = \frac{1}{2}\partial X^\mu \partial X_\mu$ is the energy–momentum tensor of the $X^\mu$ and:

$$T_{bc} = c\partial b + 2(\partial c)b\,, \tag{8.1.5}$$

which gives $b$ conformal dimension 2 and $c$ dimension $-1$. $T_{\rm tot}$ has a central charge $c_{\rm mat} - 26$, see section 2.6. Exactly the same is true for the antiholomorphic part $\bar{T}_{\rm tot}$, and we will drop the antiholomorphic symmetry generators in the rest of this chapter. As each $X^\mu$ contributes 1 to the central charge, we see that it is only in 26 dimensions that the metric decouples from the theory. When this condition is satisfied, the string theory is said to be "critical", otherwise the theory suffers from an anomaly, which can be canceled by introducing an action for the Liouville mode $\phi$ [158]. We will not study non-critical string theory here.

The gauge fixed action (8.1.3) is a sum of free field actions, and it is supplemented by a BRST operator of which we give only the holomorphic part:

$$Q = \oint c(T_{\rm mat} + \frac{1}{2}T_{bc})\,, \tag{8.1.6}$$

$Q$ is only nilpotent when the total central charge $c_{\rm mat} - 26$ vanishes. Requiring BRST invariance of the physical states implements the classical constraint $T = 0$ which arose from the equations of motion of the metric. One then identifies physical states with the elements of the cohomology of $Q$ in the complex generated by $\{\partial X^\mu, \exp(k_\mu X^\mu), b, c\}$, where the vertex operators were defined in subsection 2.6.1. As an example, using the OPEs of section 2.6, we can compute:

$$Q(b) = T_{\rm tot}\,. \tag{8.1.7}$$

This means that $T_{\rm tot}$ is BRST-trivial. Because the operators in an OPE between two BRST-trivial operators are trivial themselves, we again find that $T_{\rm tot}$ should have no central extension.

The cohomology of the bosonic string can be computed as follows. Let $\mathcal{X}$ be an operator depending only on $X^\mu$, and $\mathcal{G}$ on $(b, c)$. The action of Q (8.1.6) on the normal ordered product of $\mathcal{X}$ and $\mathcal{G}$ is given by:

$$Q(\mathcal{X}\mathcal{G}) = \sum_{n\geq 0} [T\mathcal{X}]_{n+1}\partial^n c\ \mathcal{G} + \mathcal{X}\ Q(\mathcal{G})\,. \tag{8.1.8}$$

Now, one has:

$$Q(c) = -\partial c c \quad \text{and} \quad Q(\partial c c) = 0\,. \tag{8.1.9}$$

Combining these three equations, we see that the field:

$$\mathcal{X}c \tag{8.1.10}$$

is physical when $\mathcal{X}$ is a primary field of dimension 1. Also, for a primary $\mathcal{X}$ of dimension $h$,

$$\mathcal{X}\ \partial c c \tag{8.1.11}$$

is BRST invariant. However, unless $h = 1$, it is a trivial field as it is proportional to $Q(\mathcal{X}c)$. Hence, for a primary field of dimension 1, $\mathcal{X}\partial c c$ is also a physical state. One can prove that all elements of the cohomology can be written in one of these forms, (8.1.10) or (8.1.11)[2] [88, 84, 90]. Of course, to study the spectrum, we should construct the fields of dimension 1 generated by $\{\partial X^\mu, \exp(k_\mu X^\mu)\}$. This problem was solved in [53] by introducing the spectrum generating DDF-operators. We only mention the tachyonic state $\exp(k_\mu X^\mu)$, which has mass squared $-k^2 = -2$.

A string scattering amplitude on the sphere of physical fields $\Phi_i$ is given in a pathintegral formalism by:

$$\int [dX][db][dc]\ \exp\left(-S_{gf}\right) b\,\partial b\,\partial^2 b\ \Phi_1(x_1)\Phi_2(x_2)\ldots\,, \tag{8.1.12}$$

where we only wrote the holomorphic antighosts explicitly. The insertion of the antighosts corresponds to the zero-modes of the gauge fixing determinant. This insertion restricts non-zero correlation functions to fields $\Phi_i$ where the total ghost number adds up to 3. By considering two-point functions, we see that the expression (8.1.12) vanishes unless $\Phi_1 \sim \mathcal{X}c$ and $\Phi_2 \sim \mathcal{Y}\partial c c$ (for some $X^\mu$ dependent fields $\mathcal{X}, \mathcal{Y}$), or vice versa. This leads us to identify the states in the different sectors of the cohomology:

$$\mathcal{X}c \sim \mathcal{X}\partial c c\,. \tag{8.1.13}$$

For $n$-point functions, an integration over the moduli of the worldsheet with punctures at the $x_i$ is implied in eq. (8.1.12). We do not pursue this topic here, see [105, 127].

## 8.2 $\mathcal{W}$–strings

The bosonic string can be generalised by considering a classical theory with local gauge symmetries, generated by traceless symmetric tensors. The analogues of the

[2] Note that any derivative of a physical state $\phi$ is BRST trivial. Indeed, we have that $Q([b,\phi]_1) = [T_{\rm tot}\phi]_1 = \partial\phi$.

Weyl invariance imply that the generators can be split in holomorphic and anti-holomorphic components, forming each a copy of a classical $\mathcal{W}$–algebra. The local symmetries are then gauge fixed, leading to the introduction of ghost fields. The gauge-fixed action has a nilpotent symmetry generated by the classical BRST operator. To quantise the theory, one must renormalise the symmetry transformation rules and introduce counterterms, order by order in $\sqrt{\hbar}$, such that BRST invariance of the effective action is preserved at the quantum level. The theory is quantisable if one can carry out the procedure in all orders of $\sqrt{\hbar}$. If such procedure is not possible, the theory then suffers from an anomaly.

In a bosonic string theory with 26 scalars, there is no need to add quantum counterterms or to modify the transformation rules. This is because a central charge $c = -26$ from the ghosts is cancelled by the contributions of the matter scalars. If there were $d \neq 26$ scalars in the theory, it would still be anomaly free after adding $\sqrt{\hbar}$ dependent counterterms and modifications of the transformation rules. These counterterms have the interpretation of background charges in the matter energy–momentum tensor, with the criticality condition $c = 26$ achieved by choosing appropriate background charges. Also in this case, the matter energy–momentum tensor forms a quantum Virasoro algebra with $c = 26$. Thus one can construct the quantum BRST operator directly from the quantum Virasoro algebra, eq. (8.1.6).

For the simplest nonlinear algebra, one begins with a theory with classical $\mathcal{W}_3$ symmetry generated by fields $T, W$ of dimension 2 and dimension 3. The classical OPE of the primary current $W$ is given by:

$$W \times W \;=\; \ll 2T^2 \mid \partial(T^2) \gg \;. \tag{8.2.1}$$

Despite the nonlinearity, it is straightforward to obtain the classical BRST operator. One way to realise the classical algebra is in terms of a scalar field $\varphi$[3] and an arbitrary energy–momentum tensor $T_X$ [168]:

$$\begin{aligned} T &= -\tfrac{1}{2}(\partial\varphi)^2 + T_X \\ W &= \frac{i}{\sqrt{2}}\Big(\tfrac{1}{3}(\partial\varphi)^3 + 2\partial\varphi\, T_X\Big)\,. \end{aligned} \tag{8.2.2}$$

Here, $T_X$ can be realised in terms of any system with a traceless symmetric energy–momentum tensor, *e.g.* for the multiscalar realisation of $\mathcal{W}_3$ it is the sum of a number of energy–momentum tensors of free scalars (possibly with background charge). With the realisation eq. (8.2.2), the theory can be quantised by adding counterterms and modifying the transformation rules. The corresponding quantum BRST operator is the same as the one that was constructed by Thierry-Mieg [195] from an abstract quantum $\mathcal{W}_3$ algebra with critical central charge $c = 100$. The quantum corrections of the theory can be interpreted as adding background charges to the classical currents, leading to a quantum realisation of the quantum $\mathcal{W}_3$ algebra at $c = 100$ [163]. Unlike the Virasoro algebra, the quantum modification of the classical $\mathcal{W}_3$ algebra is not merely reflected by introducing a central charge. The (quantum) OPE of the primary current $\mathcal{W}$ is given by the $\mathcal{W}_3$-algebra, which is, with explicit insertions of $\hbar$, given by:

$$\begin{aligned} \hbar^{-1} W(z)\, W(w) \;=\;& \frac{16}{(22+5c)}\Big(\frac{2\Lambda}{(z-w)^2} + \frac{\partial\Lambda}{z-w}\Big) \\ &+\hbar\Big(\frac{2T}{(z-w)^4} + \frac{\partial T}{(z-w)^3} + \frac{\frac{3}{10}\partial^2 T}{(z-w)^2} + \frac{\frac{1}{15}\partial^3 T}{z-w}\Big) + \hbar^2 \frac{c/3}{(z-w)^6}\,, \end{aligned} \tag{8.2.3}$$

with:

$$\Lambda = (TT) - \tfrac{3}{10}\hbar\partial^2 T\,. \tag{8.2.4}$$

Interactions for the critical $\mathcal{W}_3$–string were studied by analogy to the bosonic string in [142, 89]. Its spectrum was finally determined in [146].

The above considerations can be extended to more complicated $\mathcal{W}$–algebras. A discussion of the classical BRST operators for the $\mathcal{W}_N$ algebras, and the structure of the quantum BRST operators, may be found in [174, 17, 18]. Detailed results for the quantum BRST operator for $\mathcal{W}_4$ were obtained in [112, 213]. In general, the quantum $\mathcal{W}_N$ BRST operator can be regarded as the appropriate quantum renormalisation of the classical operator that arises in an anomaly-free quantisation of the theory.

The classical BRST operator is derived from the symmetry algebra and can thus be written in terms of the symmetry generators, irrespective of the model on which the $\mathcal{W}$-symmetry is realised. This is not necessarily true after quantisation. An example of this can be found in the $\mathcal{W}_3$–string. In addition to the standard multi-scalar classical realisations (8.2.2), there are four special classical realisations associated with four Jordan algebras [168]. It has been shown [168, 150, 79] that these realisations cannot be extended to realisations of the quantum $\mathcal{W}_3$ algebra. This does not preclude the possibility to build quantum-consistent $\mathcal{W}_3$–string theories based on these classical realisations of the symmetry. In other words, the possibility exists that one could still find quantum nilpotent BRST operators having the classical BRST operators built from the Jordan realisations as their classical limits. We explicitly checked that for the simplest case based on the real Jordan algebra, making a 5 scalar realisation of the classical $\mathcal{W}_3$, it is not possible to add order $\sqrt{\hbar}$ corrections to the classical $\mathcal{W}_3$ BRST operator, such that the resulting quantum BRST operator is nilpotent [148]. Thus it appears that one cannot consistently quantise $\mathcal{W}_3$–strings based on the classical Jordan realisations of the $\mathcal{W}_3$ algebra. This result was obtained for all four Jordan realisations by Vandoren *et al.* using the Batalin-Vilkovisky quantisation scheme [198].

The observation that the quantisation of a theory with gauge symmetries depends on the particular model, leads us to study string theories where no quantum $\mathcal{W}$–al-gebra corresponding to the classical gauge algebra exists. In [143, 147] quantum BRST operators were constructed for theories with a symmetry algebra formed by $T$ and a dimension $s$ current. Because the field-content of a $\mathcal{W}_{2,s}$ classical $\mathcal{W}$–algebra is the smallest possible, it seems simpler to study $\mathcal{W}_{2,s}$–strings than $\mathcal{W}_N$–strings. While the quantum $\mathcal{W}_N$ $\mathcal{W}$–algebra does exist for arbitrary values of the central charge, deformable quantum algebras with the same field-content as $\mathcal{W}_{2,s}$ exist only for $s = 3, 4, 6$. We will study a particular type of realisation of critical $\mathcal{W}_{2,s}$–strings in this chapter.

[3] Throughout this chapter we use a normalisation $\varphi \times \varphi = \ll -1 \mid 0 \gg$ to conform with the literature.

For more details about $\mathcal{W}$–strings, the reader can consult the reviews [117, 165, 206, 31].

## 8.3 $\mathcal{W}_{2,s}$–strings

In this section, we shall investigate higher-spin string theories based on the classical symmetry algebra generated by $T$ and a bosonic primary field $W$ of dimension $s$, where $s$ is an integer. Such a closed, nonlinear, $\mathcal{W}_{2,s}$ algebra exists classically for all $s \geq 3$. The classical OPEs of the generators $T$ and $W$ are given by[4]:

$$\begin{aligned} T \times T &= \ll 2\,T \mid \partial T \gg \\ T \times W &= \ll s\,W \mid \partial W \gg \\ W \times W &= \ll 2\,T^{s-1} \mid \partial T^{s-1} \gg . \end{aligned} \tag{8.3.1}$$

It is straightforward to verify that this algebra satisfies the Jacobi identity at the classical level.

In the case of a linear algebra $[T_i, T_j] = f_{ij}{}^k T_k$, one knows that the BRST charge will have the form $Q = c^i\,T_i + \frac{1}{2} f_{ij}{}^k\, c^i\, c^j\, b_k$. In our case, we may interpret the nonlinearity on the *rhs* of the OPE $W(z)W(w)$ as $T$–dependent structure constants, leading to the expectation that the BRST current should have the form:

$$J = c\,(T + T_{\beta\gamma} + \tfrac{1}{2} T_{bc}) + \gamma\,W - \partial\gamma\,\gamma\,b\,T^{s-2}\,, \tag{8.3.2}$$

where the $(b, c)$ are the antighost and ghost for $T$, and $(\beta, \gamma)$ are the antighost and ghost for $W$. They are anticommuting, and have dimensions $(2, -1)$ and $(s, 1-s)$ respectively. The ghost Virasoro operators are given by eq. (8.1.5) and:

$$T_{\beta\gamma} = -s\,\beta\,\partial\gamma - (s-1)\,\partial\beta\,\gamma\,. \tag{8.3.3}$$

[4] For even $s$, a generalisation seems possible by adding $2\alpha T^{s/2-1}W$ to the second order pole in the OPE $W(z)W(w)$ and $\alpha\partial(T^{s/2-1}W)$ to the first order pole, with $\alpha$ some arbitrary constant. In this form the algebra was called $\mathcal{W}_{s/s-2}$ in [116]. However, one can always choose generators $T, \tilde{W} = W - \alpha/s^2 T^{s/2}$ such that $\alpha$ is zero for $\tilde{W}(z)\tilde{W}(w)$.

Performing the classical OPE, we find that (8.3.2) is indeed nilpotent (the coefficient $-1$ in the last term in (8.3.2) is determined by the nilpotency requirement).

In order to construct a string theory based on the classical $\mathcal{W}_{2,s}$ symmetry, we need an explicit realisation for the matter currents. Such a realisation may be obtained in terms of a scalar field $\varphi$ and an arbitrary energy–momentum tensor $T_X$, which may itself be realised, for example, in terms of scalar fields $X^\mu$:

$$\begin{aligned} T &= -\tfrac{1}{2}(\partial\varphi)^2 + T_X \\ W &= \sum_{n=0}^{N} g_n(s)\,(\partial\varphi)^{s-2n}\,T_X^n\,, \end{aligned} \tag{8.3.4}$$

where $N = [s/2]$. The constants $g_n(s)$ are determined by demanding that $W$ satisfies (8.3.1), and we find that they are given by:

$$g_n(s) = s^{-1}(-2)^{-s/2} 2^{n+1} \binom{s}{2n}\,. \tag{8.3.5}$$

Actually, as we shall discuss later, when $s$ is even there is also a second solution for the constants $g_n(s)$, which is associated with a "trivial" string theory.

In order to discuss the quantisation of the classical $\mathcal{W}_{2,s}$–string theories, the traditional procedure would be to undertake an order-by-order computation of the quantum effective action, introducing counterterms and corrections to the transformation rules in each order in the loop-counting parameter $\sqrt{\hbar}$, such that BRST invariance is preserved. Such a procedure is cumbersome and error prone, but fortunately a more straightforward method is available to us here. We can simply parametrise all the possible quantum corrections to the BRST operator, and solve for the coefficients of these terms by demanding nilpotence at the quantum level using *OPEdefs*. Before carrying out this procedure, we shall first discuss a simplification of the structure of the BRST operator that can be achieved by performing a canonical redefinition involving the ghost and the matter fields.

We conjecture that the BRST operator in (8.3.2) can be transformed by canonical field redefinition into the following graded form:

$$Q = Q_0 + Q_1 \tag{8.3.6}$$

$$Q_0 = \oint c(T + T_{\beta\gamma} + \tfrac{1}{2} T_{bc}) \tag{8.3.7}$$

$$Q_1 = \oint \gamma\Big((\partial\varphi)^s + \tfrac{s^2}{2}(\partial\varphi)^{s-2}\beta\,\partial\gamma\Big)\,. \tag{8.3.8}$$

Here $Q_0$ has grade $(1, 0)$ and $Q_1$ has grade $(0, 1)$, with $(p, q)$ denoting the grading of an operator with ghost number $p$ for the dimension 2 ghost system and ghost number $q$ for the dimension $s$ ghost system. We have ${Q_0}^2 = {Q_1}^2 = \{Q_0, Q_1\} = 0$.

This conjecture is based on the following examples. For the case of $\mathcal{W}_{2,3}$, the field redefinition which acomplishes this was first described in [142]. At the classical level, the redefinition is given by:

$$\begin{aligned} c &\longrightarrow c - b\,\partial\gamma\,\gamma + \sqrt{2}i\,\partial\varphi\,\gamma \\ b &\longrightarrow b \\ \gamma &\longrightarrow \gamma \\ \beta &\longrightarrow \beta - \partial b\,b\,\gamma - \sqrt{2}i\,\partial\varphi\,b \\ \varphi &\longrightarrow \varphi + \sqrt{2}i\,b\,\gamma \\ T_X &\longrightarrow T_X\,. \end{aligned} \qquad (8.3.9)$$

In the case of $s=4$, we explicitly constructed the field redefinitions that turn the BRST operator in (8.3.2) into the form (8.3.6),(8.3.7),(8.3.8):

$$\begin{aligned} c &\longrightarrow c - 2\beta\partial\gamma\,\gamma - \tfrac{7}{4}(\partial\varphi)^2\,\gamma + \tfrac{21}{8}(\partial\varphi)^2 b\partial\gamma\,\gamma - \tfrac{1}{2}T_X\gamma - \tfrac{5}{4}T_X b\partial\gamma\,\gamma \\ b &\longrightarrow b \\ \gamma &\longrightarrow \gamma + 2b\,\partial\gamma\,\gamma \\ \beta &\longrightarrow \beta + 4b\,\beta\,\partial\gamma + 2b\,\partial\beta\,\gamma + \tfrac{7}{4}(\partial\varphi)^2\,b + \tfrac{49}{8}(\partial\varphi)^2\,\partial b\,b\,\gamma \\ &\quad + \tfrac{1}{2}T_X\,b - \tfrac{1}{4}T_X\,\partial b\,b\,\gamma + 4\partial b\,b\,\beta\,\partial\gamma\,\gamma + 2\partial b\,\beta\,\gamma \\ \varphi &\longrightarrow \varphi - \tfrac{7}{2}\partial\varphi\,b\,\gamma \\ T_X &\longrightarrow T_X + T_X\,b\,\partial\gamma + T_X\,\partial b\,\gamma + \tfrac{1}{2}T_X\,\partial b\,b\,\partial\gamma\,\gamma + \tfrac{1}{2}\partial T_X\,b\,\gamma\,. \end{aligned} \qquad (8.3.10)$$

The field redefinition becomes more complicated with increasing $s$. Presumably, this conjecture can be proven along the lines of [17, 18].

It is worth mentioning that for $s=2k$ there exists another solution for the realisation of $W$ given in (8.3.4) in which $W$ can be written as $\frac{1}{k}T^k$. In this case, there exists a canonical field redefinition under which the BRST operator in (8.3.2) becomes simply $Q=Q_0$. It is not surprising that the BRST operator with this realisation describes the ordinary bosonic string since in this case the constraint $W=0$ is implied by the constraint $T=0$. We shall not consider this case further.

To quantise the classical $\mathcal{W}_{2,s}$–string and obtain the quantum BRST operator, we add $\sqrt{\hbar}$-dependent counterterms to the classical BRST. In order to do this in a systematic way, it is useful to identify the $\hbar$ dimensions of the quantum fields. An assignment that is consistent with the OPEs is:

$$\{T_X, \partial\varphi, b, c, \beta, \gamma\} \sim \{\hbar, \sqrt{\hbar}, \hbar, 1, \hbar^{s/2}, \hbar^{1-s/2}\}\,. \qquad (8.3.11)$$

We shall make the assumption that the graded structure of the classical BRST operator is preserved at the quantum level. For $\mathcal{W}_{2,3}$, this has been explicitly found to be true [142]. For $s \geq 4$, there certainly exist quantum BRST operators with the graded structure, as we shall discuss below. Whether there could exist further quantum BRST operators that do not posses the grading is an open question.

For the scalar field $\varphi$, the quantum corrections that can be added to $Q_0$ simply take the form of a background-charge term proportional to a constant $\alpha$. Its energy–momentum tensor becomes:

$$T_\varphi \;\equiv\; -\tfrac{1}{2}(\partial\varphi)^2 - \alpha\,\partial^2\varphi\,. \qquad (8.3.12)$$

Similar modifications to $T_X$ can occur[5]. The equation ${Q_0}^2=0$ requires that the total central charge vanishes:

$$0 = -26 - 2(6s^2 - 6s + 1) + 1 + 12\alpha^2 + c_X\,, \qquad (8.3.13)$$

with $c_X$ the central charge of $T_X$. In $Q_1$, the possible quantum corrections amount to:

$$Q_1 = \oint dz\,\gamma\,F(\varphi,\beta,\gamma)\,, \qquad (8.3.14)$$

where $F(\varphi,\beta,\gamma)$ is a dimension $s$ operator with ghost number zero such that its leading-order (*i.e.* classical) terms are given in (8.3.8). The precise form of the operator $F(\varphi,\beta,\gamma)$ is determined by the nilpotency conditions $\{Q_0,Q_1\} = {Q_1}^2 = 0$. The quantum BRST operators for $\mathcal{W}_{2,s}$ theories with $s=4$, 5 and 6 were constructed in [143], and the results were extended to $s=7$ in [147] and $s=8$ in [148]. The conclusion of these various investigations is that there exists at least one quantum BRST operator for each value of $s$. If $s$ is odd, then there is exactly one BRST operator of the type discussed. If $s$ is even, then there are two or more inequivalent quantum BRST operators. One of these is a natural generalisation to even $s$ of the unique odd-$s$ sequence of BRST operators, see also section 8.4.

As we discussed earlier, the case $s=3$ corresponds to the $\mathcal{W}_3$=$\mathcal{W}_{DS}A_2$ algebra, which exists as a closed quantum algebra for all values of the central charge, including, in particular, the critical value $c=100$. For $s=4$, it was shown in [147] that the two $\mathcal{W}_{2,4}$ quantum BRST operators correspond to BRST operators for the $\mathcal{W}_{DS}B_2$ algebra, which again exists at the quantum level for all values of the central charge. The reason why there are two inequivalent BRST operators in this case is that $B_2$ is not simply-laced and so there are two inequivalent choices for the background charges that give rise to the same critical value $c=172$ for the central charge [147]. Two of the four $\mathcal{W}_{2,6}$ BRST operators can similarly be understood as corresponding to the existence of a closed quantum $\mathcal{W}_{DS}G_2$ algebra for all values of the central charge, including in particular the critical value $c=388$ [147]. However, the remaining quantum $\mathcal{W}_{2,s}$ BRST operators cannot be associated with any closed deformable quantum $\mathcal{W}_{2,s}$ algebras. For example, the quantum $\mathcal{W}_{2,5}$ algebra [29] only satisfies the Jacobi identities (up to null fields) for a discrete set of central-charge values, namely $c=\{-7, \frac{6}{7}, -\frac{350}{11}, 134 \pm 60\sqrt{5}\}$. Since none of these central charges includes the value $c = 268$ needed for criticality, we see that although the quantum $\mathcal{W}_{2,5}$ BRST operator can certainly be viewed as properly describing the quantised $\mathcal{W}_{2,5}$–string, it is not the case that there is a quantum $\mathcal{W}_{2,5}$ symmetry in the $\mathcal{W}_{2,5}$–string. This is an explicit example of the fact that a classical theory can be successfully quantised, without anomalies, even when a quantum version of the symmetry algebra does not exist. It appears that the existence of closed quantum $\mathcal{W}$–algebras is inessential for the existence of consistent $\mathcal{W}$–string theories.

[5] For ease of notation, we still write $T_X$ for the quantum energy–momentum tensor.

As usual, physical fields $\Phi$ are determined by the requirement that they be annihilated by the BRST operator, and that they be BRST non-trivial. In other words, $Q\Phi = 0$ and $\Phi \neq Q\Psi$ for any $\Psi$. There are two different sectors [143]. The "discrete" physical fields, with zero momentum in the $X^\mu$, will not be considered here [143]. The other sector consists of fields with continuous $X^\mu$ momentum. Both sectors have only physical fields for particular values of the $\varphi$-momentum. In this sense, $\varphi$ is considered a "frozen" coordinate, and the $X^\mu$ form coordinates in the effective spacetime.

It was conjectured in [142] and [143] that all continuous-momentum physical states for multi-scalar $\mathcal{W}_{2,s}$ string theories can be described by physical operators of the form:

$$\Phi_\Delta = c\,U(\varphi,\beta,\gamma)\,V_\Delta(X)\,, \tag{8.3.15}$$

where $\Delta$ denotes the conformal dimension of the operator $V_\Delta(X)$ which creates an effective spacetime physical state which is a highest weight field with respect to $T_X$. For simplicity, one can always take the effective-spacetime operator $V_\Delta(X)$ to be tachyonic, since the discussion of physical states with excitations in the effective spacetime proceeds identically to that of bosonic string theory. The interesting new features of the $\mathcal{W}$–string theories are associated with excitations in the $(\varphi,\beta,\gamma)$ fields. Thus we are primarily concerned with solving for the operators $U(\varphi,\beta,\gamma)$ that are highest weight under $T_\varphi + T_{\gamma,\beta}$ with conformal weights $h = 1-\Delta$, and that in addition satisfy $Q_1(U) = 0$. Solving these conditions for $U(\varphi,\beta,\gamma)$, with $V_\Delta(X)$ being highest weight under $T_X$ with conformal weight $\Delta = 1-h$, is equivalent to solving the physical-state conditions for $\Phi_\Delta$ in eq. (8.3.15).

## 8.4 Minimal models and $\mathcal{W}_{2,s}$–strings

It has been known for some time that there is a close connection between the spectra of physical states in $\mathcal{W}$–string theories, and certain Virasoro or $\mathcal{W}$ minimal models. This connection first came to light in the case of the $\mathcal{W}_3$–string [37, 140, 141, 89], where it was found that the physical states in a multi-scalar realisation can be viewed as the states of Virasoro-type bosonic strings with central charge $c_X = 25\frac{1}{2}$ and intercepts $\Delta = \{1, \frac{15}{16}, \frac{1}{2}\}$. These quantities are dual to the central charge $c_{mm} = \frac{1}{2}$ and weights $h = \{0, \frac{1}{16}, \frac{1}{2}\}$ for the $(p,q) = (3,4)$ Virasoro minimal model, the Ising model, in the sense that $26 = c_X + c_{mm}$, and $1 = \Delta + h$. In fact, the physical operators of the multi-scalar $\mathcal{W}$–string have the form eq. (8.3.15) with $V_\Delta(X)$ a dimension $\Delta$ field. Further support for this connection was found in [142, 89] by considering the scattering of physical states. Using certain identifications similar to eq. (8.1.13), it was found that the S-matrix elements of the lowest $\mathcal{W}_3$–string states obey selection rules also found in the Ising model.

If one were to look at the multi-scalar $\mathcal{W}_N$–string, one would expect that analogously the physical states would be of the form of effective Virasoro string states for a $c_X = 26 - \left(1 - \frac{6}{N(N+1)}\right)$ theory, tensored with operators $U(\vec{\varphi},\vec{\beta},\vec{\gamma})$ that are primaries of the $c_{mm} = 1 - \frac{6}{N(N+1)}$ Virasoro minimal model, *i.e.* the $(p,q) = (N, N+1)$ unitary model. Here, $\vec{\varphi}$ denotes the set of $(N-2)$ special scalars which, together with the $X^\mu$ appearing in $T_X$, provide the multi-scalar realisation of the $\mathcal{W}_N$ algebra. Similarly, $\vec{\beta}$ and $\vec{\gamma}$ denote the sets of $(N-2)$ antighosts and ghosts for the dimension $3, 4, 5, \ldots, N$ currents. The identification with a particular minimal model is in these cases based solely on the set of conformal dimensions that occur for the $U$ field in eq. (8.3.15). The rapid growth of the complexity of the $\mathcal{W}_N$ algebras with increasing $N$ means that only incomplete results are available for $N \geq 4$, but partial results and general arguments have provided supporting evidence for the above connection.

A simpler case to consider is a $\mathcal{W}_{2,s}$–string, corresponding to the quantisation of the classical theories described in the previous section. As already mentioned, there is for any $s$ a "regular" BRST operator, which has the feature that the associated minimal model, with energy–momentum tensor $T_{mm} = T_\varphi + T_{\beta\gamma}$, has central charge:

$$c_{mm} = \frac{2(s-2)}{(s+1)}\,. \tag{8.4.1}$$

This is the central charge of the lowest unitary $\mathcal{W}_{s-1}$ minimal model. We will study the case $s = 4$, 5 and 6 in further detail in the subsections which follow. We will make use of the results of [143] for the quantum BRST operators and the lowest physical states. Our calculations clarify the connection with the minimal models [144].

When $s$ is even, there are further "exceptional" BRST operators in addition to the regular one described above. When $s = 4$, there is one exceptional case, for $s = 6$ three [143] and for $s = 8$ four [148]. The spectra of these theories are studied in [147, 148]. They share the feature that a negative weight for $U(\varphi,\beta,\gamma)$ occurs. This implies correspondingly an intercept value $\Delta > 1$ for the effective spacetime Virasoro string, and hence the existence of some negative-norm physical states. For some of the exceptional theories the dimensions of the physical states point towards a correspondence with Virasoro or $\mathcal{W}_{DS}\ B_n$ minimal models, while for others no connection with minimal models has been found yet. We will not study the theories related to the exceptional BRST operators here.

We now outline the procedure which will be followed in the study of $\mathcal{W}_{2,s}$ for $s = 4, 5, 6$. The fields in a minimal model of a $\mathcal{W}$–algebra $\mathcal{W}$ are given by the $\mathcal{W}$-descendants of some highest weight fields, among which is the unit operator of the OPA. This implies that the generators of $\mathcal{W}$ are contained in the set of fields, as they are descendants of $\mathbb{1}$. Hence, if the physical states (8.3.15) are connected with the $\mathcal{W}$-minimal model, there should be physical states such that the $\varphi, \beta, \gamma$ dependent parts $U^i$ form a realisation of the generators of $\mathcal{W}$. In this respect it is important to note that we are looking for a realisation of the $\mathcal{W}$–algebra in the BRST cohomology, *i.e.* up to BRST exact terms.
The $U^i$ which generate the $\mathcal{W}$–algebra should have ghost number zero. This is because $\mathcal{W}$ has always non-zero central charge (8.4.1) and the generators generally satisfy $[U^i U^i]_{2h_i} \sim \mathbb{1}$. Furthermore, they should depend on $\varphi$ in such a way that they have well-defined OPEs, *i.e.* the OPE of $\varphi$ with any other physical state should be meromorphic. This points to zero $\varphi$-momentum states. This claim is further supported by an analysis of the spectrum which shows that $\varphi, \beta, \gamma$ dependent parts $U$ of the $\mathcal{W}$-highest weight fields and their descendants have the same $\varphi$-momentum. To summarise, we should look for operators $U^i(\varphi, \beta, \gamma)$ annihilated by $Q_1$. They have ghost number and $\varphi$-momentum zero. For the energy–momentum tensor of $\mathcal{W}$ the obvious candidate for $U$ is:

$$T_{mm} = T_\varphi + T_{\beta\gamma}\,, \tag{8.4.2}$$

where $T_\varphi$ and $T_{\beta\gamma}$ are given in (8.3.12) and (8.3.3).

### 8.4.1 The $\mathcal{W}_{2,4}$–string

Let us consider first the $\mathcal{W}_{2,4}$–string. The BRST operator is then given by (8.3.6), (8.3.7), (8.3.12), (8.3.14), with $\alpha^2 = \frac{243}{20}$ and the operator $F(\varphi, \beta, \gamma)$ given by:

$$\begin{aligned} F(\varphi,\beta,\gamma) = {} & (\partial\varphi)^4 + 4\alpha\,\partial^2\varphi\,(\partial\varphi)^2 + \tfrac{41}{5}(\partial^2\varphi)^2 + \tfrac{124}{15}\partial^3\varphi\,\partial\varphi + \tfrac{46}{135}\alpha\,\partial^4\varphi \\ & +8(\partial\varphi)^2\,\beta\partial\gamma - \tfrac{16}{9}\alpha\,\partial^2\varphi\,\beta\partial\gamma - \tfrac{32}{9}\alpha\,\partial\varphi\,\beta\partial^2\gamma - \tfrac{4}{5}\beta\,\partial^3\gamma \\ & +\tfrac{16}{3}\partial^2\beta\,\partial\gamma\,. \end{aligned} \tag{8.4.3}$$

In [143], all physical states up to and including level[6] $\ell = 9$ in $(\varphi, \beta, \gamma)$ excitations were studied for the $\mathcal{W}_{2,4}$–string. It was found that all the continous-momentum physical states fall into a set of different sectors, characterised by the value $\Delta$ of the effective spacetime intercept, eq. (8.3.15). Specifically, for the $\mathcal{W}_{2,4}$ string, $\Delta$ can take values in the set $\Delta = \{1, \frac{14}{15}, \frac{3}{5}, \frac{1}{3}, -\frac{2}{5}, -2\}$. As one goes to higher and higher levels $\ell$, one just encounters repetitions of these same intercept values, with more and more complicated operators $U(\varphi, \beta, \gamma)$. These operators correspondingly have conformal weights $h$ that are conjugate to $\Delta$, *i.e.* $h = 1 - \Delta = \{0, \frac{1}{15}, \frac{2}{5}, \frac{2}{3}, \frac{7}{5}, 3\}$. For convenience, table 8.1 reproduces the results up to level 9, giving the $(\beta, \gamma)$ ghost number $g$ of the operators $U(\varphi, \beta, \gamma)$, their conformal weights $h$, and their $\varphi$ momenta $\mu$.

[6] The level $\ell$ of a state is defined with respect to the standard ghost vacuum $c_1\gamma_1 \cdots \gamma_{s-1}\big|0\big\rangle$.

| | $g$ | | $h$ | | | $\mu$ (In units of $\alpha/27$) | |
|---|---|---|---|---|---|---|---|
| $\ell = 0$ | 3 | | $\frac{1}{15}$ | 0 | | $-26$ | $-24$ |
| $\ell = 0$ | 3 | | $\frac{1}{15}$ | 0 | | $-28$ | $-30$ |
| $\ell = 1$ | 2 | $\frac{2}{3}$ | $\frac{2}{5}$ | $\frac{1}{15}$ | $-20$ | $-18$ | $-16$ |
| $\ell = 2$ | 2 | | $\frac{7}{5}$ | $\frac{2}{3}$ | | $-18$ | $-14$ |
| $\ell = 3$ | 1 | | $\frac{2}{3}$ | $\frac{1}{15}$ | | $-10$ | $-8$ |
| $\ell = 4$ | 1 | | | $\frac{2}{5}$ | | | $-6$ |
| $\ell = 5$ | 1 | | $\frac{7}{5}$ | $\frac{2}{3}$ | | $-6$ | $-4$ |
| $\ell = 6$ | 0 | | | 0 | | | 0 |
| $\ell = 7$ | 0 | | | $\frac{1}{15}$ | | | 2 |
| $\ell = 8$ | 0 | | | $\frac{1}{15}$ | | | 4 |
| $\ell = 9$ | 0 | | 3 | 0 | | 0 | 6 |

Table 8.1: $U(\varphi, \beta, \gamma)$ operators for the $\mathcal{W}_{2,4}$ string of level $\ell$, ghostnumber $g$, dimension $h$ and $\varphi$-momentum $\mu$.

The explicit expressions for the operators $U(\varphi, \beta, \gamma)$ can be quite complicated, and we shall not give them all here. Some simple examples are as follows. We find $U = \partial^2\gamma\,\partial\gamma\,\gamma\,e^{\mu\varphi}$ at level $\ell = 0$; $U = \partial\gamma\,\gamma\,e^{\mu\varphi}$ at $\ell = 1$; $U = \Big(10\,\partial\varphi\,\partial\gamma\,\gamma - (\mu + 2\alpha)\partial^2\gamma\,\gamma\Big)e^{\mu\varphi}$ at $\ell = 2$; and $U = \mathbb{1}$ at $\ell = 6$. The values of the momentum $\mu$ are given in table 8.1.

We wish to identify the dimension 3 generator $W_{mm}$ of the associated $\mathcal{W}_3$ algebra, realised on the $(\varphi, \beta, \gamma)$ system. We observe from the results in [143] that at level $\ell = 9$ there is an operator $U(\varphi, \beta, \gamma)$ with conformal weight 3, ghost number $g = 0$, and momentum $\mu = 0$. Clearly this is the required primary dimension 3 operator. Its detailed form is:

$$\begin{aligned} W_{mm} \;=\; & \sqrt{\tfrac{2}{13}}\Big(\tfrac{5}{3}\,(\partial\varphi)^3 + 5\alpha\,\partial^2\varphi\,\partial\varphi + \tfrac{25}{4}\,\partial^3\varphi + 20\,\partial\varphi\,\beta\,\partial\gamma \\ & +12\,\partial\varphi\,\partial\beta\,\gamma + 12\,\partial^2\varphi\,\beta\,\gamma + 5\alpha\,\partial\beta\,\partial\gamma + 3\alpha\,\partial^2\beta\,\gamma\Big)\,, \end{aligned} \tag{8.4.4}$$

where we have given it the canonical normalisation in which:

$$W_{mm}(z)W_{mm}(w) \sim \frac{c_{mm}/3}{(z-w)^6} + \text{more}\,, \tag{8.4.5}$$

with the central charge $c_{mm} = \frac{4}{5}$. It is now a straightforward matter to compute the OPEs of the $T_{mm}$ and $W_{mm}$ currents with *OPEdefs* and verify that they do indeed generate the $\mathcal{W}_3$ algebra at $c_{mm} = \frac{4}{5}$. The only noteworthy point in the verification is that at the second-order pole in the OPE of $W_{mm}$ with $W_{mm}$ there is an additional dimension 4 primary, but BRST-trivial current, $\{Q_1, \beta\}$.

Having found the currents that generate the $\mathcal{W}_3$ algebra, we are now in a position to see how they act on the operators $U(\varphi,\beta,\gamma)$ occuring in the physical states of the $\mathcal{W}_{2,4}$ string. Of course, we already know that the operators $U(\varphi,\beta,\gamma)$ are primary fields under $T_{mm}$. Acting with $W_{mm}$, we find that when $h$ takes values in the set $\{0, \frac{1}{15}, \frac{2}{5}, \frac{2}{3}\}$, the corresponding operators are highest weight under $W_{mm}$, *i.e.* $(\widehat{W_{mm}})_n U = 0$ for $n > 0$, and $(\widehat{W_{mm}})_n U = w\,U$. We find that the weights are as follows:

$$\begin{aligned} T_{mm}: &\quad \{0, \tfrac{1}{15}, \tfrac{2}{5}, \tfrac{2}{3}\} \\ \tfrac{243}{\alpha}\sqrt{\tfrac{13}{8}}\,W_{mm}: &\quad \{0, \pm 1, 0, \pm 26\}\,. \end{aligned} \tag{8.4.6}$$

**Intermezzo 8.4.1**

To be precise, we find that for $h = \frac{1}{15}$ the $W_{mm}$ weight is positive for those operators $U(\varphi,\beta,\gamma)$ that have $\frac{27}{\alpha}\mu = 4 \bmod 6$, and negative when $(\frac{\alpha}{27})^{-1}\mu = 2 \bmod 6$. Similarly, for operators with $h = \frac{2}{3}$, we find the $W_{mm}$ weight is positive when $(\frac{\alpha}{27})^{-1}\mu = 2 \bmod 6$, and negative when $(\frac{\alpha}{27})^{-1}\mu = 4 \bmod 6$. These results accord with the observation in [143] that there are two independent towers of $h = \frac{1}{15}$ operators, and two independent towers of $h = \frac{2}{3}$ operators, with the screening operator $\beta\, \exp(\frac{2}{9}\alpha\varphi)$ generating each tower from its lowest-level member.

---

Comparing with the results in [66], we see that these $T_{mm}$ and $W_{mm}$ weights are precisely those for the lowest $\mathcal{W}_3$ minimal model, with $c_{mm} = \frac{4}{5}$. The remaining operators $U(\varphi,\beta,\gamma)$ in the physical states of the $\mathcal{W}_{2,4}$ string have $T_{mm}$ weights $h = \frac{7}{5}$ and 3. We find that these are not highest weight under the $W_{mm}$ current. In fact, they are $\mathcal{W}_3$ descendant fields; those with $h = \frac{7}{5}$ can be written as $(\widehat{W_{mm}})_{-1} + \cdots$ acting on operators $U(\varphi,\beta,\gamma)$ with $h = \frac{2}{5}$, and those with $h = 3$ can be written as $(\widehat{W_{mm}})_{-3} + \cdots$ acting on operators $U(\varphi,\beta,\gamma)$ with $h = 0$.

The conclusion of the above discussion is that the $U(\varphi,\beta,\gamma)$ operators appearing in the physical states of the $\mathcal{W}_{2,4}$ string are precisely those associated with the $c_{mm} = \frac{4}{5}$ lowest $\mathcal{W}_3$ minimal model. Those with $h = \{0, \frac{1}{15}, \frac{2}{5}, \frac{2}{3}\}$ are $\mathcal{W}_3$ highest weight fields, whilst those with $h = \frac{7}{5}$ and 3 are $\mathcal{W}_3$ descendants. Viewed as purely Virasoro fields, they are all primaries. In fact, what we are seeing is an explicit example of the phenomenon under which the set of highest weight fields of a $\mathcal{W}$ minimal model decomposes into a larger set of highest weight fields with respect to the Virasoro subalgebra. In this example, since $c_{mm}$ is less than 1, the $W_{mm}$ highest weight fields decompose into a finite number of Virasoro primaries (namely a subset of the primaries of the $c_{mm} = \frac{4}{5}$ 3-state Potts model). In a more generic example, where the $\mathcal{W}$ minimal model has $c_{mm} \geq 1$, the finite number of $\mathcal{W}$ highest weight fields will decompose into an infinite number of Virasoro primaries, with infinitely many of them arising as $W_{mm}$ descendants. We shall encounter explicit examples of this when we study the $\mathcal{W}_{2,s}$ strings with $s = 5$ and $s = 6$.

### 8.4.2 The $\mathcal{W}_{2,5}$ string

Let us now turn to the example of the $\mathcal{W}_{2,5}$ string. The operator $F(\varphi,\beta,\gamma)$ appearing in eq. (8.3.14) is given by [143]:

$$\begin{aligned} F(\beta,\gamma,\varphi) = {} & (\partial\varphi)^5 + 5\alpha\,\partial^2\varphi\,(\partial\varphi)^3 + \tfrac{305}{8}(\partial^2\varphi)^2\,\partial\varphi + \tfrac{115}{6}\partial^3\varphi\,(\partial\varphi)^2 \\ & + \tfrac{10}{3}\alpha\,\partial^3\varphi\,\partial^2\varphi + \tfrac{55}{48}\alpha\,\partial^4\varphi\,\partial\varphi + \tfrac{251}{576}\partial^5\varphi + \tfrac{25}{2}(\partial\varphi)^3\,\beta\,\partial\gamma + \tfrac{25}{4}\alpha\,\partial^2\varphi\,\partial\varphi\,\beta\,\partial\gamma \\ & + \tfrac{125}{16}\partial^3\varphi\,\beta\,\partial\gamma + \tfrac{325}{12}\partial^2\varphi\,\partial\beta\,\partial\gamma + \tfrac{375}{16}\partial\varphi\,\partial^2\beta\,\partial\gamma - \tfrac{175}{48}\partial\varphi\,\beta\,\partial^3\gamma \\ & + \tfrac{5}{3}\alpha\,\partial^3\beta\,\partial\gamma - \tfrac{35}{48}\alpha\,\partial\beta\,\partial^3\gamma\,, \end{aligned} \tag{8.4.7}$$

with $\alpha^2 = \frac{121}{6}$. Here, we have $c_X = 25$, and the associated minimal model, with $c_{mm} = 1$, is expected to be the lowest $\mathcal{W}_4$ minimal model [143]. Following the same strategy as before, we should begin by looking amongst the operators $U(\varphi,\beta,\gamma)$ associated with the physical states eq. (8.3.15) with zero $\varphi$ momentum, and zero ghost number, at dimensions 3 and 4. These will be the candidate dimension 3 and 4 primary fields of the $\mathcal{W}_4$ algebra. In [143], all physical states of the $\mathcal{W}_{2,5}$ string up to level $\ell = 13$ were obtained. In fact one can easily see that the required physical states associated with the dimension 3 and 4 generators will occur at levels 13 and 14 respectively. We find the following expressions for the primary dimension 3 current $W_{mm}$ and dimension 4 current $V_{mm}$ of the $\mathcal{W}_4$ algebra at $c_{mm} = 1$:

$$\begin{aligned} W_{mm} = {} & \tfrac{1}{2}(\partial\varphi)^3 + \tfrac{3}{2}\alpha\,\partial^2\varphi\partial\varphi + \tfrac{31}{12}\partial^3\varphi + \tfrac{15}{2}\partial\varphi\beta\partial\gamma \\ & + 5\,\partial\varphi\partial\beta\gamma + 5\,\partial^2\varphi\beta\gamma + \tfrac{3}{2}\alpha\,\partial\beta\partial\gamma + \alpha\,\partial^2\beta\gamma, \end{aligned} \tag{8.4.8}$$

$$\begin{aligned} V_{mm} = {} & \frac{-25}{\sqrt{864}}\Big((\partial\varphi)^4 + 4\alpha\,\partial^2\varphi(\partial\varphi)^2 + \tfrac{2317}{150}(\partial^2\varphi)^2 + \tfrac{277}{25}\partial^3\varphi\partial\varphi + \tfrac{617}{1650}\alpha\,\partial^4\varphi \\ & + \tfrac{108}{5}\partial^2\varphi\partial\varphi\beta\gamma + 20\,(\partial\varphi)^2\beta\partial\gamma + \tfrac{292}{25}(\partial\varphi)^2\partial\beta\gamma + \tfrac{62}{11}\alpha\,\partial^2\varphi\beta\partial\gamma \\ & + \tfrac{2104}{275}\alpha\,\partial^2\varphi\partial\beta\gamma + \tfrac{216}{55}\alpha\,\partial\varphi\partial^2\beta\gamma + \tfrac{378}{55}\alpha\,\partial\varphi\partial\beta\partial\gamma + \tfrac{108}{55}\alpha\,\partial^3\varphi\beta\gamma \\ & - \tfrac{44}{15}\beta\partial^3\gamma - \tfrac{132}{25}\partial\beta\partial^2\gamma + \tfrac{321}{25}\partial^2\beta\partial\gamma + \tfrac{544}{75}\partial^3\beta\gamma + \tfrac{44}{5}\partial\beta\beta\partial\gamma\gamma\Big)\,. \end{aligned} \tag{8.4.9}$$

We have normalised these currents canonically, so that the coefficient of the highest-order pole in the OPE of a dimension $s$ operator with itself is $c_{mm}/s$, where the central charge $c_{mm} = 1$ in the present case.

It is now a straightforward matter to check that $T_{mm}$, $W_{mm}$ and $V_{mm}$ indeed generate the $\mathcal{W}_4$ algebra at $c_{mm} = 1$. We find complete agreement with the algebra given in [26, 131] again modulo the appearance of certain additional primary fields that are BRST exact. Specifically, we find a dimension 5 null primary field $Q_1(\beta)$ and its Virasoro descendants in the OPE of $W_{mm}(z)V_{mm}(w)$, and a dimension 6 null primary field $Q_1(30\,\partial\varphi\,\beta+11\sqrt{6}\,\partial\beta)$ and its descendants in the OPE $V_{mm}(z)V_{mm}(w)$.

| | $g$ | $h$ | | | | $\mu$ (In units of $\alpha/22$) | | | |
|---|---|---|---|---|---|---|---|---|---|
| $\ell=0$ | 4 | | $\frac{1}{12}$ | $\frac{1}{16}$ | 0 | | $-22$ | $-21$ | $-20$ |
| $\ell=0$ | 4 | | | $\frac{1}{16}$ | 0 | | $-22$ | $-23$ | $-24$ |
| $\ell=1$ | 3 | $\frac{3}{4}$ | $\frac{9}{16}$ | $\frac{1}{3}$ | $\frac{1}{16}$ | $-18$ | $-17$ | $-16$ | $-15$ |
| $\ell=2$ | 3 | | $\frac{25}{16}$ | $\frac{4}{3}$ | $\frac{3}{4}$ | | $-17$ | $-16$ | $-14$ |
| $\ell=3$ | 2 | | 1 | $\frac{9}{16}$ | $\frac{1}{12}$ | | $-12$ | $-11$ | $-10$ |
| $\ell=4$ | 2 | | | $\frac{25}{16}$ | $\frac{9}{16}$ | | | $-11$ | $-9$ |
| $\ell=5$ | 2 | 3 | $\frac{25}{12}$ | $\frac{25}{16}$ | 1 | $-12$ | $-10$ | $-9$ | $-8$ |
| $\ell=6$ | 1 | | | $\frac{3}{4}$ | $\frac{1}{16}$ | | | $-6$ | $-5$ |
| $\ell=7$ | 1 | | | | $\frac{1}{3}$ | | | | $-4$ |
| $\ell=8$ | 1 | | | $\frac{4}{3}$ | $\frac{9}{16}$ | | | $-4$ | $-3$ |
| $\ell=9$ | 1 | | $\frac{49}{16}$ | $\frac{25}{16}$ | $\frac{3}{4}$ | | $-5$ | $-3$ | $-2$ |
| $\ell=10$ | 0 | | | | 0 | | | | 0 |
| $\ell=11$ | 0 | | | | $\frac{1}{16}$ | | | | 1 |
| $\ell=12$ | 0 | | | | $\frac{1}{12}$ | | | | 2 |
| $\ell=13$ | 0 | | | 3 | $\frac{1}{16}$ | | | 0 | 3 |
| $\ell=14$ | 0 | 4 | $\frac{49}{16}$ | $\frac{25}{12}$ | 0 | 0 | 1 | 2 | 4 |

Table 8.2: $U(\varphi,\beta,\gamma)$ operators for the $\mathcal{W}_{2,5}$ string

Having obtained the currents that generate the $\mathcal{W}_4$ algebra, we may now examine the $U(\varphi,\beta,\gamma)$ operators in the physical states of the $\mathcal{W}_{2,5}$ string, in order to compare their weights with those of the lowest $\mathcal{W}_4$ minimal model. Specifically, this model has highest weight fields with conformal weights $h=\{0,\frac{1}{16},\frac{1}{12},\frac{1}{3},\frac{9}{16},\frac{3}{4},1\}$. The results presented in [143], extended to level $\ell=14$, are given in table 8.2. One can see from the results in table 8.2 that indeed all the conformal weights of the primary fields of the lowest $\mathcal{W}_4$ minimal model occur in the $\mathcal{W}_{2,5}$ string. We find that the corresponding weights under the $\mathcal{W}_4$ currents (8.4.8) and (8.4.9) are:

$$\begin{aligned} T_{mm}: &\quad \{0,\tfrac{1}{16},\tfrac{1}{12},\tfrac{1}{3},\tfrac{9}{16},\tfrac{3}{4},1\},\\ \tfrac{352}{\alpha}\,W_{mm}: &\quad \{0,\pm1,0,0,\pm11,\pm32,0\}, \qquad (8.4.10)\\ 6912\sqrt{6}\,V_{mm}: &\quad \{0,27,-64,128,-405,1728,-6912\}\,. \end{aligned}$$

**Intermezzo 8.4.2**
If $h=\frac{1}{16}$, the $W_{mm}$–weight is positive when $(\frac{\alpha}{22})^{-1}\mu=3 \bmod 4$, and negative when $(\frac{\alpha}{22})^{-1}\mu=1 \bmod 4$. If $h=\frac{9}{16}$, the $W_{mm}$-weight is positive when $(\frac{\alpha}{22})^{-1}\mu=1 \bmod 4$, and negative when $(\frac{\alpha}{22})^{-1}\mu=3 \bmod 4$. If $h=\frac{3}{4}$, for which $(\frac{\alpha}{22})^{-1}\mu=12n-18$ or $12n-14$, with $n$ a non-negative integer (see [143]), the $W_{mm}$ weight is positive when $n$ is odd, and negative when $n$ is even.

---

We have checked that these weights agree with those that one finds using the highest weight vertex-operators of the $\mathcal{W}_N$ minimal models in the "Miura" realisations discussed in [63][7]. The remaining $U(\varphi,\beta,\gamma)$ operators obtained here and in [143], with conformal weights $h$ that lie outside the set of weights for the $\mathcal{W}_4$ minimal model, correspond to $W_{mm}$ and $V_{mm}$ descendant states. In other words, they are secondaries of the $\mathcal{W}_4$ minimal model, but they are primaries with respect to a purely Virasoro $c_{mm}=1$ model. In this more generic case, with $c_{mm}\geq1$, the number of primaries in the purely Virasoro model will be infinite. Thus if we would go on solving the physical state conditions at higher and higher levels $\ell$, we would find a set of operators $U(\varphi,\beta,\gamma)$ with conformal weights $h$ that increased indefinitely. All those lying outside the set $h=\{0,\frac{1}{16},\frac{1}{12},\frac{1}{3},\frac{9}{16},\frac{3}{4},1\}$ would be given by certain integers added to values lying in the set, corresponding to $W_{mm}$ and $V_{mm}$ descendant fields.

### 8.4.3 The $\mathcal{W}_{2,6}$ string

In [143], it was found that there are four different nilpotent BRST operators of the form (8.3.6),(8.3.7),(8.3.14), corresponding to different values of $\alpha$, and hence $c_X$. As usual, we shall be concerned with the case which seems to be associated with a unitary string theory. This is given by $\alpha^2=\frac{845}{20}$, implying $c_X=\frac{174}{7}$ and hence the $(\varphi,\beta,\gamma)$ system describes a model with $c=\frac{8}{7}$. We expect this to be the lowest $\mathcal{W}_5$ minimal model. The operator $F(\varphi,\beta,\gamma)$ in this case takes the form [143]:

$$\begin{aligned} F(\beta,\gamma,\varphi) &= (\partial\varphi)^6+6\alpha\partial^2\varphi(\partial\varphi)^4+\tfrac{765}{7}(\partial^2\varphi)^2(\partial\varphi)^2+\tfrac{256}{7}\partial^3\varphi(\partial\varphi)^3\\ &+\tfrac{174}{35}\alpha(\partial^2\varphi)^3+\tfrac{528}{35}\alpha\partial^3\varphi\partial^2\varphi\partial\varphi+\tfrac{18}{7}\alpha\partial^4\varphi(\partial\varphi)^2+\tfrac{1514}{245}(\partial^3\varphi)^2\\ &+\tfrac{2061}{245}\partial^4\varphi\partial^2\varphi+\tfrac{2736}{1225}\partial^5\varphi\partial\varphi+\tfrac{142}{6125}\alpha\partial^6\varphi+18(\partial\varphi)^4\beta\partial\gamma+\tfrac{72}{5}\alpha\partial^2\varphi(\partial\varphi)^2\beta\partial\gamma\\ &+\tfrac{48}{5}\alpha(\partial\varphi)^3\partial\beta\partial\gamma+\tfrac{216}{5}\partial^3\varphi\partial\varphi\beta\partial\gamma+\tfrac{1494}{35}(\partial^2\varphi)^2\beta\partial\gamma+\tfrac{5256}{35}\partial^2\varphi\partial\varphi\partial\beta\partial\gamma\\ &+\tfrac{324}{5}(\partial\varphi)^2\partial^2\beta\partial\gamma-\tfrac{72}{7}(\partial\varphi)^2\beta\partial^3\gamma+\tfrac{204}{175}\alpha\partial^4\varphi\beta\partial\gamma+\tfrac{192}{25}\alpha\partial^3\varphi\partial\beta\partial\gamma\\ &+\tfrac{2376}{175}\alpha\partial^2\varphi\partial^2\beta\partial\gamma-\tfrac{144}{175}\alpha\partial^2\varphi\beta\partial^3\gamma+\tfrac{1296}{175}\alpha\partial\varphi\partial^3\beta\partial\gamma-\tfrac{576}{175}\alpha\partial\varphi\partial\beta\partial^3\gamma\\ &+\tfrac{1614}{175}\partial^4\beta\partial\gamma-\tfrac{216}{35}\partial^2\beta\partial^3\gamma+\tfrac{144}{1225}\beta\partial^5\gamma+\tfrac{144}{35}\partial\beta\beta\partial^2\gamma\partial\gamma. \qquad (8.4.11) \end{aligned}$$

In [143], physical states in the theory up to and including level $\ell=6$ were studied. Here, we are primarily concerned with finding the physical states associated with the expected dimension 3, 4, 5, primary fields of the $\mathcal{W}_5$ minimal model. These should occur at levels $\ell = 18$, 19 and 20 respectively. It is a straightforward matter to solve for such physical states eq. (8.3.15) with $U(\varphi, \beta, \gamma)$ having zero ghost number and zero $\varphi$ momentum. We find the following results for the dimension 3, 4, 5 operators $W_{mm}$, $V_{mm}$ and $Y_{mm}$:

$$\begin{aligned} W_{mm} &= \sqrt{\tfrac{2}{57}}\Big(\tfrac{7}{3}(\partial\varphi)^3 + 7\alpha\partial^2\varphi\partial\varphi + \tfrac{185}{12}\partial^3\varphi + 42\partial\varphi\beta\partial\gamma \\ &\quad +30\partial\varphi\partial\beta\gamma + 30\partial^2\varphi\beta\gamma + 7\alpha\partial\beta\partial\gamma + 5\alpha\partial^2\beta\gamma\Big) \end{aligned}$$

$$\begin{aligned} V_{mm} &= -\sqrt{\tfrac{7}{60819}}\Big(\tfrac{427}{8}(\partial\varphi)^4 + \tfrac{427}{2}\alpha\partial^2\varphi(\partial\varphi)^2 + \tfrac{10619}{8}(\partial^2\varphi)^2 + 743\partial^3\varphi\partial\varphi \\ &\quad +\tfrac{3313}{156}\alpha\partial^4\varphi + 1455\partial^2\varphi\partial\varphi\beta\gamma + 1281(\partial\varphi)^2\beta\partial\gamma + 825(\partial\varphi)^2\partial\beta\gamma \\ &\quad +\tfrac{5370}{13}\alpha\partial^2\varphi\beta\partial\gamma + \tfrac{6900}{13}\alpha\partial^2\varphi\partial\beta\gamma + \tfrac{2910}{13}\alpha\partial\varphi\partial^2\beta\gamma + \tfrac{4656}{13}\alpha\partial\varphi\partial\beta\partial\gamma \\ &\quad +\tfrac{1455}{13}\alpha\partial^3\varphi\beta\gamma - 247\beta\partial^3\gamma - 494\partial\beta\partial^2\gamma + \tfrac{6891}{7}\partial^2\beta\partial\gamma \\ &\quad +\tfrac{7785}{14}\partial^3\beta\gamma + 1170\partial\beta\beta\partial\gamma\gamma\Big) \end{aligned}$$

$$\begin{aligned} Y_{mm} = \sqrt{\tfrac{7}{122}}\Big(&\tfrac{749}{165}(\partial\varphi)^5 + \tfrac{749}{33}\alpha(\partial\varphi)^3\partial^2\varphi + \tfrac{6091}{22}(\partial^2\varphi)^2\partial\varphi + \tfrac{1361}{66}\alpha\partial^3\varphi\partial^2\varphi \\ &+\tfrac{14351}{132}\partial^3\varphi(\partial\varphi)^2 + \tfrac{2330}{429}\alpha\partial^4\varphi\partial\varphi + \tfrac{4825}{1848}\partial^5\varphi + \tfrac{1498}{11}(\partial\varphi)^3\beta\partial\gamma \\ &+\tfrac{2570}{33}(\partial\varphi)^3\partial\beta\gamma + \tfrac{11382}{143}\alpha\partial^2\varphi\partial\varphi\beta\partial\gamma + \tfrac{15670}{143}\alpha\partial^2\varphi\partial\varphi\partial\beta\gamma \\ &+\tfrac{430}{11}\alpha(\partial^2\varphi)^2\beta\gamma + \tfrac{4270}{143}\alpha\partial^3\varphi\partial\varphi\eta\gamma + \tfrac{2350}{11}\partial^2\varphi(\partial\varphi)^2\beta\gamma \\ &+\tfrac{7840}{143}\alpha(\partial\varphi)^2\partial\beta\partial\gamma + \tfrac{4380}{143}\alpha(\partial\varphi)^2\partial^2\beta\gamma - 52\partial\varphi\beta\partial^3\gamma - \tfrac{624}{7}\partial\varphi\partial\beta\partial^2\gamma \\ &+\tfrac{17331}{77}\partial\varphi\partial^2\beta\partial\gamma + \tfrac{9775}{77}\partial\varphi\partial^3\beta\gamma - \tfrac{624}{7}\partial^2\varphi\beta\partial^2\gamma + \tfrac{18541}{77}\partial^2\varphi\partial\beta\partial\gamma \\ &+\tfrac{3390}{11}\partial^2\varphi\partial^2\beta\gamma + \tfrac{7957}{154}\partial^3\varphi\beta\partial\gamma + \tfrac{77705}{462}\partial^3\varphi\partial\beta\gamma + \tfrac{11575}{462}\partial^4\varphi\beta\gamma \\ &-\tfrac{26}{3}\alpha\partial\beta\partial^3\gamma - \tfrac{104}{7}\alpha\partial^2\beta\partial^2\gamma + \tfrac{6775}{1001}\alpha\partial^3\beta\partial\gamma + \tfrac{5365}{1001}\alpha\partial^4\beta\gamma \\ &+120\partial\varphi\partial\beta\beta\partial\gamma\gamma + \tfrac{120}{13}\alpha\partial^2\beta\beta\partial\gamma\gamma\Big). \end{aligned} \qquad (8.4.12)$$

[7] After converting from the non-primary basis of Miura currents to the primary basis that we are using here.

We have as usual given these currents their canonical normalisations. We have checked that they indeed, together with $T_{mm} = T_\varphi + T_{\beta\gamma}$, generate the $\mathcal{W}_5$ algebra, given in [113], with central charge $c_{mm} = \frac{8}{7}$. Again, one finds additional BRST exact fields appearing on the right-hand sides of the OPEs of the primary generators. These null fields are primaries (and their descendants) except in the case of the OPE $Y_{mm}(z)Y_{mm}(w)$. The new field occuring at the second order pole of this OPE is only primary up to BRST exact terms.

It was found in [143] that the physical states of the $\mathcal{W}_{2,6}$ string were associated with operators $U(\varphi, \beta, \gamma)$ whose conformal weights included those of the highest weight fields of the lowest $\mathcal{W}_5$ minimal model, which has $c = \frac{8}{7}$. Indeed, here we find that the highest weight fields have the weights

$$\begin{aligned} &T_{mm}: && \{0, \tfrac{2}{35}, \tfrac{3}{35}, \tfrac{2}{7}, \tfrac{17}{35}, \tfrac{23}{35}, \tfrac{4}{5}, \tfrac{6}{7}, \tfrac{6}{5}\} \\ &\tfrac{325}{\alpha}\sqrt{\tfrac{57}{8}}W_{mm}: && \{0, \pm2, \pm1, 0, \pm13, \pm39, \pm76, 0, \pm38\} \\ &25\sqrt{\tfrac{141911}{3}}\,V_{mm}: && \{0, 11, -14, 50, -74, 66, 836, -1100, -2299\} \\ &\tfrac{89375}{\alpha}\sqrt{\tfrac{427}{32}}\,Y_{mm}: && \{0, \pm11, \mp48, 0, \pm314, \mp902, \pm1452, 0, \mp16621\}\,. \end{aligned} \qquad (8.4.13)$$

Note that the $\pm$ signs for the weights under the dimension 3 operator $W_{mm}$ are correlated with those for the weights under the dimension operator $Y_{mm}$. Again we have checked that these weights agree with those calculated from the realisations of the $\mathcal{W}_N$ minimal models given in [63]. All the physical states of the $\mathcal{W}_{2,6}$ string are presumably associated with operators $U(\varphi, \beta, \gamma)$ that are either highest weight under the $\mathcal{W}_5$ algebra, as given in eq. (8.4.13), or they are $W_{mm}$, $V_{mm}$ or $Y_{mm}$ descendants of such operators. Some examples of descendant operators can be found in [143]. Again one expects, since the $\mathcal{W}_5$ minimal model has $c_{mm} = \frac{8}{7} \geq 1$, that there will be an infinite number of descendant operators.

## 8.5 Hierarchies of string embeddings

It was proposed recently [14] that as part of the general programme of looking for unifying principles in string theory, one should look for ways in which string theories with smaller worldsheet symmetries could be embedded into string theories with larger symmetries. In particular, it was shown in [14] that the bosonic string could be embedded in the $N = 1$ superstring, and that in turn, the $N = 1$ string could be embedded in the $N = 2$ superstring. In subsequent papers, it was shown by various methods that the cohomologies of the resulting theories were precisely those of the embedded theories themselves [80, 120].

The essential ingredient in the embeddings discussed in [14] is that a realisation for the currents of the more symmetric theory can be found in terms of the currents of the less symmetric theory, together with some additional matter fields whose eventual rôle for the cohomology is to supply degrees of freedom that are cancelled by the additional ghosts of the larger theory. For example, the $N = 1$ superconformal algebra, at critical central charge $c = 15$, can be realised in terms of a $c = 26$ energy–momentum tensor $T_M$ as:

$$\begin{aligned} T &= T_M - \tfrac{3}{2}b_1\,\partial c_1 - \tfrac{1}{2}\partial b_1\,c_1 + \tfrac{1}{2}\partial^2(c_1\,\partial c_1) \\ G &= b_1 + c_1\,(T_M + \partial c_1\,b_1) + \tfrac{5}{2}\partial^2 c_1\,, \end{aligned} \qquad (8.5.1)$$

where $b_1$ and $c_1$ are ghost-like dimension $(\frac{3}{2}, -\frac{1}{2})$ anticommuting matter fields. The cohomology of the BRST operator for the $N = 1$ superstring, with this realisation of the $N = 1$ superconformal algebra, is precisely that of the usual bosonic string [14, 80, 120]. This is most easily seen using the method of [120], where a unitary canonical transformation $Q \longrightarrow e^R\, Q\, e^{-R}$ is applied to the $N = 1$ BRST operator, transforming it into the BRST operator for the bosonic string plus a purely topological BRST operator. In effect, the degrees of freedom of $b_1$ and $c_1$ are cancelled out by the degrees of freedom of the commuting dimension $(\frac{3}{2}, -\frac{1}{2})$ ghosts for the dimension $\frac{3}{2}$ current $G$. The central charge of the energy–momentum tensor for $(b_1, c_1)$ is $c = 11$, which precisely cancels the $c = -11$ central charge for the dimension $\frac{3}{2}$ ghost system for the dimension $\frac{3}{2}$ current $G$.

It is natural to enquire whether some analogous sequence of embeddings for $\mathcal{W}$–strings might exist, with, for example, the usual Virasoro string contained within the $\mathcal{W}_3$–string, which in turn is contained in the $\mathcal{W}_4$–string, and so on [14]. In fact, as was observed in [145], such sequences of embeddings are already well known for $\mathcal{W}$–strings. The simplest example is provided by the $\mathcal{W}_3$–string, where the $\mathcal{W}_3$–currents $T$ and $W$ are realised in terms of an energy–momentum tensor $T_X$, and a scalar field $\varphi$. The $\varphi$ field here plays a role analogous to the $(b_1, c_1)$ matter fields in the embedding of the bosonic string in the $N = 1$ superstring. Here, however, the central charge $c = \frac{149}{2}$ for the energy–momentum tensor of $\varphi$ does not quite cancel the central charge $c = -74$ of the $(\beta, \gamma)$ ghosts for the dimension 3 current $W$, and so the nilpotence of the $\mathcal{W}_3$ BRST operator requires that $T_X$ has central charge $c = \frac{51}{2}$ rather than $c = 26$. The $\varphi$ field has no associated continuous degrees of freedom in physical states, and the cohomology of the $\mathcal{W}_3$ string is just that of a $c = 25\frac{1}{2}$ Virasoro string tensored with the Ising model.

It has also been suggested that one might be able to embed the $c = 26$ Virasoro string into, for example, the $\mathcal{W}_3$ string. However, it would, perhaps, be surprising if it were possible to embed the Virasoro string into the $\mathcal{W}_3$ string in two different ways, both for $c_X = \frac{51}{2}$ and also for $c_X = 26$. Indeed, there is no known way of realising the currents of the $\mathcal{W}_3$ algebra, with the central charge $c = 100$ needed for nilpotence of the BRST operator, in terms of a $c = 26$ energy–momentum tensor plus other fields that would contribute no continuous degrees of freedom in physical states.

**Intermezzo 8.5.1**

A very different approach was proposed in [15], where it was shown that by performing a sequence of canonical transformations on the BRST operator of the $\mathcal{W}_3$ string, it could be transformed into the BRST operator of an ordinary $c = 26$ bosonic string plus a purely topological BRST operator. However, as was shown in [147], and subsequently reiterated in [207], one step in the sequence of canonical transformations involved a nonlocal transformation that reduced the original $\mathcal{W}_3$ BRST operator to one with completely trivial cohomology. A later step in the sequence then involved another nonlocal transformation that caused the usual cohomology of the bosonic string to grow out of the previous trivial cohomology. In effect one is glueing two trivialised theories back to back, and so the physical spectra of the two theories prior to trivialisation are disconnected from one another, making the embedding quite meaningless.

---

An interesting possibility for generalising the ideas in [14] is to consider the case where the bosonic string is embedded in a fermionic higher-spin string theory. The simplest such example would be provided by looking at a theory with a dimension $\frac{5}{2}$ current in addition to the energy–momentum tensor. In order to present some results on this example, it is useful first to recast the $N = 1$ superstring, with the matter currents realised as in (8.5.1), in a simpler form. We do this by performing a canonical redefinition involving the dimension 2 ghosts $(b, c)$, the dimension $\frac{3}{2}$ ghosts $(r, s)$, and the ghost-like matter fields $(b_1, c_1)$ (which we shall refer to as pseudo-ghosts). If we transform these according to:

$$\begin{aligned} c &\longrightarrow c - s\, c_1 \\ r &\longrightarrow r - b\, c_1 \\ b_1 &\longrightarrow b_1 + b\, s\,, \end{aligned} \tag{8.5.2}$$

(with $b$, $s$ and $c_1$ suffering no transformation), then the BRST operator assumes the graded form $Q = Q_0 + Q_1$, where:

$$\begin{aligned} Q_0 &= \oint c\Big(T_M + T_{b_1 c_1} + T_{rs} + \tfrac{1}{2}T_{bc} + x\,\partial^2(\partial c_1 c_1)\Big) \\ Q_1 &= \oint s\Big(b_1 - x\, b_1 \partial c_1 c_1 + 3x\, r\partial s c_1 + x\,\partial r s c_1 + 2x^2\,\partial^2 c_1 \partial c_1 c_1\Big)\,. \end{aligned} \tag{8.5.3}$$

Here $x$ is a free constant which actually takes the value $-\frac{1}{2}$ when one transforms (8.5.1) according to (8.5.2), but can be made arbitrary by performing a constant OPE-preserving rescaling of $b_1$ and $c_1$. The reason for introducing $x$ is that it can be viewed as a power-counting parameter for a second grading of $Q_0$ and $Q_1$, under the $(b_1, c_1)$ pseudo-ghost number. Thus $Q_0$ has terms of pseudo-ghost degrees 0 and 2, whilst $Q_1$ has terms of pseudo-ghost degrees $-1$, 1 and 3. (We have dropped an overall $x^{-1}$ factor from $Q_1$ for convenience. We are free to do this owing to the first grading under $(r, s)$ degree, which implies that $Q_0^2 = Q_1^2 = \{Q_0, Q_1\} = 0$.)

Before moving on to the generalisation to higher dimensions, it is useful to present the unitary canonical transformation of ref. [120] in this language, which maps the BRST operator into that of the bosonic string plus a topological term. Thus we find that the charge:

$$R = \oint c_1\Big(-c\,\partial r - \tfrac{3}{2}\partial c\, r - x\, r\, s\, \partial c_1\Big) \tag{8.5.4}$$

acts on the BRST operator $Q = Q_0 + Q_1$ to give:

$$e^R\, Q\, e^{-R} = \oint c\,(T_M - b\,\partial c) + \oint s\, b_1\,. \tag{8.5.5}$$

The first term on the *rhs* is the usual BRST operator of the bosonic string, and the second term is purely topological, with no cohomology.

We may now seek a dimension $(2,\frac{5}{2})$ generalisation of this dimension $(2,\frac{3}{2})$ theory. Thus we now consider commuting ghosts $(r,s)$ of dimensions $(\frac{5}{2},-\frac{3}{2})$ for a dimension $\frac{5}{2}$ current, and anticommuting pseudo-ghosts $(b_1,c_1)$ of dimensions $(\frac{5}{2},-\frac{3}{2})$. We find that a graded BRST operator $Q=Q_0+Q_1$ again exists, where $Q_0$ contains terms with pseudo-ghost degrees 0, 2 and 4, whilst $Q_1$ has terms of pseudo-ghost degrees $-1$, 1, 3 and 5. The coefficients of the various possible structures in $Q_0$ and $Q_1$ are determined by the nilpotency conditions $Q_0^2=Q_1^2=\{Q_0,Q_1\}=0$. $Q_0$ takes the form:

$$Q_0=\oint c\Big(T_M+T_{b_1c_1}+T_{rs}+\tfrac{1}{2}T_{bc}+x\,\partial^2\big(3\partial^3c_1c_1+7\partial^2c_1\partial c_1\big) \\ +\,y\,\partial^2\big(\partial^3c_1\partial^2c_1\partial c_1c_1\big)\Big)\ , \qquad (8.5.6)$$

where $x$ and $y$ are arbitrary constants associated with the terms in $Q_0$ of pseudo-ghost degree 2 and 4 respectively. The form of $Q_1$ is quite complicated:

$$Q_1=\oint s\Big(b_1-6x\,b_1\partial^2c_1\partial c_1-4x\,b_1\partial^3c_1c_1-6x\,\partial b_1\partial^2c_1c_1- \\ 2x\,\partial^2b_1\partial c_1c_1+\cdots \\ +x\big(\tfrac{26}{3}x^2+\tfrac{25}{6}y\big)\partial^4c_1\partial^3c_1\partial^2c_1\partial c_1c_1\Big)\ , \qquad (8.5.7)$$

where the ellipsis represents 13 terms of pseudo-ghost degree 3.

One may again look for a charge $R$ that acts unitarily and canonically on the BRST operator to give it a simpler form. We find that the required charge is given by:

$$R=\oint c_1\Big(-c\partial r-\tfrac{5}{2}\partial cr-x\,c\partial^2c_1\partial c_1\partial r-\tfrac{5}{2}x\,\partial c\partial^2c_1\partial c_1r \\ -2x\,\partial c_1\partial^2rs-6x\,\partial^2c_1\partial rs+2x\,\partial^3c_1rs-\tfrac{1}{2}y\,\partial^3c_1\partial^2c_1\partial c_1rs\Big)\ . \qquad (8.5.8)$$

Acting on the BRST operator $Q=Q_0+Q_1$, this gives exactly eq. (8.5.5), which shows that this theory is again simply equivalent to the bosonic string.

Although the dimension $(2,\frac{5}{2})$ theory that we have described above has a BRST operator that is a natural generalisation of the $N=1$ superconformal BRST operator with the realisation (8.5.1) for the matter currents, there is one important aspect which we should consider. From the graded $(2,\frac{3}{2})$ BRST operator given by (8.5.3), one can invert the canonical transformation (8.5.2), and get back to a form in which one can replace the specific realisation (8.5.1) of the superconformal currents by an abstract realisation in terms of currents $T$ and $G$. In this sense, one can say that the realisation (8.5.1) describes an embedding of the bosonic string in the $N=1$ superstring. Let us look if the some procedure goes through for the $(2,\frac{5}{2})$ BRST operator.

Let us consider the $\mathcal{W}_{2,5/2}$ algebra in more detail. Classically, the primary dimension $\frac{5}{2}$ current $G$ satisfies the OPE:

$$G(z)G(w)\sim\frac{T^2}{z-w}\ . \qquad (8.5.9)$$

The Jacobi identity is satisfied modulo the classical null field:

$$N_1\equiv 4T\,\partial G-5\partial T\,G\ . \qquad (8.5.10)$$

Before specifying what we mean with a classical null field, we wish to show that one can realise the algebra in the following way:

$$\begin{aligned} T &= -\tfrac{1}{2}\psi\,\partial\bar\psi+\tfrac{1}{2}\partial\psi\,\bar\psi\ , \\ G &= \tfrac{1}{2}\left(\psi+\bar\psi\right)T\ , \end{aligned} \qquad (8.5.11)$$

where $\psi$ is a complex fermion satisfying the OPE $\psi(z)\bar\psi(w)\sim 1/(z-w)$. One can easily verify for this realisation that the null field $H$ vanishes.

We checked that both for the abstract algebra, and for the realisation no BRST operator can be constructed along the familiar lines, *e.g.* by introducing ghosts–antighost pairs for $T$ and $G$, and constructing a BRST charge $c^1T+c^2G+f_{ij}{}^kc^ic^jb_k$, where the structure constants $f_{ij}{}^k$ depend on $T$. To explain this surprising fact, we need to elaborate on the meaning of a null field in a Poisson algebra.

We use exactly the same definition for a null field as in the OPE case, see subsection 2.3.3. That is, all null fields form an ideal in the Poisson algebra. We can check by repeatedly computing Poisson brackets with $N_1$ that it there is an ideal in the Poisson algebra of $T$ and $G$, generated by $N_1$ and

$$N_2\equiv 4T^3-30\partial G\,G\ . \qquad (8.5.12)$$

More precisely, all other null fields can be written as:

$$f_1(T,G)N_1+f_2(T,G)N_2\ , \qquad (8.5.13)$$

with $f_i(T,G)$ a differential polynomial in $T$ and $G$. We see that the phase space of the Poisson algebra is not simply the space of differential polynomials in $T$ and $G$, but the additional constraints $N_1=N_2=0$ have to be taken into account. In such a case, the ordinary procedure of constructing a (classical) BRST charge does not work. Indeed, one should use the BRST-formalism appropriate for reducible constraints, which requires the introduction of "ghosts for ghosts", see *e.g.* [110]. This clearly explains why no "ordinary" BRST charge exists for this system.

Thus, it seems that the $\mathcal{W}_{2,\frac{5}{2}}$ string is of very different type than other strings considered up to now. It remains to be seen if the resulting BRST-charge is in any way related to the one we constructed above, eqs, (8.5.6,8.5.7).

We have explicitly checked for all higher half-integer dimensions, and we find that again a classical $\mathcal{W}_{2,n/2}$ algebra does not identically satisfy the Jacobi identity. Thus again, we expect that ghosts for ghosts should be introduced to enable a proper treatment in the BRST formalism of such algebras.

## 8.6 Conclusion and discussion

In this chapter we have looked at the quantisation of $\mathcal{W}$-string theories based on the classical $\mathcal{W}_{2,s}$ higher-spin algebras. One of the more noteworthy features of these theories is that anomaly-free quantisation is possible even when there does not exist a closed quantum extension of the classical $\mathcal{W}_{2,s}$ algebra at the critical central charge. We can identify quantum currents as the coefficients of the dimension $-1$ and $1-s$ ghosts in the BRST current, and discarding antighosts. Of course, these fields do generate a realisation of a quantum $\mathcal{W}$–algebra. However, the corresponding abstract algebra is probably infinitely generated. A previous example of this kind of phenomenon, where a BRST operator exists even when the matter system does not generate a closed algebra at the quantum level, was found in the context of the non-critical $\mathcal{W}_3$–string discussed in [19, 23, 22].

It is quite puzzling that there can be several inequivalent quantum theories that arise from the same classical theory, corresponding to different possible choices for the coefficients of the quantum corrections to the classical BRST operator. A study of the relation between the classical and quantum cohomology of the corresponding BRST operators should be able to shed some light on this point.

In a multi-scalar realisation, the spectrum of physical states for a $\mathcal{W}_{2,s}$ string turns out to be described by the tensor product of sets of bosonic-string states in the effective spacetime times certain primary operators built from the $(\varphi,\beta,\gamma)$ fields. In most cases these primary fields are conjectured to correspond to those of some Virasoro or $\mathcal{W}$ minimal model. For example, the regular sequence of $\mathcal{W}_{2,s}$ BRST operators, which exist for all $s$, corresponds to the lowest unitary $\mathcal{W}_{s-1}$ minimal model, with $c_{mm}=2(s-2)/(s+1)$. We have tested the above conjecture in detail for the cases $s=4$, 5 and 6 of the $\mathcal{W}_{2,s}$–string. We have shown, for the lowest few levels, that indeed the operators $U(\varphi,\beta,\gamma)$ that arise in the physical states, are associated with the highest weight fields of the lowest unitary $\mathcal{W}_{s-1}$ minimal models. Specifically, we find in all physical states that $U(\varphi,\beta,\gamma)$ is either a highest weight field of the corresponding $\mathcal{W}_{s-1}$ minimal model, or else it is a descendant field in the sense that it is obtained from a highest weight field by acting with the negative modes of the primary currents of the $\mathcal{W}_{s-1}$ algebra. Since the central charge $c=\frac{2(s-2)}{(s+1)}$ of the $(\varphi,\beta,\gamma)$ system satisfies $c\ge 1$ for $s\ge 5$, it follows that in these cases there are infinite numbers of such descendant fields in the models. Thus the $\mathcal{W}_{s-1}$ generators provide a strikingly powerful organising symmetry in these cases.

The original realisations of the $\mathcal{W}_N$ algebras were the $(N-1)$-scalar realisations from the Miura transformation, introduced in [66, 63]. By contrast, the realisations of the lowest unitary $\mathcal{W}_{s-1}$ minimal models that we find here are all given in terms of just one scalar field $\varphi$, and the $(\beta,\gamma)$ ghost system for dimension $s$. This ghost system can be bosonised, yielding two-scalar realisations. However even when $s=4$, our two-scalar realisation is quite different from the usual Miura realisation of $\mathcal{W}_3$. In particular, our realisations close on the $\mathcal{W}_{s-1}$ algebras modulo the appearance of certain null primary fields in the OPEs of the currents, whereas no such null fields arise in the Miura realisations. Presumably the realisations that we find here are very specific to the particular unitary minimal models that arise in these higher-spin string theories. As an example, we present the dimension 2 and dimension 3 currents eq. (8.4.2) and (8.4.4) for the $\mathcal{W}_3$ algebra at $c=\frac{4}{5}$ in the bosonised language, where $\gamma=\exp(i\rho)$ and $\beta=\exp(-i\rho)$:

$$\begin{aligned} T_{mm} &= -\tfrac{1}{2}(\partial\varphi)^2-\tfrac{1}{2}(\partial\rho)^2-\alpha\,\partial^2\varphi+\tfrac{7}{2}i\,\partial^2\rho, \\ W_{mm} &= \sqrt{\tfrac{2}{13}}\Big(\tfrac{5}{3}\,(\partial\varphi)^3+5\alpha\,\partial^2\varphi\,\partial\varphi+\tfrac{25}{4}\,\partial^3\varphi+4\,\partial\varphi\,(\partial\rho)^2-16i\,\partial\varphi\,\partial^2\rho \\ &\qquad -12i\,\partial^2\varphi\,\partial\rho-\tfrac{2}{3}i\alpha\,(\partial\rho)^3-3\alpha\,\partial^2\rho\,\partial\rho-\tfrac{11}{6}i\alpha\,\partial^3\rho\Big)\,, \end{aligned} \tag{8.6.1}$$

where $\alpha^2=\frac{243}{20}$. It is interesting to note that this realisation of the $\mathcal{W}_3$ algebra at $c=\frac{4}{5}$ is precisely the one obtained in [16] (case $I$, after an $SO(1,1)$ rotation of the two scalars), where more general scalar realisations of $\mathcal{W}_3$ modulo a null dimension 4 operator were considered.

If there is just one $X^\mu$ coordinate in the effective energy–momentum tensor $T_X$, the spectrum of physical states for the $\mathcal{W}_{2,s}$ string becomes more complicated, as observed in [143]. In particular, there are additional physical states over and above those of the form (8.3.15), which do not factorise into the product of effective-spacetime physical states times operators $U(\varphi,\beta,\gamma)$. Examples of these were found for the $\mathcal{W}_3$ string in [164], and for $\mathcal{W}_{2,s}$ strings in [143]. A general discussion of the BRST cohomology for the two-scalar $\mathcal{W}_3$ string is given in [30]. It may well be that the $\mathcal{W}_{2,s}$ strings with just one additional coordinate $X^\mu$ capture the more subtle aspects of the underlying higher-spin geometry.

We have looked also at string theories based on classical algebras involving a higher-spin fermionic current in addition to the energy–momentum tensor. These classical algebras do not satisfy the Jacobi identity identically, but only modulo null fields. When there exists a classical realisation, these null fields are identically zero. The appearance of (classical) null fields obliges one to introduce "ghosts for ghosts". This topic remains to be studied furher.

# Appendix A

# Green's function for the Laplacian in two dimensions

In this appendix we provide some useful formulas that can be derived from the Green's function of the Laplacian[1]. We will first show that in two dimensions this Green's function is given by $\frac{1}{4}\log|x-x_0|^2$. We need to prove:

$$\frac{1}{4}\nabla_0^2\int dx^2\, f(x)\log|x-x_0|^2 = \pi f(x_0)\,. \tag{A.1}$$

For this we write the integral in cylinder coordinates around $x_0$. We get for the *lhs*:

$$\frac{1}{4}\nabla_0^2\int drd\theta\, f(x+x_0)2r\log r\,.$$

We can bring the Laplacian inside the integral, and let it act on $x$:

$$\frac{1}{2}\int drd\theta\, r\log r\left(\frac{1}{r}\frac{\partial}{\partial r}r\frac{\partial f(x+x_0)}{\partial r}+\frac{1}{r^2}\frac{\partial^2 f(x+x_0)}{\partial\theta^2})\right)\,.$$

The second part of the integral is zero because of the total derivative in $\theta$. The first part can easily be evaluated to:

$$\frac{1}{2}\int drd\theta\,\frac{\partial}{\partial r}\left(r\log r\frac{\partial f(x+x_0)}{\partial r}-f(x+x_0)\right)\,.$$

Assuming that $f$ decays fast enough to zero at infinity (we call the set of these functions $\mathcal{S}$), we get the desired result (A.1).

Written in terms of the $z,\bar{z}$ coordinates, eq. (A.1) becomes:

$$\frac{\partial}{\partial\bar{z}_0}\frac{\partial}{\partial z_0}\int dx^2\, f(z,\bar{z})\log(z-z_0)(\bar{z}-\bar{z}_0) = \pi f(z_0,\bar{z}_0)\,, \tag{A.2}$$

or also

$$\int d^2x\,\frac{\partial f}{\partial\bar{z}}(z,\bar{z})\,(z-z_0)^{-(n+1)} = -\pi\frac{1}{n!}\frac{\partial^n f}{\partial z_0^n}(z_0,\bar{z}_0)\,,\ \forall n\in\mathbb{N} \tag{A.3}$$

where we have generalised the result by taking additional derivatives with respect to $z_0$ and changed the derivatives with respect to $z_0$, $\bar{z}_0$ to $z$, $\bar{z}$. This is often written as an equality between distributions:

$$\frac{\partial}{\partial\bar{z}_0}\frac{1}{(z-z_0)^{n+1}} = -\pi\frac{1}{n!}\frac{\partial^n}{\partial z_0^n}\delta^{(2)}(x-x_0)\,, \tag{A.4}$$

where the integration over $z$, $\bar{z}$ is implied.

If $f$ is analytic in an neighbourhood of the point $(z_0,\bar{z}_0)$ (it cannot be analytic in the whole complex plane as it has to be an element of $\mathcal{S}$), we can use the Cauchy residue formula to rewrite eq. (A.3):

$$\frac{1}{\pi}\int d^2x\,\frac{\partial f}{\partial\bar{z}}(z,\bar{z})\,(z-z_0)^{-(n+1)} = -\oint_{C_0}\frac{dz}{2\pi i}\,f(z)\,(z-z_0)^{-(n+1)}\,, \tag{A.5}$$

where the contour $C_0$ surrounds $z_0$ anti-clockwise and lies in the region where $f$ is analytic. This important formula is used in the subsection on Ward-identities 2.2.2.

Eq. (A.3) can be used to define an inverse derivative. Indeed, for $n=0$ we see that (using the notation $\bar{\partial}\equiv\frac{d}{d\bar{z}}$):

$$\int d^2x\, f(z,\bar{z})\,(z-z_0)^{-1} = -\pi g(z_0,\bar{z}_0)\Rightarrow\bar{\partial}g = f\,, \tag{A.6}$$

for $f$ such that (A.4) is valid for $\bar{\partial}f$, or $\bar{\partial}f\in\mathcal{S}$. We define:

$$\bar{\partial}^{-1}f(z_0,\bar{z}_0) = -\frac{1}{\pi}\int d^2x\, f(z,\bar{z})\,(z-z_0)^{-1}\,. \tag{A.7}$$

[1] To get the normalisation constants right, I checked most of the formulas for some test functions by using numeric integration.

# Appendix B

# Superconventions

### Indices

We use the conventions of [55] when working with supervectors, -matrices, and the like. Left and right indices occur, an element of a supermatrix is for instance ${}_iA^j$. Summation is implied over indices which occurs twice, where no signs are introduced when indices are near each other and one is on top, the other down, *e.g.* $b^i\ {}_iA^j$. We use the two following rules for shifting indices:

$$A^i = (-1)^{Ai}\ {}^iA \qquad A_i = (-1)^{(A+1)i}\ {}_iA\,, \tag{B.1}$$

where in exponents of $-1$, $A$ denotes the parity of the "symbol" $A$ and $i$ the parity of the index, such that an element of a supermatrix ${}_iA^j$ has parity $(-1)^{A+i+j}$. For instance one could have a bosonic matrix where the indices take values in a bosonic and fermionic sector. The rules (B.1) are such that:

$$A^i\ {}_iB = (-1)^{AB}B^i\ {}_iA\,. \tag{B.2}$$

### (Super)Lie algebras

We use in chapter 6 and 7 realisations of a super Lie algebra $\bar{g}$. The generators (which are matrices with numbers as elements) are denoted by ${}_at$, $a \in \{1, \cdots, d_B + d_F\}$, where $d_B$ ($d_F$) is the number of bosonic (fermionic) generators. We take $t$ a bosonic symbol. We will always assume that the representation is such that ${}_at$ is a matrix containing numbers (zero parity). This means that ${}_i({}_at)^j$ is zero unless $(-1)^{a+i+j} = 1$. We denote the (anti)commutation relations by:

$$[{}_at, {}_bt] = {}_at\ {}_bt - (-)^{ab}{}_bt\ {}_at = {}_{ab}f^c\ {}_ct\,, \tag{B.3}$$

where ordinary matrix multiplication is used. The structure constants are thus also numbers, and ${}_{ab}f^c$ is zero except if $(-1)^{a+b+c} = 1$.

When we have a super Lie algebra valued field $A$, its components are:

$${}_iA^j = {}_i(A^a\ {}_at)^j \equiv (-1)^{i(A+a+1)}A^a\ \ {}_i({}_at)^j\,, \tag{B.4}$$

where the last equation follows the convention (B.1). This gives us for the elements of the matrix (anti)commutator:

$${}_i([A,B])^j = {}_i\left(-(-1)^{aB}A^aB^b\ {}_{ba}f^c\ {}_ct\right)^j\,. \tag{B.5}$$

From the Jacobi identities, one shows that the adjoint representation is given by:

$${}_b({}_at)^c \equiv {}_{ba}f^c\,. \tag{B.6}$$

The Killing metric ${}_ag_b$ is defined by

$$(-1)^d\ {}_{ca}f^d\ {}_{db}f^c = -\tilde{h}\ {}_ag_b\,, \tag{B.7}$$

where $\tilde{h}$ is the dual Coxeter number. Though this is perfect for ordinary Lie algebras, this is not sufficient for super algebras as the dual Coxeter number might vanish in this case[1]. More generally, it is defined via the supertrace:

$$str({}_at\ {}_bt) \equiv {}_i({}_at)^j\ {}_j({}_bt)^i(-1)^i \equiv -x\ {}_{ab}g\,, \tag{B.8}$$

where $x$ is the index of the representation. Obviously we have $x = \tilde{h}$ in the adjoint representation. Note that using eq. (B.4), the supertrace of a product of two Lie algebra valued fields is:

$$str(AB) \equiv {}_i(AB)^i(-)^{i(A+B+1)} = -xA^a\ {}_ag_b\ {}^bB = (-1)^{AB}str(BA)\,. \tag{B.9}$$

[1] The possibility of a vanishing quadratic Casimir in the adjoint representation, has the interesting consequence that in that case, the affine Sugawara construction always gives a value for $c$ which is independent of the level of the affine super Lie algebra. E.g. for $D(2,1,\alpha)$, one always has $c = 1$.

The Killing metric is used to raise and lower indices:

$$A^a = A_b{}^b g^a , \qquad A_a = A^b{}_b g_a \, , \tag{B.10}$$

where ${}^a g^b$ is the inverse of ${}_a g_b$: ${}^a g^c{}_c g_b = {}^a\delta_b$.

Table B.1 contains some properties of the (super) Lie algebras which appear in chapter 7. We denote by $x_{\rm fun}$, the index of the fundamental (defining) representation. For $D(2,1,\alpha)$, it is not clear how to define the fundamental representation. The size of the smallest representation of $D(2,1,\alpha)$ depends on $\alpha$. The smallest representation which exists for generic values of $\alpha$ is the adjoint representation.

Table B.1: Properties of some (super) Lie algebras.

| algebra | bosonic subalgebra | $d_B$ | $d_F$ | $\tilde{h}$ | $x_{\rm fun}$ |
|---|---|---|---|---|---|
| $sl(n)$ | $sl(n)$ | $n^2-1$ | 0 | $n$ | $\frac{1}{2}$ |
| $so(n)$ | $so(n)$ | $\frac{1}{2}n(n-1)$ | 0 | $n-2$ | 1 |
| $osp(n\|2)$ | $sl(2)+so(n)$ | $\frac{1}{2}(n^2-n+6)$ | $2n$ | $\frac{1}{2}(4-n)$ | $\frac{1}{2}$ |
| $D(2,1,\alpha)$ | $sl(2)+su(2)+su(2)$ | 9 | 8 | 0 | – |
| $su(1,1\|2)$ | $sl(2)+su(2)$ | 6 | 8 | 0 | $\frac{1}{2}$ |
| $su(m\|n)$ $m\neq n$ | $su(m)+su(n)+u(1)$ | $m^2+n^2-1$ | $2mn$ | $m-n$ | $\frac{1}{2}$ |

## $sl(2)$ embeddings of (super)Lie algebras

We now fix our conventions for embeddings of $sl(2)$ in a (super) Lie algebra $\bar{g}$. We denote the $sl(2)$ generators with $\{e_0, e_\pm\}$. We use normalisations for the $sl(2)$ generators such that:

$$[e_0, e_\pm] = \pm 2 e_\pm \, , \qquad [e_+, e_-] = e_0 \, , \tag{B.11}$$

with supertraces (B.8):

$$str(e_0 e_0) = 2 str(e_+ e_-) = 4xy \, , \tag{B.12}$$

with $y$ the index of the $sl(2)$ embedding.

We define the kernels:

$$\mathcal{K}_\pm \equiv \ker \mathrm{ad}\, e_\pm \, . \tag{B.13}$$

Any highest weight representation $L(\Lambda)$ of $\bar{g}$ decomposes according to irreducible $sl(2)$ representations [61]:

$$L(\Lambda) = \bigoplus_{j\in\frac{1}{2}\mathbf{N}} n_j(\Lambda)\ \underline{2j+1} \, , \tag{B.14}$$

where $n_j(\Lambda)$ denotes the multiplicity of the $sl(2)$ representation. Taking eq. (B.14) for the adjoint representation of $\bar{g}$, we can make a choice for the generators of $\bar{g}$ which reflects the $sl(2)$ embedding: $\{t_{(jm,\alpha_j)}\}$ where $j \in \frac{1}{2}\mathbf{Z}$, $m = -j, -j+1, \cdots, j$ and $\alpha_j = 1, \cdots n_j(\text{adjoint})$. We take for the $sl(2)$ generators $e_m = t_{(1m,0)}$.[2]

The algebra $\bar{g}$ acquires a grading:

$$\bar{g} = \bigoplus_{m\in\frac{1}{2}\mathbf{Z}} \bar{g}_m \quad \text{where} \quad \bar{g}_m = \{a \in \bar{g} \,|\, [e_0, a] = 2ma\} \, . \tag{B.15}$$

We use the symbol $\Pi$ for projection operators in $\bar{g}$. $\Pi_{hw}$ denotes a projection on $\mathcal{K}_+$, and $\Pi_{lw}$ on $\mathcal{K}_-$. We also use $\Pi_{\geq n}\bar{g} = \oplus_{m\geq n}\bar{g}_m$, $\Pi_n \bar{g} = \bar{g}_n$, etc. Furthermore, $\Pi_+ \equiv \Pi_{>0}$ and $\Pi_- \equiv \Pi_{<0}$.

[2] This differs with a factor $\sqrt{2}$ from [181] for $m = \pm 1$.

# Appendix C

# A *Mathematica* primer

This appendix gives a short introduction to *Mathematica*. We refer to [210] for further details.

Symbols in *Mathematica* are case sensitive. Expressions, except for some built-in types like `Integer` and `Symbol`, have always the same structure: they consist of a head and some (possibly null) arguments. The head and arguments are again expressions. Application of a head on some arguments is denoted with square brackets. As an example:

*In[1] :=* `FullForm[1+a+b]`
*Out[1] =* `Plus[1,a,b]`

One extracts the $n$th subexpression using `Part[expr,n]`, or `expr[[n]]`.

One can attach transformation rules to any symbol which tell how to transform an expression with the symbol as head or as one of its arguments. Evaluation of an expression proceeds by evaluating first the head, then the arguments (unless some attributes are assigned to the head). After this, *Mathematica* checks if any transformation rules for the expression can be found. It first checks rules assigned to the arguments (`UpValues`), and then rules assigned to the head (`DownValues`).

Programming in *Mathematica* is effectively done by specifying a set of transformation rules. When the *lhs* of a rule matches the expression that is being evaluated (taking into account any conditions specified with `/;`), the transformation rule is applied and the result is again evaluated. Evaluation continues until no further rules apply. The most specific rules are used first[1], *i.e.* if a rule is given that any expression with head `f` is zero, and another rule specifies that `f[0]` is 1, then the latter rule will be checked first.

Before giving an example it is necessary to discuss the pattern matching which is used in *Mathematica*. The purpose of a pattern is to specify the conditions when a certain transformation rule has to be used. We give a list of some frequently occuring patterns. In this table *p1*, *p2* stand for any pattern.

| pattern | explanation | example |
|---|---|---|
| `_` | any expression | `f[1][1,2,3]` |
| `_f` | any expression with head `f` | `f[1,2,3]` |
| `__` | sequence of expressions (length $\geq 1$) | `1,2,3` |
| `___` | sequence of expressions (length $\geq 0$) | `1,2,3` |
| `f[`*p1*`,`*p2*`]` | expression with head `f` whose arguments match *p1*,*p2* | `f[x,1]` matches `f[_,_Integer]` |

One can give a pattern a name by prepending it with the name and a colon, *e.g.* `a:f[_,_]`. An abbreviation for this syntax is possible when the pattern begins with an underscore, *e.g.* `a_Integer`. Named patterns are useful first of all to name arguments of a transformation rule (see below). Additionally, when a named pattern occurs more than once, all matching items have to be identical, *e.g.* `f[a_,a_]` matches `f[1,1]`, but not `f[1,2]`.

Let us present a small example to show how all these things fit together to make a very powerful programming language. The factorial function could be defined as follows:

*In[2] :=* `factorial[n_Integer] := n factorial[n-1]`
*In[3] :=* `factorial[0] = 1;`

As explained above, the order in which these statements are given is not important. It is quite simple to make that factorials of an expression plus a small integer should be transformed into a product:

*In[4] :=* `factorial[n_+ m_Integer] :=`
`factorial[n] Product[n+i,{i,1,m}] /; 0<=m<=10`

Here the notation `/;` is used to specify a condition. These rules were all concerning transformations of expressions with head `factorial`. However, we can also attach a rule to `factorial` to handle quotients:

[1] When *Mathematica* is not able to figure out an order between two rules, it checks the rules in the order they were defined.

*In[5] :=* `factorial /: factorial[n_] / factorial[m_] :=`
`Pochhammer[m+1,n-m]`

where `Pochhammer` is an internal function corresponding to the Pochhammer symbol defined in appendix 2.A.

Note that *Mathematica* does not enforce the use of types like *Axiom*, but patterns can be used to simulate this.

It is sometimes useful to have a set of transformation rules which is not applied automatically. Such local rules are normal *Mathematica* expressions `Rule[`*pat*`,`*expr*`]`, with alternative notation *pat* `->` *expr*. They are used as follows:

*In[6] :=* `x + y /. x -> z`
*Out[6] =* `y + z`

Two different assignments are possible in *Mathematica*. With `Set` (or `=`) the *rhs* is evaluated when the assignment is done (useful for assignment of results), while with `SetDelayed` (or `:=`) the *rhs* is evaluated when the transformation rule is used (useful for function definitions). A similar difference exists between `Rule` (`->`) and `RuleDelayed` (`:>`).

We also need some *lisp* and *APL*-like functions which are heavily used in *OPEdefs*.

| function | abbreviation | example in | example out |
|---|---|---|---|
| `Map` | `/@` | `f /@ {1,2,3}` | {f[1],f[2],f[3]} |
| `Apply` | `@` | `f @ g[1,2,3]` | f[1,2,3] |

`Scan` is like `Map` but has only side-effects, *i.e.* the function is applied on the elements of the list, but the results are discarded.

Finally, when defining a more complicated transformation rule, we will need local variables. This is done with the `Block` statement which has the syntax `Block[{`*vars*`},` *statement*`]`, where *vars* is a list of local variables (possibly with assignments) and *statement* can be a compound statement, *i.e.* statements separated with a semicolon. The value of the block is result of the statement. Hence a function definition could be:

*In[7] :=* `f[x_] := Block[a = g[x], a + a^2]`

Note that *Mathematica* 2.0 introduced a similar statement `Module`. It makes sure that there is no overlap between globally defined symbols and the local variables. However, this introduces considerable run-time overhead compared to `Block`. When a function is defined in the `Private‘` section of a *Mathematica* package, no conflict is possible, and `Block` is to be preferred.

## An example : generating tube plots

We will first define a general purpose function `TubePlot` which generates a three dimensional plot of a tube, which is specified by a parametric curve in three dimensions, like in `ParametricPlot3D`. Then, we will use this function to generate a surface representing a second order Feynman “diagram” for the interaction of two closed strings, which can be found at the start of this book.

The algorithm for `TubePlot` is to construct at some points a circle perpendicular to the tangent vector of the curve (`TubeCircle`). These circles are then sampled and the samples are returned by `TubePlot`. The function `ListSurfaceGraphics3D` can then be used to visualise the plot.

```
Needs["Graphics‘Graphics3D‘"]
rotMatrix[{a_,b_,c_}] :=
   Block[{sqrab=Sqrt[a^2+b^2],sqrabc=Sqrt[a^2+b^2+c^2]},
     If[N[sqrab / sqrabc] < 10^-6,
       {{-1,1,0}/Sqrt[2],{-1,-1,0}/Sqrt[2],{0,0,1}},
       {{-b,a,0}/sqrab,
        {-a c,-b c,a^2+b^2}/sqrab/sqrabc,
        {a,b,c}/sqrabc}//Transpose
     ]
   ]
TubeCircle[r0_,rp0_,R_] :=
   Evaluate[rotMatrix[rp0] . {R Cos[#], R Sin[#],0} + r0]&
Options[TubePlot] = {PlotPoints->15, Radius->.5,PointsOnCircle->5};
TubePlot[f_,{t_,start_,end_},opts___Rule] :=
   TubePlot[f,D[f,t], {t,start,end},opts]
TubePlot[f_,fp_,{t_,start_,end_},opts___Rule] :=
   Block[{ nrtpoints,nrcpoints, R, tcircle, circlepoints, tval},
     {nrtpoints, nrcpoints, R} =
         {PlotPoints, PointsOnCircle, Radius}/.{opts}/.Options[TubePlot];
     circlepoints = N[Range[0,nrcpoints]/nrcpoints 2Pi];
     Table[
        tcircle = N[TubeCircle[f/.t->tval, fp/.t->tval,R]];
        tcircle /@ circlepoints,
        {tval,N[start], end, N[(end-start)/nrtpoints]}
     ]
   ]
```

We wish to use `TubePlot` with a smooth curve through some points. The curve can be constructed using cubic spline interpolation, which is defined in the `NumericalMath‘SplineFit‘` package. Unfortunately, this package does not define the derivative of the fitted spline. We can do this ourselves fairly easily because `SplineFunction[Cubic,__]` simply stores the coefficients of the cubic polynomials used in the interpolation.

```
Needs["NumericalMath‘SplineFit‘"];
Derivative[1][SplineFunction[Cubic,se_,pts_,internal_]]:=
   Block[{int = Map[{#[[2]], 2 #[[3]], 3 #[[4]], 0}&, internal, {2}]},
      SplineFunction[Cubic,se,
            Append[ {#[[1,1]],#[[2,1]]}& /@ int,
                    Apply[Plus,int[[-1]],1]
            ],
            int]
   ]
```

Finally, we need a suitable list of points. A last catch is that `SplineFit` returns a parametric curve in the plane, while `TubePlot` expects a curve in 3D.

```
pts={{3,0},{2,0},{2,1},{2.6,1},{2,1},{2,2},{1,2},{1,1},{.8,1.8},{0,2},
     {0,0},{.7,0},{1,1},{1,-.2},{1.1,1},{1.1,1.1},{2,1},{2,2},{3,2}};
sf=SplineFit[pts,Cubic];             dsf=sf';
f1[t_?NumberQ]:= Append[sf[t], 1]    f2[t_?NumberQ]:= Append[dsf[t], 0]
tplt=TubePlot[f1[t],f2[t], {t,0,Length[pts]-1.1},
      PointsOnCircle->13,Radius->.25,PlotPoints->70];
ListSurfacePlot3D[tplt,ViewPoint->{0,0,2},
  AmbientLight->GrayLevel[.1],RenderAll->False,Boxed->False];
```

**Warning:** The CPU time to generate this plot is fairly small. However, rendering the plot can take several hours depending on your platform.

# Bibliography

[1] M. Ademollo et al., *Supersymmetric strings and colour confinement*, Phys. Lett. **B62** (1976) 105.
M. Ademollo et al., *Dual string with U(1) colorsymmetry*, Nucl. Phys. **B111** (1976) 77.
M. Ademollo et al., *Dual string models with non-abelian color and flavor symmetry*, Nucl. Phys. **B114** (1976) 29.

[2] A. Alekseev, S. Shatashvili, *Path integral quantization of the coadjoint orbits of the Virasoro group and 2D gravity*, Nucl. Phys. **B323** (1989) 719.

[3] O. Alvarez, *Fermion determinants, chiral symmetry, and the Wess-Zumino anomaly*, Nucl. Phys. **B238** (1984) 61.

[4] G.E. Andrews, R.J. Baxter, P.J. Forrester, *8-Vertex SOS model and generalized Roger-Ramanujan-type identities*, J. Stat. Phys. **35** (1984) 193.

[5] F.A. Bais, P. Bouwknegt, M. Surridge, K. Schoutens, *Extensions of the Virasoro algebra constructed from Kač-Moody algebras using higher order Casimir invariants*, Nucl. Phys. **B304** (1988) 348.

[6] F.A. Bais, P. Bouwknegt, M. Surridge, K. Schoutens, *Coset construction for extended Virasoro algebras*, Nucl. Phys. **304**(1988) 371.

[7] F.A. Bais, T. Tjin, P. van Driel, *Coupled chiral algebras obtained by reduction of WZNW theories*, Nucl. Phys. **B357** (1991) 632.

[8] I. Bakas, E. Kiritsis, *Bosonic realization of a universal $\mathcal{W}$-algebra and $\mathbb{Z}_\infty$ parafermions*, Nucl. Phys. **343** (1990) 185.

[9] J. Balog, L. Fehér, P. Forgács, L. O'Raifeartaigh, A. Wipf, *Toda theory and $\mathcal{W}$-algebra from a gauged WZNW point of view*, Ann. Phys. **B203** (1990) 76.
J. Balog, L. Fehér, P. Forgács, L. O'Raifeartaigh, A. Wipf, *Kač-Moody realization of $\mathcal{W}$-algebras*, Phys. Lett. **B44** (1990) 435.

[10] J.A. Batalin, G.A. Vilkovisky, *Quantization of gauge theories with linearly dependent generators*, Phys. Rev. **D28** (1983) 2567.
J.A. Batalin, G.A. Vilkovisky, *Closure of the gauge algebra, generalized Lie equations and Feynmann rules*, Nucl. Phys. **B234** (1984) 106.
J.A. Batalin, G.A. Vilkovisky, *Correction*, Phys. Rev. **D30** (1984) 508.

[11] J.A. Batalin, G.A. Vilkovisky, *Existence theorem for gauge algebras*, J. Math. Phys. **26** (1985) 172.

[12] C. Becchi, A.Rouet, R. Stora, *Renormalisation of gauge theories*, Ann. Phys. **98** (1976) 287.

[13] A.A. Belavin, A.M. Polyakov, A.B. Zamolodchikov, *Infinite conformal symmetry in two dimensional quantum field theory*, Nucl. Phys. **B241** (1984) 333.

[14] N. Berkovits, C. Vafa, *On the uniqueness of string theory* , Mod. Phys. Lett. **A9** (1994) 653.

[15] N. Berkovits, M.D. Freeman, P.C. West, *A $\mathcal{W}$-string realisation of the bosonic string*, hep-th/9312013.

[16] E. Bergshoeff, H.J. Boonstra, M. de Roo, *Realisations of $\mathcal{W}_3$ symmetry*, Phys. Lett. **B292** (1992) 307.

[17] E. Bergshoeff, H.J. Boonstra, M. de Roo, S. Panda, A. Sevrin, *On the BRST operator of $\mathcal{W}$-strings*, Phys. Lett. **B308** (1993) 34.

[18] E. Bergshoeff, H.J. Boonstra, M. de Roo, S. Panda, *A BRST analysis of $\mathcal{W}$ symmetries*, Nucl. Phys. **B411** (1994) 717.

[19] E. Bergshoeff, A. Sevrin, X. Shen, *Noncritical $\mathcal{W}$-strings*, Phys. Lett. **B296** (1992) 95.

[20] M. Bershadsky, *Superconformal algebras in two dimensions with arbitrary N*, Phys. Lett. **B174** (1986) 285.

[21] M. Bershadsky, *Conformal field theories via hamiltonian reduction*, Comm. Math. Phys. **139** (1991) 71.

[22] M. Bershadsky, W. Lerche, D. Nemeschansky, N.P. Warner, *Extended $N = 2$ superconformal structure of gravity and $\mathcal{W}$-gravity coupled to matter*, preprint USC-92/021, CERN-TH.6694/92.

[23] M. Bershadsky, W. Lerche, D. Nemeschansky, N.P. Warner, *A BRST operator for non-critical $\mathcal{W}$-strings*, Phys. Lett. **B292** (1992) 35.

[24] M. Bershadsky, H. Ooguri, *Hidden $SL(n)$ symmetry in conformal field theories*, Comm. Math. Phys. **126** (1989) 49.

[25] M. Bershadsky, H. Ooguri, *Hidden $OSp(N,2)$ symmetries in superconformal field theories*, Phys. Lett. **B229** (1989) 374.

[26] R. Blumenhagen, M. Flohr, A. Kliem, W. Nahm, A. Recknagel, R. Varnhagen, *$\mathcal{W}$-Algebras with two and three Generators*, Nucl. Phys. **B361** (1991) 255.

[27] R. Blumenhagen, W. Eholzer, A. Honecker, K. Hornfeck, R. Hübel, *Unifying $\mathcal{W}$–algebras*, preprint DFTT-15/94, BONN-TH-94-01, hep-th/9404113.

[28] R.E. Borcherds, *Vertex algebras, Kač-Moody algebras, and the Monster*, Proc.Nat.Acad.Sci. (1986) 3068.

[29] P. Bouwknegt, *Extended conformal algebras*, Phys. Lett. **B207** (1988) 295.

[30] P. Bouwknegt, J. McCarthy, K. Pilch, *Semi-infinite cohomology of $\mathcal{W}$ algebras*, USC-93/11, hep-th9302086.

[31] P. Bouwknegt, K. Schoutens, *$\mathcal{W}$-symmetry in conformal field theory*, Phys. Rep. **223** (1993) 183.

[32] R. Bott, L.W. Tu, *Differential Forms in Algebraic Topology*, Springer Verlag, 1986.

[33] P. Bowcock, *Quasi-primary fields and associativity of chiral algebras*, Nucl. Phys. **B356** (1991) 367.

[34] P. Bowcock, G.M.T. Watts, *On the classification of quantum $\mathcal{W}$-algebras*, Nucl. Phys. **B379** (1992) 63.

[35] J. L. Cardy, *Operator content of two dimensional conformally invariant theories*, Nucl. Phys. **270** [FS16] (1986) 186.

[36] Char, Geddes, Leong, Monagan, Watt, *Maple V Language reference material*, *Maple V Library Reference Manual*, and *First Leaves: A tutorial introduction to Maple V*.

[37] S.R. Das, A. Dhar, S.K. Rama, *Physical properties of $\mathcal{W}$-gravities and $\mathcal{W}$-strings*, Mod. Phys. Lett. **A6** (1991) 3055.
S.R. Das, A. Dhar, S.K. Rama, *Physical states and scaling properties of $\mathcal{W}$-gravities and $\mathcal{W}$-strings*, Int. J. Mod. Phys. **A7** (1992) 2295.

[38] J. de Boer, J. Goeree, *The covariant $\mathcal{W}$ gravity and its moduli space from gauge theory*, Nucl. Phys. **B401** (1993) 369.

[39] J. de Boer, J. Goeree, *The effective action of $\mathcal{W}_3$ gravity to all orders*, Nucl. Phys. **B401** (1993) 348.

[40] J. de Boer, L. Fehér, A. Honecker, *A Class of $\mathcal{W}$-algebras with Infinitely Generated Classical Limit*, preprint BONN-HE-93-49, ITP-SB-93-84, hep-th/9312094, to appear in Nucl. Phys. **B**.

[41] J. de Boer, T. Tjin, *Quantization and representation theory of finite $\mathcal{W}$–algebras*, Comm. Math. Phys. **158** (1993) 485.

[42] J. de Boer, *Extended Conformal Symmetry in Non-Critical String Theory*, PhD. thesis, RU Utrecht (1993).

[43] J. de Boer, T. Tjin, *The Relation between Quantum $\mathcal{W}$–algebras and Lie Algebras*, Comm. Math. Phys. **160** (1994) 317.

[44] A. Deckmyn, *On the generalized Miura transformation*, Phys. Lett. **298B** (1993) 318.

[45] A. Deckmyn, *Extended Conformal Algebras and their Realizations*, PhD. thesis, KU Leuven (1994).

[46] A. Deckmyn, A. Sevrin, R. Siebelinck, W. Troost, in preparation.

[47] A. Deckmyn, K. Thielemans, *Factoring out free fields*, preprint KUL-TF-93/26, hep-th/9306129.

[48] F. Defever, Z. Hasiewicz, W. Troost, *Superconformal algebras with $N = 5,6,7,8$, (I), (II)*, Class. Quantum Grav. **8** (1991) 253, Class. Quantum Grav. **8** (1991) 257.

[49] F. Defever, W. Troost, Z. Hasiewicz, *Superconformal algebras with quadratic nonlinearity*, Phys. Lett. **B273** (1991) 51.

[50] F. Defever, S. Schrans, K. Thielemans, *Moding of superconformal algebras*, Phys. Lett. **B212** (1988) 467.

[51] F. De Jonghe, *The Batalin-Vilkovisky Lagrangian quantisation scheme, with applications to the study of anomalies in gauge theories*, PhD. thesis, KU Leuven (1994).

[52] F. Delduc, E. Ragoucy, P. Sorba, *Rational $\mathcal{W}$ algebras from composite operators*, preprint ENSLAPP-AL-429/93, NORDITA-93/47-P, June 1993.

[53] E. del Giudice, P. Di Vecchia, S. Fubini, *General properties of the dual resonance model*, Ann. Phys. **70**, (1972) 378.

[54] G.W. Delius, M.T. Grisaru, P. van Nieuwenhuizen, *Induced $(N,0)$ supergravity as a constrained $OSp(N|2)$ WZNW model and its effective action*, Nucl. Phys. **B389** (1993) 25.

[55] B. de Witt, *Supermanifolds*, Cambridge University Press.

[56] A. Diaz, J.M. Figueroa-O'Farrill, *A new explicit construction of $\mathcal{W}_3$ from the affine algebra $A_2(1)$*, Nucl. Phys. **B349** (1991) 237.

[57] P. Dirac, *Lectures on Quantum Mechanics*, Belfer Graduate School of Science, Yeshiva Univ., NY (1964).

[58] P. Di Vecchia, P. Rossi, *On the equivalence between the Wess-Zumino action and the free Fermi theory in 2 dimensions*, Phys. Lett. **B140** (1984) 344.
P. Di Vecchia, B. Durhuus and J. L. Petersen, *The Wess-Zumino action in 2 dimensions and the non-abelian bosonization*, Phys. Lett. **B144** (1984) 245.

[59] Vl.S. Dotsenko, V.A. Fateev, *Conformal algebra and multi-point correlation functions in 2D statistical models*, Nucl. Phys. **B240** (1984) 312.
Vl.S. Dotsenko, V.A. Fateev, *Four-point correlation functions and the operator algebra in 2D conformal invariant theories with central charge $c \leq 1$*, Nucl. Phys. **B251** [FS13] (1985) 691.

[60] V.G. Drinfeld, V.V. Sokolov, *Lie algebras and equations of Korteweg-de Vries type*, J. Sov. Math. **30** (1984) 1975.

[61] E. B. Dynkin, *Semi-simple subalgebras of semi-simple Lie algebras*, Amer. Math. Soc. Transl. Ser. 2 **6** (1967) 111.

[62] W. Eholzer, A. Honecker, R. Hübel,*How complete is the classification of $\mathcal{W}$-symmetries?*, Phys. Lett. **B308** (1993) 42.

[63] V.A. Fateev, S.L. Lukyanov, *The models of two dimensional conformal quantum field theory with $Z_N$-symmetry*, Int. J. Mod. Phys. **A3** (1988) 507.

[64] V.A. Fateev, S.L. Lukyanov, *Conformally invariant models of two-dimensional quantum field theory with $Z_N$-symmetry*, Sov. Phys. JETP **67** (1988) 447.

[65] V.A. Fateev, S.L. Lukyanov, *Additional Symmetries and Exactly Soluble Models in Two-Dimensional Conformal Field Theory, I. Quantization of Hamiltonian Structures, II. $\mathcal{W}$-Algebra Representation Theory, III. Minimal Models*, Moscow preprints 1988-1989.

[66] V.A. Fateev, A.B. Zamolodchikov , *Conformal quantum field theory models in two dimensions having $Z_3$ symmetry*, Nucl. Phys. **B280** [FS18] (1987) 644.

[67] V.A. Fateev, A.B. Zamolidchikov, *Nonlocal (parafermion) currents in two-dimensional conformal quantum field theory and self-dual critical points in $Z_N$-symmetric statistical systems*, Sov. Phys. JETP **62** (1985) 215.

[68] V.A. Fateev, A.B. Zamolidchikov, *Operator algebra and correlation functions in two dimensional $SU(2) \times SU(2)$ chiral Wess-Zumino mode*, Sov. J. Nuc. R. 43 (1986) 657.

[69] L. Fehér, L. O'Raifeartaigh, P. Ruelle, I. Tsutsui, A. Wipf, *Generalized Toda theories and $\mathcal{W}$–algebras associated with integral gradings*, Ann. Phys. **213** (1992) 1.

[70] L. Fehér, L. O'Raifeartaigh, I. Tsutsui, *The Vacuum Preserving Lie Algebra of a Classical $\mathcal{W}$–algebra*, Phys. Lett. **B316** (1993) 275.

[71] L. Fehér, L. O'Raifeartaigh, P. Ruelle, I. Tsutsui, *On the Classification of the Set of Classical $\mathcal{W}$–algebras obtained from DS reductions*, Preprint BONN-HE-93-14, DIAS-STP-93-02, hep-th/9304125, to appear in Comm. Math. Phys. .

[72] B.L. Feigin, E. Frenkel, *Quantization of the Drinfeld-Sokolov reduction*, Phys. Lett. **B246** (1990) 75.

[73] B.L. Feigin, D.B.Fuchs, *Invariant skew-symmetric differential operators on the line and Verma modules over the Virasoro algebra*, Funct. Anal. Appl. **16** (1982) 114.

[74] J.M. Figueroa-O'Farrill, S. Schrans, *The spin-6 extended conformal algebra*, Phys. Lett. **B245** (1990) 471.

[75] J.M. Figueroa-O'Farrill, S. Schrans, unpublished.

[76] J.M. Figueroa-O'Farrill, S. Schrans, *The Conformal Bootstrap and Super $\mathcal{W}$-Algebras*, Int. J. Mod. Phys. **A7** (1992) 591.

[77] J. M. Figueroa-O'Farrill, S. Schrans, *Extended superconformal algebras*, Phys. Lett. **B257** (1991) 69.

[78] J.M. Figueroa-O'Farrill, S. Schrans, K. Thielemans, *On the Casimir Algebra of B2*, Phys. Lett. **B263** (1991) 378.

[79] J. Figueroa-O'Farrill, *A comment on the magical realisations of $\mathcal{W}_3$*, QMW-PH-94-1, hep-th/9401108.

[80] J. Figueroa-O'Farrill, *On the universal string theory*, hep-th/9310200.

[81] J. Figueroa-O'Farrill, S. Stanciu, *Nonsemisimple Sugawara constructions*, preprint QMW-PH-94-2, hep-th/9402035.

[82] J. Fisch, M. Henneaux, J. Stasheff, C. Teitelboim, *Existence, uniqueness and cohomology of the classical BRST charge with ghosts of ghosts*, Comm. Math. Phys. **120** (1989) 379.

[83] M. Flohr, *$\mathcal{W}$-Algebren Quasiprimäre Felder and Nicht-Minimale Modelle*, Preprint Diplomarbeit BONN-IR-91-30 (1991).

[84] E.S. Fradkin, T.E Fradkina, *Quantization of realistic systems with constraints*, Phys. Lett. **B72** (1975) 343.

[85] E.S. Fradkin, V. Linetsky, *Result of the classification of superconformal algebras in two dimensions*, Phys. Lett. **B291** (1992) 71.

[86] L. Frappat, E. Ragoucy, P. Sorba, *$\mathcal{W}$–algebras and superalgebras from constrained WZW models: a group theoretical classification*, Comm. Math. Phys. **157** (1993) 499.

[87] L. Frappat, E. Ragoucy, P. Sorba, *$\mathcal{W}$-algebras and superalgebras from constrained WZW models: a group theoretical classification*, Comm. Math. Phys. **157** (1993) 499.

[88] M.D. Freedman, D.I. Olive, *BRS cohomology in string theory and the no-ghost theorem*, Phys. Lett. **B175** (1986) 151.

[89] M. Freeman, P. West, *$\mathcal{W}_3$-string scattering*, Int. J. Mod. Phys. **A8** (1993) 4261.

[90] I.B. Frenkel, H. Garland, G. Zuckerman, *Semi-infinite cohomology and string theory*, Yale University preprint (1986).

[91] D. Friedan, E. Martinec, S. Shenker, *Conformal invariance, supersymmetry and string theory*, Nucl. Phys. **B271** (1986) 93.

[92] S. Fubini, A.J. Hanson, R. Jackiw, *New approach to field theory*, Phys. Rev. **D7** (1973) 7.

[93] A. Fujitsu, *ope.math: operator product expansions in free field realizations on conformal field theory*, Computer Phys. Comm. **79** (1994) 78.

[94] S. J. Gates, M. T. Grisaru, W. Siegel, M. Roček, *Superspace*, Benjamin/Cummings pub. comp. (1983).

[95] D. Gepner, Z. Qiu, *Modular invariant partition functions for parafermionic field theories*, Nucl. Phys. **B285** [FS19] (1977) 423.

[96] E. Getzler, *Manin triples and $N = 2$ superconformal field theory*, hep-th/9307041.

[97] P. Ginsparg, *Applied conformal field theory*, in Les Houches session XLIX, eds. E. Brézin and J. Zin–Justin, Elsevier (1988).

[98] P. Goddard, *Meromorphic conformal field theory*, in 'Infinite dimensional Lie algebras and Lie groups', ed. V. Kač, Proc. CIRM-Luminy conf. (1988), World Scientific.

[99] P. Goddard, A. Kent, D. Olive, *Virasoro algebras and coset space models*, Phys. Lett. **B152** (1985) 88.

[100] P. Goddard, W. Nahm, D. Olive, A. Schwimmer, *Vertex operators for non-simply-laced algebras*, Comm. Math. Phys. **107** (1986) 179.

[101] P. Goddard, D. Olive, *Kač-Moody and Virasoro algebras in relation to quantum physics*, Int. J. Mod. Phys. **A1** (1986) 303.

[102] P. Goddard, D. Olive, G. Waterson, *Superalgebras, symplectic bosons and the Sugawara construction*, Comm. Math. Phys. **112** (1987) 591.

[103] P. Goddard, A. Schwimmer, *Factoring out free fermions and superconformal algebras*, Phys. Lett. **B214** (1988) 209.
[104] T. Goto, *Relativistic quantum mechanics of one-dimensional mechanical continuum and subsidiary condition of dual resonance model*, Prog. Theor. Phys. **46** (1971) 1560.
[105] M. Green, J. Schwarz, E. Witten, *Superstring Theory*, Vol. I& II, Cambridge University Press (1987).
[106] M.T. Grisaru, W. Siegel, M. Roček, *Improved methods for supergraphs*, Nucl. Phys. **B159** (1979) 429.
[107] M.T. Grisaru, R.M. Xu, *Quantum supergravities in 2 dimensions*, Phys. Lett. **B205** (1988) 486.
[108] K. Hamada and M. Takao, *Spin-4 current algebra*, Phys. Lett. **B209** (1988) 247. Erratum Phys. Lett. **B213** (1988) 564.
[109] Z. Hasiewicz, K. Thielemans, W. Troost, *Superconformal algebras and Clifford algebras*, J. Math. Phys. **31**, (1989) 744.
[110] M. Henneaux, *Hamiltonian form of the path integral for theories with gauge freedoms*, Phys. Rep. **126** (1985) 1.
[111] A. Honecker, *A note on the algebraic evaluation of correlators in local chiral conformal field theory*, preprint BONN-HE-92-25 (1992), hep-th/ 9209029.
[112] K. Hornfeck, *Explicit construction of the BRST charge for $\mathcal{W}_4$*, Phys. Lett. **B315** (1993) 287.
[113] K. Hornfeck, *$\mathcal{W}$–algebras with a Set of Primary Fields of Dimensions* $(3, 4, 5)$ *and* $(3, 4, 5, 6)$, Nucl. Phys. **B407** (1993) 237.
[114] K. Hornfeck, *Classification of Structure Constants for $\mathcal{W}$–algebras from Heighest Weights*, Nucl. Phys. **B411** (1994) 307.
[115] C.M. Hull, *Chiral $\mathcal{W}$-gravities for general extended conformal algebras*, Phys. Lett. **B259** (1991) 68.
[116] C.M. Hull, *Higher-spin extended conformal algebras and $\mathcal{W}$-gravities*, Nucl. Phys. **B353** (1991) 707.
[117] C.M. Hull, *Lectures on $\mathcal{W}$-Gravity, $\mathcal{W}$-Geometry and $\mathcal{W}$-Strings*, Invited talk at Lectures given at Trieste Summer School on High Energy Physics and Cosmology, Trieste, Italy, 15 Jun - 14 Aug 1992. In "Trieste 1992, Proceedings, High energy physics and cosmology" 76-142 and London Queen Mary and Westfield Coll, (1993) QMW-93-2, hep-th/9302110.
[118] D.A. Huse, *Exact exponents for the infinitely many new multicritical points*, Phys. Rev. **B30** (1984) 3908.
[119] T. Inami, Y. Matsuo, I. Yamanaka, *Extended conformal algebras with $N = 1$ supersymmetry*, Phys. Lett. **B215** (1988) 701.
[120] H. Ishikawa, M. Kato, *Note on $N = 0$ string as $N = 1$ string*, UT-KOMABA/93-23, hep-th/9311139.
[121] K. Ito, *Quantum Hamiltonian Reduction and $\mathcal{W}B$ Algebra*, Int. J. Mod. Phys. **A7** (1992) 4885.
[122] C. Itzykson, *Invariance conforme et modèles critiques bidimensionelles*, Cours de C. Itzykson au DEA de Physique Théorique de Marseille, preprint CNRS Marsèille CTP-86/P.1915.
[123] R. D. Jenks, R. S. Sutor, *Axiom: The Scientific Computation system*, NAG, Springer Verlag (1992).
[124] V.G. Kač, Funct. Anal. Appl. 1 (1967) 328.
[125] V.G. Kač, in *Lecture Notes in Physics* **94**, Springer-Verlag (1979).
[126] V. Kač, *Infinite Dimensional Lie Algebras*, Birkhauser (1983).
[127] M. Kaku, *Introduction to Superstrings*, Springer-Verlag (1989).
[128] M. Kato, S. Matsuda, *Construction of singular vertex operators as degenerate primary conformal fields*, Phys. Lett. **B172** (1986) 216.
[129] H.G. Kausch, *Extended conformal algebras generated by a multiplet of primary fields*, Phys. Lett. **B259** (1991) 448.
[130] H.G. Kausch, *Chiral algebras in conformal field theory*, Ph.D. thesis, Cambridge university, (1991).
[131] H.G. Kausch, G.M.T. Watts, *A study of $\mathcal{W}$-Algebras using Jacobi Identities*, Nucl. Phys. **B354** (1991) 740.
[132] H.G. Kausch, G.M.T. Watts, *Quantum Toda theory and the Casimir algebra of $B_2$ and $C_2$*, Int. J. Mod. Phys. **A7** (1992) 4175.
[133] V.G. Knizhnik, *Superconformal algebras in two dimensions*, Theor. Math. Phys. **66** (1986) 68.
[134] V.G. Knizhnik, A.M. Polyakov, A.B. Zamolodchikov, *Fractal structure of 2d-quantum gravity*, Mod. Phys. Lett. **A3** (1988) 819.
[135] V.G. Knizhnik, A.B. Zamolodchikov, *Current algebra and Wess-Zumino model in two dimensions*, Nucl. Phys. **B247** (1984) 83.
[136] S.O. Krivonos, A. Sorin, *Linearizing $\mathcal{W}$–algebras*, ICTP preprint, hep-th/9406005.
[137] S.O. Krivonos, K. Thielemans, *A Mathematica package for super OPEs*, in preparation.
[138] T. Kugo, I. Ojima, *Manifestly covariant canonical formulation of Yang-Mills theories, physical state subsidiary conditions and physical S-matrix unitarity*, Phys. Lett. **B73** (1978) 459.
[139] W. Lerche, *Chiral rings in topological W-gravity*, $26^{th}$ workshop *From superstrings to supergravity*, Erice, (1992), preprint CERN-TH.6812/93.
[140] H. Lu, C.N. Pope, S. Schrans, K.-W. Xu, *The complete spectrum of the $\mathcal{W}(N)$ string*, Nucl. Phys. **B385** (1992) 99.
[141] H. Lu, C.N. Pope, S. Schrans, X.J. Wang, *The interacting $\mathcal{W}_3$- string*, Nucl. Phys. **B403** (1993) 351.
[142] H. Lu, C.N. Pope, S. Schrans, X.J. Wang, *On the spectrum and scattering of $\mathcal{W}_3$ strings*, Nucl. Phys. **B408** (1993) 3.
[143] H. Lu, C.N. Pope, X.J. Wang, *On higher-spin generalizations of string theory*, Int. J. Mod. Phys. **A9** (1994) 1527.

[144] H. Lu, C.N. Pope, K. Thielemans, X.J. Wang, *Higher-spin string and $\mathcal{W}$ minimal models*, Class. Quantum Grav. **11** (1994) 119.

[145] H. Lu, C.N. Pope, X.J. Wang, S.C. Zhao, *Critical and noncritical $\mathcal{W}_{2,4}$ strings*, Class. Quantum Grav. **11** (1994) 939.

[146] H. Lu, C.N. Pope, X.J. Wang, K.-W. Xu, *The Complete Cohomology of the $\mathcal{W}_3$ String*, Class. Quantum Grav. **11** (1994) 967.

[147] H. Lu, C.N. Pope, X.J. Wang, S.C. Zhao, *A note on $\mathcal{W}_{2,s}$ strings*, Phys. Lett. **B327** (1994) 241.

[148] H. Lu, C.N. Pope, K. Thielemans, X.J. Wang, K.-W. Xu, *Quantising Higher-spin String Theories*, in preparation.

[149] K.A. Meissner, J. Pawełczyk, *Some perturbative results for two-dimensional gravity*, Mod. Phys. Lett. **A5** (1990) 763.

[150] N. Mohammedi, *General super-Virasoro construction on affine-G*, Mod. Phys. Lett. **A6** (1991) 2977.

[151] R.V. Moody, *Lie algebras associated with generalized Cartan matrices*, Bull. Am. Mat. Soc. **73** (1974) 217.

[152] Y. Nambu, *Lectures at the Copenhagen Summer Symposium* (1970).

[153] W. Nahm,*Conformal quantum field theories in two dimensions*, Proceedings of the Trieste Conference on Recent Developments in Conformal Field Theories, Trieste, October 1989.

[154] C. R. Nappi, E. Witten, *Wess-Zumino-Witten model based on a nonsemisimple group*, Phys. Rev. Lett. **71** 371.

[155] F.J. Narganes-Quijano, *Bosonization of Parafermions and Related Conformal Models*, Brussels preprint, ULB-TH 89/09, Ann. Phys. to be published.

[156] S.P. Novikov, Usp. Mat. Nauk. **37** (1982) 3.

[157] H. Ooguri, K. Schoutens, A. Sevrin, P. van Nieuwenhuizen, *The induced action of $\mathcal{W}_3$ gravity*, Comm. Math. Phys. **145** (1992) 515.

[158] A.M. Polyakov, *Quantum geometry of bosonic strings*, Phys. Lett. **B103** (1981) 207.

[159] A.M. Polyakov, *Quantum gravity in two dimensions*, Mod. Phys. Lett. **A2** (1987) 893.

[160] A.M. Polyakov, in *Physics and Mathematics of Strings*, ed. L. Brink, D. Friedan, A.M. Polyakov, World Scientific (1990), p. 13.

[161] A.M. Polyakov, P.B. Wiegmann, *Theory of nonabelian goldstone bosons in two dimensions*, Phys. Lett. **B131** (1983) 121
A.M. Polyakov, P.B. Wiegmann, *Goldstone fields in 2 dimensions with multivalued actions*, Phys. Lett. **B141** (1984) 223.

[162] A.M. Polyakov, A.B. Zamolodchikov, *Fractal structure of two dimensional supergravity*, Mod. Phys. Lett. **A3** (1988) 1213.

[163] C.N. Pope, L.J. Romans, K.S. Stelle, *Anomaly-free $\mathcal{W}_3$ gravity and critical $\mathcal{W}_3$ strings*, Phys. Lett. **B268** (1991) 167.

[164] C.N. Pope, E. Sezgin, K.S. Stelle, X.J. Wang, *Discrete states in the $\mathcal{W}_3$ string*, Phys. Lett. **B299** (1993) 247.

[165] C.N. Pope, *$\mathcal{W}$-Strings 93*, Talks given at Spring Workshop on High Energy Physics, Trieste, Italy, April 1993, and at the International Conference on Strings 93, Berkeley, CA, 24-29 May 1993, (1993) CTP-TAMU-55-93.

[166] K. Popper, *The logic of scientific discovery*, Hutchinson (London) (1960).

[167] E. Ragoucy, *$OSp(1|2)$ and $Sl(2)$ reductions in generalized supertoda models and factorization of spin $1/2$ fields*, Nucl. Phys. **B411** (1994) 778.

[168] L.J. Romans, *Realizations of classical and quantum $\mathcal{W}_3$ symmetry*, Nucl. Phys. **B352** (1991) 829.

[169] M.V. Savelev, A.M. Vershik, *Continuum analogues of contragedient Lie algebras. (Lie algebras with a Cartan operator and nonlinear dynamical systems)*, Comm. Math. Phys. **126** (1989) 367.

[170] K. Schoutens, *A non-linear representation of the $D = 2$ $SO(4)$-extended superconformal algebra*, Phys. Lett. **B194** (1987) 75.

[171] K. Schoutens, *$O(N)$-extended superconformal field theory in superspace*, Nucl. Phys. **B295** [FS21] (1988) 634.

[172] K. Schoutens, A. Sevrin, P. van Nieuwenhuizen, *Quantum $\mathcal{W}_3$ gravity in the chiral gauge*, Nucl. Phys. **B364** (1991) 584.

[173] K. Schoutens, A. Sevrin, P. van Nieuwenhuizen, *On the effective action of chiral $\mathcal{W}_3$ gravity*, Nucl. Phys. **B371** (1992) 315.

[174] K. Schoutens, A. Sevrin, P. van Nieuwenhuizen, *Quantum BRST charge for quadratically nonlinear Lie algebras*, Comm. Math. Phys. **124** (1989) 87.

[175] K. Schoutens, A. Sevrin, P. van Nieuwenhuizen, in Proceedings of the Jan. 1991 Miami Workshop on *Quantum Field Theory, Statistical Mechanics, Quantum Groups and Topology*, Plenum (1991).

[176] K. Schoutens, A. Sevrin, P. van Nieuwenhuizen, *Induced gauge theories and $\mathcal{W}$-gravity*, in the proceedings of the Stony Brook conference *Strings and Symmetries 1991*, World Scientific (1992).

[177] S. Schrans, *Uitbreidingen van Conforme Invariantie in Tweedimensionale Quantumveldentheorie*, PhD. thesis (in English), KU Leuven (1991).

[178] A. Schwimmer, N. Sieberg, *Comments on the $N = 2, 3, 4$ superconformal algebras in two dimensions*, Phys. Lett. **B184** 191.

[179] W.M. Seiler, *SUPERCALC, a REDUCE Package for commutator calculations*, Karlsruhe preprint KA-THEP-20/90.

[180] A. Sevrin, R. Siebelink, W. Troost, *Regularization of nonlocal actions in two-dimensional field theories*, Nucl. Phys. **B413** (1994) 185.

[181] A. Sevrin, K. Thielemans, W. Troost, *Induced and effective gravity theories in $D = 2$*, Nucl. Phys. **B407** (1993) 459.

[182] A. Sevrin, K. Thielemans, W. Troost, *The relation between linear and nonlinear $N = 3, 4$ supergravity theories*, Phys. Rev. **D48** (1993) 1768.

[183] A. Sevrin, K. Thielemans, W. Troost, *A systematic approach to extensions of the Virasoro algebra and 2D gravities*, in Proc. of the Strings '93 conference, Berkeley, World Scientific (1994).

[184] A. Sevrin, W. Troost, *Extensions of the Virasoro algebra and gauged WZW models*, Phys. Lett. **B315** (1993) 304.
[185] A. Sevrin, W. Troost, A. Van Proeyen, *Superconformal algebras in two dimensions with $N = 4$*, Phys. Lett. **B208** (1988) 447.
[186] A. Sevrin, W. Troost, A. Van Proeyen, P. Spindel, *Extended supersymmetric $\sigma$-models on group manifolds II. current algebras*, Nucl. Phys. **B311** (1988) 465.
[187] R. Siebelink,*Regularisation of two-dimensional induced models and non-critical strings*, PhD. thesis, KU Leuven (1994).
[188] L.J. Slater, *Generalized hypergeometric functions*, Cambridge University Press (1966).
[189] I.V. Tyupin, Lebedev preprint, FIAN No. 39 (1975).
[190] Y. Saint-Aubin, *Phénomènes critiques en deux dimensions et invariance conforme*, Montréal preprint CRM-1247.
[191] K. Symanzik, *Small distance behaviour in field theory and power counting*, Comm. Math. Phys. **18** (1970) 227.
[192] K. Thielemans, *A Mathematica package for computing operator product expansions*, Int. J. Mod. Phys. C Vol. **2**, No. 3, (1991) 787.
[193] K. Thielemans, *New computing techniques in Physics Research II*, proceedings of the Second International Workshop on Software Engineering, Artificial Intelligence and Expert Systems in High Energy and Nuclear Physics, ed. D. Perret-Gallix, World Scientific (1992).
[194] K. Thielemans, in preparation.
[195] J. Thierry-Mieg, *BRS-analysis of Zamolodchikov's spin 2 and 3 current algebra*, Phys. Lett. **B197** (1987) 368.
[196] W. Troost, P. van Nieuwenhuizen, A. Van Proeyen, *Anomalies and the Batalin-Vilkovisky Lagrangian-formalism*, Nucl. Phys. **B333** (1990) 727.
[197] W. Troost, A. Van Proeyen, *An introduction to Batalin-Vilkovisky Lagrangian Quantisation*, Leuven University Press, in preparation.
[198] S. Vandoren, A. Van Proeyen, *Simplifications in Lagrangian BV quantization exemplified by the anomalies of chiral $\mathcal{W}_3$ gravity*, Nucl. Phys. **B411** (1994) 257.
[199] A. Van Proeyen, preprint KUL-TF-91/35, in the proceedings of "Strings and Symmetries 1991", World Scientific.
[200] H. Verlinde, *conformal field-theory, 2 dimensional quantum gravity and quantization of Teichmuller space*, Nucl. Phys. **B337** (1990) 652.
[201] D. Verstegen, *On the classification of $\mathcal{W}$-algebras*, Internal Report KU Leuven (1992).
[202] M. Wakimoto, *Fock representations of the affine Lie algebra $A_1$(1)*, Comm. Math. Phys. **104** (1986) 605.
A. Gerasimov, A. Marshakov, A. Morozov, M. Olshanetsky, S. Shatashvili, *Wess-Zumino-Witten model as a theory for free fields*, Int. J. Mod. Phys. **A5** (1990) 2495.
[203] G.M.T. Watts, *$\mathcal{W}B$ algebra representation theory*, Nucl. Phys. **339** (1990) 177.
[204] G.M.T. Watts, *$\mathcal{W}$-algebras and coset models*, Phys. Lett. **B245** (1990) 65.
[205] G.M.T. Watts, *$\mathcal{W}B_n$ symmetry, Hamiltonian Reduction and $B(0,n)$ Toda Theory*, Cambridge preprint, DAMTP-90/23.
[206] P. West, *A Review of $\mathcal{W}$ Strings*, preprint Göteborg-ITP-93-40.
[207] P.C. West, *$\mathcal{W}$ strings and cohomology in parafermionic theories*, hep-th/9403185.
[208] K.G. Wilson, *Non-Lagrangian models of current algebra*, Phys. Rev. **179** (1969) 1499.
[209] E. Witten, *Nonabelian bosonization in two dimensions*, Comm. Math. Phys. **92** (1984) 455.
[210] S. Wolfram, *Mathematica, A system for Doing Mathematics by Computer* 2nd ed. (1991), Addison-Wesley Publishing Company, Inc.
[211] A.B. Zamolodchikov, *Infinite additional symmetries in two dimensional conformal quantum field theory*, Theor. Math. Phys. **63** (1985) 1205.
[212] A.B. Zamolodchikov, *Vacuum Ward identities for higher generalisation*, Nucl. Phys. **B316** (1989) 573.
[213] C.J. Zhu, *The BRST quantisation of the nonlinear $\mathcal{W}B_2$ and $\mathcal{W}_4$ algebras*, SISSA/77/93/EP.